\documentclass[manuscript]{acmart}
\AtBeginDocument{ 
  \providecommand\BibTeX{{ 
    \normalfont B\kern-0.5em{\scshape i\kern-0.25em b}\kern-0.8em\TeX}}}

\setcopyright{acmcopyright}
\copyrightyear{2020}
\acmMonth{April}
\acmYear{2022}
\acmDOI{10.1145/xxxxxx}

\acmConference[CSUR]{ACM Computing Surveys}{ACM Computing Surveys}{April 2022}
\acmPrice{0.00}
\acmISBN{TBD}
\renewcommand\footnotetextcopyrightpermission[1]{} 

\usepackage[utf8x]{inputenc}
\usepackage{xspace}
\usepackage{makecell}
\usepackage{xcolor}
\usepackage[position=bottom]{subfig}
\usepackage{float}
\usepackage{graphics}
\usepackage{graphicx}
\usepackage{color, colortbl}
\usepackage{caption}
\usepackage{wrapfig}
\usepackage{paralist}
\usepackage{booktabs} 
\usepackage{array}
\usepackage{balance}
\usepackage{lipsum}
\usepackage{optidef} 
\usepackage{dsfont} 
\usepackage[ruled,norelsize]{algorithm2e}
\usepackage{algorithmic}
\usepackage{amsmath}
\makeatletter
\newcommand{\removelatexerror}{\let\@latex@error\@gobble}
\makeatother

\usepackage{amsmath}

\usepackage{tikz}
\usepackage{physics}
\usepackage{cancel}
\usepackage{siunitx}
\usepackage{multirow}
\usepackage{tabularx}

\usetikzlibrary{arrows.meta}
\usetikzlibrary{positioning}
\usetikzlibrary{fadings}
\usetikzlibrary{patterns}
\usetikzlibrary{shadows.blur}
\usetikzlibrary{shapes}
\usetikzlibrary{shapes.symbols} 

\usepackage{adjustbox}
\usepackage{bbm}

\tikzset{%
  >={Latex[width=2mm,length=2mm]},
            base/.style = {rectangle, rounded corners, draw=black, fill=gray!10,
                           minimum width=2.3cm, minimum height=1cm,
                           text centered, font=\large\sffamily},
            data/.style = {rectangle, draw=black, fill=gray!10, minimum width=2.3cm, minimum height=1cm, text centered, node distance=1cm and 1cm, font=\large\sffamily},
            corpus/.style={tape, tape bend top=none,
                draw=black, fill=gray!10, minimum width=2.3cm, minimum height=1cm, text centered, font=\large\sffamily},
            bias/.style= {draw=none, 
                          fill=none,
                          minimum width=1.1cm,
                          text=red,
                   font=\bf\large\sffamily},
            plaintext/.style= {draw=none, 
                          fill=none,
                          minimum width=1.1cm,
                          text=black,
                 font=\bf\large\sffamily},
            process/.style= {base,
                             node distance=1cm and 1cm,
                             fill=gray!25, font=\large\sffamily},
            node/.style = {base, 
                           node distance=1cm and 1cm, font=\large\sffamily}
}

\definecolor{Lightgray}{gray}{0.9}
\newcolumntype{L}[1]{>{\raggedright\let\newline\\\arraybackslash\hspace{0pt}}m{#1}}
\newcolumntype{C}[1]{>{\centering\let\newline\\\arraybackslash\hspace{0pt}}m{#1}}
\newcolumntype{R}[1]{>{\raggedleft\let\newline\\\arraybackslash\hspace{0pt}}m{#1}}

\DeclareMathOperator*{\argmax}{arg\,max}

\newcommand{\eop}{EO\xspace}

\newcommand*{\kl}{\textsf{rKL}\xspace}
\newcommand*{\nd}{\textsf{rND}\xspace}
\newcommand*{\rd}{\textsf{rRD}\xspace}
\newcommand*{\agg}{\textsf{IGF-Aggregated}\xspace}
\newcommand*{\ratio}{\textsf{IGF-Ratio}\xspace}

\def\btau{\boldsymbol{\tau}}

\newcommand{\oracle}{\mathcal O}
\newcommand{\candidateSet}[1]{\mathcal C_{\mathit{#1}}}
\newcommand{\userSet}[1]{\mathcal U_{\mathit{#1}}}
\newcommand{\featureSet}[0]{\mathbf X}
\newcommand{\sensFeatSet}[0]{\mathbf A}
\newcommand{\sensAttr}[0]{A}
\newcommand{\sensAttrSet}[0]{\mathbf A}
\newcommand{\setOfQueries}[0]{\mathbf Q}
\newcommand{\query}[0]{Q}
\newcommand{\score}[1]{Y_{#1}}
\newcommand{\queryScore}[1]{Y_{#1}}
\newcommand{\predScore}[0]{\hat{Y}}
\newcommand{\featOfCand}[1]{\mathbf{X}_{#1}} 
\newcommand{\featOfUser}[1]{\mathbf{U}_{#1}}
\newcommand{\fairFeatOfCand}[1]{\widetilde{\mathbf{X}}_{#1}}
\newcommand{\npFeatOfCand}[1]{\mathbf{X}^*_{#1}}
\newcommand{\utilityTwoPara}[2]{U^{#1}(#2)}
\newcommand{\utilityThreePara}[3]{U^{#1}(#2, #3)}
\newcommand{\yUtilLoss}[3]{L^{#1}_{\score{}}(#2, #3)}
\newcommand{\unfairnessThreePara}[3]{D\left(#1, #2 | #3\right)}
\newcommand{\unfairnessTwoPara}[2]{D\left(#1, #2\right)}
\newcommand{\unfairnessOnePara}[1]{D\left(#1\right)}
\newcommand{\exposure}[1]{\operatorname{Exposure}\left(#1\right)}
\newcommand{\group}[1]{\mathcal G_{#1}}
\newcommand{\groupSet}[0]{\boldsymbol{\mathcal G}}
\newcommand{\groupFunc}[0]{g}
\newcommand{\attention}[2]{\mathit{att}(#1, #2)}
\newcommand{\posBias}[1]{\mathbf v(#1)}
\newcommand{\recsysTrainSet}[1]{\mathcal{T}_{\text{#1}}}
\newcommand*{\ie}{i.e.,\xspace}
\newcommand*{\eg}{e.g.,\xspace}

\newcommand*{\ltr}{LtR\xspace}

\newcommand{\deltr}{\textsc{DELTR}\xspace}
\newcommand{\algofair}{\textsc{FA*IR}\xspace}
\newcommand{\spara}[1]{\smallskip\noindent{\bf #1}}
\newcommand{\cifrank}{\textsc{CIF-Rank}\xspace}
\newcommand{\fairpg}{\textsc{Fair-PG-Rank}\xspace}

\newcommand{\rev}[1]{\textcolor{black}{#1}}

\newcommand{\revv}[1]{\textcolor{black}{#1}}

\newcommand*{\ranking}[1]{\langle #1 \rangle}

\newcommand*{\lst}[2][n] {#2_1, \ldots, #2_{#1}}

\def\e#1{{\em #1}}
\def\val#1{\texttt{#1}}
\def\angs#1{\mathord{\langle #1 \rangle}}

\definecolor{LightCyan}{rgb}{0.88,1,1}

\newcommand*{\cs}{{\it CS}\xspace}
\newcommand*{\os}{{\it OS}\xspace}
\newcommand*{\ds}{{\it DS}\xspace}

\begin{document}

\title{Fairness in Ranking: A Survey}

\author{Meike Zehlike}
\email{zehlike@gmail.com}
\affiliation{%
  \institution{Humboldt University of Berlin,}
  \institution{Max Planck Institute for Software Systems, and}
  \institution{Zalando Research}
  \country{Germany}
}

\author{Ke Yang}
\email{ky630@nyu.edu}
\affiliation{ 
  \institution{New York University, NY}
  \institution{University of Massachusetts, Amherst, MA}
  \country{USA}
}

\author{Julia Stoyanovich}
\email{stoyanovich@nyu.edu}
\affiliation{ 
  \institution{New York University, NY}
  \country{USA}
}

\renewcommand{\shortauthors}{Zehlike, Yang, Stoyanovich}

\begin{abstract}
   In the past few years, there has been much work on incorporating fairness requirements into algorithmic rankers, with contributions coming from the data management, algorithms,  information retrieval, and recommender systems communities.  In this survey we give a systematic overview of this work, offering a broad perspective that connects formalizations and algorithmic approaches across subfields.  An important contribution of our work is in developing a common narrative around the value frameworks that motivate specific fairness-enhancing interventions in ranking.  This allows us to unify the presentation of mitigation objectives and of algorithmic techniques to help meet those objectives or identify trade-offs.
  
   In this survey, we describe four classification frameworks for fairness-enhancing interventions, along which we relate the technical methods surveyed in this paper, discuss evaluation datasets, and present technical work on fairness in score-based ranking. Then, we present methods that incorporate fairness in supervised learning, and also give representative examples of recent work on fairness  in recommendation and matchmaking systems.  We also discuss evaluation frameworks for fair score-based ranking and fair learning-to-rank, and draw a set of recommendations for the evaluation of fair ranking methods.
\end{abstract}

\setcopyright{acmcopyright}
\acmJournal{CSUR}
\acmYear{2022} \acmVolume{1} \acmNumber{1} \acmArticle{1} \acmMonth{1} \acmPrice{15.00}\acmDOI{10.1145/3533379}



\maketitle

\section{Introduction}
\label{sec:intro}
The research community recognizes several important normative dimensions of information technology including privacy, transparency, and fairness. In this survey we focus on fairness --- a  broad and inherently interdisciplinary topic of which the social and philosophical foundations are still unresolved~\cite{DBLP:journals/cacm/ChouldechovaR20}.    

Research on fair machine learning has mainly focused on classification and prediction tasks~\cite{fairMLbook,DBLP:journals/cacm/ChouldechovaR20}, while we focus on ranking.  As is customary in fairness research, we assume that input data describes \emph{individuals} --- natural persons seeking education, employment, or financial opportunities, or being prioritized for access to goods and services.   While some of the algorithmic techniques described here can be applied to entities other than people, we believe that the concept of fairness, along with the corresponding normative frameworks, applies predominantly to scenarios where data describes people.  For consistency, we will refer to the set of individuals in the input to a ranking task as \emph{candidates}. 

We consider two types of ranking tasks: score-based and supervised learning. In score-based ranking, a given set of candidates is sorted on the score attribute, which may itself be computed on the fly, and returned in sorted order.  In  supervised learning\rev{-to-rank}, a preference-enriched training set of candidates is given, with preferences among them stated in the form of scores, preference pairs, or lists; this training set is used to train a model that predicts the ranking of unseen candidates.  For both score-based and supervised learning tasks, we typically return the best-ranked $k$ candidates, the top-$k$.  Set selection is a special case of ranking that ignores the relative order among the top-$k$, returning them as a set.  

\rev{While supervised learning-to-rank appears to be similar to classification, there is one crucial difference.  The goal of classification is to assign a class label to each item, and this  assignment is made independently for each item.  In contrast, learning-to-rank positions items relative to each other, and so the outcome for one item is not independent of the outcomes for the other items.  This lack of independence has profound implications for the design of learning-to-rank methods in general, and for fair learning-to-rank in particular.}

To make our discussion concrete, we now present our running example from university admissions, a domain in which ranking and set selection are very natural and are broadly used. 

\subsection{Running example: university admissions}
\label{sec:intro:example}

Consider an admissions officer at a university who selects candidates from a large applicant pool.  When making their decision, the officer pursues some or all of the goals listed below.  Some of these goals may be legally mandated, while others may be based on the policies adopted by the university, and include admitting students who:
\begin{itemize}
    \item are likely to succeed: complete the program with high marks and graduate on time;
    \item show strong interest in specific majors like computer science, art, or literature;  and
    \item form a demographically diverse group in terms of their demographics, both overall and in each major.
\end{itemize}

\newcommand{\dataquerytab}{
    \small
	\begin{tabular}{|c||c|c||c|c|c||c||c|c|c|}
		\hline
		\rowcolor[HTML]{C0C0C0} 
		candidate & $A_1$ & $A_2$ & $X_1$ & $X_2$ & $X_3$ & $X_4$ & $\queryScore{1}$ & $\queryScore{2}$ & $\queryScore{3}$ \\ \hline
		\val{b}  & \val{male}    & \val{White}    & 4 & 5 & 5  & $\{$\val{cs}:0.9; \val{art}:0.2$\}$ & 14 & 9 & 1 \\ \hline
		\val{c}  & \val{male}    & \val{Asian}    & 5 & 3 & 4 & $\{$\val{math}:0.9; \val{cs}:0.5$\}$ & 12 & 9 & 1 \\ \hline
		\val{d}  & \val{female}  & \val{White}    & 5 & 4 & 2 & $\{$\val{lit}:0.8; \val{math}:0.8$\}$ & 11 & 4 & 6 \\ \hline
		\val{e}  & \val{male}    & \val{White}    & 3 & 3 & 4 & $\{$\val{math}:0.8; \val{econ}:0.4$\}$ & 10 & 7 & 6 \\ \hline
		\val{f}  & \val{female}  & \val{Asian}    & 3 & 2 & 3 & $\{$\val{econ}:0.9; \val{math}:0.5$\}$ & 8 & 5 & 8 \\ \hline
		\val{k}  & \val{female}  & \val{Black}    & 2 & 2 & 3 & $\{$\val{lit:0.9;\val{art}:0.8}$\}$ & 7 & 1 & 9 \\ \hline
		\val{l}  & \val{male}    & \val{Black}    & 1 & 1 & 4 & $\{$\val{lit}:0.5; \val{math}:0.7$\}$ & 6 & 6 & 2  \\ \hline
		\val{o}  & \val{female}  & \val{White}    & 1 & 1 & 2  & $\{$\val{econ}:0.9; \val{cs}:0.8$\}$ & 4 & 7 & 8 \\ \hline
	\end{tabular}
}
\newcommand{\rankscoretab}{
    \small
	\begin{tabular}{|c|}
		\hline
		\rowcolor[HTML]{C0C0C0} 
		$\btau_1$ \\ \hline
		\rowcolor[HTML]{EFEFEF} 
		b                    \\ \hline
		\rowcolor[HTML]{EFEFEF} 
		c                    \\ \hline
		\rowcolor[HTML]{EFEFEF} 
		d                    \\ \hline
		\rowcolor[HTML]{EFEFEF} 
		e                    \\ \hline
		f                    \\ \hline
		k                    \\ \hline
		l                    \\ \hline
		o                    \\ \hline
	\end{tabular}
}
\newcommand{\rankstemtab}{
    \small
	\begin{tabular}{|c|}
	   
		\hline
		\rowcolor[HTML]{C0C0C0} 
		$\btau_2$ \\ \hline
		\rowcolor[HTML]{EFEFEF} 
		c                    \\ \hline
		\rowcolor[HTML]{EFEFEF} 
		b                    \\ \hline
		\rowcolor[HTML]{EFEFEF} 
		e                    \\ \hline
		\rowcolor[HTML]{EFEFEF} 
		f                    \\ \hline
		d                    \\ \hline
		o                    \\ \hline
		l                    \\ \hline
		k                    \\ \hline
	\end{tabular}
}
\newcommand{\ranknonstemtab}{
    \small
	\begin{tabular}{|c|}
		\hline
		\rowcolor[HTML]{C0C0C0} 
		$\btau_3$ \\ \hline
		\rowcolor[HTML]{EFEFEF} 
		k                    \\ \hline
		\rowcolor[HTML]{EFEFEF} 
		l                    \\ \hline
		\rowcolor[HTML]{EFEFEF} 
		b                   \\ \hline
		\rowcolor[HTML]{EFEFEF} 
		d                    \\ \hline
		e                    \\ \hline
		f                    \\ \hline
		c                    \\ \hline
		o                    \\ \hline
	\end{tabular}
}

\begin{figure*}[t!]
    \centering
    \setlength{\tabcolsep}{0.3em}
	\subfloat[]
	{\dataquerytab
	 \label{fig:ad_example_queries:data}
	}
	\hfill
	\subfloat[]{
		\rankscoretab
		\label{fig:ad_example_queries:ranking_score}
	}
	\hfill
	\subfloat[]{
		\rankstemtab
		\label{fig:ad_example_queries:ranking_stem}
	}
	\hfill
	\subfloat[]{
		\ranknonstemtab
		\label{fig:ad_example_queries:ranking_non_stem}
	}
    \caption{(a) dataset $\candidateSet{}$ of college applicants, with demographic attributes $A_1$ (sex) and $A_2$ (race), numerical attributes $X_1$ (high school GPA), $X_2$ (verbal SAT), and $X_3$ (math SAT), and attribute $X_4$ (choice) that is a vector extracted from the applicants' essays;~(b) is a ranking $\btau_1$ on $\queryScore{1}$, computed as the sum of $X_1$, $X_2$, and $X_3$;~(c) is a ranking on $\queryScore{2}$, predicted based on historical performance of STEM  (\val{cs}, \val{econ},  \val{math}) majors; ~(d) is a ranking on $\queryScore{3}$, predicted based on historical performance of humanities  (\val{art}, \val{lit}) majors.  In all cases, the top-4 candidates will be interviewed in score order, and potentially admitted.}
    \label{fig:admissions}
\end{figure*}

Figure~\ref{fig:admissions} shows a dataset $\candidateSet{}$ of applicants and illustrates the admissions process.  Each applicant submits several quantitative scores, all transformed here to a discrete scale of 1 (worst) through 5 (best) for ease of exposition: $X_1$ is the high school GPA (grade point average), $X_2$ is the verbal portion of the SAT (Scholastic Assessment Test) score, and $X_3$ is the mathematics portion of the SAT score. Attribute $X_4$ (choice) is a weighted feature vector extracted from the applicant's essay, with weight ranging between 0 and 1, and with a higher value corresponding to stronger interest in a specific major.  For example, candidate \val{b} is a White male with a high GPA (4 out of 5), perfect SAT verbal and SAT math scores (5 out of 5), a strong interest in studying computer science (feature weight 0.9), and some interest in studying art (weight 0.2).

The admissions officer uses a suite of tools to sift through the applications and identify promising candidates.  Many of these tools are \emph{rankers}, illustrated in Figure~\ref{fig:ranker_general}. A ranker takes a dataset of candidates, described by structured features, text, or both, as input and produces a permutation of these candidates, also called a \emph{ranking}.  The admissions officer will take the order in which the candidates appear in a ranking under advisement when deciding whom to consider more closely, interview, and admit.

These tools include score-based rankers (Sections~\ref{sec:intro:score-based}) that compute the score of each candidate based on a formula that the admissions officer gives, and then return some number of highest-scoring applicants in ranked order.  This \emph{scoring formula} may, for example, specify the score as a linear combination of the applicant's high-school GPA and the two components of their SAT score, each carrying an equal weight.  This is done in Figure~\ref{fig:admissions}(a), where a candidate's score is computed as $\queryScore{1} = X_1 + X_2 + X_3$  and then ranking $\btau_1$ in Figure~\ref{fig:admissions}(b) is produced. 

\newcommand{\datatab}{
    \small
	\begin{tabular}{|c||c|c||c|c||c||c|}
		\hline
		\rowcolor[HTML]{C0C0C0} 
		candidate & $A_1$ & $A_2$ & $X_1$ & $X_2$ & $X_3$ & $Y$   \\ \hline
		\rowcolor[HTML]{FFFFFF}
		\val{b}  & \val{male}      & \val{White}     & 4   &  5 & 5  & 14   \\ \hline
		\val{c}  & \val{male}      & \val{Asian}    & 5 & 3 &  4 & 12 \\ \hline
		\val{d}  & \val{female}      & \val{White}     & 5 & 4 & 2 & 11 \\ \hline
		\val{e}  & \val{male}      & \val{White}    & 3 & 3 & 4 & 10 \\ \hline
		\val{f}  & \val{female}       & \val{Asian}    & 3 & 2 & 3 & 8 \\ \hline
		\val{k}  & \val{female}       & \val{Black}    & 2 & 2 & 3 & 7   \\ \hline
		\val{l}  & \val{male}      & \val{Black}    & 1 & 1 & 4 & 6   \\ \hline
		\val{o}  & \val{female}       & \val{White}    & 1 & 1 & 2 & 4 \\ \hline
	\end{tabular}
}
\newcommand{\rankrawtab}{
    \small
	\begin{tabular}{|c|}
		\hline
		\rowcolor[HTML]{C0C0C0} 
		$\btau_1$ \\ \hline
		\rowcolor[HTML]{EFEFEF} 
		b                    \\ \hline
		\rowcolor[HTML]{EFEFEF} 
		c                    \\ \hline
		\rowcolor[HTML]{EFEFEF} 
		d                    \\ \hline
		\rowcolor[HTML]{EFEFEF} 
		e                    \\ \hline
		\rowcolor[HTML]{FFFFFF}
		f                    \\ \hline
		k                    \\ \hline
		l                    \\ \hline
		o                    \\ \hline
	\end{tabular}
}
\newcommand{\rankproptab}{
    \small
	\begin{tabular}{|c|}
		\hline
		\rowcolor[HTML]{C0C0C0} 
		$\btau_2$ \\ \hline
		\rowcolor[HTML]{CBCEFB} 
		\val{b}   \\ \hline
		\rowcolor[HTML]{CBCEFB} 
		\val{c}   \\ \hline
		\rowcolor[HTML]{FFCE93} 
		\val{d}   \\ \hline
		\rowcolor[HTML]{FFCE93} 
		\val{f}   \\ \hline
		\rowcolor[HTML]{FFFFFF}
		\val{e}   \\ \hline
		\val{k}   \\ \hline
		\val{l}   \\ \hline
		\val{o}   \\ \hline
	\end{tabular}
}
\newcommand{\rankpreftab}{
    \small
	\begin{tabular}{|c|}
		\hline
		\rowcolor[HTML]{C0C0C0} 
		$\btau_3$ \\ \hline
		\rowcolor[HTML]{CBCEFB} 
		\val{b}   \\ \hline
		\rowcolor[HTML]{FFCE93} 
		\val{d}   \\ \hline
		\rowcolor[HTML]{CBCEFB} 
		\val{c}   \\ \hline
		\rowcolor[HTML]{FFCE93} 
		\val{f}   \\ \hline
		\rowcolor[HTML]{FFFFFF}
		\val{e}   \\ \hline
		\val{k}   \\ \hline
		\val{l}   \\ \hline
		\val{o}   \\ \hline
	\end{tabular}
}
\setlength{\tabcolsep}{0.3em}
\begin{table*}
    \begin{tabular}{cc}
        \begin{minipage}{0.6\textwidth}
            \centering
           \begin{minipage}{0.65\textwidth}
                \datatab
           \end{minipage}
           \begin{minipage}{0.05\textwidth}
                \rankrawtab
           \end{minipage}
           \hspace{2em}
           \begin{minipage}{0.05\textwidth}
                \rankproptab
           \end{minipage}
           \hspace{2em}
           \begin{minipage}{0.05\textwidth}
                \rankpreftab
           \end{minipage}
        \end{minipage}
        &
        \begin{minipage}{0.4\textwidth}
            \centering
\begin{tikzpicture}[scale=0.8, every node/.style={fill=white, font=\small\sffamily, transform shape}, align=left]
  \node (TableData) [data] {Structured data};
  \node (Ranker) [process, below= of TableData] {Ranker};
  \node (TextData) [corpus, right= of TableData] {Text data};
  \node (OutputRanking) [data, below= of Ranker] {Ranking $\btau$};

  \draw [->] (TableData) -- (Ranker) node [midway, right] (TextNode) {1. };
  \draw [->] (TextData) -- (Ranker) node [midway, right] (TextNode) {2. };
  \draw [->] (Ranker) -- (OutputRanking) node [midway, right] (TextNode) {3. };
\end{tikzpicture}
        \end{minipage}
     \\
    \begin{minipage}[t]{0.6\textwidth}
        \captionof{figure}{A dataset $\candidateSet{}$ of college applicants. Score $\score{}$ is computed by a score-based ranker $f(X)= X_1+ X_2+X_3$. Ranking $\btau_1$ of $\score{}$ in $\candidateSet{}$.  Ranking $\btau_2$ with proportional representation by sex at the top-$4$. Ranking $\btau_3$ with proportional representation by sex in every prefix of the top-$4$. The top-4 candidates will be interviewed in score order and potentially admitted.
        }
        \label{fig:ad_example}
    \end{minipage}
     & 
    \begin{minipage}[t]{0.35\textwidth}
        \centering
          
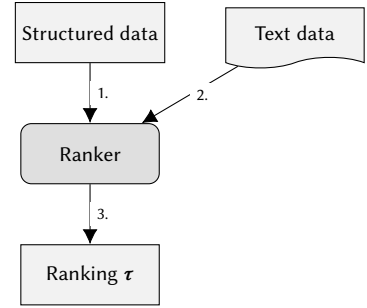
\captionof{figure}{Functional principle of rankers:
            (1. and 2.) Structured and text data that correspond to candidates serve as inputs to a ranker; 
            (3.) The ranker outputs a ranking of the candidates $\btau$.}
         \label{fig:ranker_general}
    \end{minipage}
    \end{tabular}
\end{table*}

Predictive analytics are also among the admissions officer's toolkit.  For example, multiple ranking models may be trained, one per undergraduate major or set of majors, on features $X_1, X_2, X_3, X_4$ of the successful applicants from the past years, to predict applicant's standing upon graduation (based, \eg on their GPA in the major). 
These ranking models are then used to predict a ranking of this year's applicants.  In our example in Figure~\ref{fig:admissions}(a), feature $\queryScore{2}$ predicts performance in a STEM (Science, Technology, Engineering, Mathematics) major such as computer science (\val{cs}), economics (\val{econ}), or mathematics (\val{math}) and leads to ranking $\btau_2$ in Figure~\ref{fig:admissions}(c), while feature $\queryScore{3}$ predicts performance in a humanities major such as literature (\val{lit}) or fine arts (\val{art}) and leads to ranking  $\btau_3$ in Figure~\ref{fig:admissions}(d).

The promising applicants identified in this way---with the help of either a score-based ranker or a predictive analytic---will then be considered more closely, \emph{in ranked order}: invited for an interview and potentially admitted.

Let us recall that, in addition to incorporating quantitative scores and students' choice, an admissions officer also aims to admit a demographically diverse group of students to the university and to each major.  Further, the admissions officer is increasingly aware that the data on which their decisions are based may be biased, in the sense that this data may carry results of historical discrimination or disadvantage, and that the computational tools at their disposal may be exacerbating or introducing new forms of bias, or even creating a kind of a self-fulfilling prophecy. (See discussion of the types of bias in Section~\ref{sec:frame:bias}.)   For this reason, the officer may elect to incorporate one or several fairness objectives into the ranking process. 

For example, they may assert, for legal or ethical reasons, that the proportion of the female applicants among those selected for further consideration should match their proportion in the input.  Applying this requirement to ranking $\btau_1$ in Figure~\ref{fig:ad_example} (in which we elaborate on the already familiar example in Figure~\ref{fig:admissions}) yields ranking $\btau_2$ in Figure~\ref{fig:ad_example}. Further, the admissions officer may assert that, because applicants are interviewed in ranked order, it is important to achieve proportional representation by sex in \emph{every prefix} of the produced ranking, which yields ranking $\btau_3$ in Figure~\ref{fig:ad_example}.  In this survey we give an overview of the technical work that would allow an admissions officer to compute ranked results under these and other fairness requirements. 

\subsection{Scope and contributions of the survey} 
\label{sec:intro:scope}

In the past few years, there has been much work on incorporating fairness requirements into algorithmic rankers.  \rev{And while several surveys on fairness in classification have been published (\eg~\cite{DBLP:journals/csur/MehrabiMSLG21,DBLP:journals/corr/abs-2001-09784}, ranking has not yet received systematic attention. Giving an overview of this large and growing body of work, and the underlying value frameworks that serve as a basis for classification, is the primary goal of our survey.}  Which specific fairness requirements an admissions officer will assert depends on the values they are operationalizing and, thus, on the mitigation objectives.  An important goal of this survey is to create an explicit mapping between mitigation objectives, which we will characterize in Section~\ref{sec:frame:mit_goal}. Without such a mapping, an admissions officer in our running example would have a difficult time selecting an appropriate fairness-enhancing intervention, and would not know which interventions are mutually comparable and which are not. 

In our survey we will present a selection of approaches for fairness in ranking that were developed in several subfields of computer science, including data management, algorithms,  information retrieval, and recommender systems.   We are aware of several recent tutorials on fairness in ranking at \rev{SIGIR 2019~\cite{castillo2019fairness}}, RecSys 2019~\cite{DBLP:conf/recsys/EkstrandBD19}, and VLDB 2020~\cite{vldb20tutorial}, pointing to the need to systematize the work in this area and motivating our survey. Our goal is to offer a broad perspective, connecting work across subfields. We discuss existing technical methods for fairness in score-based ranking in Section~\ref{sec:fair_db}. Technical work on fairness in supervised learning, \rev{with a focus on information retrieval, is covered in Section~\ref{sec:fair_ir}. In Section~\ref{sec:fair_recsys} we highlight representative examples of fairness in recommender systems and matching.} 

The primary focus of this survey is on associational fairness measures for ranking, although we do include one recently proposed causal framework.

\subsection{Survey roadmap}
This survey is organized as follows:
\begin{itemize}
 \item We gave a general introduction in Section~\ref{sec:intro}.
 \item We start with the preliminaries and fix notation in Section~\ref{sec:prelim}. 
 \item We present value classification frameworks along which we relate all surveyed  technical methods in Section~\ref{sec:02-four-frameworks}. 
 \item We present the evaluation datasets that are used by the surveyed technical methods  in Section~\ref{sec:datasets}.  
 \item We describe technical work on fairness in score-based ranking in Section~\ref{sec:fair_db}.
    \item We describe technical work on fair supervised learning in Section~\ref{sec:fair_ir}.
    \item We highlight representative work on fairness in recommender systems and matching in Section~\ref{sec:fair_recsys}.
    \item We discuss evaluation frameworks for fair score-based ranking and fair learning-to-rank in Section~\ref{sec:eval}.
    \item We draw a set of recommendations for the evaluation of fair ranking methods in Section~\ref{sec:discuss}.
    \item We conclude the survey, and identify directions for future work, in Section~\ref{sec:conc}.
\end{itemize}

\section{Preliminaries and notation}
\label{sec:prelim}
In this section we will build on our running example to discuss score-based and supervised learning-based rankers more formally, and fix the necessary notation.  We summarize notation in Table~\ref{tbl:notation} and illustrate it throughout this section.

\subsection{Score-based ranking}
\label{sec:intro:score-based}

Formally, we are given a set $\candidateSet{}$ of candidates; each candidate is described by a set of features  $\featureSet$ and a score attribute $\score{}$. Additionally we are given a set of sensitive attributes $\sensFeatSet \subseteq \featureSet$, which are categorical, denoting membership of a candidate in demographic groups. \rev{Sensitive attributes like age or degree of disability may be drawn from a continuous domain, and several fairness-in-classification methods for continuous sensitive attributes have been proposed~\cite{grari2019fairnessaware,pmlr-v97-mary19a}.  However, we are not aware of any work of this kind that applies to fairness in ranking, and so will assume that sensitive attributes are categorical in the remainder of this survey.}  A sensitive attribute $\sensAttr \in \sensFeatSet$ may be binary, with one of the values (\eg $\sensAttr=1$ or $\sensAttr=\val{female}$, as in Figure~\ref{fig:admissions}) denoting membership in a minority or historically disadvantaged group (often called ``protected group'') and with the other value (\eg $\sensAttr=0$ or $\sensAttr=\val{male}$) denoting membership in a majority or privileged group.  Alternatively, a sensitive attribute may take on three or more values, for example, to represent ethnicity or (non-binary) gender identity of candidates.  

A \e{ranking} $\btau$ is a permutation over the candidates in $\candidateSet{}$. 
Letting $n = \left| \candidateSet{} \right|$, we denote by $\btau = \ranking{\lst{\tau}}$ a ranking that places candidate $\tau_i$ at rank $i$.  We denote by $\btau(i)$ the candidate at rank $i$, and by $\btau^{-1}(a)$ the rank of candidate $a$ in $\btau$.   We are often interested in a sub-ranking of $\btau$ containing its best-ranked $k$ elements, for some integer $k \leq n$; this sub-ranking is called the top-$k$ and is denoted $\btau_{1\ldots k}$.  For example, given a ranking $\btau=\angs{b,c,d,e,f,k,l,o}$, $\btau(3)=d$, $\btau^{-1}(l)=7$, and the top-$4$ is $\btau_{1\ldots 4}=\angs{b,c,d,e}$.

\begin{table}[t!]
\caption{Summary of notation used throughout the survey.}
\label{tbl:notation}
\small
\begin{adjustbox}{max width=\textwidth,center}
\begin{tabular}{p{.1\textwidth}p{.4\textwidth}p{.1\textwidth}p{.45\textwidth}l}\toprule
$\candidateSet{}$  & A set of candidates to be ranked & $\score{}$ & the score feature and ground truth for supervised learning\\
$a, b, c$ &  Candidates in $\candidateSet{}$ & $\predScore{}$ & the scores predicted by $\hat{f}$\\
$n$ & Number of candidates $\left| \candidateSet{} \right|$ & $\score{a}$ & the score of candidate $a$\\
$\featureSet$ & a set of features of the candidates in $\candidateSet{}$ & $\btau$ & Ranking: permutation of candidates from $\candidateSet{}$\\
$\featOfCand{a}$ & Features of candidate $a$ & $\btau(i)$ & The candidate at position $i$ in $\btau$\\
$\sensFeatSet$ & A set of sensitive features, $\sensFeatSet \subseteq \featureSet$ & $\posBias{i}$ & the position bias of rank $i$\\
$\group{}$ & A group (subset) of candidates, $\group{} \subseteq \candidateSet{}$ & $\utilityTwoPara{k}{\btau}$ & Utility of the top-$k$ candidates in $\btau$ \\
$\group{1}$ & A protected group (subset) of candidates, $\group{1} \subseteq \candidateSet{}$ & $\utilityThreePara{k}{\btau}{\group{}}$ & Utility of the top-$k$ candidates of group $\group{}$ in $\btau$ \\
$\userSet{}$ & A set of users that use the ranking system & $\utilityThreePara{}{\btau}{a}$ & Utility of candidate $a$ in $\btau$ \\
$\setOfQueries$ & A set of queries & $\unfairnessTwoPara{a}{b}$ &Disparity in visibility between candidates $a$ and $b$ \\

$f,\hat{f}$ & a ranker, a ranker learned from training data & $\unfairnessTwoPara{\group{1}}{\group{2}}$ & Disparity in visibility between groups $\group{1}$ and $\group{2}$\\

\bottomrule
\end{tabular}
\end{adjustbox}
\end{table}
\paragraph{Utility} Because score $Y$ is assumed to encode a candidate's appropriateness, quality, or \emph{utility}, a score-based ranking usually satisfies: 
\begin{equation}
    \score{\btau(1)} \geq \score{\btau(2)} \geq \ldots \geq \score{\btau(n)}
    \label{eq:sort}
\end{equation}

We will find it convenient to denote by $\utilityTwoPara{k}{\btau}$ the utility of $\btau_{1\ldots k}$.  Different methods surveyed in this paper adopt different notions of utility, and we will make their formulations precise as appropriate.  The simplest method is to treat $\btau_{1\ldots k}$ as a set (disregarding candidate positions), and to compute the utility of the set as the sum of scores of its elements: 
\begin{equation}
    \utilityTwoPara{k}{\btau} = \sum_{i=1}^{k} \score{\btau(i)} 
    \label{eq:agg_utility}
\end{equation}

Another common method incorporates position-based discounts, following the observation that it is more important to present high-quality items at the top of the ranked list, since these items are more likely to attract the attention of the consumer of the ranking.   For example, we may compute position-discounted utility of a ranking as:

\begin{equation}
    \utilityTwoPara{k}{\btau} = \sum_{i=1}^{k} \frac{\score{\btau(i)}}{\log_{2} (i+1)}
    \label{eq:disc_agg_utility}
\end{equation}

For example, the utility at top-$4$ of $\btau_1$ in Figure~\ref{fig:ad_example} is
$47$ based on Equation~\ref{eq:agg_utility} and $31.4$ based on Equation~\ref{eq:disc_agg_utility}. Note that the base of the logarithm in the denominator of Equation~\ref{eq:disc_agg_utility} is empirically determined, and it can be set to some value $b>1$ other than 2.

For these variants of utility and for others, it is often useful to quantify utility realized by candidates belonging to a particular demographic group $\group{} \subseteq \candidateSet{}$, defined by an assignment of values to one or several sensitive attributes. For example, $\group{}$ may contain female candidates, or Asian female candidates.  We can then compute to the utility of  $\btau_{1\ldots k}$ (per Equation~\ref{eq:agg_utility}) for group $\group{}$ as:

\begin{equation}
    \utilityThreePara{k}{\btau}{\group{}} = \sum_{i=1}^{k} Y_{\btau(i)} \times \mathbbm{1}{[\btau(i) \in \group{}]}
    \label{eq:agg_utility_g}
\end{equation}

Here, $\mathbbm{1}$ is an indicator variable that returns 1 when $\btau(i) \in \group{}$ and 0 otherwise.  Position-discounted utility (per Equation~\ref{eq:disc_agg_utility}) for group $\group{}$ can be defined analogously.  For the ranking $\btau_1$ in Figure~\ref{fig:ad_example}, $\utilityThreePara{4}{\btau_1}{sex=\val{male}} = 36$, $\utilityThreePara{4}{\btau_1}{sex=\val{male} \wedge race=\val{White}} = 24$, and 
$\utilityThreePara{4}{\btau_1}{sex=\val{male} \wedge race=\val{Black}} = 0$.

\paragraph{Fairness} To satisfy objectives other than utility, such as \emph{fairness}, we may output a ranking $\hat \btau$ that is not simply sorted based on the observed values of $\score{}$ as in Equation~\ref{eq:sort}.   As is the case for classification and prediction, numerous fairness measures have been defined for rankings.  These measures can be used both to assess the fairness of a ranking and to intervene on unfairness, for example, by serving as basis for constraints.

A prominent class of fairness measures corresponds to \emph{proportional representation} in the top-$k$ treated as a set, or in every prefix of the top-$k$.  These measures are motivated by the need to mitigate different types of bias, based on assumptions about its origins and with a view of specific objectives (to be discussed  in Section~\ref{sec:02-four-frameworks}).  For example, ranking $\btau_2$ in Figure~\ref{fig:ad_example} re-ranks candidates to satisfy proportional representation by gender at the top-$4$ (treating it as a set), swapping candidates $e$ and $f$.  The ranking $\btau_3$ in Figure~\ref{fig:ad_example} additionally reorders candidates $c$ and $d$ to achieve proportional representation by gender in every prefix of the top-$4$.

In addition to fairness measures, \emph{diversity} measures have also been proposed in the literature~\cite{drosou2017diversity}.  In this survey we will \rev{discuss} coverage-based diversity that is most closely related to fairness, and requires that members of multiple, possibly overlapping, groups, be sufficiently well-represented among the top-$k$, treated either as a set or as a ranked list.  Diversity constraints may, for example, be stated to require that members of each ethnic group, each gender group, and of selected intersectional groups on ethnicity and gender, all be represented at the top-$k$ in proportion to their prevalence in the input. \rev{The terminology we adopt in this paper is that ``coverage-based diversity'' is a technical notion that can be used to express several fairness objectives.  In contrast, fairness is never purely technical: it is always associated with a value framework and with a socio-technical context of use.}

When candidates are re-ranked to meet objectives other than score-based utility, we may be interested to compute \emph{$\score{}$-utility loss}, denoted $L_{\score{}}(\btau, \hat \btau)$.  We can use a variety of metrics that quantify the distance between ranked lists for this purpose, including, for example, the Kendall distance that counts the number of pairs that appear in the opposite relative order in $\btau$ and $\hat{\btau}$, or one in a family of generalized distances between rankings~\cite{DBLP:conf/www/KumarV10}.  However, loss functions that compare rankings $\btau$ and $\hat \btau$ in their entirety are uncommon.  Rather, $\score{}$-utility loss is usually specified over the top-$k$.  The simplest formulation is:

\begin{equation}
    \yUtilLoss{k}{\btau}{\hat{\btau}} =  \utilityTwoPara{k}{\btau} - \utilityTwoPara{k}{\hat{\btau}}
\end{equation}

Alternatively, we may normalize this quantity:

\begin{equation}
    \yUtilLoss{k}{\btau}{\hat{\btau}} =  1 - \frac{\utilityTwoPara{k}{\hat{\btau}}}{\utilityTwoPara{k}{\btau}}
\end{equation}

Further, we may be interested to quantify utility loss for a particular demographic group $\group{}$.  In that case, we define the utility of $\btau$ and $\hat \btau$ for group $\group{}$, as was done in Equation~\ref{eq:agg_utility_g}, or analogously for other utility formulations.  Interestingly, underrepresented groups may see a gain, rather than a loss, in $\score{}$-utility, because they may receive better representation at the top-$k$ when a fairness objective is applied.

\subsection{Supervised learning to rank}
\label{sec:intro:learned}
In supervised learning to rank (\ltr), we are given a set  $\candidateSet{}$ of candidates; each candidate is described by a set $\featureSet$ of features, including also sensitive features $\sensFeatSet \subseteq \featureSet$.  
Each candidate $a \in \candidateSet{}$ has an associated score attribute $\score{}$, which describes their quality with respect to a given task ( \eg college admissions).
Every such association forms an instance of either the training  dataset $\candidateSet{train}$ or the test dataset $\candidateSet{test}$.
Like score-based rankers, \ltr rankers compute candidate scores and return a ranking $\btau$ with the highest-scoring candidates appearing closer to the top (per Eq.~\ref{eq:sort}).   The difference between score-based and \ltr rankers is in how the score is obtained: in score-based ranking, a function is given to calculate the scores $Y$, while in supervised learning, the ranking function $\hat{f}$ is learned from a set of training examples and the score  $\predScore$ is estimated.\footnote{Note that the literature distinguishes point-wise, pair-wise and list-wise \ltr methods and that $\score{}$ has a slightly different meaning for each of them~\cite{DBLP:series/synthesis/2014Li}.
However, because the overall procedure remains the same, we will focus on point-wise LtR in the remainder of this section, and give technical details for pair-wise and list-wise methods in later sections, as appropriate.} 

Figure~\ref{fig:supervised_ranker} describes the \ltr process.   
We are given two datasets $\candidateSet{train}$ and $\candidateSet{test}$.\footnote{To follow machine learning best practices, we may also produce a separate validation dataset, used to tune model hyperparameters.  We leave this out from our discussion for brevity.}   
We use $\candidateSet{train}$ to train an \ltr model, learning a ranking function $\hat{f}(\featureSet)$ that minimizes the prediction errors on $\score{\text{train}}$. This is usually done by minimizing the sum of the individual errors $\hat{f}$ makes between the ground truth $\score{}$ and its prediction $\predScore$ for $\candidateSet{train}$.  To evaluate the performance of the model $\hat{f}$, we apply it to $\candidateSet{test}$, and then compare ground truth scores and predictions.  If model testing succeeds, meaning that the ranker's predictions are deemed sufficiently accurate, then $\hat{f}$ is deployed: a new set of candidates is ranked by predicting their scores $\predScore = \hat{f}(\featureSet)$, and ranking the candidates according to these predictions.
\newcommand{\datatabIRexample}{
    \small
	\begin{tabular}{|c||c|c||c|c|c||c||c|c|}
		\hline
		\rowcolor[HTML]{C0C0C0} 
		candidate & $A_1$ & $A_2$ & $X_1$ & $X_2$ & $X_3$ & $X_4$ & $\queryScore{1}$ & $\queryScore{2}$  \\ \hline
		\val{b}  & \val{male}    & \val{White}    & 4 & 5 & 5  & $\{$\val{cs}:0.9; \val{art}:0.2$\}$ & 9 & 1 \\ \hline
		\val{c}  & \val{male}    & \val{Asian}    & 5 & 3 & 4 & $\{$\val{math}:0.9; \val{cs}:0.5$\}$ & 9 & 1 \\ \hline
		\rowcolor[HTML]{ADD8E6}
		\val{d}  & \val{female}  & \val{White}    & 5 & 4 & 2 & $\{$\val{lit}:0.8; \val{math}:0.8$\}$ & 8 & 1 \\ \hline
		\val{e}  & \val{male}    & \val{White}    & 3 & 3 & 4 & $\{$\val{math}:0.8; \val{econ}:0.4$\}$ & 7 & 6 \\ \hline
		\val{f}  & \val{female}  & \val{Asian}    & 3 & 2 & 3 & $\{$\val{econ}:0.9; \val{math}:0.5$\}$ & 5 & 8 \\ \hline
		\val{k}  & \val{female}  & \val{Black}    & 2 & 2 & 3 & $\{$\val{lit:0.9;\val{art}:0.8}$\}$  & 1 & 9 \\ \hline
		\rowcolor[HTML]{ADD8E6}
		\val{l}  & \val{male}    & \val{Black}    & 1 & 1 & 4 & $\{$\val{lit}:0.5; \val{math}:0.7$\}$ & 6 & 7  \\ \hline
		\val{o}  & \val{female}  & \val{White}    & 1 & 1 & 2  & $\{$\val{econ}:0.9; \val{cs}:0.8$\}$ & 7 & 8 \\ \hline
	\end{tabular}
}

\newcommand{\rankrawtabIRexample}{
    \small
	\begin{tabular}{|c|}
		\hline
		\rowcolor[HTML]{C0C0C0} 
		$\btau$ \\ \hline
		d                    \\ \hline
		l                    \\ \hline
	\end{tabular}
}
\newcommand{\rankproptabIRexample}{
    \small
	\begin{tabular}{|c|}
		\hline
		\rowcolor[HTML]{C0C0C0} 
		$\hat{\btau}$ \\ \hline
		\val{l}   \\ \hline
		\val{d}   \\ \hline
	\end{tabular}
}

\begin{figure*}[t!]
    \centering
	\subfloat[]
	{\datatabIRexample
	 \label{fig:ir-example_data}
	}
	\hfill
	\subfloat[]{
		\rankrawtabIRexample
		\label{fig:ir-example_ranking}
	}
	\hfill
	\subfloat[]{
		\rankproptabIRexample
		\label{fig:ir-example_rank_prop}
	}
    \caption{(a) Dataset $\candidateSet{}$ of college applicants, with demographic attributes $A_1$ (sex) and $A_2$ (race), numerical attributes $X_1$ (high school GPA), $X_2$ (verbal SAT), and $X_3$ (math SAT), and attribute $X_4$ (choice) that is a vector extracted from the applicants' essays. Scores $\score{1}$ and $\score{2}$ are the respective ground truth scores for queries $\query_1, \query_2 \in \setOfQueries$. We randomly distribute $\candidateSet{}$ into a training dataset $\candidateSet{train} = \{b, c, e, f, k, o\}$ and a test dataset $\candidateSet{test} = \{d, l\}$ (blue lines);~(b) The ground truth ranking of $\candidateSet{test}$ for query $\query_1$. The ranking model $\hat f$ should reproduce this ordering of candidates $d$ and $l$ when presented $\candidateSet{test}$;~(c) A ranking predicted by a model with a bias against women. Note that by randomly choosing candidates $d$ and $l$ for $\candidateSet{test}$ we accidentally injected a bias against women into our training data (all women are now ranked below men). A learning model is likely to pick up this bias and wrongly assign feature $\sensAttr_1$ a high weight. (We remark that this is a very simple example on data bias for illustrative purposes).
    }
    \label{fig:ir_example}
\end{figure*}
\begin{figure}[t]
    \centering
    \begin{tikzpicture}[scale=0.8, every node/.style={fill=white, font=\small\sffamily, transform shape}, align=center]
      \node (TrainData) [data] {$\featureSet_{\text{train}}, \score{\text{train}}$};
      \node (Learning) [process, right= of TrainData] {Model Training};
      \node (Model) [data, below= of Learning] {Model  $\hat{f}(\featureSet{})$};
      \node (TestData) [data, left= of Model] {$\featureSet_{\text{test}}$};
      \node (ModelOutput) [data, right= of Model] {$\predScore$};
      \node (TestModel) [process, above= of ModelOutput] {Model Testing};
      \node (TestLabels) [data, right= of TestModel] {$\score{\text{test}}$};
    
      \draw [->] (TrainData) -- (Learning) node [midway, above] (TextNode) {1. };
      \draw [->] (Learning) to [loop above] node {2. minimize errors for $\candidateSet{train}$} (Learning);
      \draw [->] (Learning) -- (Model) node [midway, above] (TextNode) {3. };
      \draw [->] (TestData) -- (Model) node [midway, above] (TextNode) {4. };
      \draw [->] (Model) -- (ModelOutput) node [midway, above] (TextNode) {5. };
      \draw [->] (TestLabels) -- (TestModel) node [midway, above] (TextNode) {6. };
      \draw [->] (ModelOutput) -- (TestModel) node [midway] (TextNode) {6. };
    \end{tikzpicture}
    \caption{Functional principle of supervised learning in ranking, commonly denoted learning-to-rank, or \ltr. 
    (1.) Training data $\candidateSet{train}$, consisting of tuples $(\featureSet, \score{})$, is given as input to an \ltr algorithm that (2.) trains a ranking function $\hat{f}(\featureSet)$ (3.). 
    This is done by minimizing the errors $\hat{f}(\featureSet)$ makes when predicting the scores $\predScore$ for $\candidateSet{train}$. 
    (4.) To test  predictive accuracy of the model, the features $\featureSet_{\text{test}}$ of $\candidateSet{\text{test}}$ are given as input to $\hat{f}$ to (5.) predict scores $\predScore$. 
    (6.) Then, $\predScore$ is compared to the ground truth $\score{\text{test}}$. 
    }
    \label{fig:supervised_ranker}
\end{figure}

As an example, consider Figure~\ref{fig:ir_example} that revisits our college admissions example from Figure~\ref{fig:admissions} in a supervised learning setting. 
We are given six features, as previously described, and two ground truth scores $\score{1}$ and $\score{2}$ for each candidate. 

First, training data is prepared as input:  We divide the data into a training set $\candidateSet{train} = \{b, c, e, f, k, o\}$ and a test set $\candidateSet{test} = \{d, l\}$ (blue lines).  Then a model is trained and tested using any available \ltr method, such as RankNet~\cite{burges2010ranknet} or ListNet~\cite{cao2007learning}. Ranking $\btau$ in Figure~\ref{fig:ir-example_ranking} depicts the ground truth ranking of $\candidateSet{test}$ based on score~$\score{1}$. 

\paragraph{Prediction accuracy.}  In traditional supervised learning, the term \emph{utility} is often used to refer to prediction accuracy of $\hat{f}$.  A common measure of prediction accuracy in \ltr is the Normalized Discounted Cumulative Gain (\texttt{NDCG})~\cite{jarvelin2002cumulated}, which compares a ranking generated by model $\hat{f}$ to a ground-truth ranking (sometimes called the ``ideal'' ranking).  \texttt{NDCG} measures the usefulness, or \emph{gain}, of the candidates based both on their scores and on their positions in the ranking.  NDCG incorporates position-based discounts, capturing the intuition that it is more important to retrieve high-quality (according to score) candidates at higher ranks, and is hence closely related to Eq.~\ref{eq:disc_agg_utility}.
\texttt{NDCG} of a predicted ranking is computed relative to the gain of the ground-truth ranking, \texttt{IDCG}, and thus \texttt{NDCG} measures the extent to which the model is able to reproduce the ground-truth ranking from $\candidateSet{train}$ in its predictions $\predScore$.  We are usually interested in \texttt{NDCG} at the top-$k$ (denoted $\texttt{NDCG}^k$), and so normalize the position-discounted gain of the top-$k$ in the predicted ranking by the position-discounted gain of the top-$k$ in the ideal ranking (denoted $\texttt{IDCG}^k$), per Equation~\ref{eq:ndcg}.
\begin{equation}
    \texttt{NDCG}^k = \frac{1}{\texttt{IDCG}^k} \cdot \sum_{i=1}^{k} \frac{\predScore{\btau(i)}}{log_2 (i+1)}
    \label{eq:ndcg}
\end{equation}

An important application of \ltr are information retrieval systems, where users issue search queries, and expect the system to find relevant information and rank the results by decreasing relevance to their queries.  Consider again our example in Figure~\ref{fig:ir_example}, and suppose that we are additionally given a set $\setOfQueries$ of queries, each associated with $\candidateSet{}$ via a score.

In our example, two queries are given, $\query_1 = $ ``What are the most promising candidates to admit to a STEM major?'' associated with score  $\score{1}$, and $\query_2 = $ ``What are the most promising candidates to admit to a humanities or arts major?'' associated with  $\score{2}$. 
The training and test sets are formed by assigning the candidate features $\featureSet$ and their respective scores $\score{}$ per query: $\candidateSet{train} = \left\{ (\featureSet_{\textit{train}}, \score{\query}) \right\}_{\query \in \setOfQueries}$ and $\candidateSet{test} = \left\{ (\featureSet_{\textit{test}}, \score{\query}) \right\}_{\query \in \setOfQueries}$.
With these sets as input, we can use the \ltr procedure shown in Figure~\ref{fig:supervised_ranker} to train a single model. 

To evaluate model performance, its accuracy measures need to be extended to handle multiple queries. Commonly used measures are \texttt{NDCG} (Eq.~\ref{eq:ndcg}) averaged over all queries and Mean Average Precision (\texttt{MAP}). \texttt{MAP}~\cite{manning2008evaluation} consists of several parts: first,  precision at position~$k$ ($P@k$) is calculated as the proportion of query-relevant candidates in the top-$k$ positions of the predicted ranking $\hat{\btau}$. This proportion is computed for all positions in $\hat{\btau}$, and then averaged by the number of relevant candidates for a given query to compute average precision (\texttt{AP}).
Finally, \texttt{MAP} is calculated as the mean of AP values across all queries.  \texttt{MAP} enables a performance comparison between models irrespective of the number of queries that were given at training time.  
\paragraph{Fairness.} 
As is the case in score-based ranking, \ltr methods may incorporate fairness objectives in addition to utility.
Fairness interventions in \ltr are warranted because the procedure described in Figure~\ref{fig:supervised_ranker} is prone to pick up and amplify different types of bias (see Section~\ref{sec:frame:bias} for details).  For example, let us return to Figure~\ref{fig:ir_example} and note that, by randomly selecting $\candidateSet{train} = \{b, c, e, f, k, o\}$, in which all men are ranked above all women, we may have accidentally injected a strong gender bias into the learned model.  This, in turn, may result in an estimated ranking $\hat{\btau}$ in Figure~\ref{fig:ir-example_rank_prop} that places the male candidate $l$ above the female candidate $d$, although their ground truth scores would place them in the opposite relative order. 
This may lead to biased future predictions for rankings that systematically disadvantage women~($\hat{\btau}$ in Fig.~\ref{fig:ir-example_rank_prop}).\footnote{We denote that this is a very simplified example for technical bias which we use for illustrative purposes.}

As a remedy, two main lines of work on measuring fairness in rankings, and enacting fairness-enhancing interventions, have emerged over the past several years: probability-based  and exposure-based. Both interpret fairness as a requirement to provide a predefined share of visibility for one or more protected groups throughout a ranking.

\emph{Probability-based fairness} is defined by means of statistical significance tests that ask how likely it is that a given ranking was created by a fair process, such as by tossing a coin to decide whether to put a protected-group or a privileged-group candidate at position $i$~\cite{yang2017measuring, zehlike2017fa}.  

\emph{Exposure-based fairness} is defined by quantifying the expected attention received by a candidate, or a group of candidates, typically by comparing their average \e{position bias}~\cite{joachims2017accurately} to that of other candidates or groups.
\begin{equation}
    \exposure{\btau(i)} = \mathbb{E}_{\btau \sim \pi}\left[\posBias{\btau(i)}\right]
    \label{eq:exp}
\end{equation}
\noindent Here, $\pi: rnk(\candidateSet{}) \rightarrow [0,1]$ is the probability mass function over the ranking space, and position bias $\posBias{\btau(i)}$ refers to the observation that users of a ranking system tend to prefer candidates at higher positions, and that their attention decreases either geometrically or logarithmically with increasing rank~\cite{joachims2017accurately,DBLP:journals/cacm/Baeza-Yates18}. 
Logarithmic position-based discounting when computing exposure is in-line with position-based discounting of utility for score-based rankers (Eq.~\ref{eq:disc_agg_utility}) and with the NDCG measure for supervised \ltr (Eq.~\ref{eq:ndcg}). 

The algorithmic fairness community is familiar with the distinction between individual fairness, a requirement that individuals who are similar with respect to a task are treated similarly by the algorithmic process, and group fairness, a requirement that outcomes of the algorithmic process be in some sense equalized across groups.  Probability-based fairness definitions are designed to express strict group fairness goals.  
Thus, they do not allow later compensation for unfairness in higher ranking positions, since a ranking has to pass the statistical significance test at every position to be declared fair. If a ranking fails the fairness test at any point, it is immediately declared unfair, in contrast to exposure-based definitions.
Exposure-based fairness can serve the goals of either individual fairness or group fairness, depending on the specific formalization.  
Individual unfairness in exposure can be expressed as the discrepancy~$\unfairnessOnePara{.}$ in exposure between two candidates $a$ and $b$:
\begin{equation}
    \unfairnessTwoPara{a}{b} = \left| \exposure{a} - \exposure{b} \right|
\end{equation}

Group unfairness can be expressed as the discrepancy~$\unfairnessOnePara{.}$ in the average exposure between two groups $\group{1}$ and $\group{2}$:
\begin{equation}
    \unfairnessTwoPara{\group{1}}{\group{2}} = \left| \frac{1}{\left|\group{1}\right|} \sum_{a \in \group{1}} \exposure{a} - \frac{1}{\left|\group{2}\right|} \sum_{b \in \group{2}} \exposure{b} \right|
\end{equation}

Note that consensus on a definition of exposure has not yet been found and, while many measures feature position bias in some way, they disagree on its importance.  An additional distinctive characteristic of fairness definitions is that some of them consider a notion of a candidate's merit when measuring disparities in exposure, while others explicitly leave it out. Most of the former understand merit as the utility score $\score{}$ at face value.  However, as we will discuss in Section~\ref{sec:frame:mit_goal}, the understanding of merit depends on worldviews and on one's conception of equal opportunity. In Sections~\ref{sec:fair_ir} and ~\ref{sec:fair_recsys} we will present different interpretations of merit and exposure that have been used by \ltr methods in information retrieval and recommender systems.
\section{Four Classification Frameworks for Fairness-Enhancing Interventions}
\label{sec:02-four-frameworks}

\begin{figure}[h]
    \centering
    \begin{tikzpicture}[scale=0.9,align=center]
    \tikzset{
        root/.style = {rectangle, rounded corners, draw=black, fill=gray!30, font=\small\sffamily},
        child/.style = {rectangle, rounded corners,draw=black, font=\small\sffamily},
        orangec/.style = {rectangle, rounded corners, draw=black, fill=orange!70, font=\small\sffamily},
        greenc/.style = {rectangle, rounded corners, draw=black, fill=green!40, font=\small\sffamily},
        bluec/.style = {rectangle, rounded corners, draw=black, fill=blue!30, font=\small\sffamily},
        yellowc/.style = {rectangle, rounded corners, draw=black, fill=yellow!70, font=\small\sffamily},
        pinkc/.style = {rectangle, rounded corners, draw=black, fill=pink!70, font=\small\sffamily},
    }
    \node (ND) at (-0.2,0) [root] {Normative Dimensions};    
    \node (GS) at (1.8,1.75) [orangec] {Group Structure};    
    \node (BT) at (1.8,-1.75) [greenc] {Bias Type};    
    \node (EO) at (-1.8,1.75) [bluec] {Equal Opportunity};    
    \node (WV) at (-1.8,-1.75) [yellowc] {Worldview};  
    \node (IT) at (3.3,0.0) [pinkc] {Intersectional};
    
    \node (AC) at (3.9,2.5) [child] {Attribute Cardinality};
    \node (ACB) at (6.3,3.2) [child] {Binary};
    \node (ACM) at (6.3,1.8) [child] {Multinary};
    
    \node (AN) at (3.9,1.0) [child] {Attribute Number};
    \node (ANO) at (6.9,1.15) [child] {One};
    \node (ANM) at (6.9,0.5) [child] {Multiple};
    
    \node (ANMI) at (8.9,0.8) [child] {Independent};
    \node (ANMC) at (8.9,0.0) [child] {Combination};
    
    \node (ITY) at (5.3,0.2) [child] {Yes};
    \node (ITN) at (5.3,-0.4) [child] {No};
    
    \node (BTP) at (4.0,-1.2) [child] {Pre-existing};
    \node (BTT) at (4.0,-2.0) [child] {Technical};
    \node (BTE) at (4.0,-2.75) [child] {Emergent};
    
    \node (EOF) at (-3.8,2.8) [child] {Formal};
    \node (EOFP) at (-4.6,2.1) [child] {Formal plus};
    \node (EOS) at (-3.8,1.0) [child] {Substantive};
    
    \node (EOSR) at (-5.8,1.4) [child] {Rawlsian};
    \node (EOSL) at (-5.4,0.1) [child] {Luck-egalitarian};

    \node (WVI) at (-3.8,-0.8) [child] {WYSIWYG};
    \node (WVE) at (-3.8,-1.8) [child] {WAE};
    \node (WVC) at (-3.8,-2.8) [child] {Continuous};
    
    \draw (ND.east) -- (GS);
    \draw (ND.east) -- (BT);
    \draw (ND.east) -- (IT);
    \draw (ND.west) -- (EO);
    \draw (ND.west) -- (WV);
    
    \draw (GS) -- (AC);
    \draw (GS) -- (AN);
    
    \draw (AC) -- (ACB);
    \draw (AC) -- (ACM);
    
    \draw (AN) -- (ANO);
    \draw (AN) -- (ANM);
    
    \draw (IT) -- (ITY);
    \draw (IT) -- (ITN);
    
    \draw (ANM) -- (ANMI);
    \draw (ANM) -- (ANMC);
    
    \draw (BT) -- (BTP);
    \draw (BT) -- (BTT);
    \draw (BT) -- (BTE);
    
    \draw (EO) -- (EOF);
    \draw (EO) -- (EOFP);
    \draw (EO) -- (EOS);
    
    \draw (EOS) -- (EOSR);
    \draw (EOS) -- (EOSL);
    
    \draw (WV) -- (WVI);
    \draw (WV) -- (WVE);
    \draw (WV) -- (WVC);
\end{tikzpicture}
    \caption{\rev{A mind map summary of the structure of the four classification frameworks.}}
    \label{fig:normative-dimensions}
\end{figure}
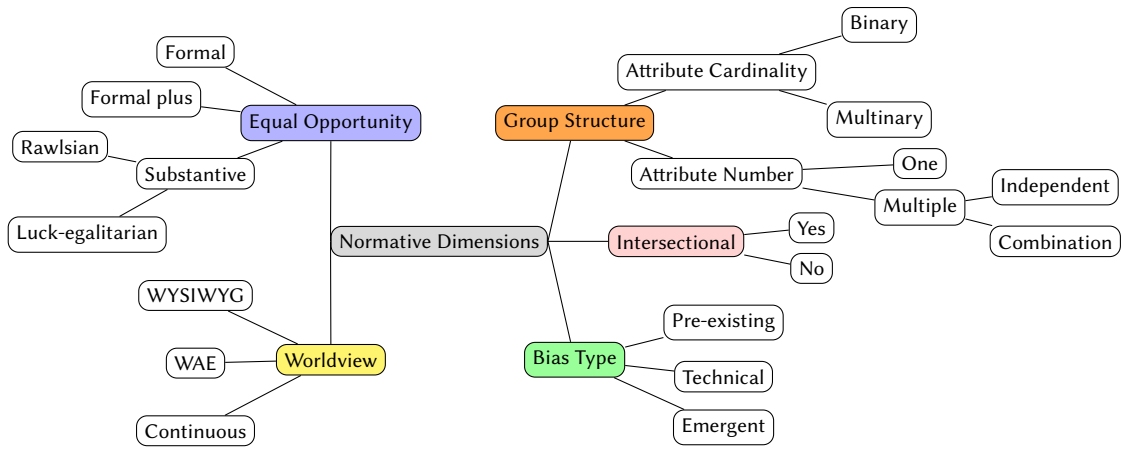

Operationally, algorithmic approaches surveyed in this paper differ in how they represent candidates (e.g., whether they support one or multiple sensitive attribute, and whether these are binary), in the type of bias they aim to surface and mitigate, in what fairness measure(s) they adopt, in how they navigate the trade-offs between fairness and utility during mitigation, and, for supervised learning methods, at what stage of the pipeline a mitigation is applied.  Conceptually, these operational choices correspond to normative statements about the types of bias being observed and mitigated, and the objectives of the mitigation.  In this section we give four classification frameworks that allow us to relate the technical choices with the normative judgments they encode, and to identify the commonalities and the differences between the many algorithmic approaches.
\rev{Figure~\ref{fig:normative-dimensions} gives a structural overview of the frameworks and their sub-categories in the form of a \emph{mind map}. For each method, we will highlight which normative choices they make based on this mind map. }

\subsection{Group structure}
\label{sec:frame:group}

Recall that fairness of a method is stated with respect to a set of categorical sensitive attributes (or features).  Individuals who have the same value of a particular sensitive attribute, such as gender or race, are called a \emph{group}. In this survey, we consider several orthogonal dimensions of group structure, based on the handling of sensitive attributes.

\paragraph*{Cardinality of sensitive attributes} Some methods consider only \emph{binary} sensitive attributes (\eg binary gender, majority or minority ethnic group), while other methods handle higher-cardinality (\emph{multinary}) domains of values for sensitive attributes. If a multinary domain is supported, methods differ in whether they consider one of the values to be protected (corresponding to a designated group that has been experiencing discrimination), or if they treat all values of the sensitive attribute as potentially being subject to discrimination.

\paragraph*{Number of sensitive attributes} Some methods are designed to handle a \emph{single sensitive attribute} at a time (
\eg they handle gender or race, but not both), while other methods handle \emph{multiple sensitive attributes} simultaneously (\eg they handle both gender and race).  

\paragraph*{Handling of multiple sensitive attributes} Methods that support multiple sensitive attributes differ in whether they handle these \emph{independently} (\eg by asserting fairness constraints w.r.t. the treatment of both women and Blacks) or \emph{in combination} (\eg by requiring fairness w.r.t. Black women).  Note that any method that supports a single multinary attribute can be used to represent multiple sensitive attributes with the help of a computed high-cardinality sensitive attribute.  For example, a computed sensitive attribute \emph{gender-race-disability} can represent the Cartesian product $\{male,female,non\text{-}binary\}\cross\{White,Black,Asian\}\cross\{disabled,non\text{-}disabled\}$.
We may be tempted to say that such methods take the point of view of intersectional discrimination~\cite{crenshaw1990mapping, makkonen2002multiple}.  However, as we will discuss in Section~\ref{sec:frame:mit_goal}, detecting and mitigating intersectional discrimination is  more nuanced, and so it is in general not true that if a method  takes a Cartesian product of sensitive attribute values then handles intersectional discrimination, and if a method  treats sensitive attributes independently then it does not. 

\subsection{Type of bias}
\label{sec:frame:bias}

We study ranking systems with respect to the types of bias that they attempt to mitigate, namely, pre-existing bias, technical bias, and emergent bias, as defined by~\cite{DBLP:journals/tois/FriedmanN96}. 

\paragraph*{Pre-existing bias} This type of bias includes all biases that exist independently of an algorithm itself and has its origins in society. For an example of pre-existing bias in rankings, consider the Scholastic Assessment Test (SAT).  College applicants in the US are commonly ranked on their SAT score, often in combination with other features.  It has been documented that the mean score of the math section of the SAT differs across racial groups, as does the shape of the score distribution.  According to a Brookings report that analyzed 2015 SAT test results, ``The mean score on the math section of the SAT for all test-takers is 511 out of 800, the average scores for blacks (428) and Latinos (457) are significantly below those of whites (534) and Asians (598). The scores of black and Latino students are clustered towards the bottom of the distribution, while white scores are relatively normally distributed, and Asians are clustered at the top''~\cite{brookings_race}.  This disparity is often attributed to racial and class inequalities encountered early in life, and presenting persistent obstacles to upward mobility and opportunity. 

\paragraph{Technical bias} This type of bias arises from technical constraints or considerations, such as the screen size or a ranking's inherent position bias --- the geometric drop in visibility for items at lower ranks compared to those at higher ranks.  Position bias arises because in Western cultures we read from top to bottom, and from left to right, and so items appearing in the top-left corner of the screen attract more attention~\cite{DBLP:journals/cacm/Baeza-Yates18}.  A practical implication of position bias in rankings that do not admit ties is that, even if two items are equally suitable for a searcher, only one of them can be placed above the other in a ranking, suggesting to the searcher that it is better and should be prioritized.  

Note that, as all rankings carry an inherent position bias, any method that produces rankings with equalized candidate visibility implicitly addresses this technical bias. However, we will only assign a method to technical bias mitigation, if the paper is explicitly concerned with it, such as~\cite{biega2018equity}. 

\paragraph{Emergent bias} This type of bias arises in a context of use and may be present if a system was designed with different users in mind or when societal concepts shift over time.  In the context of ranking and recommendation it arises most notably because searchers tend to trust the systems to indeed show them the most suitable items at the top positions~\cite{pan2007google}, which in turn shapes a searcher's idea of a satisfactory answer for their search. 
These feedback loops can create a ``the-winner-takes-it-all'' situation in which consumers increasingly prefer one majority product over everything else.

\subsection{Mitigation objectives} 
\label{sec:frame:mit_goal}

\subsubsection{Worldviews} Friedler et al. ~\cite{friedler2016possibility} reflect on the impossibility of a purely objective interpretation of algorithmic fairness (in the sense of a lack of bias):  ``In order to make fairness mathematically precise, we tease out the difference between beliefs and mechanisms to make clear what aspects of this debate are opinions and which choices and policies logically follow from those beliefs.''  They model the decision pipeline of a task as a sequence of mappings between three metric spaces: construct space (\cs), observed space (\os), and decision space (\ds),  and define worldviews (belief systems) as assumptions about the properties of these mappings.  

\begin{figure*}[t]
    \centering
    \subfloat[WYSIWYG]{
        \includegraphics[height=80pt,width=200pt]{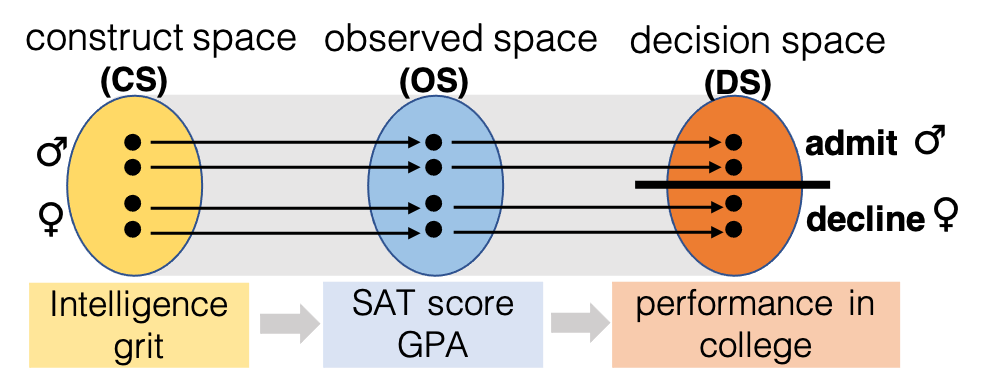}
    }
    \subfloat[WAE]{
        \includegraphics[height=80pt,width=200pt]{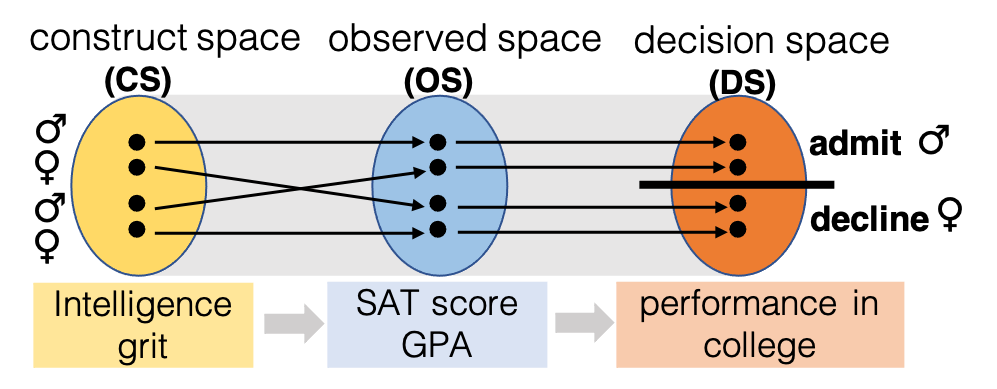}
    }
    \caption{An illustration of the worldviews from Frieder et al.~\cite{friedler2016possibility}: ``What you see is what you get'' (WYSIWYG) vs. ``We are all equal'' (WAE). 
    WYSIWYG assumes that the mapping from the construct space (\cs) to the observed space (\os) shows very low distortion.  In contrast, WAE assumes that the mapping from \cs to \os distorts the structure of the groups in \cs, leading to structural bias. }
   \label{fig:wae_wys}
\end{figure*}

The spaces and the mappings between them are illustrated in Figure~\ref{fig:wae_wys} for the college admissions example.  Individuals are represented by points. \cs represents the ``true'' unobservable properties of an individual (\eg intelligence and grit), while \os represents the properties that we can measure (\eg SAT score as a proxy for intelligence, high school GPA as a proxy for grit) and serves as the  feature space of an algorithmic ranker.  An observation process $g(p) = \hat{p}$ maps from an individual $p \in \cs$ to an entity $\hat{p} \in OS$. An example of such a process is an SAT test. The decision space \ds maps from \os to a metric space of decisions, which for rankings represent the degree of relevance of an entity $\hat{p}$ by placing it at a particular position in the ranking.

Note that the mappings between the spaces are prone to distortions, of which those that map from \cs to either \os or \ds are by definition unobservable.  Because the properties of these mapping cannot be independently verified, a belief system has to be postulated.  \citet{friedler2016possibility} describe two extreme cases: WYSIWYG (``what you see is what you get'') and WAE (``we are all equal''). The former assumes that \cs and \os are essentially the same and any distortion between the two is at most $\epsilon$. The latter assumes that any differences between the utility distributions of different groups are due to an erroneous or biased observation process $g$. 
In our college admissions example this would mean that any differences in the GPA or IQ distributions across different groups are solely caused by biased school systems and IQ tests. 
It is also assumed that $g$ shows different biases across groups, to which the authors refer as \emph{group skew}.

The authors further define different terms from the Fairness, Accountability, Transparency, and Ethics (FATE) literature in terms of the underlying group skew. 
Their \emph{fairness} definition is inspired by~\citet{dwork2012fairness} and says that items that are close in construct space shall also be close in decision space, which is widely known as individual fairness: \rev{similar individuals should receive similar outcomes}.
Group fairness, however, is defined indirectly through the terms \emph{direct discrimination} and \emph{non-discrimination}, \rev{requiring that an individual's treatment should not depend on their group membership}. 
\rev{More formally, d}irect discrimination is absent if the group skew of a mapping between $OS$ and $DS$ is less than $\epsilon$. 
Non-discrimination is present, if the group skew of a mapping between $CS$ and $DS$ is less than $\epsilon$. 
Note that the last definition requires a choice of world view beforehand in order to be evaluated.
If WYSIWYG is chosen, group fairness is given as soon as there is no direct discrimination, because $CS \approx OS$.

We will classify the investigated algorithms in terms of which worldview they choose and which of the three terms (fairness, direct discrimination, non-discrimination) they aim to optimize.

When categorizing surveyed methods with respect to worldview, we consider whether their fairness objective aims for equality of outcome or equality of treatment.  If the goal of a method is to achieve equality of outcome, and if it is asserted that OS is not trustworthy because of biased or erroneous distortion $g$ between CS and OS, then we consider this method to fall under the WAE worldview.  If, on the other hand, the goal is to achieve equality of treatment and it is asserted that the mapping between CS and OS shows low distortion, then the method falls under the WYSIWYG worldview.

\subsubsection{Equality of Opportunity} 
\rev{Equality of Opportunity (\revv{\eop}) is a philosophical doctrine that aims to remove morally irrelevant and arbitrary barriers to the attainment of desirable positions.}  \citet{heidari2019moral} show an application of equality of opportunity (\revv{\eop}) frameworks to algorithmic fairness:  
``At a high level, in these models an individual’s outcome/position is assumed to be affected by two main factors: his/her circumstance $c$ and effort $e$. Circumstance $c$ is meant to capture all factors that are deemed irrelevant, or for which the individual should not be held morally accountable; for instance $c$ could specify the socio-economic status they were born into. Effort $e$ captures all accountability factors---those that can morally justify inequality.'' Several conceptions of \revv{\eop} have been proposed, differing \rev{in what features they consider morally relevant, and in how the relationship between circumstance and effort is modeled.}

\emph{Formal \revv{\eop}} considers a competition to be fair when candidates are evaluated on the basis of their qualifications, and the most qualified candidate wins. This view rejects any qualifications that are irrelevant, such as hereditary privileges or social status, but it makes no attempt to correct for arbitrary privileges and disadvantages leading up to the competition that can lead to disparities in qualifications at the time of competition.  Formal \revv{\eop} is typically understood in the algorithmic fairness literature as fairness-through-blindness --- disallowing the direct impact from sensitive attributes (\eg gender and race) on the outcome but allowing them to impact the outcome through proxies.  

Limiting formal \revv{\eop} to fairness through blindness has been challenged in recent work by~\citet{fairfriends}, who argue for a broader interpretation: ``For example, think of the SAT as a predictor of college success: when students can  afford to do a  lot of test  prep, scores are an inflated reflection of students’ college potential.  When students don’t have access to test prep, the  SAT underestimates students’ college potential. The SAT systematically overestimates more privileged students, while  systematically underestimating less  privileged  students.  The test’s validity as a predictor of  college potential varies across groups.  That’s also a violation of formal \revv{\eop}.  After all, in the college admissions contest, applicants should only be judged by ‘college-relevant’ qualifications--but this test’s accuracy as a yardstick for college potential varies with students’ irrelevant privilege'.''  
\revv{\emph{Formal-plus \revv{\eop}}, due to Fishkin~\cite{Fishkin2014Bottlenecks}, addresses this important shortcoming of formal \eop, capturing the desideratum that test performance should not skew along the lines of morally irrelevant factors.  Tests that satisfy formal-plus \eop include those that aim to balance error rates~\cite{kleinberg_et_al:LIPIcs:2018:8323}, as well as equalized odds~\cite{hardt2016equality}.}

\revv{\emph{Substantive \eop} doctrines take a broader view of Equal Opportunity --- one that is not limited to fair competitions. Instead, they consider whether people have comparable opportunities over the course of a lifetime, including crucial developmental opportunities such as access to education. In order to make such a determination, substantive doctrines attempt to mitigate the effect of morally arbitrary factors such as gender, race, and socio-economic status, on people’s relevant qualifications, which are the basis for attaining desirable positions.  Importantly, in contrast to formal and formal plus \eop that focus on the \emph{current competition}, substantive \eop aims to make people's \emph{future prospects} comparable.} 

\revv{There are several prominent conceptions of substantive \eop. 
\emph{Luck-egalitarian \eop} (see~\citet{dworkin_1981} and~\citet{Roemer2002}) would distribute outcomes after conditioning people’s morally relevant qualification score on their morally irrelevant circumstances. Such an approach may, for example, rank individuals separately by group, and then take the specified number of top-ranked individuals from each list.}

\revv{An alternative iterative approach to equalizing people’s life chances could follow \emph{Rawls’ Fair \eop}, and distribute outcomes in a way that improves the parity in people’s  future prospects of success, setting them up to be competitive in future competitions, even if it means ``unfairness'' in the outcomes of the current competition~\cite{Rawls}. In this paper, we will interpret fairness interventions that attempt to model what an individuals' qualifications \emph{would have looked like}, in a world where \emph{equally talented people have equal prospects of success}, as Rawls's Fair \eop. Once this has been satisfied, we can look more broadly at improving people's life prospects by applying the Difference Principle, which explicitly focuses on improving outcomes for the most disadvantaged (\ie maximizing the minimum).}

We will classify surveyed approaches with respect to the \revv{\eop} framework based on how their fairness definition compares individuals according to some qualifications (\eg test scores, credit amount, and times of being arrested).  However, such a mapping is elusive if a paper does not clearly state its \revv{assumptions about how morally arbitrary factors affect} an individual's \revv{relevant qualifications}. 
Note also that we explicitly map the \textit{fairness definitions}, not the overall approaches. This is because many methods \textit{combine} a fairness and a utility objective into a single optimization problem and, by doing so, lose a clear association with a particular  framework. As a result, many of the methods we survey fall between the WAE and WYSIWYG worldviews, and do not cleanly map to a single \revv{\eop} category. Some are even designed to allow a continuous shift between the frameworks, by providing a tuning parameter~\cite{zehlike2017matching}.

\rev{Some authors~\cite{Arneson2018FourCO,heidari2019moral} categorize the \emph{libertarian} view as an \revv{\eop} framework.  
According to this view, any information about an individual that was legally obtained can be used to make a decision.  Because there is no attempt to equalize access to opportunity, this view corresponds to a narrow notion of procedural fairness, and we do not categorize it under \revv{\eop}~\cite{fairfriends}.  If a fairness definition assumes that all individuals are comparable in all dimensions, as long as there is no gross violations of their privacy during the comparison, then we map this definition to the libertarian view. }

\paragraph{Worldviews vs. Equality of Opportunity} 

\revv{The different worldviews~\cite{friedler2016possibility} give us an intuitive way of thinking about the sufficiency of different \eop doctrines.  The WYSIWYG worldview assumes that there is no distortion between the construct space and the observed space, and, in such a setting, formal and formal-plus \eop conceptions that focus on adjudicating outcomes fairly based on observable qualifications are sufficient. The WAE worldview, on the other hand, models structural bias that leads to the mis-measurement of qualifications of certain demographic groups. One way to correct for this is by adopting a formal-plus \eop approach that attempts to eliminate skew in test performance between groups \emph{at the point of competition}.  Alternatively, the conditions modelled by WAE may be mitigated by interventions consistent with substantive \eop, which seeks to equalize opportunities \emph{over a lifetime} by modeling and controlling for \emph{causes of the skew}.}

\subsubsection{\revv{Intersectional discrimination}}

Intersectional Discrimination~\cite{crenshaw1990mapping, makkonen2002multiple} states that individuals who belong to several protected groups simultaneously (\eg Black women) experience stronger discrimination compared to individuals who belong to a single protected group (\eg White women or Black men), and that this disadvantage compounds more than additively.  This effect has been demonstrated by numerous case studies, and by theoretical and empirical work~\cite{collins2002black,shields2008gender,d2020data,noble2018algorithms}.  
The most immediate interpretation for ranking is that, if fairness is taken to mean proportional representation among the top-$k$, then it is possible to achieve proportionality for each gender subgroup (\eg men and women) and for each racial subgroup (\eg Black and White), while still having inadequate representation for a subgroup defined by the intersection of both attributes (\eg Black women).   

Intersectional concerns also arise in more subtle ways.  For example,~\citet{DBLP:conf/ijcai/YangGS19} observed that when representation constraints are stated on individual attributes, like race and gender, and when the goal is to maximize score-based utility subject to these constraints, then a particular kind of unfairness can arise, namely,  utility loss can be particularly severe in historically disadvantaged intersectional groups.  When discussing specific technical methods, we will speak to whether they consider intersectional discrimination and, if so, which specific concerns they aim to address.

\subsection{Mitigation method}
\label{sec:frame:mit_where}

\begin{table}
\begin{tabular}{cc}
    \begin{minipage}{0.48\textwidth}
        \begin{adjustbox}{width=0.9\textwidth}
    \begin{tikzpicture}[every node/.style={fill=white, font=\sffamily}, align=center]
      \node at (-1, -0.1) (bias) [bias] {Bias};
      \node (scoreDistribution) [process, right of=bias, node distance=1.5cm] {candidates $\candidateSet{}$};
      \node (scoreFunction) [process, right= of scoreDistribution] {ranking function $f$};
      \node (rankingProcess) [process, right= of scoreFunction]   {ranking $\btau$};
      \node (scoreIntervention) [node, below= of scoreDistribution] {intervene on scores};
      \node (functionInvertion) [node, below= of scoreFunction] {intervene on $f$};
      \node (rankingIntervention) [node, below= of rankingProcess] {intervene on $\btau$};
       
      \draw[->]             (scoreDistribution) -- (scoreFunction);
      \draw [->]     (scoreFunction) -- (rankingProcess);
      \draw[thick, dashed, red]
            (-1,0) sin (0,1) cos (1,0) sin (2,-1) cos (3,0) sin (4,1) cos (5,0) sin (6,-1) cos (7,0)
          sin (8,1);
      \draw[->] (scoreIntervention) -- (scoreDistribution);
      \draw[->] (functionInvertion) -- (scoreFunction);
      \draw[->] (rankingIntervention) -- (rankingProcess);
    \end{tikzpicture}
\end{adjustbox}
    
    \end{minipage}
     &  
    \begin{minipage}{0.48\textwidth}
        \begin{adjustbox}{width=0.9\textwidth}
    \begin{tikzpicture}[every node/.style={fill=white, font=\sffamily}, align=center]
      \node at (-1, -0.1) (bias) [bias] {Bias};
      \node (trainingData) [process, right of=bias, node distance=1.5cm] {Training Data};
      \node (modelTraining) [process, right= of trainingData] {Model Training};
      \node (modelOutput)      [process, right= of modelTraining]   {Output Rankings};
      \node (pre) [node, below= of trainingData] {Pre-Processing};
      \node (in) [node, below= of modelTraining] {In-Processing};
      \node (post) [node, below= of modelOutput] {Post-Processing};
       
      \draw[->]             (trainingData) -- (modelTraining);
      \draw[->]     (modelTraining) -- (modelOutput);
      \draw[thick, dashed, red]
            (-1,0) sin (0,1) cos (1,0) sin (2,-1) cos (3,0) sin (4,1) cos (5,0) sin (6,-1) cos (7,0) sin (8,1);
      \draw[->] (pre) -- (trainingData);
      \draw[->] (in) -- (modelTraining);
      \draw[->] (post) -- (modelOutput);
    \end{tikzpicture}
\end{adjustbox}
    \end{minipage}
     \\
    \begin{minipage}[t]{0.48\textwidth}
        
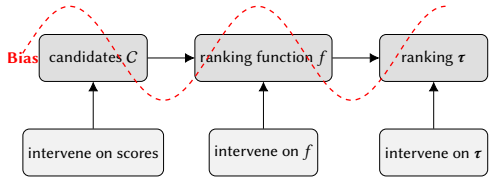
\captionof{figure}{Bias mitigation in score-based ranking: intervening on the score distribution of the candidates in $\candidateSet{}$,  on the ranking function $f$, or on the ranked outcome.}
        \label{fig:db-flowchart}
    \end{minipage}
     & 
    \begin{minipage}[t]{0.47\textwidth}
        
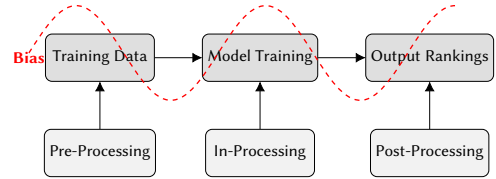
\captionof{figure}{Bias mitigation at different stages of supervised learning-to-rank: pre-processing, in-processing, and post-processing.}
        \label{fig:ir-flowchart}
    \end{minipage}
\end{tabular}
\end{table}

Score-based and supervised learning based rankers use different types of bias mitigation methods.  

In score-based ranking, we categorize mitigation methods into those that intervene on the score distribution, or on the scoring function, or on the ranked outcome, as illustrated in Figure~\ref{fig:db-flowchart}. Methods that \emph{intervene on the score distribution} aim to mitigate disparities in candidate scores, either before these candidates are processed by an algorithmic ranker or during ranking.  Methods that \emph{intervene on the ranking function}  identify a function that is similar to the input function but that produces a ranked outcome that meets the specified fairness criteria.    Methods that \emph{intervene on the ranked outcome} impose constraints to require a specific level of diversity or representation among the top-$k$ as a set, or in every prefix of the top-$k$. 

In supervised learning, we categorize mitigation methods into pre-processing, in-processing, and post-processing. This is illustrated in Figure~\ref{fig:ir-flowchart}, which is analogous to Figure~\ref{fig:db-flowchart}. \emph{Pre-processing} methods seek to mitigate discriminatory bias in training data, and have the advantage of early intervention on pre-existing bias.  \emph{In-processing} methods aim to learn a bias-free model. Finally, \emph{post-processing} methods re-rank candidates in the output  subject to given fairness constraints~\cite{hajian2016algorithmic}.    

The advantage of post-processing methods in supervised learning is that they provide a guaranteed share of visibility for protected groups.  In contrast, in-processing methods only consider fairness at training time and make no guarantees about fairness of the test set.  However, post-processing methods may be subject to legal challenges because of due process concerns that may make it illegal to intervene at the decision stage (\eg Ricci v. DeStefano~\cite{ricci}).  Thus, like all technical choices, the choice of whether to use a pre-, in-, or post-processing fairness-enhancing intervention is not purely technical, but must also consider the social and legal context of use of the algorithmic ranker. 

\begin{table*}[thb!]
\caption{Experimental datasets used in the surveyed papers.}
\centering
\small
\begin{tabular}{p{0.22\textwidth}p{0.17\textwidth}p{0.2\textwidth}p{0.18\textwidth}p{0.12\textwidth}}
\toprule 
 \textbf{Dataset} & \textbf{Size} & \textbf{Sensitive attrs} & \textbf{Score} & \textbf{Used in} \\
\midrule 
\rowcolor{Lightgray}
AirBnB~\cite{AirBnBData} & 10,201 houses & gender of host & rating, price & \cite{biega2018equity,lahoti2019ifair} \\ 
COMPAS~\cite{COMPASData} & 7,214 people & gender, race & risk scores &  \cite{yang2017measuring,asudeh2019designing,zehlike2017fa} \\
\rowcolor{Lightgray} 
CS departments~\cite{CSData} & 51 departments &\makecell[l]{department size, \\geographic region} & number of publications in different CS areas& \cite{DBLP:conf/ijcai/YangGS19}  \\
DOT~\cite{DOTData} & 1.3 million flights & airline name &  \makecell[l]{departure delay,  arrival \\delay, \\taxi-in time} & \cite{asudeh2019designing} \\
\rowcolor{Lightgray}
Engineering students~\cite{EngineeringData} & \makecell[l]{5 queries, 650 stu-\\dents per query} & \makecell[l]{gender, \\high school type} & \makecell[l]{academic performance \\after first year} &  \cite{zehlike2018reducing}  \\
Forbes richest U.S.~\cite{ForbesRichesData} & 400 people & gender & net worth & \cite{DBLP:conf/edbt/StoyanovichYJ18} \\
\rowcolor{Lightgray}
German credit~\cite{GermanCreditData} & 1,000 people & gender, age  & \makecell[l]{credit amount, \\duration} & \cite{yang2017measuring,DBLP:conf/kdd/WuZW18,singh2019policy,zehlike2017fa} \\
IIT-JEE~\cite{IITJEEData} & 384,977 students & birth category, gender, disability status & test scores & \cite{celis2020interventions}  \\
\rowcolor{Lightgray}
LSAC~\cite{LSACData} & 21,792 students & gender, race & LSAT scores & \cite{zehlike2017matching} \\
MEPS~\cite{MEPSData} & 15,675 people & gender, race, age & \makecell[l]{number of trips \\requiring medical care} & \cite{DBLP:conf/ijcai/YangGS19}  \\
\rowcolor{Lightgray}
NASA astronauts~\cite{NASAData} & 357 astronauts & major in college & flight hours & \cite{DBLP:conf/edbt/StoyanovichYJ18} \\
Pantheon~\cite{PantheonData} & 11,341 people & occupation & popularity of Wikipedia page & \cite{DBLP:conf/edbt/StoyanovichYJ18}   \\
\rowcolor{Lightgray}
SAT~\cite{SATData} & 1.6M students & gender & SAT score & \cite{zehlike2017fa}  \\
StackExchange~\cite{StackData} & \makecell[l]{253,000 queries\\6M documents} & domains & document relevance & \cite{biega2018equity} \\
\rowcolor{Lightgray}
SSORC~\cite{SSORCData} & 8,975,360 papers & \makecell[l]{gender of authors \\(inferred)}& number of citations &  \cite{celis2020interventions}  \\
W3C experts~\cite{W3CData} & \makecell[l]{60 queries, 200 \\experts per query} & gender & probability of being an expert & \cite{zehlike2018reducing}  \\
\rowcolor{Lightgray}
XING~\cite{XINGData} & 40 candidates & gender & \makecell[l]{years of experience, \\education} & \cite{lahoti2019ifair,zehlike2017fa}   \\
Yahoo LTR~\cite{YahooData} & \makecell[l]{26,927 queries \\638,794 documents} & N/A & relevance & \cite{singh2019policy} \\
\rowcolor{Lightgray}
Yow news~\cite{YowData} & unknown & source of news & relevance & \cite{singh2018fairness} \\
\bottomrule
\end{tabular}
\label{tab:datasets}
\end{table*}

\section{Datasets}
\label{sec:datasets}
Before diving into a description of the fair ranking methods, we present the experimental datasets used by them.  We summarize the datasets in Table~\ref{tab:datasets}, where we highlight the following aspects: size (usually the number of candidates), sensitive attributes, scoring attributes, and the surveyed papers that use this dataset in their evaluation. We then briefly describe each dataset, and refer the reader to the description of each method for details about that dataset's use: score-based ranking in Section~\ref{sec:fair_db}, supervised learning in Section~\ref{sec:fair_ir}, and recommender systems in Section~\ref{sec:fair_recsys}.  All datasets are publicly available under the referenced links unless otherwise indicated.

The papers surveyed here rarely substantiate their choice of an experimental dataset, other than that by the fact that the dataset was available, and that items in it have scores on which to rank. 
Both of these reasons can be seen as purely technical (or even syntactic) rather than conceptual.
Unfortunately little explicit attention has been paid to explaining whether a particular dataset was collected with a ranking task in mind, and \emph{why} it is deemed appropriate for the specific fairness definition, that is, whether and to what extent the task for which the dataset was collected or can plausibly be used exhibits unfairness of the kind that the proposed fairness definition is designed to address.  We see this as an important limitation of empirical studies in fairness in ranking and, more generally, in algorithmic fairness research, and posit that the use of a dataset must be explicitly justified. 

\paragraph{AirBnB~\cite{AirBnBData}} This dataset consists of 10,201 house listings from three major cities: Hong Kong (4,529 items), Boston (3,944 items), and Geneva (1,728 items). The gender of the hosts is used as the sensitive attribute, and the ranking score is computed as the ratio of the rating and the price.

\paragraph{COMPAS (Correctional Offender Management Profiling for Alternative Sanctions)~\cite{COMPASData}} This dataset is derived based on a recidivism risk assessment tool called COMPAS. 
The dataset contains the COMPAS scores from the Broward County Sheriff’s Office in Florida in 2013 and 2014, and the profile of each person’s criminal history collected by ProPublica~\cite{angwin2016machine}. In total there are 7,214 data points, with sensitive attributes gender and race. 

\paragraph{CS department rankings~\cite{CSData}} This dataset contains information about 51 computer science departments in the U.S.. The methods in \cite{DBLP:conf/ijcai/YangGS19,yang2020causal} use the number of publications as the ranking score. Two categorical attributes are treated as sensitive: department size (with values ``large'' and ``small'') and geographic area (with values ``North East'', ``West'', ``Middle West'', ``South Center'', and ``South Atlantic''). 

\paragraph{DOT (Department of Transportation)~\cite{DOTData}} This dataset consists of about 1.3 million records of  flights conducted by 14 U.S. airlines in the first three months of 2016. The dataset was collected by~\citet{asudeh2019designing} from the flight on-time
database that is published by the U.S. Department of Transportation. Three scoring attributes are used in~\cite{asudeh2019designing}: departure delay, arrival delay, and taxi-in time. The name of the airline conducting the flight is treated as the sensitive attribute. 

\paragraph{Engineering students~\cite{EngineeringData}} This dataset contains the results of a Chilean university admissions test from applicants to a large engineering school in five consecutive years. The task in~\cite{zehlike2018reducing} is to predict the students' academic performance after the first year based on their admissions test results and school grades. The sensitive attributes are gender and whether the applicants graduated from a private or public high school. 
This dataset is only accessible upon request, see referenced link for details.

\paragraph{Forbes richest Americans~\cite{ForbesRichesData}} This dataset consists of 400 individuals from the 2016 Forbes US Richest list \footnote{\url{https://www.forbes.com/forbes-400/list/}}, ranked by their net worth. Gender is the sensitive attribute, with 27 female vs. 373 male individuals in the dataset. 

\paragraph{German credit~\cite{GermanCreditData}} This dataset, hosted by the UCI machinle learning repository~\cite{lichman_2013_uci}, contains financial information of 1,000 individuals, and is associated with a binary classification task that predicts whether an individual's credit is good or bad. The sensitive attributes are gender and age, where age is categorized into younger or older based on a threshold (25 or 35 years old is variably used as the threshold). 
Attributes credit amount and duration (how long an individual has had a line of credit) have been used as scoring attributes in fair ranking papers. 

\paragraph{IIT-JEE (The Joint Entrance Exam of Indian Institutes of Technology)~\cite{IITJEEData}} This dataset consists of scores of 384,977 students in the Mathematics, Physics, and Chemistry sections of IIT-JEE 2009, along with their gender, birth category (see~\cite{DBLP:journals/corr/abs-1904-06698}), disability status, and zip code.  The students are scored on a scale from −35 to +160 points in all three sections, with an average total score of +28.36, a maximum score of +424, and a minimum score of −86.

\paragraph{LSAC~\cite{LSACData}} This dataset consists of a U.S. national longitudinal bar exam passage data gathered from the class that started law school in Fall 1991. Data is provided by the students, their law schools, and state boards of bar examiners over a 5-year period~\cite{wightman1998lsac}. The dataset consists of 21,791 students, with the sensitive attributes sex and race. Rankings are produced based on LSAT scores. 

\paragraph{MEPS (Medical Expenditure Panel Survey)~\cite{MEPSData}} This dataset consists of 15,675 people and their information regarding the amount of health expenditures~\cite{cohen2009medical,coston2019fair}. The sensitive attributes are gender, race, and age of each individual, where age is categorized into younger or older based on a threshold (35 years old) in~\cite{DBLP:conf/ijcai/YangGS19,yang2020causal}. 
The ranking score is based on utilization, defined by the IBM AI Fairness 360 toolkit~\cite{bellamy2018ai} as the total number of trips requiring medical care. Utilization is computed as the sum of the number of office-based visits, the number of outpatient visits, the number of ER visits, the number of inpatient nights, and the number of home health visits. 

\paragraph{NASA astronauts~\cite{NASAData}} This dataset consists of 357 astronauts with their demographic information.  The method in~\cite{DBLP:conf/edbt/StoyanovichYJ18} ranks this dataset by the number of space flight hours, and assigns individuals to categories based on their undergraduate major, treating is a the sensitive attribute. A total of 83 majors are represented in the dataset, the 9 most frequent are assigned to their individual categories --- Physics (35), Aerospace Engineering (33), Mechanical Engineering (30), etc, and the remaining 141 individuals are combined into the category ``Other'', resulting in 10 groups.

\paragraph{SAT~\cite{SATData}} This dataset contains about 1.6 million data points, in which the score column corresponds to an individual's results in the US Scholastic Assessment Test (SAT) in 2014~\cite{sat_2014}. The sensitive attribute is gender.

\paragraph{SSORC~\cite{SSORCData}} The Semantic Scholar Open Research Corpus contains the meta-data of 46,947,044 published research papers in computer science, neuroscience, and biomedicine from 1936 to 2019 on Semantic Scholar. The meta-data for each paper includes the list of authors of the paper, the year of publication, the list of papers citing it, and the journal of publication, along with other details. The sensitive attribute is the gender of the authors,  collected by~\citet{celis2020interventions}. The ranking score is the number of citations of each paper.

\paragraph{StackExchange~\cite{StackData}} This dataset contains a query log and a document collection using the data from the Stack Exchange Q\&A community (dump as of 13-06-2016)~\cite{biega2017privacy}. It consists of about 6 million posts inside the type ``Question'' or ``Answer'' in 142 diverse subforums (\eg Astronomy, Security, Christianity, Politics, Parenting, and Travel). The questions are translated into about 253,000 queries, and the respective answers serve as the documents for the queries. The sensitive attribute is the query domain.

\paragraph{W3C experts~\cite{W3CData}} The task behind this dataset corresponds to a search of experts for a given topic based on a corpus of e-mails written by possible candidates.  The sensitive attribute is the gender of the expert. The experimental setup in~\cite{zehlike2018reducing} investigates situations in which bias is unrelated to relevance: expertise has been judged correctly, but ties have been broken in favor to the privileged group (\ie all male experts are followed by all female experts, followed by all male non-experts, followed finally by all female non-experts). 

\paragraph{Xing~\cite{XINGData}} This dataset was collected by~\citet{zehlike2017fa} from a German online job market website\footnote{\url{https://www.xing.com}}. The authors collected the top-40 profiles returned for 54 queries, and computed an ad-hoc score based on educational features, job experience and profile popularity. The sensitive attribute is gender, which was inferred based on the first name associated with the profile and the profile picture, when available. Items are ranked based on an ad-hoc score. 

\paragraph{Yahoo! LTR~\cite{YahooData}} This dataset consists of 19,944 training queries and 6,983 test set queries. Each query has a variable sized candidate set of documents that needs to be ranked. There are 473,134 training and 165,660 test documents. The query-document pairs are represented by a 700-dimensional feature vector. For supervision, each query-document pair is assigned an integer relevance judgments from 0 (bad) to 4 (perfect). The dataset is used to evaluate the effectiveness of Learning to Rank methods in~\cite{singh2019policy}, thus no sensitive attribute is specified. 

\paragraph{Yow news~\cite{YowData}} This dataset contains explicit and implicit feedback from a set of users for news articles in the ``people'' topic produced by Yow~\cite{zhang2005bayesian}. The ranking score is the explicitly given relevance field. The source of news is treated as the sensitive attribute.

\section{Score-based Ranking}
\label{sec:fair_db}

\begin{table*}[ht!]
     \caption{Classification of score-based ranking methods according to the frameworks in Section~\ref{sec:02-four-frameworks}. }
    \small
    \begin{tabular}{p{0.2\textwidth}p{0.2\textwidth}p{0.12\textwidth}p{0.1\textwidth}p{0.11\textwidth}l}
        \toprule
         \textbf{Method} & \textbf{Group structure} & \textbf{Bias} & \textbf{Worldview} & \textbf{\revv{\eop}} & \textbf{Intersectional} \\ \midrule 
          \makecell[l]{Rank-aware proportional\\ representation~\cite{yang2017measuring}} & one binary sensitive attr. & pre-existing & WAE & luck-egalitarian & no\\
          \rowcolor{Lightgray}
          \makecell[l]{Constrained ranking\\ maximization~\cite{celis2018ranking}} & \makecell[l]{multiple sensitive attrs.;\\ multinary; \\handled independently} & pre-existing & WAE & \makecell[l]{\revv{luck-}\\\revv{egalitarian}\\ \revv{(1 sensitive}\\ \revv{attr. only) }} & no \\
          \makecell[l]{Balanced diverse\\ ranking~\cite{DBLP:conf/ijcai/YangGS19}} & \makecell[l]{multiple sensitive attrs.;\\ multinary; \\handled independently} & \makecell[l]{pre-existing;\\ technical} & WAE & luck-egalitarian & yes \\
         \rowcolor{Lightgray}
         \makecell[l]{Diverse $k$-choice\\ secretary~\cite{DBLP:conf/edbt/StoyanovichYJ18}} & one multinary sensitive attr. & pre-existing & WAE & luck-egalitarian & no \\
         \makecell[l]{Utility of selection with\\ implicit bias~\cite{kleinberg_et_al:LIPIcs:2018:8323}} & 
         one binary sensitive attr. &\makecell[l]{pre-existing;\\ implicit} & WAE & N/A & no\\
          \rowcolor{Lightgray}
         \makecell[l]{Utility of ranking with\\ implicit bias~\cite{celis2020interventions}} &  \makecell[l]{multiple sensitive attrs.;\\ multinary;\\ handled independently} & \makecell[l]{pre-existing;\\ implicit} & WAE & N/A & yes\\
         \makecell[l]{Causal intersectionally\\ fair ranking~\cite{yang2020causal}} &  \makecell[l]{multiple sensitive attrs.;\\ multinary;\\ handled independently} & pre-existing & WAE & \makecell[l]{\revv{Rawlsian}} & yes\\
         \rowcolor{Lightgray}
         \makecell[l]{Designing fair ranking\\ functions~\cite{asudeh2019designing}} & any & pre-existing & any & any & yes\\
         \hline 
    \end{tabular}
    \label{tbl:method-summary_score}
\end{table*}

In this section we present several methods for fairness in score-based ranking.  \rev{Rather than giving a purely technical comparison, we re-iterate that the choice of a method should be based on assumptions about the nature of unfairness, and on the fundamental modeling choices.} Table~\ref{tbl:method-summary_score} summarizes the methods presented in this section according to the frameworks of Section~\ref{sec:02-four-frameworks}. \rev{Additionally, every technical methods is placed on the mind map in Figure~\ref{fig:normative-dimensions}, to give a visual summary and as a means to compare the methods.}

Recall that, in score-based ranking we categorize mitigation methods into those that intervene on the ranking process, on the score distribution, or on the scoring function.  In Section~\ref{sec:fair_db:prop}, we describe methods that \emph{intervene on the ranked outcome} by ensuring  proportional representation across groups.   Next, in Section~\ref{sec:fair_db:bounds}, we discuss several methods that formulate \emph{fairness} and \emph{coverage-based diversity} constraints by specifying bounds on the number of candidates from groups of interest to be present in prefixes of a ranked list. These methods also intervene on the ranked outcome.  Then, in Section~\ref{sec:fair_db:latent}, we describe methods that \emph{intervene on the score distributions}.  Finally, in Section~\ref{sec:fair_db:geo}, we present a method that treats the fairness objective as a black-box and proposes a geometric interpretation of score-based ranking to reach the objective by \emph{intervening on the ranking function}.

\subsection{Intervening on the Ranked Outcome: Rank-aware Proportional Representation}
\label{sec:fair_db:prop}

To the best of our knowledge,~\citet{yang2017measuring} were the first formalize rank-aware fairness, under the assumption that the scores based on which the ranking is produced encode pre-existing bias.

Consider a ranking in which candidates are assigned to one of two groups, $\group{1}$ or $\group{2}$, according to a single binary sensitive attribute (\eg, binary gender), and with one of these groups, $\group{1}$, corresponding to the protected group (\eg the female gender).   The fairness measures proposed in this paper are based on the following intuition: Because it is more beneficial for an item to be ranked higher, it is also more important to achieve proportional representation at higher ranks. The idea, then, is to take several well-known proportional representation measures and to make them \emph{rank-aware}, by placing them within a framework that applies position-based discounts. 

\spara{Fairness definition and problem formalization.}  Recall from Section~\ref{sec:prelim} that position-based discounting is commonly used to quantify utility (Eq.~\ref{eq:disc_agg_utility}) or prediction accuracy in a ranking (Eq.~\ref{eq:ndcg}). In a similar vein, the use of position-based discounting in \citet{yang2017measuring} is a natural way to make set-wise proportional representation requirements rank-aware.  Specifically, the idea is to consider a series of prefixes of a ranking $\btau$, for $k=10,20,\dots$, to treat each top-$k$ prefix $\btau_{1 \ldots k}$ as a set,  to compute \emph{statistical parity} at top-$k$, and to compare that value to the proportion of the protected group in the entire ranking. (Naturally,  perfect statistical parity is achieved when $k=n$.) The values computed at each cut-off point are summed up with a position-based discount.  Based on this idea, the authors propose three fairness measures that differ in the specific interpretation of statistical parity:  normalized discounted difference (\nd), ratio (\rd), and KL-divergence (\kl). 

Normalized discounted difference (\nd) (Equation~\ref{eq:nd}) computes the difference between the  proportions of the protected group $\group{1}$ at the top-$k$ and in the over-all population. Normalizer $Z$ is computed as the highest possible value of \nd for the given number of items $n$ and protected group size $|\group{1}|$. 

\begin{equation}
\nd(\btau)=\frac{1}{Z} \sum_{k=10,20,...}^{n}{ \frac{1}{\log_{2}{k}} 
\left(\frac{|\btau_{1\ldots k} \cap \group{1}|}{k} - \frac{|\group{1}|}{n} \right)}
\label{eq:nd}
\end{equation}

Normalized discounted ratio (\rd) is defined analogously, as follows:

\begin{equation}
\rd(\btau)=\frac{1}{Z} \sum_{k=10,20,...}^{n}{ \frac{1}{\log_{2}{k}} 
\left(\frac{|\btau_{1\ldots k} \cap \group{1}|}{|\btau_{1\ldots k} \cap \group{2}|} - \frac{|\group{1}|}{|\group{2}|} \right)}
\label{eq:rd}
\end{equation}

When either the numerator or the denominator of a term in Eq.~\ref{eq:rd} is 0, the value of the term is set to 0.  

Finally, normalized discounted KL-divergence (\kl) uses Kullback-Leibler (KL) divergence to quantify the expectation of the logarithmic difference between two discrete probability distributions, $P_k$ that quantifies the proportions in which groups are represented at the top-$k$:

\begin{equation}
P_k = \left( \frac{|\btau_{1\ldots k} \cap \group{1}|}{k}, \frac{|\btau_{1\ldots k} \cap \group{2}|}{k} \right)
\label{eq:kl_pk}
\end{equation}

and $Q$ that quantifies the proportions in which groups are represented in the over-all ranking: 

\begin{equation}
Q = \left( \frac{|\group{1}|}{n}, \frac{|\group{2}|}{n} \right)
\label{eq:kl_q}
\end{equation}

KL-divergence between $P_k$ and $Q$, denoted  $D_{KL}(P_k||Q)$, is computed at every cut-off point $k$, and position-based discounting is applied as the values are compounded, with normalizer $Z$ defined analogously as for \nd:

\begin{equation}
\kl (\btau)=\frac{1}{Z} \sum_{k=10,20,...}^{n}{ \frac{1}{\log_{2}{k}} D_{KL}(P_k||Q)}
\label{eq:kl}
\end{equation}

Note that, unlike \nd and \rd, which are limited to a binary sensitive attribute, \kl can handle a multinary sensitive attribute and so is more flexible.  

\spara{Experiments and observations.} The authors evaluate the empirical behavior of the proposed fairness measures using real and synthetic datasets.  Real datasets used are COMPAS~\cite{COMPASData} and German Credit~\cite{GermanCreditData}, see Section~\ref{sec:datasets} for details. Synthetic datasets are generated using an intuitive data generation procedure described below.  This procedure was later used in the work of~\citet{zehlike2017fa} and~\citet{,DBLP:conf/kdd/WuZW18}, and is of independent interest. 

Recall that $\group{1}$ represents the protected group and $\group{2}$ represents the privileged group, and suppose for simplicity that each group constituted one half of the candidates $\candidateSet{}$.  An example is given in Figure~\ref{fig:exm_measure_fair_rank_data}, in which $\candidateSet{}$ contains 8 candidates, 4 female ($\group{1}$) and 4 male ($\group{2}$).  The data generation procedure, presented in Algorithm~\ref{alg:rankgen}, takes two inputs: a ranking $\btau$ of $\candidateSet{}$ and a ``fairness probability'' $p$, and it produces a ranking $\Tilde{\btau}$. The input ranking $\btau$ is assumed to be generated by the vendor according to their usual process (\eg based on candidate scores, as in Figure~\ref{fig:exm_measure_fair_rank_input}).  Algorithm~\ref{alg:rankgen} splits up $\btau$ into two rankings: $\btau_1$ of candidates in $\group{1}$ and $\btau_2$ of candidates in $\group{2}$.  It then repeatedly considers pairs of candidates at the top of the lists, $\btau_1(1)$ and $\btau_2(1)$, and decides which of these should be ranked above the other, selecting  $\btau_1(1)$ with probability $p$ and $\btau_2(1)$ with probability $1-p$, and appending the selected candidate to $\Tilde(\btau)$.

The parameter $p$ specifies the relative preference between candidates in $\group{1}$ and in $\group{2}$. When $p = 0.5$, groups are mixed in approximately equal proportion for as long as there are items in both groups. This is illustrated in Figure~\ref{fig:fig:alg1:gender_05} for the sensitive attribute $A_1$ (gender) and in Figure~\ref{fig:alg1:race_05} for the sensitive attribute $A_2$ (race).  When $p > 0.5$, members of the protected group $\group{1}$ are preferred, and when $p < 0.5$ members of the privileged group $\group{2}$ are preferred. In extreme cases, when $p=0$, all (or most) members of $\group{2}$ will be placed before any members of $\group{1}$, as shown in Figure~\ref{fig:exm_male_then_female} for the sensitive attribute $A_1$ (gender).   Note that candidates within a group always remain in the same relative order in $\Tilde{\btau}$ as in $\btau$ (that is, there is \emph{no reordering within a group}), but there may be \emph{reordering between groups}.  

\begin{algorithm}[ht!]
	\caption{FairGen}
	\begin{algorithmic}[1]
		\REQUIRE Ranking $\btau$, fairness probability $p$.\\
		\COMMENT {Initialize the output ranking $\Tilde{\btau}$.}\\
		\STATE $\Tilde{\btau} \leftarrow \emptyset$\\
		\STATE $\btau_{1} = \btau \cap \group{1}$\\
		\STATE $\btau_{2} = \btau \cap \group{2}$\\
		\WHILE {$(\btau_{1} \neq \emptyset) \wedge (\btau_{2} \neq \emptyset)$}
		\STATE $r=random([0, 1])$\\
		\COMMENT {Append the next selected item to $\Tilde{\btau}$}\\
		\IF {$r<p$} 
		\STATE $\Tilde{\btau} \leftarrow pop(\btau_{1})$\\
		\ELSE
		\STATE $\Tilde{\btau} \leftarrow pop(\btau_{2})$\\
		\ENDIF
		\ENDWHILE\\
		\COMMENT {If any items remain in $\btau_{1}$ or $\btau_{2}$, append them to $\Tilde{\btau}$}
		\STATE $\Tilde{\btau} \leftarrow \btau_{1}$\\
		\STATE $\Tilde{\btau} \leftarrow \btau_{2}$\\
		\RETURN $\Tilde{\btau}$\\
	\end{algorithmic}
	\label{alg:rankgen}
\end{algorithm}

\begin{figure}[t!]
    \centering
    \small
    \setlength{\tabcolsep}{0.3em}
    	\subfloat[]
    	{
        	\begin{tabular}{|c||c|c||c|}
        		\hline
        		\rowcolor[HTML]{C0C0C0} 
        		candidate & $A_1$ & $A_2$ & $Y$   \\ \hline
    		    \val{b}  & \cellcolor[HTML]{CBCEFB} \val{male} & \cellcolor[HTML]{CBCEFB}  \val{White} & 9   \\ \hline
    		    \val{c}  & \cellcolor[HTML]{CBCEFB} \val{male} & \cellcolor[HTML]{FFCE93} \val{Black} & 8 \\ \hline
    		    \val{d}  & \cellcolor[HTML]{FFCE93} \val{female} & \cellcolor[HTML]{CBCEFB} \val{White} & 7 \\ \hline
    		    \val{e}  & \cellcolor[HTML]{CBCEFB} \val{male} & \cellcolor[HTML]{CBCEFB} \val{White} & 6 \\ \hline
    		    \val{f}  & \cellcolor[HTML]{FFCE93} \val{female} & \cellcolor[HTML]{CBCEFB} \val{White} & 5 \\ \hline
    		    \val{k}  & \cellcolor[HTML]{FFCE93} \val{female} & \cellcolor[HTML]{CBCEFB} \val{White} & 4   \\ \hline
    		    \val{l}  & \cellcolor[HTML]{CBCEFB} \val{male} & \cellcolor[HTML]{CBCEFB} \val{White} & 3   \\ \hline
    		    \val{o}  & \cellcolor[HTML]{FFCE93} \val{female} & \cellcolor[HTML]{FFCE93} \val{Black} & 2 \\ \hline
    	    \end{tabular}
        	\label{fig:exm_measure_fair_rank_data}
    	}
    	\hspace{5mm}
    	\subfloat[]{
    		\begin{tabular}{|c|}
        		\hline
        		\rowcolor[HTML]{C0C0C0} 
        		$\btau$ \\ \hline
        		\rowcolor[HTML]{CBCEFB} 
        		\val{b}                    \\ \hline
        		\rowcolor[HTML]{CBCEFB} 
        		\val{c}                    \\ \hline
        		\rowcolor[HTML]{FFCE93} 
        		\val{d}                    \\ \hline
        		\rowcolor[HTML]{CBCEFB} 
        		\val{e}                    \\ \hline
        		\rowcolor[HTML]{FFCE93} 
        		\val{f}                   \\ \hline
        		\rowcolor[HTML]{FFCE93} 
        		\val{k}                    \\ \hline
        		\rowcolor[HTML]{CBCEFB} 
        		\val{l}                    \\ \hline
        		\rowcolor[HTML]{FFCE93} 
        		\val{o}                    \\ \hline
        	\end{tabular}
    	    \label{fig:exm_measure_fair_rank_input}
    	}
    	\hspace{5mm}
    	\subfloat[]{
    		\begin{tabular}{|c|}
        		\hline
        		\rowcolor[HTML]{C0C0C0} 
        		$\Tilde{\btau}$ \\ \hline
        		\rowcolor[HTML]{CBCEFB} 
        		\val{b}   \\ \hline
        		\rowcolor[HTML]{FFCE93} 
        		\val{d}   \\ \hline
        		\rowcolor[HTML]{CBCEFB} 
        		\val{c}   \\ \hline
        		\rowcolor[HTML]{FFCE93} 
        		\val{f}   \\ \hline
        		\rowcolor[HTML]{CBCEFB} 
        		\val{e}   \\ \hline
        		\rowcolor[HTML]{FFCE93} 
        		\val{k}   \\ \hline
        		\rowcolor[HTML]{CBCEFB} 
        		\val{l}   \\ \hline
        		\rowcolor[HTML]{FFCE93} 
        		\val{o}   \\ \hline
        	\end{tabular}
    	    \label{fig:fig:alg1:gender_05}
    	}
    	\hspace{5mm}
    	\subfloat[]{
    		\begin{tabular}{|c|}
        		\hline
        		\rowcolor[HTML]{C0C0C0} 
        	    $\Tilde{\btau}$ \\ \hline
        		\rowcolor[HTML]{CBCEFB} 
        		\val{b}                    \\ \hline
        		\rowcolor[HTML]{CBCEFB} 
        		\val{c}                    \\ \hline
        		\rowcolor[HTML]{CBCEFB} 
        		\val{e}                    \\ \hline
        		\rowcolor[HTML]{CBCEFB} 
        		\val{l}                    \\ \hline
        		\rowcolor[HTML]{FFCE93} 
        		\val{d}                    \\ \hline
        		\rowcolor[HTML]{FFCE93} 
        		\val{f}                   \\ \hline
        		\rowcolor[HTML]{FFCE93} 
        		\val{k}                    \\ \hline
        		\rowcolor[HTML]{FFCE93} 
        		\val{o}                    \\ \hline
        	\end{tabular}
    	    \label{fig:exm_male_then_female}
    	}
    	\hspace{5mm}
    	\subfloat[]{
    		\begin{tabular}{|c|}
        		\hline
        		\rowcolor[HTML]{C0C0C0} 
        	    $\Tilde{\btau}$ \\ \hline
        		\rowcolor[HTML]{CBCEFB} 
        		\val{b}                    \\ \hline
        		\rowcolor[HTML]{FFCE93} 
        		\val{c}                    \\ \hline
        		\rowcolor[HTML]{CBCEFB} 
        		\val{d}                    \\ \hline
        		\rowcolor[HTML]{FFCE93} 
        		\val{o}                    \\ \hline
        		\rowcolor[HTML]{CBCEFB} 
        		\val{e}                    \\ \hline
        		\rowcolor[HTML]{CBCEFB} 
        		\val{f}                   \\ \hline
        		\rowcolor[HTML]{CBCEFB} 
        		\val{k}                    \\ \hline
        		\rowcolor[HTML]{CBCEFB} 
        		\val{l}                    \\ \hline
        	\end{tabular}
    	    \label{fig:alg1:race_05}
    	}
    \caption{{\bf (a)} A set of applicants for college admissions $\candidateSet{}$, with two binary sensitive attributes: $A_1$ (gender), with protected group $\group{\val{F}}=\{d,f,k,o\}$ and privileged group $\group{\val{M}}=\{b,c,e,l\}$; and $A_2$ (race), with protected group $\group{\val{B}}=\{c,o\}$ and privileged  group $\group{\val{W}}=\{b,d,e,f,k,l\}$. Protected values of $A_1$ and $A_2$ are shown in orange, and privileged values---in blue. {\bf (b)} Ranking $\btau$ sorts the applicants in descending order of their score $Y$, as shown in Figure~\ref{fig:exm_measure_fair_rank_input}, with male candidates appearing in higher proportion at the top ranks.  {\bf (c)} Ranking  $\Tilde{\btau}$ for $A_1$ mixes candidates in approximately equal proportion by gender, with $p=0.5$ in Algorithm~\ref{alg:rankgen}, and is expected to achieve statistical parity for this attribute, since gender groups are represented in equal proportion in $\candidateSet{}$. {\bf (d)} Ranking  $\Tilde{\btau}$ for $A_1$, with $p=0$ in Algorithm~\ref{alg:rankgen},  places all, or most, male applicants about the female applicants. {\bf (e)} Ranking  $\Tilde{\btau}$ for $A_2$ (race), with $p=0.5$ in Algorithm~\ref{alg:rankgen}, is expected to achieve equal representation by race at top ranks, but not statistical parity, since $\candidateSet{}$ is not balanced by race.}
    \label{fig:exm_measure_fair_rank}
\end{figure}

The proposed fairness measures ---normalized discounted difference (\nd), ratio (\rd), and KL-divergence (\kl)--- are evaluated on rankings produced by Algorithm~\ref{alg:rankgen} with a range of values for $p$ and with different relative proportions of $\group{1}$ and  $\group{2}$ in $\candidateSet{}$.  The authors conclude that \kl is the most promising measure both because it is smooth and because it naturally generalized to multinary sensitive attributes.

This paper also proposes a bias mitigation methodology, inspired by~\citet{DBLP:conf/icml/ZemelWSPD13}, that integrates fairness objectives into an optimization framework that balance fairness against utility, with an experimental evaluation on the German credit dataset~\cite{GermanCreditData} (see details in Sec.~\ref{sec:datasets}).

\begin{figure}[h]
    \centering
      \begin{tikzpicture}[scale=0.9,align=center]
    \tikzset{
        root/.style = {rectangle, rounded corners, draw=black, fill=gray!30, font=\small\sffamily},
        child/.style = {rectangle, rounded corners,draw=black, font=\small\sffamily},
        orangec/.style = {rectangle, rounded corners, draw=black, fill=orange!70, font=\small\sffamily},
        greenc/.style = {rectangle, rounded corners, draw=black, fill=green!40, font=\small\sffamily},
        bluec/.style = {rectangle, rounded corners, draw=black, fill=blue!30, font=\small\sffamily},
        yellowc/.style = {rectangle, rounded corners, draw=black, fill=yellow!70, font=\small\sffamily},
        pinkc/.style = {rectangle, rounded corners, draw=black, fill=pink!70, font=\small\sffamily},
    }
    \node (ND) at (-0.2,0) [root] {Normative Dimensions};    
    \node (GS) at (1.8,1.75) [orangec] {Group Structure};    
    \node (BT) at (1.8,-1.75) [greenc] {Bias Type};    
    \node (EO) at (-1.8,1.75) [bluec] {Equal Opportunity};    
    \node (WV) at (-1.8,-1.75) [yellowc] {Worldview};  
    \node (IT) at (3.3,0.0) [pinkc] {Intersectional};
    
    \node (AC) at (3.9,2.5) [child] {Attribute Cardinality};
    \node (ACB) at (6.3,3.2) [orangec] {Binary};
    \node (ACM) at (6.3,1.8) [child] {Multinary};
    
    \node (AN) at (3.9,1.0) [child] {Attribute Number};
    \node (ANO) at (6.9,1.15) [orangec] {One};
    \node (ANM) at (6.9,0.5) [child] {Multiple};
    
    \node (ANMI) at (8.9,0.8) [child] {Independent};
    \node (ANMC) at (8.9,0.0) [child] {Combination};
    
    \node (ITY) at (5.3,0.2) [child] {Yes};
    \node (ITN) at (5.3,-0.4) [pinkc] {No};
    
    \node (BTP) at (4.0,-1.2) [greenc] {Pre-existing};
    \node (BTT) at (4.0,-2.0) [child] {Technical};
    \node (BTE) at (4.0,-2.75) [child] {Emergent};
    
    \node (EOF) at (-3.8,2.8) [child] {Formal};
    \node (EOFP) at (-4.6,2.1) [child] {Formal plus};
    \node (EOS) at (-3.8,1.0) [child] {Substantive};
    
    \node (EOSR) at (-5.8,1.4) [child] {Rawlsian};
    \node (EOSL) at (-5.4,0.1) [bluec] {Luck-egalitarian};

    \node (WVI) at (-3.8,-0.8) [child] {WYSIWYG};
    \node (WVE) at (-3.8,-1.8) [yellowc] {WAE};
    \node (WVC) at (-3.8,-2.8) [child] {Continuous};
    
    \draw (ND.east) -- (GS);
    \draw (ND.east) -- (BT);
    \draw (ND.east) -- (IT);
    \draw (ND.west) -- (EO);
    \draw (ND.west) -- (WV);
    
    \draw (GS) -- (AC);
    \draw (GS) -- (AN);
    
    \draw (AC) -- (ACB);
    \draw (AC) -- (ACM);
    
    \draw (AN) -- (ANO);
    \draw (AN) -- (ANM);
    
    \draw (IT) -- (ITY);
    \draw (IT) -- (ITN);
    
    \draw (ANM) -- (ANMI);
    \draw (ANM) -- (ANMC);
    
    \draw (BT) -- (BTP);
    \draw (BT) -- (BTT);
    \draw (BT) -- (BTE);
    
    \draw (EO) -- (EOF);
    \draw (EO) -- (EOFP);
    \draw (EO) -- (EOS);
    
    \draw (EOS) -- (EOSR);
    \draw (EOS) -- (EOSL);
    
    \draw (WV) -- (WVI);
    \draw (WV) -- (WVE);
    \draw (WV) -- (WVC);
\end{tikzpicture}
    \caption{\revv{Summary of the normative values encoded by Rank-aware proportional representation (\citet{yang2017measuring}).}}
    \label{fig:mind:prop_represent}
\end{figure}
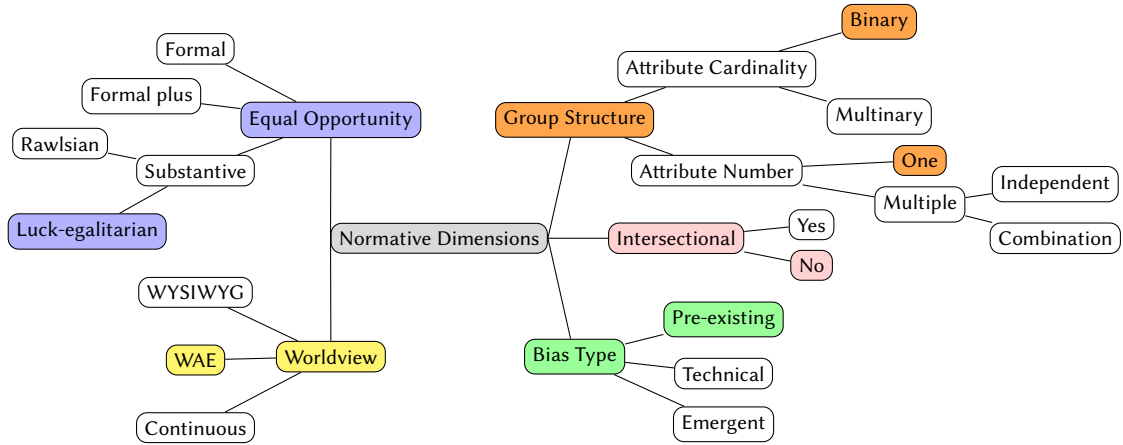


\spara{Insights.} The fairness definitions of this paper aim to address pre-existing bias, per classification in Section~\ref{sec:frame:bias}. 

Fairness is interpreted as equality of outcomes, suggesting an underlying assumption of WAE, per classification in Section~\ref{sec:frame:mit_goal}.  Assuming the existence of indirect discrimination in candidate scores (\ie that the observation process between construct space \cs and observable space \os is biased), the paper aims to ensure a similar representation of groups in the ranked outcomes.

The approach is designed around \revv{conditioning qualification scores on morally-irrelevant circumstances}:
 candidates are ranked according to score within a demographic group, and a ranked outcome is considered fair if the groups are mixed in equal proportion when the input is balanced, as in Figure~\ref{fig:exm_measure_fair_rank_data} by $A_1$ (gender), or, more generally, when statistical parity is achieved at high ranks.  \revv{Assuming that the goal of the competition is to make future prospects comparable, this is consistent with luck-egalitarian \eop}, per classification in Section~\ref{sec:frame:mit_goal}. \rev{Figure~\ref{fig:mind:prop_represent} and Table~\ref{tbl:method-summary_score} summarize our analysis.}

\subsection{Intervening on the Ranked Outcome: Diversity Constraints}
\label{sec:fair_db:bounds}

In Section~\ref{sec:fair_db:prop} we discussed how fairness measures that are based on (set-wise) proportional representation  can be made rank-aware.  The methods described in this section start with the observation that if the total number of candidates in $\candidateSet{}$, and the number of candidates in each demographic group of interest, is available as input (\ie that these quantities are known a priori or can be estimated), then any measure that aims to equalize or bound the difference in proportions can be equivalently re-formulated with the help of counts.  Specifically, proportional representation constraints and coverage-based diversity constraint~\cite{drosou2017diversity} for \emph{set selection tasks} can be expressed by specifying a lower-bound $L_k^{\group{}}$ and an upper-bound $U_k^{\group{}}$ on the representation of group $\group{} \subseteq \candidateSet{}$ among the top-$k$ set of a ranking. Such constraints can be formulated for one or several demographic groups of interest, and also for their intersections, and a score-based ranker can then optimize utility under such constraints.  Generalizing beyond set selection, constraints $L_p^{\group{}}$ and $U_p^{\group{}}$ can be specified over every prefix of the top-$k$ of a ranked list, with $p \in [k]$, or, more practically, at some specific cut-off points within the top-$k$.

Similarly to the methods of Section~\ref{sec:fair_db:prop}, the methods described in this section are designed to enforce fairness and diversity in the sense of representation.  In contrast Section~\ref{sec:fair_db:prop}, these methods are designed to handle multiple sensitive attributes simultaneously---individually or in combination. 

\subsubsection{\citet{celis2018ranking}} \label{sec:fair_db:celis2018}
\spara{Fairness definition and problem formalization.} The authors formulate the \emph{constrained ranking maximization problem}: Consider a set of $n$ candidates $\candidateSet{}$, and the integer $k \ll n$,  along with 1) the utility of placing a candidate in a particular position in the ranking, 2) the collection of sensitive attributes (\eg  gender, race, or disability status) that map candidates to groups $\groupSet$, and 3) a collection of lower-bound constraints $L_p^{\group{}}$ and upper-bound constraints $U_p^{\group{}}$ that, for each prefix $p \in [k]$ and for each group $\group{} \in \groupSet$, bound the number of candidates from that group that are allowed to appear in the top-$p$ positions of the ranking. The goal is to output a ranking that maximizes overall utility with respect to the original utility metric, while respecting the constraints.  Note that this problem formulation has the flexibility to explicitly associate a utility with an assignment of candidate $a \in \candidateSet{}$ to rank position $j \in [k]$, and may already incorporate position-based discounting (per Equation.~\ref{eq:disc_agg_utility}).  However, for consistency and ease of exposition, we will assume that utility score $Y$ is fixed per candidate.

An example of the constrained ranking maximization problem is given in Figure~\ref{fig:crm}, where the goal is to select $k=4$ candidates, with at least two of each gender ($L_4^{\val{M}}=2$, $L_4^{\val{F}}=2$) and at least one of each race ($L_4^{\val{W}}=1$, $L_4^{\val{B}}=1$, $L_4^{\val{A}}=1$) among the top-$k$, and with no further constraints on the prefixes of the top-$k$.  (For convenience, we are referring to each groups by the first letter of the attribute value that defines it, such as \val{M} for \val{male} and \val{A} for \val{Asian}). Ranking $\btau_1$ in Figure~\ref{fig:crm} is a ranked outcome of the top-$4$ candidates selected based on utility $Y$: two of them are male and two are female, and all are White. Applying diversity constraints on gender and race yields $\btau_2$, a ranking of the top-$4$ in Figure~\ref{fig:crm}, selecting the top-scoring White male candidates $a$ and $b$, and two lower-scoring female candidates, $g$ and $k$.   Computing total utility as the sum of scores of selected candidates (for simplicity), we observe that $U(\btau_1)=68$ and $U(\btau_2)=53$ in this example.

Note that the example in Figure~\ref{fig:crm} is deliberately constructed to highlight disparities in scores due to pre-existing bias on gender and race: all male candidates are ranked above all female candidates of a given race, and all Whites  are ranked above all Black, who are in turn ranked above all Asians.  For this reason, imposing diversity constraints leads to a substantial drop in score-utility of $\btau_2$ in Figure~\ref{fig:crm}.
\newcommand{\celisdata}{
    \small
    \begin{tabular}{|c||c|c||c|}
		\hline
		\rowcolor[HTML]{C0C0C0} 
		candidate & $A_1$  & $A_2$  & $Y$   \\ \hline
		\rowcolor[HTML]{FFFFFF}
    	\val{a}  & \val{male} & \val{White} & 19   \\ \hline
    	\val{b}  & \val{male} & \val{White} & 18 \\ \hline
    	\val{c}  & \val{female} & \val{White}  & 16 \\ \hline
    	\val{d}  & \val{female}  & \val{White} & 15 \\ \hline
    	\val{e}  & \val{male} & \val{Black} & 11 \\ \hline
    	\val{f}  & \val{male} & \val{Black} & 11   \\ \hline
    	\val{g}  & \val{female} & \val{Black} & 10   \\ \hline
    	\val{h}  & \val{female} & \val{Black}& 9 \\ \hline
    	\val{i}  & \val{male} & \val{Asian}& 7 \\ \hline
    	\val{j}  & \val{male} & \val{Asian}& 7 \\ \hline
    	\val{k}  & \val{female} & \val{Asian}& 6 \\ \hline
    	\val{l}  & \val{female} & \val{Asian}& 3 \\ \hline
    \end{tabular}
}
\newcommand{\celisrankuti}{
    \small
    \begin{tabular}{|c|}
		\hline
		\rowcolor[HTML]{C0C0C0} 
		$\btau_1$ \\ \hline
		\rowcolor[HTML]{FFFFFF}
		\val{a}                    \\ \hline
		\val{b}                    \\ \hline
		\val{c}                    \\ \hline
		\val{d}                    \\ \hline
	\end{tabular}
}
\newcommand{\celisrankdiv}{
    \small
    \begin{tabular}{|c|}
		\hline
		\rowcolor[HTML]{C0C0C0} 
		$\btau_2$ \\ \hline
		\rowcolor[HTML]{FFFFFF}
		\val{a}                    \\ \hline
		\val{b}                    \\ \hline
		\val{g}                    \\ \hline
		\val{k}                    \\ \hline
	\end{tabular}
}
\newcommand{\celisrankbal}{
    \small
    \begin{tabular}{|c|}
		\hline
		\rowcolor[HTML]{C0C0C0} 
		$\btau_3$ \\ \hline
		\rowcolor[HTML]{FFFFFF}
		\val{a}                    \\ \hline
		\val{c}                    \\ \hline
		\val{e}                    \\ \hline
		\val{k}                    \\ \hline
	\end{tabular}
}
\newcolumntype{f}{>{\columncolor[HTML]{FFCE93}}c}
\newcolumntype{o}{>{\columncolor[HTML]{CBCEFB}}c}
\newcolumntype{t}{>{\columncolor[HTML]{C0C0C0}}c}

\newcommand{\secretaryone}{
    \small
    \begin{tabular}{|t||o|f||o|f|o|f|}

  	  \multicolumn{7}{l}{(a)} \\
      \hline
      candidate & \val{a} & \val{e} & {\bf b} & \val{f} & \val{c} & {\bf d}  \\
  	  \hline
	  $Y$ & 7 & 4 & 8 & 5 & 9 & 3 \\
      \hline
   \end{tabular}
}
\newcommand{\secretarytwo}{
    \small
    \begin{tabular}{|t||o|f||o|f|o|f|}
  	  \multicolumn{7}{l}{(b)} \\
  	  \hline
      candidate & \val{a} & \val{e} & {\bf b} & {\bf f} & \val{c} & \val{d}   \\
  	  \hline
	  $Y$ & 7 & 4 & 8 & 5 & 9 & 3 \\
      \hline
   \end{tabular}
}
\setlength{\tabcolsep}{0.25em}
\begin{table*}
    \begin{tabular}{cc}
        \begin{minipage}{0.55\textwidth}
            \centering
           \begin{minipage}{0.5\textwidth}
                \celisdata
           \end{minipage}
           \begin{minipage}{0.05\textwidth}
                \celisrankuti
           \end{minipage}
           \hspace{0.6em}
           \begin{minipage}{0.05\textwidth}
                \celisrankdiv
           \end{minipage}
           \hspace{0.6em}
           \begin{minipage}{0.05\textwidth}
                \celisrankbal
           \end{minipage}
        \end{minipage}
        &
        \begin{minipage}{0.25\textwidth}
            \centering
            \begin{minipage}{\textwidth}
                \secretaryone
           \end{minipage}
           \begin{minipage}{\textwidth}
                \vspace{4em}
           \end{minipage}
           \begin{minipage}[b]{\textwidth}
                \secretarytwo
           \end{minipage}
        \end{minipage}
     \\
    \begin{minipage}[t]{0.5\textwidth}
        \captionof{figure}{A set of applicants for college admissions $\candidateSet{}$, with two sensitive attributes: $A_1$ (gender), with groups $\group{\val{M}}=\{a,b,e,f,i,j\}$ and $\group{\val{F}}=\{c,d,g,h,k,l\}$, and $A_2$ (race), with groups $\group{\val{W}}=\{a,b,c,d\}$, $\group{\val{B}}=\{e,f,g,h\}$, and $\group{\val{A}}=\{i, j, k, l\}$. Top-$4$ ranking $\btau_1$ selects the highest-scoring candidates according to $Y$; all selected candidates are White, two of them are female and two are male. The utility of $\btau_1$, computed as the sum of scores, is 68. Top-$4$ ranking $\btau_2$ selects highest-scoring candidates subject to constraints to select at least two candidates of each gender, $L_4^{\val{M}}=2$, $L_4^{\val{F}}=2$, and at least one candidate of each race, $L_4^{\val{W}}=1$, $L_4^{\val{B}}=1$, $L_4^{\val{A}}=1$. The utility of $\btau_1$, computed as the sum of scores, is 53. Top-$4$ ranking $\btau_3$, subject to the same diversity constrains as $\btau_2$, but additionally balancing utility loss within each group. This ranking has utility 52, and it returns the highest-scoring male, female, White, and Black candidates.}
        \label{fig:crm}
    \end{minipage}
     & 
    \begin{minipage}[t]{0.5\textwidth}
        \captionof{figure}{An instance of the diverse $k$-choice secretary problem. A set of $n=6$ college applicants $\candidateSet{}$, are arriving for in-person interviews.  The order of interviews is from left to right,   with $a$ arriving first, followed by $e$, etc.  A candidate's score $Y$ is revealed when they are interviewed. $\candidateSet{}$ is partitioned into two groups based on (binary) gender,  $\group{\val{M}}=\{a,b,c\}$ with $n_{\val{M}} = 3$ candidates, and $\group{\val{F}}=\{d,e,f\}$ with $n_{\val{F}} = 3$ candidates. The goal is to select $k=2$ candidates, with one of every gender ($L_k^{\val{M}} = 1$,  $L_k^{\val{F}} = 1$), and to maximize the expected sum of $Y$-scores subject to these diversity constraints.  {\bf (a)} using a common warm-up period yields candidates \val{b} and \val{d}, selecting the 2nd best male candidate but the lowest-scoring female candidate {\bf (b)} separating warm-up per-group yields candidates \val{b} (as before) and \val{f}, the top-scoring female candidate.}
        \label{fig:sec}
    \end{minipage}
    \end{tabular}
\end{table*}

In the example we constructed, diversity constraints are satisfiable.  However, as was shown by~\citet{celis2018ranking}, the constrained ranking maximization problem can be seen to generalize various NP-hard problems such as independent set, hypergraph matching and set packing, and so is hard in the general case.  It turns out that even checking if there is a complete feasible ranking is NP-hard.  The authors show that a special case of the problem, in which each candidate is assigned to (at most) one group, and so the assignment induces a partitioning on $\candidateSet{}$, can be solved in polynomial time.  In this case, diversity constraints can only be specified with respect to a single sensitive attribute, which may be binary or multinary, and so can represent multiple sensitive attributes in combination (see discussion on group structure in Section~\ref{sec:frame:group}).  

Recall that the problem formulation allows to associate a utility with an assignment of candidate $a \in \candidateSet{}$ to rank position $j \in [k]$.  While the nature of these assignments can in principle be arbitrary, many reasonable utility metrics, including NDCG, Bradley-Terry~\cite{bradley1952rank} or Spearman’s rho~\cite{spearman1961proof}, are non-increasing with increasing rank position, and with decreasing utility score $Y$, which is intuitively interpreted to mean that, if $Y_a \geq Y_b$ then placing $a$ above $b$ cannot decrease the utility of the overall ranking.   Such metrics are said to be monotone and to satisfy the Monge property.  For this family of utility metrics, the authors propose an exact dynamic programming algorithm that solves the constrained ranking optimization problem in time polynomial in the number of candidates $m$ and size of the selected set $k$, and exponential in the number of possible assignments of candidates to groups (typically the product of domain cardinalities of the sensitive attributes, $card(A_1) \times card(A_2) = 6$ in our example in Figure~\ref{fig:crm}).  The authors also propose approximation algorithms that allow violations of diversity constraints, and study the quality of these approximations.

\spara{Insights.} The focus of this work is on the formal  properties of the constrained ranking maximization problem, including its hardness and approximability under different assumptions about the sensitive attributes, the diversity constraints, and the properties of the utility metric.  The paper does not include an experimental evaluation. 
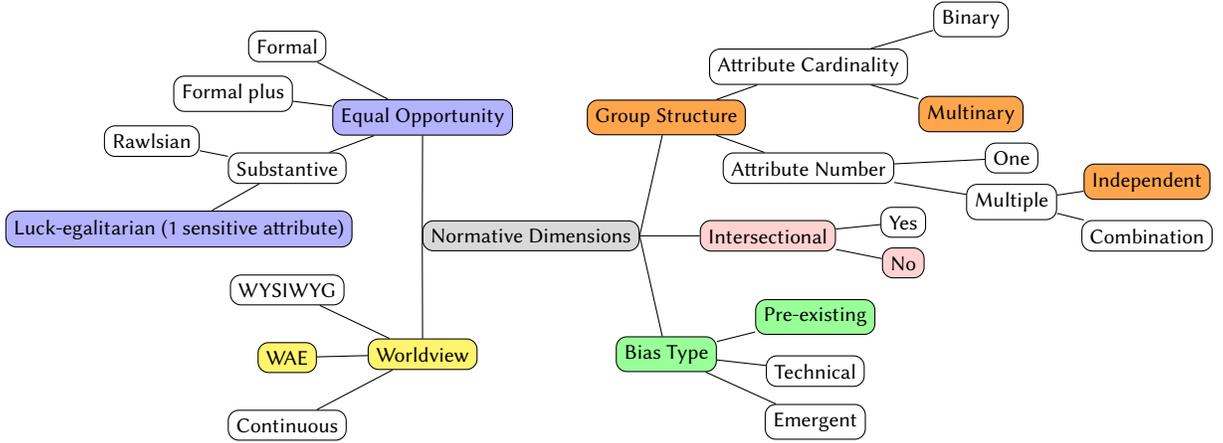
\begin{figure}[h]
    \centering
    \begin{tikzpicture}[scale=0.9,align=center]
    \tikzset{
        root/.style = {rectangle, rounded corners, draw=black, fill=gray!30, font=\small\sffamily},
        child/.style = {rectangle, rounded corners,draw=black, font=\small\sffamily},
        orangec/.style = {rectangle, rounded corners, draw=black, fill=orange!70, font=\small\sffamily},
        greenc/.style = {rectangle, rounded corners, draw=black, fill=green!40, font=\small\sffamily},
        bluec/.style = {rectangle, rounded corners, draw=black, fill=blue!30, font=\small\sffamily},
        yellowc/.style = {rectangle, rounded corners, draw=black, fill=yellow!70, font=\small\sffamily},
        pinkc/.style = {rectangle, rounded corners, draw=black, fill=pink!70, font=\small\sffamily},
    }
    \node (ND) at (-0.2,0) [root] {Normative Dimensions};    
    \node (GS) at (1.8,1.75) [orangec] {Group Structure};    
    \node (BT) at (1.8,-1.75) [greenc] {Bias Type};    
    \node (EO) at (-1.8,1.75) [bluec] {Equal Opportunity};    
    \node (WV) at (-1.8,-1.75) [yellowc] {Worldview};  
    \node (IT) at (3.3,0.0) [pinkc] {Intersectional};
    
    \node (AC) at (3.9,2.5) [child] {Attribute Cardinality};
    \node (ACB) at (6.3,3.2) [child] {Binary};
    \node (ACM) at (6.3,1.8) [orangec] {Multinary};
    
    \node (AN) at (3.9,1.0) [child] {Attribute Number};
    \node (ANO) at (6.9,1.15) [child] {One};
    \node (ANM) at (6.9,0.5) [child] {Multiple};
    
    \node (ANMI) at (8.9,0.8) [orangec] {Independent};
    \node (ANMC) at (8.9,0.0) [child] {Combination};
    
    \node (ITY) at (5.3,0.2) [child] {Yes};
    \node (ITN) at (5.3,-0.4) [pinkc] {No};
    
    \node (BTP) at (4.0,-1.2) [greenc] {Pre-existing};
    \node (BTT) at (4.0,-2.0) [child] {Technical};
    \node (BTE) at (4.0,-2.75) [child] {Emergent};
    
    \node (EOF) at (-3.8,2.8) [child] {Formal};
    \node (EOFP) at (-4.6,2.1) [child] {Formal plus};
    \node (EOS) at (-3.8,1.0) [child] {Substantive};
    
    \node (EOSR) at (-5.8,1.4) [child] {Rawlsian};
    \node (EOSL) at (-5.4,0.1) [bluec] {Luck-egalitarian (1 sensitive attribute)};

    \node (WVI) at (-3.8,-0.8) [child] {WYSIWYG};
    \node (WVE) at (-3.8,-1.8) [yellowc] {WAE};
    \node (WVC) at (-3.8,-2.8) [child] {Continuous};
    
    \draw (ND.east) -- (GS);
    \draw (ND.east) -- (BT);
    \draw (ND.east) -- (IT);
    \draw (ND.west) -- (EO);
    \draw (ND.west) -- (WV);
    
    \draw (GS) -- (AC);
    \draw (GS) -- (AN);
    
    \draw (AC) -- (ACB);
    \draw (AC) -- (ACM);
    
    \draw (AN) -- (ANO);
    \draw (AN) -- (ANM);
    
    \draw (IT) -- (ITY);
    \draw (IT) -- (ITN);
    
    \draw (ANM) -- (ANMI);
    \draw (ANM) -- (ANMC);
    
    \draw (BT) -- (BTP);
    \draw (BT) -- (BTT);
    \draw (BT) -- (BTE);
    
    \draw (EO) -- (EOF);
    \draw (EO) -- (EOFP);
    \draw (EO) -- (EOS);
    
    \draw (EOS) -- (EOSR);
    \draw (EOS) -- (EOSL);
    
    \draw (WV) -- (WVI);
    \draw (WV) -- (WVE);
    \draw (WV) -- (WVC);
\end{tikzpicture}
    \caption{\revv{Summary of the normative values encoded by Constrained ranking maximization (\citet{celis2018ranking}).}}
    \label{fig:mind:cons_maximize}
\end{figure}

The paper states that ``[...] left unchecked, the output of ranking algorithms can result in decreased diversity in the type of content presented, promote stereotypes, and polarize opinions.'' The goal of imposing diversity constrains is to counteract pre-existing bias.  Further, since these constraints enforce equality of outcomes, the method relates to the WAE worldview.

When the candidates are partitioned by a single sensitive attribute (either binary or multinary), the method will select the highest-scoring members of each group to satisfy diversity constraints.  \revv{Under the assumption that the goal of the competition is to make future prospects comparable, this is consistent with substantive \eop, and the mechanism is consistent with luck-egalitarian \eop.}  However, when candidates are associated with two or more sensitive attributes, as is the case in the example in Figure~\ref{fig:crm}, a single utility distribution is assumed, in the sense that a higher-scoring candidate will be preferred to a lower-scoring one irrespective of their group membership, whenever constraints permit.  This was illustrated ranking $\btau_2$ as shown in Figure~\ref{fig:crm}, where the highest-scoring White and male candidates were selected among the top-$k$, but the  highest-scoring female, Black, and Asian candidates were skipped.  Based on this observation, \revv{the method falls short of satisfying the desiderata of \eop for multiple sensitive attributes.}  \rev{Figure~\ref{fig:mind:cons_maximize} and Table~\ref{tbl:method-summary_score} summarize our analysis.} We will elaborate on this specific concern in the next sub-section.

\subsubsection{\citet{DBLP:conf/ijcai/YangGS19}}
\spara{Fairness definition and problem formalization.} The authors further investigate the constrained ranking maximization problem with two or more sensitive attributes, and observe that members of multiple historically disadvantaged groups may still be treated unfairly by this process.  For an intuition, consider again Figure~\ref{fig:crm}, and recall that the goal is to select the top-$4$ candidates, with at least two  of each gender ($L_4^{\val{M}}=2$, $L_4^{\val{F}}=2$) and at least one of each race ($L_4^{\val{W}}=1$, $L_4^{\val{B}}=1$, $L_4^{\val{A}}=1$).  Maximizing utility subject to these constraints being met yields, ranking $\btau_2$ in Figure~\ref{fig:crm} that selects the best (according to score) White and male candidates $a$ and $b$, but it does not select the best Black, Asian, or female candidates. 

Our example was deliberately constructed to highlight the following: if some population groups have systematically lower scores, then it costs less to skip their best-scoring members in the name of diversity.  This runs contrary to the nature of the diversity objective, which is to equalize access to opportunity.  This also represents unfairness, under the luck-egalitarian view.  To see why, suppose that scores represent effort (\eg how hard someone studied to do well on a test), and that we consider it important to reward effort.  We may then take a relative view of effort, and assert that scores are more informative \emph{within} a group than \emph{across groups}.   Taken together, this means that the best-scoring individuals from historically disadvantaged groups should have a chance to be selected among the top-$k$. Ranking $\btau_3$ in Figure~\ref{fig:crm} represents a ranked outcome that gets closer to this objective; it presents $\btau_3$, a top-$4$ ranking that contains the highest-scoring male, female, White and Black candidates.

\citet{DBLP:conf/ijcai/YangGS19} formalize this intuition by stating that, when multiple sensitive attributes (\eg gender and race) are considered simultaneously, it is crucial to consider the utility loss that is incurred within each group, and to balance that loss across groups.  The authors propose two measures to quantify in-group utility, \ratio and \agg, both taking on values from the range (0, 1], with 1 corresponding to perfect utility within a group (no loss), and with high loss corresponding to values close to 0.  Both \ratio and \agg can be computed over the top-$k$ as a set, or in rank-aware manner, by considering every prefix of length $p \in [k]$.  In what follows, we will illustrate one of these measures, \ratio, taking the set interpretation for simplicity.

\ratio, quantifies the utility within a group (\eg female or Black) by computing the ratio of the utility score of the highest-scoring skipped candidate from that group and the lowest-scoring selected candidate.  Consider again ranking $\btau_2$ in Figure~\ref{fig:crm}.  We compute  $\ratio(\btau_2, \group{M}) = \ratio(\btau_2, \group{W}) = 1$, since the highest-scoring male and White candidates were selected.  For the female, Black, and Asian groups, we compute  $\ratio(\btau_2, \group{F}) = \frac{Y_g}{Y_c} = \frac{10}{16}$; $\ratio(\btau_2, \group{B}) = \frac{Y_g}{Y_e} = \frac{10}{11}$; and $\ratio(\btau_2, \group{A}) = \frac{Y_k}{Y_i} = \frac{6}{7}$.

\agg is based on similar intuition as \ratio, but rather than comparing the utility due to a pair of items for each group --- one selected and one skipped --- it compares the sum of scores of all items from a group up to a particular position with the sum of scores of all selected items from that group (up to the same position).

The authors go on to use \ratio and \agg to state that loss in these measures should be balanced across groups.  They implement this requirement as an additional set of constraints, and formalize the induced optimization problem that (1) meets diversity constraints for each group, (2) balances utility loss across groups, and (3) maximizes over-all utility subject to (1) and (2),  as integer linear programs.

\spara{Experiments and observations} The authors conduct experiments on two real datasets, CS departments~\cite{CSData} and MEPS~\cite{MEPSData} (see details in Section~\ref{sec:datasets}). They use these datasets to quantify the feasible trade-offs between, diversity, overall utility, and utility loss across groups.  Further, they show that utility loss can be balanced effectively, and that the over-all utility cost of such interventions is low.

\spara{Insights}  Similarly to papers surveyed earlier in this section, the work of~\citet{DBLP:conf/ijcai/YangGS19} aims to address pre-existing bias by equalizing outcomes, and so relates to the WAE worldview.  Further, because of an explicit focus on ensuring that the best-qualified candidates from each group have an opportunity to be selected, or to appear at higher ranks, this work 
\revv{conditions qualification score on morally irrelevant attributes (group membership)}, and so is firmly in the luck-egalitarian \revv{\eop} camp, \revv{under the assumption that the goal of the competition is to make individuals' future life prospects comparable.}  
\begin{figure}[h]
    \centering
    \begin{tikzpicture}[scale=0.9,align=center]
    \tikzset{
        root/.style = {rectangle, rounded corners, draw=black, fill=gray!30, font=\small\sffamily},
        child/.style = {rectangle, rounded corners,draw=black, font=\small\sffamily},
        orangec/.style = {rectangle, rounded corners, draw=black, fill=orange!70, font=\small\sffamily},
        greenc/.style = {rectangle, rounded corners, draw=black, fill=green!40, font=\small\sffamily},
        bluec/.style = {rectangle, rounded corners, draw=black, fill=blue!30, font=\small\sffamily},
        yellowc/.style = {rectangle, rounded corners, draw=black, fill=yellow!70, font=\small\sffamily},
        pinkc/.style = {rectangle, rounded corners, draw=black, fill=pink!70, font=\small\sffamily},
    }
    \node (ND) at (-0.2,0) [root] {Normative Dimensions};    
    \node (GS) at (1.8,1.75) [orangec] {Group Structure};    
    \node (BT) at (1.8,-1.75) [greenc] {Bias Type};    
    \node (EO) at (-1.8,1.75) [bluec] {Equal Opportunity};    
    \node (WV) at (-1.8,-1.75) [yellowc] {Worldview};  
    \node (IT) at (3.3,0.0) [pinkc] {Intersectional};
    
    \node (AC) at (3.9,2.5) [child] {Attribute Cardinality};
    \node (ACB) at (6.3,3.2) [child] {Binary};
    \node (ACM) at (6.3,1.8) [orangec] {Multinary};
    
    \node (AN) at (3.9,1.0) [child] {Attribute Number};
    \node (ANO) at (6.9,1.15) [child] {One};
    \node (ANM) at (6.9,0.5) [child] {Multiple};
    
    \node (ANMI) at (8.9,0.8) [orangec] {Independent};
    \node (ANMC) at (8.9,0.0) [child] {Combination};
    
    \node (ITY) at (5.3,0.2) [pinkc] {Yes};
    \node (ITN) at (5.3,-0.4) [child] {No};
    
    \node (BTP) at (4.0,-1.2) [greenc] {Pre-existing};
    \node (BTT) at (4.0,-2.0) [greenc] {Technical};
    \node (BTE) at (4.0,-2.75) [child] {Emergent};
    
    \node (EOF) at (-3.8,2.8) [child] {Formal};
    \node (EOFP) at (-4.6,2.1) [child] {Formal plus};
    \node (EOS) at (-3.8,1.0) [child] {Substantive};
    
    \node (EOSR) at (-5.8,1.4) [child] {Rawlsian};
    \node (EOSL) at (-5.4,0.1) [bluec] {Luck-egalitarian};

    \node (WVI) at (-3.8,-0.8) [child] {WYSIWYG};
    \node (WVE) at (-3.8,-1.8) [yellowc] {WAE};
    \node (WVC) at (-3.8,-2.8) [child] {Continuous};
    
    \draw (ND.east) -- (GS);
    \draw (ND.east) -- (BT);
    \draw (ND.east) -- (IT);
    \draw (ND.west) -- (EO);
    \draw (ND.west) -- (WV);
    
    \draw (GS) -- (AC);
    \draw (GS) -- (AN);
    
    \draw (AC) -- (ACB);
    \draw (AC) -- (ACM);
    
    \draw (AN) -- (ANO);
    \draw (AN) -- (ANM);
    
    \draw (IT) -- (ITY);
    \draw (IT) -- (ITN);
    
    \draw (ANM) -- (ANMI);
    \draw (ANM) -- (ANMC);
    
    \draw (BT) -- (BTP);
    \draw (BT) -- (BTT);
    \draw (BT) -- (BTE);
    
    \draw (EO) -- (EOF);
    \draw (EO) -- (EOFP);
    \draw (EO) -- (EOS);
    
    \draw (EOS) -- (EOSR);
    \draw (EOS) -- (EOSL);
    
    \draw (WV) -- (WVI);
    \draw (WV) -- (WVE);
    \draw (WV) -- (WVC);
\end{tikzpicture}
    \caption{\revv{Summary of the normative values encoded by Balanced diverse ranking (\citet{DBLP:conf/ijcai/YangGS19}).}}
    \label{fig:mind:balance}
\end{figure}
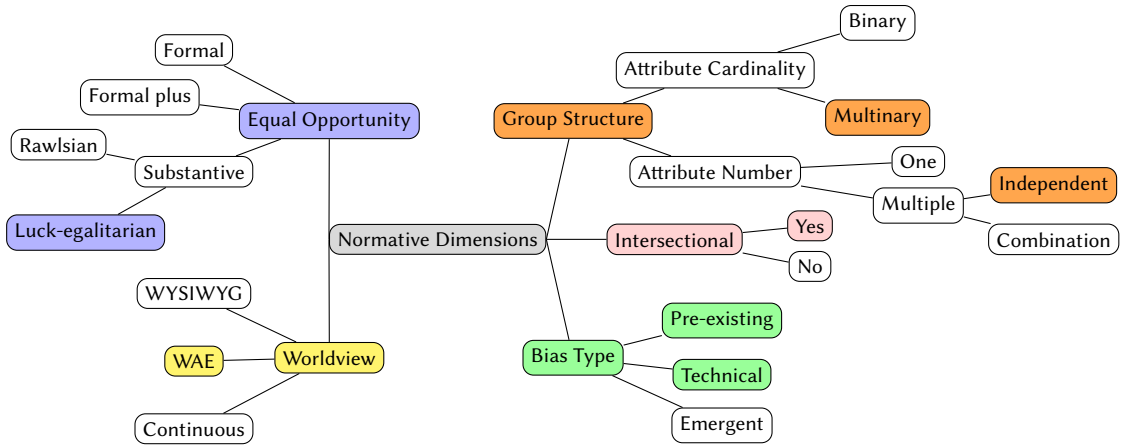
The main insight on which this work is based is that membership in multiple sensitive groups can lead to unfair treatment, and that the effects can be particularly pronounced for individuals who are multiply marginalized and who may, for example, be denied opportunity along the dimensions of both race and gender.  This insight is surfacing an important dimension of intersectional discrimination in algorithmic rankers and is, to the best of our knowledge, the first approach in this area to have observed and proposed ways to counteract intersectional effects. \rev{Figure~\ref{fig:mind:balance} and Table~\ref{tbl:method-summary_score} summarize our analysis.}

\subsubsection{\citet{DBLP:conf/edbt/StoyanovichYJ18}}
\label{subsubsec:Stoyanovichetal}
\spara{Fairness definition and problem formalization}  The final method we discuss in this section aims to incorporate diversity constraints of the kind that were used by~\citet{celis2018ranking} and~\citet{DBLP:conf/ijcai/YangGS19} into online set selection.   This setting models a sequence of job or college admissions interviews: candidates arrive one-by-one and their qualification score $Y$ is revealed at the time of the interview.  Candidates are assumed to arrive in random order according to score, and their total number $n$ is known or can be estimated.  The decision maker must hire or to reject the candidate being considered as soon as their score $Y$ is revealed, before advancing to the next candidate in the sequence. 

The classic version of this problem, known as the Secretary problem~\cite{Dynkin:1963,Lindley:1961}, aims to select a single candidate with the highest score $Y$.  It was shown by~\citet{Lindley:1961} and by~\citet{Dynkin:1963} that the optimal hiring strategy is to interview $s = \lfloor \frac{n}{e} \rfloor$ candidates without making any offers (this is called the ``warm-up period''), and make an offer to the first candidate whose score is better than the best score of the of the first $s$ candidates (or accept the last candidate if no  better candidate is seen).  This strategy yields the highest-scoring candidate with probability $\frac{1}{e}$, and is said to have ``competitive ratio'' $e$.  Further, this is the best such strategy for the secretary problem (\ie with the highest competitive ratio)~\cite{ferguson1989}.
This problem has been extended by~\citet{DBLP:journals/sigecom/BabaioffIKK08} to select $k$ candidates, maximizing the expected sum of their scores. \citet{DBLP:conf/edbt/StoyanovichYJ18} postulate the diverse $k$-choice secretary problem that enriches the $k$-choice secretary problem of~\citet{DBLP:journals/sigecom/BabaioffIKK08} with diversity constraints.

The diverse $k$-choice secretary problem is formalized as follows: In addition to a qualification score $Y$, each candidate is associated with one of $i \geq 2$ groups $\groupSet$ based on the value of a single multinary sensitive attribute (\eg gender, race, or disability status).  Both the total number of candidates $n$, and the number of candidates in each group $n_1 \ldots n_i$, are known ahead of time or can be estimated.  The goal of the decision maker is to select $k$ candidates, maximizing the expected sum of their scores, subject to diversity constraints, stated in the form of per-group lower-bounds $L_k^{\group{}}$ and upper-bounds $U_k^{\group{}}$.  Figure~\ref{fig:sec} gives an example: a set of $n=6$ college applicants, of whom $n_{\val{M}}=3$ are male and $n_{\val{F}}=3$ are female, are being interviewed in the order shown in Figure~\ref{fig:sec}, left-to-right.  The admissions officer wishes to select $k=2$ applicants, with one of each gender, specified by the lower-bound constraints $L_k^{\val{M}} = 1$ and  $L_k^{\val{F}} = 1$.

The key idea in~\citet{DBLP:conf/edbt/StoyanovichYJ18} is that, if score distributions are expected to differ between the groups, then separate warm-up periods should be conducted for each group to better estimate the scores of that group's desirable candidates.  As illustrated in the outcome (a) in Figure~\ref{fig:sec}, measuring the higher-scoring male candidates and the lower-scoring female candidates against the same (higher-scoring) standard will allow high-scoring male candidates to be selected.  However, the female candidates selected in this way are those that happen to be at the end of the interview queue: they were chosen ``at the last minute'' to satisfy the diversity constraint.   This is problematic for the reasons we outlined when discussing~\citet{DBLP:conf/ijcai/YangGS19}  earlier in this section --- it withholds opportunity from the relatively better-qualified candidates of a historically disadvantaged group, and it can build bad precedent if the lesser-qualified candidates from that group are selected but do not perform well on the task.  Outcome (b) in Figure~\ref{fig:sec} shows the result of a selection in which warm-up was conducted separately per group, yielding a higher-scoring female candidate.

The authors propose additional techniques to handle cases where the sum of the per-group lower bound is less than $k$, leaving the freedom to select high-scoring candidates from any group.  Finally, they consider the case where a constant-size waiting list of candidates is allowed, showing that it can lead to higher-utility outcomes.

\spara{Experiments and observations}
The experimental evaluation of the proposed algorithms for variants of the diverse $k$-choice secretary problems is conducted using three real datasets: Forbes richest Americans~\cite{ForbesRichesData}, NASA astronauts~\cite{NASAData}, and Pantheon~\cite{PantheonData} (see details in Section~\ref{sec:datasets}). Additional results on synthetic datasets are provided, to simulate differences in score distributions between groups. 
The evaluation on real datasets shows that the algorithms can select candidates that meet the desired diversity constraints while paying a very small cost in terms of the utility loss. The evaluation on synthetic datasets shows that if a difference in the observed scores is expected between groups, then these groups must be treated separately during processing. Otherwise, a solution may be derived that meets diversity constraints, but that results in lower utility for the disadvantaged groups.

\spara{Insights}
This work focuses on pre-existing bias that exhibits itself through differences in expected scores between groups of candidates.  Diversity constraints, and the mechanism used to enact them, aims to equalize outcomes across groups, and so this method clearly links to the WAE worldview.  
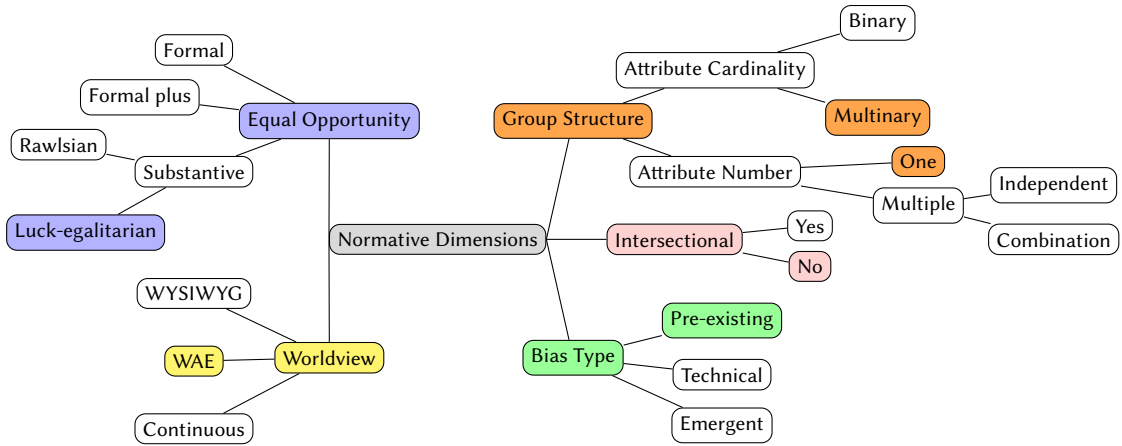
\begin{figure}[h]
    \centering
    \begin{tikzpicture}[scale=0.9,align=center]
    \tikzset{
        root/.style = {rectangle, rounded corners, draw=black, fill=gray!30, font=\small\sffamily},
        child/.style = {rectangle, rounded corners,draw=black, font=\small\sffamily},
        orangec/.style = {rectangle, rounded corners, draw=black, fill=orange!70, font=\small\sffamily},
        greenc/.style = {rectangle, rounded corners, draw=black, fill=green!40, font=\small\sffamily},
        bluec/.style = {rectangle, rounded corners, draw=black, fill=blue!30, font=\small\sffamily},
        yellowc/.style = {rectangle, rounded corners, draw=black, fill=yellow!70, font=\small\sffamily},
        pinkc/.style = {rectangle, rounded corners, draw=black, fill=pink!70, font=\small\sffamily},
    }
    \node (ND) at (-0.2,0) [root] {Normative Dimensions};    
    \node (GS) at (1.8,1.75) [orangec] {Group Structure};    
    \node (BT) at (1.8,-1.75) [greenc] {Bias Type};    
    \node (EO) at (-1.8,1.75) [bluec] {Equal Opportunity};    
    \node (WV) at (-1.8,-1.75) [yellowc] {Worldview};  
    \node (IT) at (3.3,0.0) [pinkc] {Intersectional};
    
    \node (AC) at (3.9,2.5) [child] {Attribute Cardinality};
    \node (ACB) at (6.3,3.2) [child] {Binary};
    \node (ACM) at (6.3,1.8) [orangec] {Multinary};
    
    \node (AN) at (3.9,1.0) [child] {Attribute Number};
    \node (ANO) at (6.9,1.15) [orangec] {One};
    \node (ANM) at (6.9,0.5) [child] {Multiple};
    
    \node (ANMI) at (8.9,0.8) [child] {Independent};
    \node (ANMC) at (8.9,0.0) [child] {Combination};
    
    \node (ITY) at (5.3,0.2) [child] {Yes};
    \node (ITN) at (5.3,-0.4) [pinkc] {No};
    
    \node (BTP) at (4.0,-1.2) [greenc] {Pre-existing};
    \node (BTT) at (4.0,-2.0) [child] {Technical};
    \node (BTE) at (4.0,-2.75) [child] {Emergent};
    
    \node (EOF) at (-3.8,2.8) [child] {Formal};
    \node (EOFP) at (-4.6,2.1) [child] {Formal plus};
    \node (EOS) at (-3.8,1.0) [child] {Substantive};
    
    \node (EOSR) at (-5.8,1.4) [child] {Rawlsian};
    \node (EOSL) at (-5.4,0.1) [bluec] {Luck-egalitarian};

    \node (WVI) at (-3.8,-0.8) [child] {WYSIWYG};
    \node (WVE) at (-3.8,-1.8) [yellowc] {WAE};
    \node (WVC) at (-3.8,-2.8) [child] {Continuous};
    
    \draw (ND.east) -- (GS);
    \draw (ND.east) -- (BT);
    \draw (ND.east) -- (IT);
    \draw (ND.west) -- (EO);
    \draw (ND.west) -- (WV);
    
    \draw (GS) -- (AC);
    \draw (GS) -- (AN);
    
    \draw (AC) -- (ACB);
    \draw (AC) -- (ACM);
    
    \draw (AN) -- (ANO);
    \draw (AN) -- (ANM);
    
    \draw (IT) -- (ITY);
    \draw (IT) -- (ITN);
    
    \draw (ANM) -- (ANMI);
    \draw (ANM) -- (ANMC);
    
    \draw (BT) -- (BTP);
    \draw (BT) -- (BTT);
    \draw (BT) -- (BTE);
    
    \draw (EO) -- (EOF);
    \draw (EO) -- (EOFP);
    \draw (EO) -- (EOS);
    
    \draw (EOS) -- (EOSR);
    \draw (EOS) -- (EOSL);
    
    \draw (WV) -- (WVI);
    \draw (WV) -- (WVE);
    \draw (WV) -- (WVC);
\end{tikzpicture}
    \caption{\revv{Summary of the normative values encoded by Diverse $k$-choice secretary problem (\citet{DBLP:conf/edbt/StoyanovichYJ18}).}}
    \label{fig:mind:diversek}
\end{figure}
The core idea in this work is that effort, as represented by scores, should be seen as relative:  scores are estimated per group, and individuals from a particular group are evaluated against that group's score threshold.  \revv{Thus, under the assumption that the goal of the fainess intervention is to equalize opportunities over a lifetime, this method is consistent with luck-egalitarian \eop.} 
\rev{Figure~\ref{fig:mind:diversek} and Table~\ref{tbl:method-summary_score} summarize our analysis.}

\subsection{Intervening on the Score Distribution}
\label{sec:fair_db:latent}

The methods in this subsection work under the assumption that the scores on which candidates are ranked are subject to pre-existing bias, such that members of minority or historically disadvantaged groups have lower scores, and thus are ranked less favorably.  The approach these methods take is based on correcting for the bias by adjusting the score distribution before it is given as input to a ranker.  

\subsubsection{\citet{kleinberg_et_al:LIPIcs:2018:8323} and~\citet{celis2020interventions}}
\label{subsubsec:kleinberg}
\spara{Problem formalization} The papers discussed in this section study set selection and ranking in presence of \emph{implicit bias}; they investigate under what conditions the utility of the selected set or the top-$k$ would be improved by imposing representation constraints.  \citet{kleinberg_et_al:LIPIcs:2018:8323}  consider a score-based set selection task motivated by hiring, in which a set of $n$ candidates $\candidateSet{}$ applies for an open job position, and some $k \ll n$ of them are selected as finalists to  interview.  The size of the selected set $k$ is assumed to be a small constant, with the case $k=2$ studied closely in the paper. %
Candidates in $\candidateSet{}$ belong to one of two groups, $\group{1}$ or $\group{2}$, according to a single binary sensitive attribute (\eg, binary gender), and with one of these groups, $\group{1}$, corresponding to the protected group (\eg the female gender).   It is assumed that $\group{1}$ constitutes a minority of the applicant pool, as quantified by the parameter $\alpha \in (0, 1]$, with $\left| \group{1} \right| = \alpha \cdot \left| \group{2} \right |$.  Further, it is assumed that the true qualification scores (called ``potentials'' in~\citet{kleinberg_et_al:LIPIcs:2018:8323}) are drawn from the same score distribution for the candidates in both groups, and that this distribution follows the power law, parameterized by $\delta > 0$, such that $Pr[Y\geq t]=t^{-(1+\delta)}$.

Candidates are not hired according to their true qualification scores $Y$, but rather according to their perceived  scores $\Tilde{Y}$, which are, in turn, subject to \emph{implicit bias}: hiring committee members ``downgrade'' the true scores of the candidates from $\group{1}$ by dividing them by a factor $\beta > 1$.  

The question being asked in this papers is: Under what conditions does including \emph{a single candidate} from the protected group $\group{1}$ among the $k$ finalists improve the utility of the selected set according to the true score $Y$?  (Utility is quantified as the sum of true scores of the selected candidates.) This intervention is known as the Rooney Rule~\cite{rooney}, and while its goal is to improve diversity in hiring, \citet{kleinberg_et_al:LIPIcs:2018:8323} study it explicitly from the point of view of utility rather than diversity or fairness.  The requirement of including a single protected group candidate among the finalists is a basic coverage-based diversity requirement~\cite{drosou2017diversity}.

The authors study the problem under different settings of $\alpha$ (relative proportion of the minority group), $\beta$ (bias factor), and $\delta$ (the parameter of the power law distribution of true scores).  They find that, for every $\alpha$, there exists a sufficiently small  $\delta > 0$ for which the Rooney Rule will produce a set of $k$ finalists with higher expected utility, compared to when candidates are selected according to their perceived --- and biased --- scores $\Tilde{Y}$.  Put another way, with a power law exponent $1+\delta$ that is sufficient close to 1, it is a better strategy, \emph{in terms of utility}, to commit one of the $k$ offers to the candidates from group $\group{1}$, even when $k$ is as low as 2 and $\group{1}$ forms an extremely small fraction of the population.

Figure~\ref{fig:exm_utility_kleinberg} shows an example of the selection process, where the goal is to select $k=2$ finalists from a pool of 6, with 2 male candidates for each female candidate ($\alpha=1/2$), and with females candidates being perceived as half as qualified as what their true score would suggest ($\beta=2$).  The Rooney Rule would select a top-scoring candidate from each gender group, leading to higher expected utility than if the top-two candidates were selected, both of them male.
\newcommand{\kleinbergdata}{
    \small
    \begin{tabular}{|c||c||c|c|}
		\hline
		\rowcolor[HTML]{C0C0C0} 
		candidate & $A$ (sex) & $Y$ & $\Tilde{Y}$   \\ \hline
		\rowcolor[HTML]{FFFFFF}
	    \val{b}  & \val{male} & 9 & 9  \\ \hline
	    \val{c}  & \val{male} & 9 & 6 \\ \hline
	    \rowcolor[HTML]{ADD8E6}
	    \val{d}  & \val{female}  & 8 & 4 \\ \hline
	    \rowcolor[HTML]{FFFFFF}
	    \val{e}  & \val{male} & 7 & 5 \\ \hline
	    \rowcolor[HTML]{ADD8E6}
	    \val{f}  & \val{female} & 6 & 3  \\ \hline
	    \rowcolor[HTML]{FFFFFF}
	    \val{g}  & \val{male} & 5 & 5 \\ \hline
    \end{tabular}
}
\newcommand{\kleinbergrankraw}{
    \small
	\begin{tabular}{|c|}
		\hline
		\rowcolor[HTML]{C0C0C0} 
		$\btau$ \\ \hline
		\rowcolor[HTML]{FFFFFF}
		\val{b}   \\ \hline
		\val{c}   \\ \hline
		\rowcolor[HTML]{ADD8E6}
		\val{d}   \\ \hline
		\rowcolor[HTML]{FFFFFF}
		\val{e}   \\ \hline
		\rowcolor[HTML]{ADD8E6} 
		\val{f}   \\ \hline
		\rowcolor[HTML]{FFFFFF}
		\val{g}   \\ \hline
	\end{tabular}
}
\newcommand{\kleinbergrankbiased}{
    \small
	\begin{tabular}{|c|}
		\hline
		\rowcolor[HTML]{C0C0C0} 
		$\Tilde{\btau}$ \\ \hline
		\rowcolor[HTML]{FFFFFF}
		\val{b}                   \\ \hline
		\val{c}                   \\ \hline
		\val{e}                    \\ \hline
		\rowcolor[HTML]{ADD8E6} 
		\val{d}                     \\ \hline
		\rowcolor[HTML]{FFFFFF}
		\val{g}                     \\ \hline
		\rowcolor[HTML]{ADD8E6} 
		\val{f}                     \\ \hline
	\end{tabular}
}
\newcommand{\kleinbergrankrooney}{
    \small
	\begin{tabular}{|c|}
		\hline
		\rowcolor[HTML]{C0C0C0} 
		$\btau^{R}$ \\ \hline
		\rowcolor[HTML]{FFFFFF}
		\val{b}                   \\ \hline
		\rowcolor[HTML]{ADD8E6} 
		\val{d}                     \\ \hline
	\end{tabular}
}

\setlength{\tabcolsep}{0.3em}
\begin{table*}
    \begin{tabular}{cc}
        \begin{minipage}{0.7\textwidth}
            \centering
           \begin{minipage}{0.4\textwidth}
                \kleinbergdata
           \end{minipage}
           \begin{minipage}{0.05\textwidth}
                \kleinbergrankraw
           \end{minipage}
           \begin{minipage}{0.05\textwidth}
                \kleinbergrankbiased
           \end{minipage}
           \begin{minipage}{0.05\textwidth}
                \kleinbergrankrooney
           \end{minipage}
        \end{minipage}
        &
        \begin{minipage}{0.2\textwidth}
            \centering
            \includegraphics[height=60pt,width=70pt]{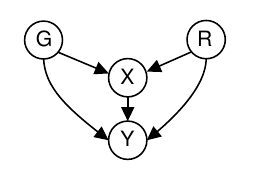}
        \end{minipage}
     \\
    \begin{minipage}[t]{0.7\textwidth}
        \captionof{figure}{Consider a set of applicants for college admissions. Observed scores $\Tilde{Y}$ of the applicants are affected by implicit bias: for the male candidates, $\Tilde{Y}=Y$, while for the female candidates, $\Tilde{Y}=Y/\beta$, with $\beta=2$.   Female candidates constitute a minority, with $\alpha=1/2$ --- there are 2 male candidates for each female one.  Ranking $\btau$ sorts candidates on their true (unobserved) scores $Y$; ranking $\Tilde{\btau}$ sorts them on scores $\Tilde{Y}$ that are subject to implicit bias; ranking $\btau^{R}$ applies to Rooney Rule to include the top-scoring female candidate among the top-2.}
        \label{fig:exm_utility_kleinberg}
    \end{minipage}
     & 
    \begin{minipage}[t]{0.2\textwidth}
        \captionof{figure}{A causal model that include sensitive attributes $G$ (gender), $R$ (race), utility score $Y$, and other covariates $\mathbf X$.}
	    \label{fig:cm}
    \end{minipage}
    \end{tabular}
\end{table*}

The results of are extended by~\citet{celis2020interventions}, who consider arbitrary utility distributions (beyond the power law) and support a richer group structure, including multiple sensitive attributes handled independently, with multinary domains.  They show that, for any (assumed) distribution of utilities and any level of implicit bias, representation constraints can lead to optimal latent utility.

\spara{Experiments and insights}
~\citet{kleinberg_et_al:LIPIcs:2018:8323} give a tight characterization of the conditions on $\alpha$, $\beta$, and $\delta$, under which applying the Rooney Rule, with its most basic representation constraint, produces a positive  change in expected utility. Proposed techniques can be used to estimate parameters of a biased decision-making process.  The paper focuses on theoretical analysis and does not provide any experimental results.
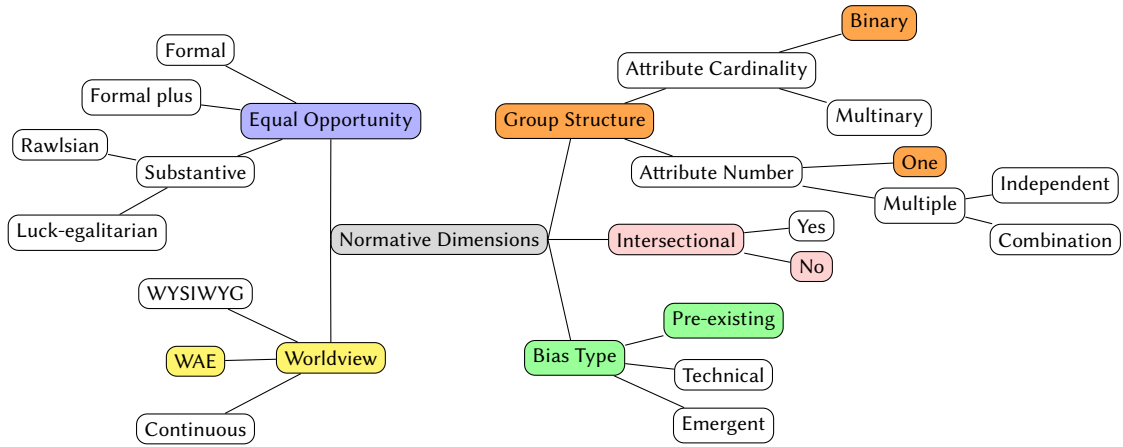
\begin{figure}[h]
    \centering
    \begin{tikzpicture}[scale=0.9,align=center]
    \tikzset{
        root/.style = {rectangle, rounded corners, draw=black, fill=gray!30, font=\small\sffamily},
        child/.style = {rectangle, rounded corners,draw=black, font=\small\sffamily},
        orangec/.style = {rectangle, rounded corners, draw=black, fill=orange!70, font=\small\sffamily},
        greenc/.style = {rectangle, rounded corners, draw=black, fill=green!40, font=\small\sffamily},
        bluec/.style = {rectangle, rounded corners, draw=black, fill=blue!30, font=\small\sffamily},
        yellowc/.style = {rectangle, rounded corners, draw=black, fill=yellow!70, font=\small\sffamily},
        pinkc/.style = {rectangle, rounded corners, draw=black, fill=pink!70, font=\small\sffamily},
    }
    \node (ND) at (-0.2,0) [root] {Normative Dimensions};    
    \node (GS) at (1.8,1.75) [orangec] {Group Structure};    
    \node (BT) at (1.8,-1.75) [greenc] {Bias Type};    
    \node (EO) at (-1.8,1.75) [bluec] {Equal Opportunity};    
    \node (WV) at (-1.8,-1.75) [yellowc] {Worldview};  
    \node (IT) at (3.3,0.0) [pinkc] {Intersectional};
    
    \node (AC) at (3.9,2.5) [child] {Attribute Cardinality};
    \node (ACB) at (6.3,3.2) [orangec] {Binary};
    \node (ACM) at (6.3,1.8) [child] {Multinary};
    
    \node (AN) at (3.9,1.0) [child] {Attribute Number};
    \node (ANO) at (6.9,1.15) [orangec] {One};
    \node (ANM) at (6.9,0.5) [child] {Multiple};
    
    \node (ANMI) at (8.9,0.8) [child] {Independent};
    \node (ANMC) at (8.9,0.0) [child] {Combination};
    
    \node (ITY) at (5.3,0.2) [child] {Yes};
    \node (ITN) at (5.3,-0.4) [pinkc] {No};
    
    \node (BTP) at (4.0,-1.2) [greenc] {Pre-existing};
    \node (BTT) at (4.0,-2.0) [child] {Technical};
    \node (BTE) at (4.0,-2.75) [child] {Emergent};
    
    \node (EOF) at (-3.8,2.8) [child] {Formal};
    \node (EOFP) at (-4.6,2.1) [child] {Formal plus};
    \node (EOS) at (-3.8,1.0) [child] {Substantive};
    
    \node (EOSR) at (-5.8,1.4) [child] {Rawlsian};
    \node (EOSL) at (-5.4,0.1) [child] {Luck-egalitarian};

    \node (WVI) at (-3.8,-0.8) [child] {WYSIWYG};
    \node (WVE) at (-3.8,-1.8) [yellowc] {WAE};
    \node (WVC) at (-3.8,-2.8) [child] {Continuous};
    
    \draw (ND.east) -- (GS);
    \draw (ND.east) -- (BT);
    \draw (ND.east) -- (IT);
    \draw (ND.west) -- (EO);
    \draw (ND.west) -- (WV);
    
    \draw (GS) -- (AC);
    \draw (GS) -- (AN);
    
    \draw (AC) -- (ACB);
    \draw (AC) -- (ACM);
    
    \draw (AN) -- (ANO);
    \draw (AN) -- (ANM);
    
    \draw (IT) -- (ITY);
    \draw (IT) -- (ITN);
    
    \draw (ANM) -- (ANMI);
    \draw (ANM) -- (ANMC);
    
    \draw (BT) -- (BTP);
    \draw (BT) -- (BTT);
    \draw (BT) -- (BTE);
    
    \draw (EO) -- (EOF);
    \draw (EO) -- (EOFP);
    \draw (EO) -- (EOS);
    
    \draw (EOS) -- (EOSR);
    \draw (EOS) -- (EOSL);
    
    \draw (WV) -- (WVI);
    \draw (WV) -- (WVE);
    \draw (WV) -- (WVC);
\end{tikzpicture}
    \caption{\revv{Summary of the normative values encoded by Selection with implicit bias (\citet{kleinberg_et_al:LIPIcs:2018:8323}).}}
    \label{fig:mind:implicit_selection}
\end{figure}

~\citet{celis2020interventions} extend these results, and also include an experimental evaluation on the \emph{IIT-JEE} dataset~\cite{IITJEEData} (see Section~\ref{sec:datasets} for details).  These results give the flavor of the utility of the proposed intervention, although experimental evaluation is substantially more limited than the problem set-up warrants, focusing on a single binary protected attribute and leaving empirically unsubstantiated the claim that proposed approach generalizes to multiple sensitive attributes and handles intersectional discrimination.
\begin{figure}[h]
    \centering
    \begin{tikzpicture}[scale=0.9,align=center]
    \tikzset{
        root/.style = {rectangle, rounded corners, draw=black, fill=gray!30, font=\small\sffamily},
        child/.style = {rectangle, rounded corners,draw=black, font=\small\sffamily},
        orangec/.style = {rectangle, rounded corners, draw=black, fill=orange!70, font=\small\sffamily},
        greenc/.style = {rectangle, rounded corners, draw=black, fill=green!40, font=\small\sffamily},
        bluec/.style = {rectangle, rounded corners, draw=black, fill=blue!30, font=\small\sffamily},
        yellowc/.style = {rectangle, rounded corners, draw=black, fill=yellow!70, font=\small\sffamily},
        pinkc/.style = {rectangle, rounded corners, draw=black, fill=pink!70, font=\small\sffamily},
    }
    \node (ND) at (-0.2,0) [root] {Normative Dimensions};    
    \node (GS) at (1.8,1.75) [orangec] {Group Structure};    
    \node (BT) at (1.8,-1.75) [greenc] {Bias Type};    
    \node (EO) at (-1.8,1.75) [bluec] {Equal Opportunity};    
    \node (WV) at (-1.8,-1.75) [yellowc] {Worldview};  
    \node (IT) at (3.3,0.0) [pinkc] {Intersectional};
    
    \node (AC) at (3.9,2.5) [child] {Attribute Cardinality};
    \node (ACB) at (6.3,3.2) [child] {Binary};
    \node (ACM) at (6.3,1.8) [orangec] {Multinary};
    
    \node (AN) at (3.9,1.0) [child] {Attribute Number};
    \node (ANO) at (6.9,1.15) [child] {One};
    \node (ANM) at (6.9,0.5) [child] {Multiple};
    
    \node (ANMI) at (8.9,0.8) [orangec] {Independent};
    \node (ANMC) at (8.9,0.0) [child] {Combination};
    
    \node (ITY) at (5.3,0.2) [child] {Yes};
    \node (ITN) at (5.3,-0.4) [pinkc] {No};
    
    \node (BTP) at (4.0,-1.2) [greenc] {Pre-existing};
    \node (BTT) at (4.0,-2.0) [child] {Technical};
    \node (BTE) at (4.0,-2.75) [child] {Emergent};
    
    \node (EOF) at (-3.8,2.8) [child] {Formal};
    \node (EOFP) at (-4.6,2.1) [child] {Formal plus};
    \node (EOS) at (-3.8,1.0) [child] {Substantive};
    
    \node (EOSR) at (-5.8,1.4) [child] {Rawlsian};
    \node (EOSL) at (-5.4,0.1) [child] {Luck-egalitarian};

    \node (WVI) at (-3.8,-0.8) [child] {WYSIWYG};
    \node (WVE) at (-3.8,-1.8) [yellowc] {WAE};
    \node (WVC) at (-3.8,-2.8) [child] {Continuous};
    
    \draw (ND.east) -- (GS);
    \draw (ND.east) -- (BT);
    \draw (ND.east) -- (IT);
    \draw (ND.west) -- (EO);
    \draw (ND.west) -- (WV);
    
    \draw (GS) -- (AC);
    \draw (GS) -- (AN);
    
    \draw (AC) -- (ACB);
    \draw (AC) -- (ACM);
    
    \draw (AN) -- (ANO);
    \draw (AN) -- (ANM);
    
    \draw (IT) -- (ITY);
    \draw (IT) -- (ITN);
    
    \draw (ANM) -- (ANMI);
    \draw (ANM) -- (ANMC);
    
    \draw (BT) -- (BTP);
    \draw (BT) -- (BTT);
    \draw (BT) -- (BTE);
    
    \draw (EO) -- (EOF);
    \draw (EO) -- (EOFP);
    \draw (EO) -- (EOS);
    
    \draw (EOS) -- (EOSR);
    \draw (EOS) -- (EOSL);
    
    \draw (WV) -- (WVI);
    \draw (WV) -- (WVE);
    \draw (WV) -- (WVC);
\end{tikzpicture}
    \caption{\revv{Summary of the normative values encoded by Ranking with implicit bias (\citet{celis2020interventions}).}}
    \label{fig:mind:implicit_ranking}
\end{figure}
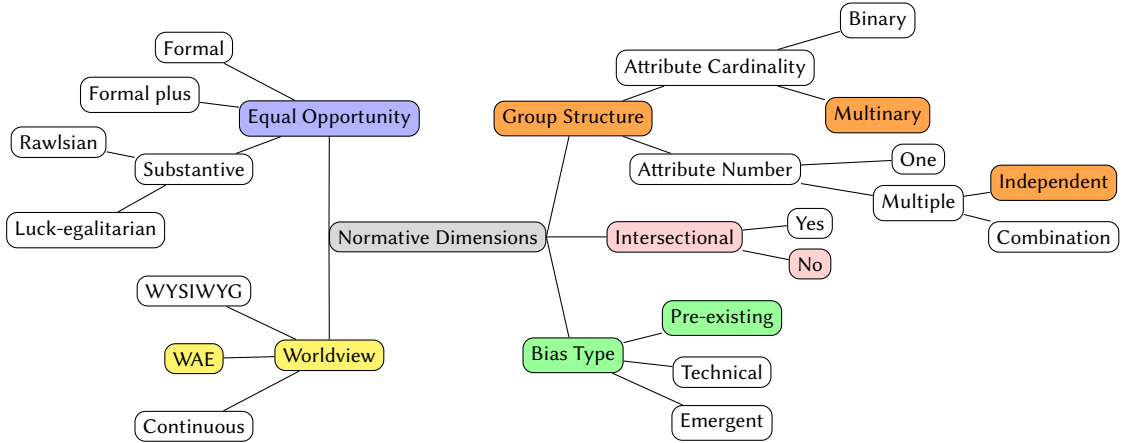

\spara{Insights} Both papers consider utility rather than diversity or fairness, and so cannot be classified according to one of our \revv{\eop} frameworks.  That said, the assumption made in the papers --- that candidates' true (unobserved) qualifications are drawn from the same score distribution --- is consistent with the WAE worldview. \rev{Figure~\ref{fig:mind:implicit_selection} and~\ref{fig:mind:implicit_ranking} summarize our analysis for the above two methods~\cite{kleinberg_et_al:LIPIcs:2018:8323} and~\cite{celis2020interventions}, respectively. The summary can be also found in Table~\ref{tbl:method-summary_score}.}

\subsubsection{\citet{yang2020causal}} \spara{Fairness definition and problem formalization} The authors define \emph{intersectional fairness for ranking} by modelling the causal effects of sensitive attributes on other variables, and then making rankers fairer by removing these effects. Their method, \cifrank, computes model-based counterfactuals to answer the question: ``What would this person's data look like if they had (or had not) been a Black woman (for example)?'' Counterfactual scores are computed by treating every candidate as though they had belonged to one specific intersectional subgroup.  Candidates are then ranked on counterfactual scores (for score-based rankers), or these scores are used to train a fair model (for rankers based on supervised learning).

Consider the hiring process of a moving company that has a dataset of applicants including their gender $G$, race $R$, weight-lifting test score $X$, and an overall qualification score $Y$ by which job candidates are ranked.  Figure~\ref{fig:cm} presents the structural causal model (SCM) that describes the data generation process. An SCM is a directed acyclic graph, where vertices represent (observed or latent) variables and edges indicate causal relationships from source to target vertices.  
The arrows pointing from $G$ (gender) and $R$ (race) directly to $Y$ encode the effect of ``direct'' discrimination.  Additionally, the SCM can encode indirect discrimination: note that $G$ and $R$ both impact $Y$ through weight-lifting ability $X$, called a ``mediator variable.''   A mediator may be designates as \emph{resolving} with respect to a sensitive variable, which means that we allow that mediator to carry the effect from the sensitive variable to the outcome.  For example, we may consider $X$ as a resolving on the path from gender $G$ to score $Y$.  Alternatively, a mediator may be designates as \emph{non-resolving}, which means that we consider the influence to be due to discrimination.  For example, we may consider $X$ as non-resolving on the path from race $R$ to score $Y$.  

The SCM, together with the information about which mediators are considered resolving, is given as input; it encodes the fairness objectives of the ranker. \cifrank will use the SCM to produce a ranking that is fair with respect to race, gender, and the intersectional subgroups of these categories. 

Let $\mathbf A$ denote the vector of sensitive attributes and let $\mathbf a$ denote a possible value. The counterfactual $Y_{\mathbf A \gets \mathbf a'}$ is computed by replacing the observed value of $\mathbf A$ with $\mathbf a'$ and then propagating this change through the DAG: any directed descendant of $\mathbf A$ has its value changed by computing the expectation for the new value of $\mathbf a'$, and this operation is iterated until it reaches all the terminal nodes that are descendants of any of the sensitive attributes $\mathbf A$. If a mediator variable is non-resolving, then its value will be set to its counterfactual value in the process.  If, however, it is designated as resolving, then we keep its observed value.

\cifrank considers a ranking $\hat \btau$ is counterfactually fair if, for all possible $x$ and pairs of vectors of actual and counterfactual sensitive attributes $a \neq a'$, respectively,
\begin{equation*}
\label{eq:cfranking}
    \begin{split}
        \mathbb P (\hat \btau(Y_{\mathbf A \gets \mathbf a}(U)) = k \mid \mathbf X = \mathbf x, \mathbf A = \mathbf a) \\
        = \mathbb P (\hat \btau(Y_{\mathbf A \gets \mathbf a'}(U)) = k \mid \mathbf X = \mathbf x, \mathbf A = \mathbf a)
    \end{split}
\end{equation*}
for any rank $k$, and with suitably randomized tie-breaking.

The causal model can be used to compute counterfactual scores $Y$ --- the scores that would have been assigned to the individuals if they belonged to one particular subgroup defined by fixed values of $R$ and $G$, while holding the weight lifting score $X$ fixed in the resolving case --- and then rank the candidates based on these scores.   The moving company can then interview or hire the highly ranked candidates, and this process would satisfy a causal and intersectional definition of fairness that corresponds to the hiring manager's explicitly stated goals.  

\spara{Experiments and observations}  The authors evaluated the performance of \cifrank on several real and synthetic datasets, including \emph{CSRankings}, \emph{COMPAS} and \emph{MEPS} (see details in Section~\ref{sec:datasets}). Results on synthetic datasets are provided to simulate different structural assumptions of the underlying causal model. The evaluation is done on two types of ranking tasks: score-based and supervised learning. 
The evaluation of score-based ranking tasks on real and synthetic datasets shows that \cifrank can be flexibly applied to different scenarios, including ones with mediating variables and numerical sensitive attributes. Counterfactually fair rankings that are produced by \cifrank compare reasonably to intuitive expectations we may have about intersectional fairness for those examples, while paying a small cost in terms of the utility loss.
The evaluation of rankers based on supervised learning on synthetic datasets shows that \cifrank can be used as a preprocessing fairness intervention to produce counterfactually fair training and test data.

\spara{Insights} \cifrank admits \emph{multiple sensitive attributes}, and is specifically designed for intersectional concerns, and so is appropriate when it is important to account for potential discrimination along two or more features.  The method supports \emph{multinary sensitive attributes}, such as non-binary gender and ethnic group membership.  The method is concerned with \emph{pre-existing bias} that in turn leads to disparities in outcomes.
\begin{figure}[h]
    \centering
    \begin{tikzpicture}[scale=0.9,align=center]
    \tikzset{
        root/.style = {rectangle, rounded corners, draw=black, fill=gray!30, font=\small\sffamily},
        child/.style = {rectangle, rounded corners,draw=black, font=\small\sffamily},
        orangec/.style = {rectangle, rounded corners, draw=black, fill=orange!70, font=\small\sffamily},
        greenc/.style = {rectangle, rounded corners, draw=black, fill=green!40, font=\small\sffamily},
        bluec/.style = {rectangle, rounded corners, draw=black, fill=blue!30, font=\small\sffamily},
        yellowc/.style = {rectangle, rounded corners, draw=black, fill=yellow!70, font=\small\sffamily},
        pinkc/.style = {rectangle, rounded corners, draw=black, fill=pink!70, font=\small\sffamily},
    }
    \node (ND) at (-0.2,0) [root] {Normative Dimensions};    
    \node (GS) at (1.8,1.75) [orangec] {Group Structure};    
    \node (BT) at (1.8,-1.75) [greenc] {Bias Type};    
    \node (EO) at (-1.8,1.75) [bluec] {Equal Opportunity};    
    \node (WV) at (-1.8,-1.75) [yellowc] {Worldview};  
    \node (IT) at (3.3,0.0) [pinkc] {Intersectional};
    
    \node (AC) at (3.9,2.5) [child] {Attribute Cardinality};
    \node (ACB) at (6.3,3.2) [child] {Binary};
    \node (ACM) at (6.3,1.8) [orangec] {Multinary};
    
    \node (AN) at (3.9,1.0) [child] {Attribute Number};
    \node (ANO) at (6.9,1.15) [child] {One};
    \node (ANM) at (6.9,0.5) [child] {Multiple};
    
    \node (ANMI) at (8.9,0.8) [orangec] {Independent};
    \node (ANMC) at (8.9,0.0) [child] {Combination};
    
    \node (ITY) at (5.3,0.2) [pinkc] {Yes};
    \node (ITN) at (5.3,-0.4) [child] {No};
    
    \node (BTP) at (4.0,-1.2) [greenc] {Pre-existing};
    \node (BTT) at (4.0,-2.0) [child] {Technical};
    \node (BTE) at (4.0,-2.75) [child] {Emergent};
    
    \node (EOF) at (-3.8,2.8) [child] {Formal};
    \node (EOFP) at (-4.6,2.1) [child] {Formal plus};
    \node (EOS) at (-3.8,1.0) [child] {Substantive};
    
    \node (EOSR) at (-5.8,1.4) [bluec] {Rawlsian};
    \node (EOSL) at (-5.4,0.1) [child] {Luck-egalitarian};

    \node (WVI) at (-3.8,-0.8) [child] {WYSIWYG};
    \node (WVE) at (-3.8,-1.8) [yellowc] {WAE};
    \node (WVC) at (-3.8,-2.8) [child] {Continuous};
    
    \draw (ND.east) -- (GS);
    \draw (ND.east) -- (BT);
    \draw (ND.east) -- (IT);
    \draw (ND.west) -- (EO);
    \draw (ND.west) -- (WV);
    
    \draw (GS) -- (AC);
    \draw (GS) -- (AN);
    
    \draw (AC) -- (ACB);
    \draw (AC) -- (ACM);
    
    \draw (AN) -- (ANO);
    \draw (AN) -- (ANM);
    
    \draw (IT) -- (ITY);
    \draw (IT) -- (ITN);
    
    \draw (ANM) -- (ANMI);
    \draw (ANM) -- (ANMC);
    
    \draw (BT) -- (BTP);
    \draw (BT) -- (BTT);
    \draw (BT) -- (BTE);
    
    \draw (EO) -- (EOF);
    \draw (EO) -- (EOFP);
    \draw (EO) -- (EOS);
    
    \draw (EOS) -- (EOSR);
    \draw (EOS) -- (EOSL);
    
    \draw (WV) -- (WVI);
    \draw (WV) -- (WVE);
    \draw (WV) -- (WVC);
\end{tikzpicture}
    \caption{\revv{Summary of the normative values encoded by Causal intersectionally fair ranking (\citet{yang2020causal}).}}
    \label{fig:mind:causal}
\end{figure}
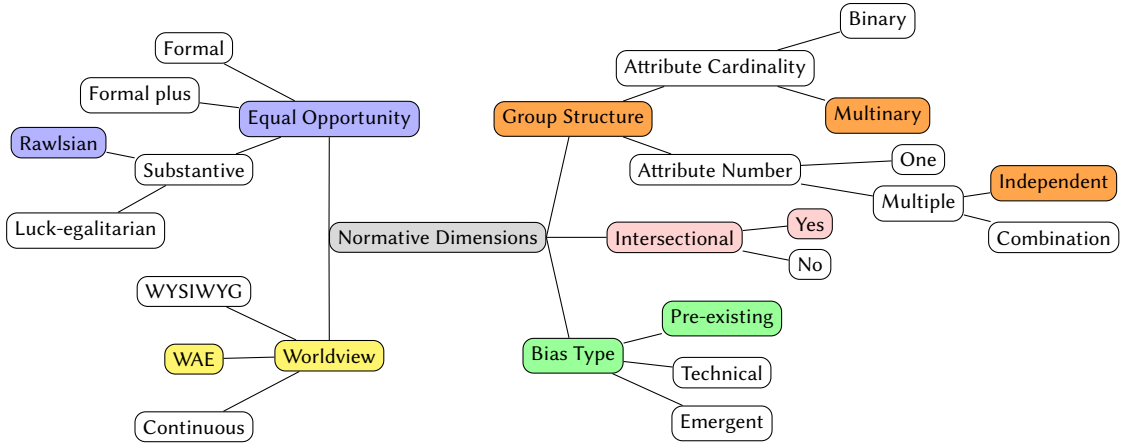
The method focuses on \emph{equality of outcome} and takes the \emph{WAE} worldview. \revv{Under the assumption that the goal is to equalize opportunity over a lifetime, the method is consistent with \emph{substantive \eop}}.  It gives the decision maker the flexibility to specify which impacts of which sensitive attribute to mitigate, and which to allow to persist.  This is done through the mediator mechanism.  A mediator $X$ may be considered resolving or not; this decision can be made separately for different sensitive attributes, and the relative strengths of causal influences of sensitive attributes on both $X$ and $Y$ can vary, creating potential for explanatory nuance.  \revv{This method supports fairness interventions that attempt to model what an individuals' qualifications \emph{would have looked like}, in a world where equally talented people have equal prospects of success.  For this reason, we classify it as Rawls's Fair \eop}. 
\rev{Figure~\ref{fig:mind:causal} and Table~\ref{tbl:method-summary_score} summarize our analysis.}

\subsection{Intervening on the Ranking Function}
\label{sec:fair_db:geo}

To motivate the methods discussed in this section, let us return to our running example described in Section~\ref{sec:intro:example}, and shown in Figure~\ref{fig:admissions}, and consider a college admissions officer who is designing a ranking scheme to evaluate a pool of applicants, each with several potentially relevant attributes.   For simplicity, let us focus on two of these attributes, high school GPA $X_1$, and verbal SAT $X_2$, and assume that they are appropriately normalized and standardized. Suppose that our fairness criterion is that the admitted class comprise at least 40\% women.  The admissions officer may believe \emph{a priori} that $X_1$ and $X_2$ should carry an approximately equal weight, computing the score of an applicant $a \in \candidateSet{}$ as $f(a) = 0.5 X_1 + 0.5 X_2$,  ranking the applicants, and returning the top 500 individuals. Upon inspection, it may be determined that an insufficient number of women is returned among the top-$k$: at least 200  were expected  and only 150 were returned, violating the fairness constraint. 

A possible mitigation is to identify an alternative scoring function $\Tilde{f}$  that, when applied to $\candidateSet{}$, meets the fairness constraint and is close to the original function $f$ in terms of attribute weights, thereby reflecting the admission officer’s notion of quality.  To arrive at such a function, the admissions officer would try a new scoring function, check whether the result meets the fairness criterion, and, if necessary, repeat.  After a few cycles of such interaction, the admissions officer may choose $f(a) = 0.45 X_1 + 0.55 X_2$ as the final scoring function.  The work of~\citet{asudeh2019designing} automates this process; the authors use a combinatorial geometry approach to efficiently explore the search space and identify a fair scoring function $\Tilde{f}$ in the neighborhood of $f$, if one exists.

\spara{Fairness definition and problem formalization} Let us assume that a dataset of candidates $\candidateSet{}$ is given, along with a linear ranking function $f$, specified by a weight vector $\Vec{w}$.  The goal is to find a ranking function $\Tilde{f}$ that is both close to $f$ in terms of the angular distance between the weight vectors of $f$ and $\Tilde{f}$, and fair according to a fairness oracle $\oracle$. 

The main technical contribution of the work is in establishing a correspondence between the space of linear ranking functions and the rankings of items from a given dataset $\candidateSet{}$ induced by these functions.  This characterization is based on the notion of an \emph{ordering exchange} that partitions the space of linear functions into disjoint regions.  Intuitively, while there is an infinite number of linear ranking functions to explore, only those of them that change the relative order among some pair of items $a, b \in \candidateSet{}$ need to be considered, because if a ranking is unchanged, then the fairness oracle $\oracle$ will not change its answer from \emph{false} to \emph{true}.  Based on this observation, the authors develop exact algorithms to determine boundaries that partition the space into regions where the desired fairness constraint is satisfied, called satisfactory regions, and regions where the constraint is not satisfied. They also develop approximation algorithms to efficiently identify and index satisfactory regions, and introduce sampling heuristics for  on-the-fly  processing in cases where the size of $\candidateSet{}$ or the number of scoring attributes are large.

\spara{Experiments and observations}  While the fairness model is general, the authors focus their experimental evaluation on proportional representation constraints that bound the number of items belonging to a particular group at the top-$k$, for some given value of $k$.  Proposed methods are evaluated on the COMPAS~\cite{COMPASData} and DOT~\cite{DOTData} datasets (see Section~\ref{sec:datasets} for details), and with two sets of fairness measures: (1) proportional representation on a single multinary protected attribute and (2) proportional representation on multiple, possibly overlapping, protected attributes. They study both how intuitive the results are --- how close a fair ranking function is to the original --- and how efficiently results can be computed in this computationally challenging setting. 

\begin{figure}[h]
    \centering
    \begin{tikzpicture}[scale=0.9,align=center]
    \tikzset{
        root/.style = {rectangle, rounded corners, draw=black, fill=gray!30, font=\small\sffamily},
        child/.style = {rectangle, rounded corners,draw=black, font=\small\sffamily},
        orangec/.style = {rectangle, rounded corners, draw=black, fill=orange!70, font=\small\sffamily},
        greenc/.style = {rectangle, rounded corners, draw=black, fill=green!40, font=\small\sffamily},
        bluec/.style = {rectangle, rounded corners, draw=black, fill=blue!30, font=\small\sffamily},
        yellowc/.style = {rectangle, rounded corners, draw=black, fill=yellow!70, font=\small\sffamily},
        pinkc/.style = {rectangle, rounded corners, draw=black, fill=pink!70, font=\small\sffamily},
    }
    \node (ND) at (-0.2,0) [root] {Normative Dimensions};    
    \node (GS) at (1.8,1.75) [orangec] {Group Structure};    
    \node (BT) at (1.8,-1.75) [greenc] {Bias Type};    
    \node (EO) at (-1.8,1.75) [bluec] {Equal Opportunity};    
    \node (WV) at (-1.8,-1.75) [yellowc] {Worldview};  
    \node (IT) at (3.3,0.0) [pinkc] {Intersectional};
    
    \node (AC) at (3.9,2.5) [child] {Attribute Cardinality};
    \node (ACB) at (6.3,3.2) [child] {Binary};
    \node (ACM) at (6.3,1.8) [child] {Multinary};
    
    \node (AN) at (3.9,1.0) [child] {Attribute Number};
    \node (ANO) at (6.9,1.15) [child] {One};
    \node (ANM) at (6.9,0.5) [child] {Multiple};
    
    \node (ANMI) at (8.9,0.8) [child] {Independent};
    \node (ANMC) at (8.9,0.0) [child] {Combination};
    
    \node (ITY) at (5.3,0.2) [child] {Yes};
    \node (ITN) at (5.3,-0.4) [child] {No};
    
    \node (BTP) at (4.0,-1.2) [greenc] {Pre-existing};
    \node (BTT) at (4.0,-2.0) [child] {Technical};
    \node (BTE) at (4.0,-2.75) [child] {Emergent};
    
    \node (EOF) at (-3.8,2.8) [child] {Formal};
    \node (EOFP) at (-4.6,2.1) [child] {Formal plus};
    \node (EOS) at (-3.8,1.0) [child] {Substantive};
    
    \node (EOSR) at (-5.8,1.4) [child] {Rawlsian};
    \node (EOSL) at (-5.4,0.1) [child] {Luck-egalitarian};

    \node (WVI) at (-3.8,-0.8) [child] {WYSIWYG};
    \node (WVE) at (-3.8,-1.8) [child] {WAE};
    \node (WVC) at (-3.8,-2.8) [child] {Continuous};
    
    \draw (ND.east) -- (GS);
    \draw (ND.east) -- (BT);
    \draw (ND.east) -- (IT);
    \draw (ND.west) -- (EO);
    \draw (ND.west) -- (WV);
    
    \draw (GS) -- (AC);
    \draw (GS) -- (AN);
    
    \draw (AC) -- (ACB);
    \draw (AC) -- (ACM);
    
    \draw (AN) -- (ANO);
    \draw (AN) -- (ANM);
    
    \draw (IT) -- (ITY);
    \draw (IT) -- (ITN);
    
    \draw (ANM) -- (ANMI);
    \draw (ANM) -- (ANMC);
    
    \draw (BT) -- (BTP);
    \draw (BT) -- (BTT);
    \draw (BT) -- (BTE);
    
    \draw (EO) -- (EOF);
    \draw (EO) -- (EOFP);
    \draw (EO) -- (EOS);
    
    \draw (EOS) -- (EOSR);
    \draw (EOS) -- (EOSL);
    
    \draw (WV) -- (WVI);
    \draw (WV) -- (WVE);
    \draw (WV) -- (WVC);
\end{tikzpicture}
    \caption{\revv{Summary of the normative values encoded by Designing fair ranking functions (\citet{asudeh2019designing}).}}
    \label{fig:mind:design}
\end{figure}
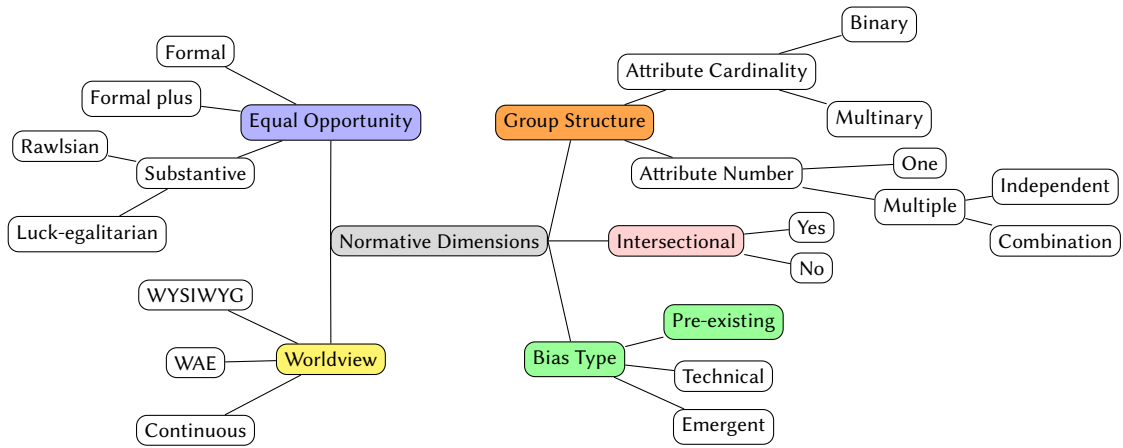

\spara{Insights} The fairness oracle $\oracle$ is treated as a black box: given a dataset $\candidateSet{}$ and a ranking function $f$, it returns \emph{true} if the ranking of $\candidateSet{}$ by $f$ meets fairness criteria and so is satisfactory, and returns \emph{false} otherwise.  The oracle is deterministic, and no further assumptions are made about the type of fairness criteria it encodes.  Because of this black-box treatment of the fairness objective, the method makes no commitment to worldview (WYSWYG or WAE) or \revv{\eop} framework, and it is not restricted in terms of group structure: the number of sensitive attributes, their cardinality, and the method by which multiple sensitive attributes are handled.  Despite this flexibility, the authors target their approach specifically at pre-existing bias. \rev{Figure~\ref{fig:mind:design} and Table~\ref{tbl:method-summary_score} summarize our analysis.}

\section{Fair Supervised learning}
\label{sec:fair_ir}

In this section, we present several methods for fairness in learning-to-rank and information retrieval. We will continue to use the notation that we introduced in Section~\ref{sec:intro:learned} wherever appropriate to illustrate the commonalities of the fields.

As we did in the previous section, we will categorize the technical methods according to four normative frameworks, discussed in detail in Section~\ref{sec:02-four-frameworks}.  Recall Figure~\ref{fig:normative-dimensions}, which gives a structural overview of these frameworks and their sub-categories in the form of a \emph{mind map}. For each method, we will highlight which normative choices they make by representing them on this mind map. Table~\ref{tbl:method-summary_ir} summarizes the categorization of all methods that are surveyed in the remainder of this document.

%

\begin{table*}[ht!]
    \caption{Summary of method classification. ``Pre-, in-'' and ``post-proc'' refer to whether a method can be classified as pre-, in- or post-processing. ``Binary'' vs. ``multinary'' tells whether the method can handle two or more protected groups per one attribute at a time (e.g. young/old is a binary manifestation of the attribute ``age'', while kid/teen/adult/old is a multinary manifestation of said attribute). Since none of the methods in this section of the survey handle intersectional discrimination, and we omit that column here.   
    See Section~\ref{sec:02-four-frameworks} for a detailed description of the classification framework.}
    \small
    \begin{tabular}{p{0.15\textwidth}p{.1\textwidth}p{0.18\textwidth}p{0.12\textwidth}p{0.1\textwidth}p{0.16\textwidth}l}
        \toprule
         \textbf{Method} & \textbf{Mitigation Point} & \textbf{Group structure} & \textbf{Bias} & \textbf{Worldview} & \textbf{\eop Framework} \\ \midrule 
         iFair~\cite{lahoti2019ifair} & pre-proc. & \makecell[l]{multiple multinary attr.;\\  \revv{independent}} & \revv{technical} & \revv{WYSWYG} & \revv{formal} \\
         \rowcolor{Lightgray}
         DELTR~\cite{zehlike2018reducing} & in-proc. &  \revv{one binary attr.} & pre-existing & WAE & luck-egalitarian \\
         Fair-PG-Rank~\cite{singh2019policy} & in-proc & \revv{one binary attr.} & technical & WYSIWYG & formal \\
         \rowcolor{Lightgray}
         \makecell[l]{Pairwise Ranking\\ Fairness~\cite{Beutel:2019:FRR:3292500.3330745}} & in-proc. & \revv{one binary attr.} & \revv{?} & \revv{WYSIWYG} & \revv{formal-plus} \\
         \algofair~\cite{zehlike2017fa} \&~\cite{zehlike2022fair} & post-proc. & \makecell[l]{one multinary attr.;\\ \revv{combination}} & pre-existing & continuous & \makecell[l]{formal /\\ luck-egalitarian} \\
         \rowcolor{Lightgray}
         \makecell[l]{Fair Ranking\\ at LinkedIn~\cite{geyik2019fairness}} & post-proc. & \makecell[l]{one multinary attr.;\\ \revv{combination}} & \makecell[l]{pre-existing; \revv{technical}} & \revv{continuous} & \makecell[l]{\revv{none} /\\ \revv{luck-egalitarian}\\ \revv{(1 sensitive attr.)}} \\
         CFA$\theta$~\cite{zehlike2017matching} & post-proc. & \makecell[l]{multiple binary attr.;\\ \revv{combination}} & pre-existing & continuous & \makecell[l]{formal /\\ \revv{substantive}} \\
         \rowcolor{Lightgray}
         \makecell[l]{Fairness of\\Exposure~\cite{singh2018fairness}}  & post-proc. & one binary attr. & \makecell[l]{pre-existing/\\ technical} & \makecell[l]{WYSIWYG /\\ WAE} & \makecell[l]{formal /\\ luck-egalitarian} \\
         \makecell[l]{Equity of\\ Attention~\cite{biega2018equity}} & post-proc. & \makecell[l]{one multinary attr.;\\ independent} & \makecell[l]{technical /\\  emergent} & WYSIWYG & formal\\
         \bottomrule
    \end{tabular}
    \label{tbl:method-summary_ir}
\end{table*}

\subsection{Pre-Processing Methods: Learning Fair Training Data}
Pre-processing approaches are usually concerned with biases in the training data which they try to mitigate. 
Those biases can be of all three types: pre-existing biases appear in any data collection procedure in various ways.
It is the way we interrogate, the decision which information we collect and which not, etc. 
Technical bias makes its way into data as rounding errors, different number and category encoding or the strategy choice how to handle missing values.
Emergent bias arises when data is used in a different way than intended during collection.

\noindent General advantages of pre-processing methods are:
\begin{itemize}
	\item Pre-processing methods consider fairness as first concern in the machine learning pipeline.
	\item Most in- and post-processing methods rely on the availability of group labels during or after training, respectively.
	Pre-processing approaches instead commonly operate on a distance measure between individuals which allows to be agnostic to group membership. 
	It is sufficient to define who should be similar to whom, based on the features that are available.
	\item Additionally it is possible to control for certain types of fairness across groups, even if only sparse information about group membership is available~\cite{lahoti2019operationalizing}.
\end{itemize}
General disadvantages are: 
\begin{itemize}
	\item Machine learning methods that rely on a separate feature engineering step are not applicable because the features identified by domain experts may be rendered meaningless, if fair representations are learned from the raw data.
	\item Current methods only operationalize individual fairness and treat group fairness as a special case of it.
\end{itemize}

\subsubsection{iFair \cite{lahoti2019ifair}} 
\newcommand{\datatabLahotiexample}{
	\centering
	 \small
	\begin{tabular}{|c||c|c||c|c|}
		\hline
		\rowcolor[HTML]{C0C0C0} 
		candidate & $\sensAttr_1$ (sex) & $\sensAttr_2$ (race) & $X_1$ (GPA) & $X_2$ (SAT)  \\ \hline
		$\val{e}$  & male      & White     & 5   &  4   \\ \hline
		$\val{f}$  & male      & Asian    & 5 & 4 \\ \hline
		$\val{g}$  & female      & Black     & 5 & 3\\ \hline
		$\val{h}$  & female      & White    & 5 & 3   \\ \hline
	\end{tabular}
}
\begin{figure}[t!]
    \datatabLahotiexample
    \caption[Intuitive functioning of a pre-processing algorithm]{\rev{Dataset of college applicants.  The goal is to find a fair feature representation and to ensure that a learning-to-rank method considers candidates based only on their non-sensitive features (here, $X_1$ and $X_2$). Assuming Manhattan distance as the distance measure $d$ between two candidates, candidates $\val{e}$ and $\val{f}$, as well as $\val{g}$ and $\val{h}$ should have the same fair feature representation $\fairFeatOfCand{}$, where $\sensAttr_1$ and $\sensAttr_2$ are irrelevant for the ranked outcome.}
    }
    \label{fig:lahoti_example}
\end{figure}
\spara{Fairness Definition. }
This work operates on an individual fairness objective to learn fair representations of training data points. 
It is based on the fairness definition by~\citet{dwork2012fairness}, which states that similar individuals should be treated similarly.
The goal is to transform an input record $\featureSet_a$ (the feature vector for candidate $a$) into fairer data representations $\fairFeatOfCand{a}$ using a mapping $\phi$, such that two individuals $a$ and $b$, who are indistinguishable on their non-sensitive attributes $\featureSet \setminus \sensAttrSet$ (marked by $\featureSet^*$) should also be nearly indistinguishable in their fair representations $\phi(\featureSet_{a})$ (where sensitive attributes are included):
\begin{equation*}
\left|d\left(\phi(\featureSet_{a}{}), \phi(\featureSet_{b}{})\right) - d(\npFeatOfCand{a}, \npFeatOfCand{b})\right| \leq \epsilon
\end{equation*}
Note that this definition assumes that a similarity measure $d$ is available that can correctly (and free of bias) capture the differences between two individuals.
As the paper uses the family of Minkowsky $p$-metrics, let us assume we choose the Minkowsky metric with $p=1$  (\ie the Manhattan distance as $d$).
\rev{Reconsidering our college admission example, in Figure~\ref{fig:lahoti_example} we see that candidates $\val{e}$ and $\val{f}$, as well as $\val{g}$ and $\val{h}$ have a distance of 0 in their non-protected features:
$ d(\npFeatOfCand{\val{e}}, \npFeatOfCand{\val{f}}) = d(\npFeatOfCand{\val{g}}, \npFeatOfCand{\val{h}}) = 0$.
When comparing candidates across groups we see that the female group shows a Manhattan distance of 1 to the male group:
$ d(\npFeatOfCand{\val{e}}, \npFeatOfCand{\val{g}}) = d(\npFeatOfCand{\val{f}}, \npFeatOfCand{\val{g}}) = d(\npFeatOfCand{\val{e}}, \npFeatOfCand{\val{h}}) = d(\npFeatOfCand{\val{f}}, \npFeatOfCand{\val{h}}) = 1$.
The proposed algorithm, \emph{iFair}, would create a new feature set $\phi(\featureSet)$ that preserves those distances \emph{and} includes sensitive attributes \revv{ in such as way as to break correlations with the non-sensitive attributes}.
In our example, $\sensAttr_1$ correlates with $X_{2}$, which may be picked up by a ranking model.
To avoid this, iFair would assign non-correlating values to $\sensAttr_1$, \revv{for example, by swapping the values of $\sensAttr_1$ for candidates $\val{b}$ and $\val{d}$}.}

\begin{figure}[b!]
    \centering
    \begin{tikzpicture}[scale=0.9,align=center]
    \tikzset{
        root/.style = {rectangle, rounded corners, draw=black, fill=gray!30, font=\small\sffamily},
        child/.style = {rectangle, rounded corners,draw=black, font=\small\sffamily},
        orangec/.style = {rectangle, rounded corners, draw=black, fill=orange!70, font=\small\sffamily},
        greenc/.style = {rectangle, rounded corners, draw=black, fill=green!40, font=\small\sffamily},
        bluec/.style = {rectangle, rounded corners, draw=black, fill=blue!30, font=\small\sffamily},
        yellowc/.style = {rectangle, rounded corners, draw=black, fill=yellow!70, font=\small\sffamily},
        pinkc/.style = {rectangle, rounded corners, draw=black, fill=pink!70, font=\small\sffamily},
    }
    \node (ND) at (-0.2,0) [root] {Normative Dimensions};    
    \node (GS) at (1.8,1.75) [orangec] {Group Structure};    
    \node (BT) at (1.8,-1.75) [greenc] {Bias Type};    
    \node (EO) at (-1.8,1.75) [bluec] {Equal Opportunity};    
    \node (WV) at (-1.8,-1.75) [yellowc] {Worldview};  
    \node (IT) at (3.3,0.0) [pinkc] {Intersectional};
    
    \node (AC) at (3.9,2.5) [child] {Attribute Cardinality};
    \node (ACB) at (6.3,3.2) [child] {Binary};
    \node (ACM) at (6.3,1.8) [orangec] {Multinary};
    
    \node (AN) at (3.9,1.0) [child] {Attribute Number};
    \node (ANO) at (6.9,1.15) [child] {One};
    \node (ANM) at (6.9,0.5) [child] {Multiple};
    
    \node (ANMI) at (8.9,0.8) [orangec] {Independent};
    \node (ANMC) at (8.9,0.0) [child] {Combination};
    
    \node (ITY) at (5.3,0.2) [child] {Yes};
    \node (ITN) at (5.3,-0.4) [pinkc] {No};
    
    \node (BTP) at (4.0,-1.2) [child] {Pre-existing};
    \node (BTT) at (4.0,-2.0) [greenc] {Technical};
    \node (BTE) at (4.0,-2.75) [child] {Emergent};
    
    \node (EOF) at (-3.8,2.8) [bluec] {Formal};
    \node (EOFP) at (-4.6,2.1) [child] {Formal plus};
    \node (EOS) at (-3.8,1.0) [child] {Substantive};
    
    \node (EOSR) at (-5.8,1.4) [child] {Rawlsian};
    \node (EOSL) at (-5.4,0.1) [child] {Luck-egalitarian};

    \node (WVI) at (-3.8,-0.8) [yellowc] {WYSIWYG};
    \node (WVE) at (-3.8,-1.8) [child] {WAE};
    \node (WVC) at (-3.8,-2.8) [child] {Continuous};
    
    \draw (ND.east) -- (GS);
    \draw (ND.east) -- (BT);
    \draw (ND.east) -- (IT);
    \draw (ND.west) -- (EO);
    \draw (ND.west) -- (WV);
    
    \draw (GS) -- (AC);
    \draw (GS) -- (AN);
    
    \draw (AC) -- (ACB);
    \draw (AC) -- (ACM);
    
    \draw (AN) -- (ANO);
    \draw (AN) -- (ANM);
    
    \draw (IT) -- (ITY);
    \draw (IT) -- (ITN);
    
    \draw (ANM) -- (ANMI);
    \draw (ANM) -- (ANMC);
    
    \draw (BT) -- (BTP);
    \draw (BT) -- (BTT);
    \draw (BT) -- (BTE);
    
    \draw (EO) -- (EOF);
    \draw (EO) -- (EOFP);
    \draw (EO) -- (EOS);
    
    \draw (EOS) -- (EOSR);
    \draw (EOS) -- (EOSL);
    
    \draw (WV) -- (WVI);
    \draw (WV) -- (WVE);
    \draw (WV) -- (WVC);
\end{tikzpicture}
    \caption{\revv{Summary of the normative values encoded by iFair (\citet{lahoti2019ifair}). This translates into normative choices that are implicitly taken when applying the method. Note, however, that this analysis is highly dependent on the choice of the metric $d$, which measures the similarity of two individuals.}}
    \label{fig:pre-proc:lahoti}
\end{figure}
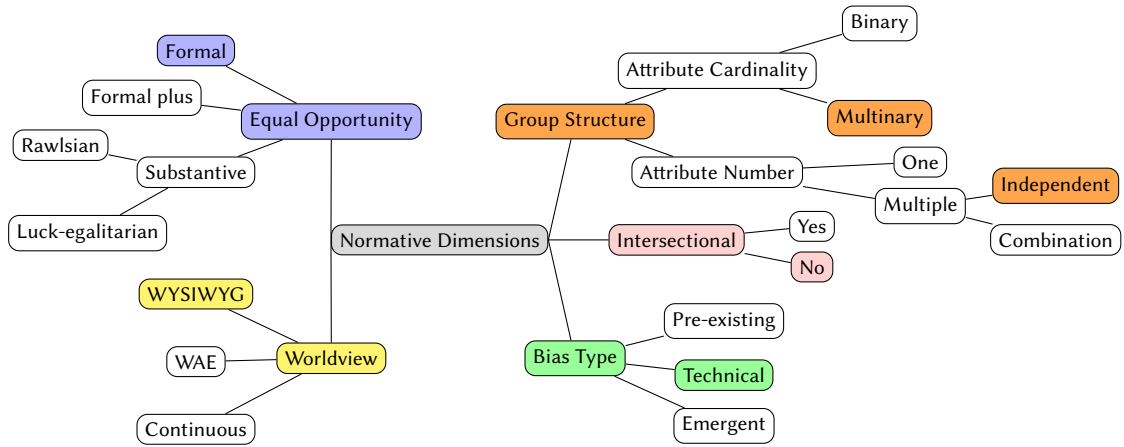
\spara{Insights.} Though not clearly stated, the wording suggests that attributes are measured in observable space \os and the definition seeks to reduce technical bias.
Depending on the choice of distance metric $d$, the method would potentially be capable of learning representations that ignore \emph{all} information about group membership, even if it is encoded partly in the non-protected features. 
In this case it would assign low weights to those non-protected features that indirectly encode protected ones, thus suggesting a leaning towards WAE.
However, the choice of Minkowsky metrics, where distances are measured in terms of absolute numbers,
suggests that the authors assume construct space $CS \sim OS$ and, hence, a leaning to a WYSIWYG worldview and to formal equality of opportunity.
As stated above, the actual normative values that are incorporated into this method depend on the choice of $d$.

Note also that the authors define fairness based on individuals, and group notions are not mentioned at all.
While we can assume that individuals may have more than one protected attribute, it is not clear whether and how intersectional discrimination is a concern to the authors.
Figure~\ref{fig:pre-proc:lahoti} and Table~\ref{tbl:method-summary_ir} summarize our analysis.

The authors do not address the question \revv{of whether Minkowsky metrics are prone to reproducing biases from training data, which may arise from biased observation processes.}
They tackle this problem in a follow-up work~\cite{lahoti2019operationalizing}, where the distance metric is replaced by a \emph{fairness graph} that captures pairwise similarities between individuals.
In the graph, a node represents an individual $a$, and an edge between two nodes indicates that these individuals are to be considered similar. 
This approach has two advantages: it allows a comparison of an individual's non-protected attributes across different domains (\eg the h-index of a successful researcher in programming languages is typically lower than that of a successful researcher in machine learning), and it allows for sparse similarity judgments, as individuals can be grouped into clusters based on their in-group relevance scores (\eg the top-10\% in the female group).

We do not further present~\citet{lahoti2019operationalizing} here, because the paper focuses on classification tasks in its experimental section.
\revv{Pre-processing methods claim to be application-agnostic, however, to the best of our knowledge,~\citet{lahoti2019ifair} is the only work to-date that has been shown to work for ranking tasks.}

\spara{Algorithm.}
The problem is formalized as a probabilistic clustering problem: given $K$ clusters of similar individuals (with $K < n$), each is represented by a prototype vector $v_k$.
A candidate record $\featureSet_{a}{}$ is assigned to one of the $v_k$ based on a record-specific probability distribution $P_a$ that reflects the distances of the record from each of the prototype vectors, and thus forms the fair representation:
\begin{equation*}
\phi(\featureSet_{a}{}) = \fairFeatOfCand{a} = \sum_{k=1..K} P_{ak} \cdot v_k
\end{equation*}
This is used to formalize a utility objective that ensures a low reconstruction loss, and a fairness objective that demands that $\phi$ should preserve pair-wise distances on non-protected attributes between data records:
\begin{equation*}
L_{\operatorname{util}} (\mathbf X, \widetilde{\mathbf X}) = \sum_{a \in \candidateSet{}} || \featureSet_{a}{}- \fairFeatOfCand{a} ||_2
\end{equation*}
\begin{equation*}
L_{\operatorname{fair}} (\mathbf X, \widetilde{\mathbf X}) = \sum_{a, b \in \candidateSet{}} \left(d(\fairFeatOfCand{a}, \fairFeatOfCand{b}) - d(\npFeatOfCand{a}, \npFeatOfCand{b}) \right)^2
\end{equation*}
The two objectives are combined into an objective function that the algorithm optimizes \revv{using} gradient descent:
\begin{equation*}
L = \lambda \cdot L_{\operatorname{util}} (\mathbf X, \widetilde{\mathbf X}) + \mu \cdot L_{\operatorname{fair}} (\mathbf X, \widetilde{\mathbf X})
\end{equation*}
The algorithm supports multiple groups. 

\spara{Experiments.} 
\revv{Experiments are performed on five real-world datasets and one synthetic dataset.  Of these, XING~\cite{XINGData} and AirBnB~\cite{AirBnBData} are used for ranking tasks.
%
%
Experiment on synthetic data} show that representations learned by iFair remain nearly the same for all configurations, irrespective of changes in group membership.
\revv{This means that changing the value of the sensitive attribute does not influence the learned representation, and so a model trained on such a representation will not learn any correlations between the sensitive attributes and other attributes.} 

The results show that applying learning algorithms on representations learned by iFair leads to more consistent decisions w.r.t. the distribution of items across a ranking than when applying the same algorithm to the original data. 
This means that two items with similar non-protected features will receive similar visibility in the resulting ranking.

\subsection{In-Processing Methods: Learning a Fair Model}
In-processing fair ranking methods extend the objective function of a learning-to-rank algorithm by a fairness term. 
Thus, the algorithm's optimization problem consists of an accuracy objective \emph{and} a fairness objective, and the method learns to find the best balance between these two. 
General advantages are:
\begin{itemize}
    \item In-processing methods yield better trade-offs between accuracy and fairness than post-processing \revv{methods}, because finding this balance is at the heart of their learning objective~\cite{zehlike2018reducing}.
    \item \revv{In-processing methods} are capable of handling different types of underlying biases without knowing which particular type is present (see Section~\ref{subsubsec:DELTR}).
\end{itemize}
General disadvantages are:
\begin{itemize}
    \item The impact of the fairness objective on the resulting ranking \revv{computed by an in-processing method} is less apparent than \revv{for} a post-processing method. The latter can make changes directly visible, while in the former case, separate models would have to be trained.
    \item Because of their goal to balance between fairness and accuracy, it is less clear what philosophical framework and worldview underlies fairness-aware models.
\end{itemize}

\subsubsection{\deltr \cite{zehlike2018reducing}.}
\label{subsubsec:DELTR}
\spara{Fairness Definition.} 
This method perceives unfairness as disparities in exposure, \revv{represented by the average visibility of a group} (see Section~\ref{sec:intro:learned} and Eq.~\ref{eq:exp}). 
The exposure of a document is defined as its probability~$P_{\predScore^{\query}} \left(\featOfCand{a}\right)$ to appear in the top position of a ranking for query $\query$:
\begin{equation}
\operatorname{Exposure}\left(\featOfCand{a}|P_{\predScore^{\query}}\right) = P_{\predScore^{\query}}\left(\featOfCand{a}\right) \cdot v_1
\end{equation}
where $ v_1 $ is the \emph{position bias} of position 1, indicating its relative importance for users of a ranking system~\cite{jarvelin2002cumulated}.
The exposure of group $\group{}$ is hence the average probability of its members to appear in the top position:
\begin{equation}
\operatorname{Exposure}(\group{}|P_{\predScore^{\query}}) = \frac{1}{|\group{}|} \sum_{\featOfCand{a} \in \group{}} \operatorname{Exposure}\left(\featOfCand{a}|P_{\predScore^{\query}}\right)
\end{equation}

\begin{figure}[b!]
    \centering
    \begin{tikzpicture}[scale=0.9,align=center]
    \tikzset{
        root/.style = {rectangle, rounded corners, draw=black, fill=gray!30, font=\small\sffamily},
        child/.style = {rectangle, rounded corners,draw=black, font=\small\sffamily},
        orangec/.style = {rectangle, rounded corners, draw=black, fill=orange!70, font=\small\sffamily},
        greenc/.style = {rectangle, rounded corners, draw=black, fill=green!40, font=\small\sffamily},
        bluec/.style = {rectangle, rounded corners, draw=black, fill=blue!30, font=\small\sffamily},
        yellowc/.style = {rectangle, rounded corners, draw=black, fill=yellow!70, font=\small\sffamily},
        pinkc/.style = {rectangle, rounded corners, draw=black, fill=pink!70, font=\small\sffamily},
    }
    \node (ND) at (-0.2,0) [root] {Normative Dimensions};    
    \node (GS) at (1.8,1.75) [orangec] {Group Structure};    
    \node (BT) at (1.8,-1.75) [greenc] {Bias Type};    
    \node (EO) at (-1.8,1.75) [bluec] {Equal Opportunity};    
    \node (WV) at (-1.8,-1.75) [yellowc] {Worldview};  
    \node (IT) at (3.3,0.0) [pinkc] {Intersectional};
    
    \node (AC) at (3.9,2.5) [child] {Attribute Cardinality};
    \node (ACB) at (6.3,3.2) [orangec] {Binary};
    \node (ACM) at (6.3,1.8) [child] {Multinary};
    
    \node (AN) at (3.9,1.0) [child] {Attribute Number};
    \node (ANO) at (6.9,1.15) [orangec] {One};
    \node (ANM) at (6.9,0.5) [child] {Multiple};
    
    \node (ANMI) at (8.9,0.8) [child] {Independent};
    \node (ANMC) at (8.9,0.0) [child] {Combination};
    
    \node (ITY) at (5.3,0.2) [child] {Yes};
    \node (ITN) at (5.3,-0.4) [pinkc] {No};
    
    \node (BTP) at (4.0,-1.2) [greenc] {Pre-existing};
    \node (BTT) at (4.0,-2.0) [child] {Technical};
    \node (BTE) at (4.0,-2.75) [child] {Emergent};
    
    \node (EOF) at (-3.8,2.8) [child] {Formal};
    \node (EOFP) at (-4.6,2.1) [child] {Formal plus};
    \node (EOS) at (-3.8,1.0) [child] {Substantive};
    
    \node (EOSR) at (-5.8,1.4) [child] {Rawlsian};
    \node (EOSL) at (-5.4,0.1) [bluec] {Luck-egalitarian};

    \node (WVI) at (-3.8,-0.8) [child] {WYSIWYG};
    \node (WVE) at (-3.8,-1.8) [yellowc] {WAE};
    \node (WVC) at (-3.8,-2.8) [child] {Continuous};
    
    \draw (ND.east) -- (GS);
    \draw (ND.east) -- (BT);
    \draw (ND.east) -- (IT);
    \draw (ND.west) -- (EO);
    \draw (ND.west) -- (WV);
    
    \draw (GS) -- (AC);
    \draw (GS) -- (AN);
    
    \draw (AC) -- (ACB);
    \draw (AC) -- (ACM);
    
    \draw (AN) -- (ANO);
    \draw (AN) -- (ANM);
    
    \draw (IT) -- (ITY);
    \draw (IT) -- (ITN);
    
    \draw (ANM) -- (ANMI);
    \draw (ANM) -- (ANMC);
    
    \draw (BT) -- (BTP);
    \draw (BT) -- (BTT);
    \draw (BT) -- (BTE);
    
    \draw (EO) -- (EOF);
    \draw (EO) -- (EOFP);
    \draw (EO) -- (EOS);
    
    \draw (EOS) -- (EOSR);
    \draw (EOS) -- (EOSL);
    
    \draw (WV) -- (WVI);
    \draw (WV) -- (WVE);
    \draw (WV) -- (WVC);
\end{tikzpicture}
    \caption{\rev{Summary of the normative values encoded by \deltr \revv{(\citet{zehlike2018reducing})}. This translates into normative choices that are implicitly taken when applying the method. Note however, that because \deltr operates in a binary setting with only two groups, it can not handle multiple attributes.}}
    \label{fig:in-proc:deltr}
\end{figure}
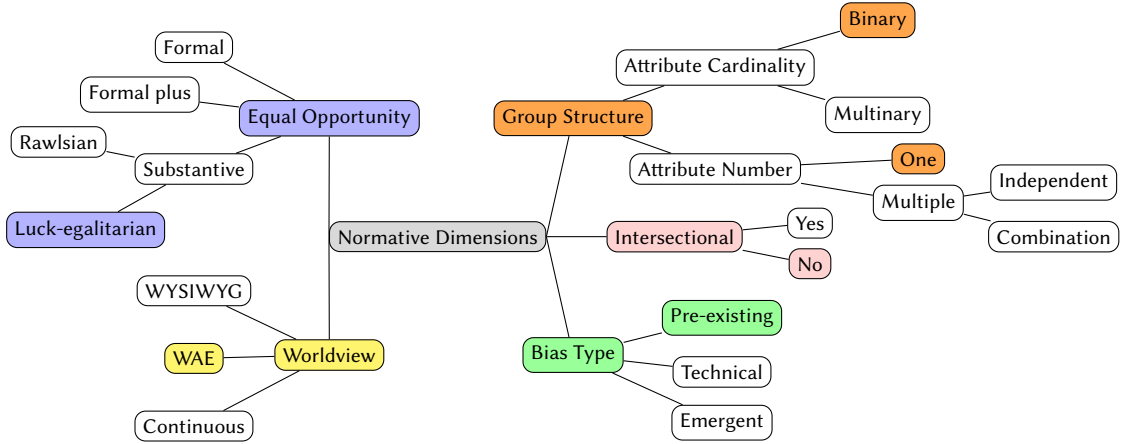
\spara{Insights. } As the definition optimizes for equality of exposure (exposure is the outcome), rather than equity, this means that the (potentially biased) qualification of a candidate is not taken into account, which suggests that its underlying assumption is a WAE worldview.
We denote however, that because exposure is defined through the probability to appear in the top position, it indirectly contains a measurement of document relevance. 
We further denote that by setting $\gamma$ to 0, the WYSIWYG worldview can also be adopted.

The method is concerned with pre-existing biases that lead to a biased observation process and\revv{, ultimately, }to disparate exposure distributions. 
%
\revv{Under the assumption that the goal of the competition is to make future prospects comparable, this is consistent with luck-egalitarian \eop. }
%
This method is agnostic to ``true'' score distributions and optimizes for equal exposure in decision space $DS$ irrespective of whether score distributions in $CS$ are different for demographic groups.  \revv{For this reason, we argue that the mechanism is consistent with conditioning the qualification score on morally irrelevant characteristics (\ie group membership), and place it into the category of luck-egalitarian \eop.}
Figure~\ref{fig:in-proc:deltr} and Table~\ref{tbl:method-summary_ir} summarize our analysis.

\spara{Algorithm.} 
The algorithm incorporates its unfairness measure into the objective function of a list-wise learning algorithm, namely ListNet~\cite{cao2007learning}, \revv{to simultaneously optimize} for an accuracy metric $L$ and an unfairness metric for two groups $\unfairnessOnePara{\predScore}$: 
    \begin{equation}
    	L_{\operatorname{\deltr}} \left( \score{}, \predScore \right) = L \left( \score{}, \predScore \right) + \gamma \unfairnessOnePara{\predScore} 
    \end{equation} with
    \begin{equation*}
        \unfairnessOnePara{\predScore} = \max \left(0, \operatorname{Exposure}\left(\group{0}|P_{\predScore{\query}}\right) - \operatorname{Exposure}\left(\group{1}|P_{\predScore{\query}}\right)\right)^2
        \label{eq:exposure}
    \end{equation*}
where $\group{0}$ denotes the non-protected group and $\group{1}$ the protected one. The squared hinge loss \revv{is asymmetric to detect unfairness only} if the protected group receives less exposure than the non-protected group, but not vice versa.
The optimization problem is solved using gradient descent.

\spara{Experiments. } To illustrate how the method works, let us \revv{return} to our running example in Figure~\ref{fig:ir_example} on page~\pageref{fig:ir_example}: 
the algorithm gets as input the sensitive feature that forms a protected group, in this case $A_1$.
As the training data $\candidateSet{train}$ shows disparities in exposure for \revv{women} (they are all ranked below \revv{the men}), a standard learning-to-rank algorithm is likely to pick up $A_1$ as a predominant criterion for its model and assign a high weight \revv{to this attribute}.
\deltr instead will learn to ignore $A_1$ as a decision criterion because the unfairness metric penalizes ranking predictions that show high disparities in exposure for the different groups of $A_1$.
\revv{In this way,} \deltr can also compensate for systematic errors in the data or in the relevance measures (\eg if the SAT test design favors male applicants, hence \revv{female applicants systematically receive} lower scores).

Experiments are performed on three \revv{datasets, each exhibiting different types of bias, which are automatically handled by the proposed method, without explicit knowledge about what  particular type of bias is present}. 

\begin{itemize}
    \item \textbf{W3C experts:}~\cite{W3CData} The experimental setup investigates situations in which bias is unrelated to relevance: expertise has been judged correctly, but ties have been broken in favor \revv{of the privileged group}.
    In this case, including the sensitive feature during training yields very bad results in terms of disparate exposure and relevance, because all experts from the protected group are ranked at the bottom of the list.
    
    \item \textbf{Engineering students:}~\cite{EngineeringData} The task is to predict a student's academic performance after the first year based on their admissions test results and school grades.
    Here, the same admissions test score relates to different levels of academic performance across groups: a score of 500 in the protected group relates to better academic performance than a score of 500 in the privileged group.
    This experiment investigates situations in which bias is coming from different score distributions among groups, and shows \revv{that a fair ranking can be derived without a degradation in accuracy. }
    In this case, including the sensitive feature during training yields \emph{better} results in terms of \revv{both} exposure and relevance.
    \item \textbf{Law School Admission Council:}~\cite{LSACData} These experiments show that disparities in exposure due to differences in academic performance can be reduced \revv{using a simple approach, but at a cost} of accuracy.
    However, the trade-off is usually \revv{better (with lower accuracy loss)} for in-processing methods than for post-processing \revv{methods}. 
\end{itemize}

\subsubsection{Fair-PG-Rank \cite{singh2019policy}}
\label{sec:in-proc:fairpgrank}
\spara{Fairness Definition.} 
The approach by~\citet{singh2019policy} addresses technical bias \revv{that may be introduced by the ranking system itself due to \emph{position bias} --- giving candidates beyond the first few positions significantly less visibility compared to those who appear in the first positions.}
\revv{Like \deltr, Fair-PG-Rank operates based on a notion of document exposure.  However, unlike \deltr,  Fair-PG-Rank defines exposure as \emph{expected attention}, which the authors consider to be equivalent} to the expected position bias (as defined in Equation~\ref{eq:exp} in Section~\ref{sec:intro:learned} on page~\pageref{eq:exp}).
\revv{Similarly to Equity of Attention by~\citet{biega2018equity}, which we will discuss in Section~\ref{subsubsec:equityOfAttention}, Fair-PG-Rank operates under a merit-based constraint: E}ach candidate $\btau(i) = a$ in ranking $\btau$ should receive exposure proportional to their utility $\utilityThreePara{}{\btau}{a}$:
\begin{equation}
    \utilityThreePara{}{\btau}{a} \geq \utilityThreePara{}{\btau}{b} \rightarrow 
    \frac{\operatorname{Exposure}(a)}{\utilityThreePara{}{\btau}{a}} \leq 
    \frac{\operatorname{Exposure}(b)}{\utilityThreePara{}{\btau}{b}}
\end{equation} 
This work further proposes a definition of individual fairness, per query $\query$, that measures the disparities in visibility $\posBias{.}$ for two candidates $a, b$ in $\btau$:
\begin{equation}\label{eq:fpgr:ind}
    \unfairnessThreePara{a}{b}{\query} = \frac{1}{|H^{\query}|} \sum_{(a, b) \in H^{\query}} \max 
    \left[0, \frac{\posBias{a}}{\utilityThreePara{}{\btau}{a}} - \frac{\posBias{b}}{\utilityThreePara{}{\btau}{b}}\right]
\end{equation}
with $H^{\query} = \left\{(a, b) \; \text{s.t.} \; \utilityThreePara{}{\btau}{a} \geq \utilityThreePara{}{\btau}{b}\right\}$. 
The authors also propose a definition of group fairness for two groups, \revv{in which the notion of individual  visibility of a document per Eq.~\ref{eq:fpgr:ind} is replaced with group visibility:}
\begin{equation}\label{eq:fpgr:grp}
    \unfairnessThreePara{\group{0}}{\group{1}}{\query} = \max \left[0, \frac{\posBias{\group{0}}}{\utilityThreePara{}{\btau}{\group{0}}} - \frac{\posBias{\group{1}}}{\utilityThreePara{}{\btau}{\group{1}}}\right]
\end{equation}
with $\posBias{\group{}} = \frac{1}{|\group{}|} \sum_{a \in \group{}} \posBias{a}$ being the average exposure of group $\group{}$.

\spara{Algorithm.}
Using the proposed fairness definitions, the authors extend the ListNet~\cite{cao2007learning} ranking function to incorporate their disparity measures.
The algorithm Fair-PG-Rank is learning an optimal ranking $\btau^*$ via empirical risk minimization using the following learning objective:
\begin{equation}\label{eq:fpgr:opt}
    \btau^*_\delta = \argmax_{\btau} \frac{1}{N} \sum^{N}_{\query=1} \left[ L(\btau) \right] - \lambda \frac{1}{N} \sum^{N}_{\query=1} \left[ \unfairnessOnePara{.|\query} \right]
\end{equation}
with $L$ being a loss function that measures the utility of $\btau$ for the user.
The optimization is done using gradient descent.

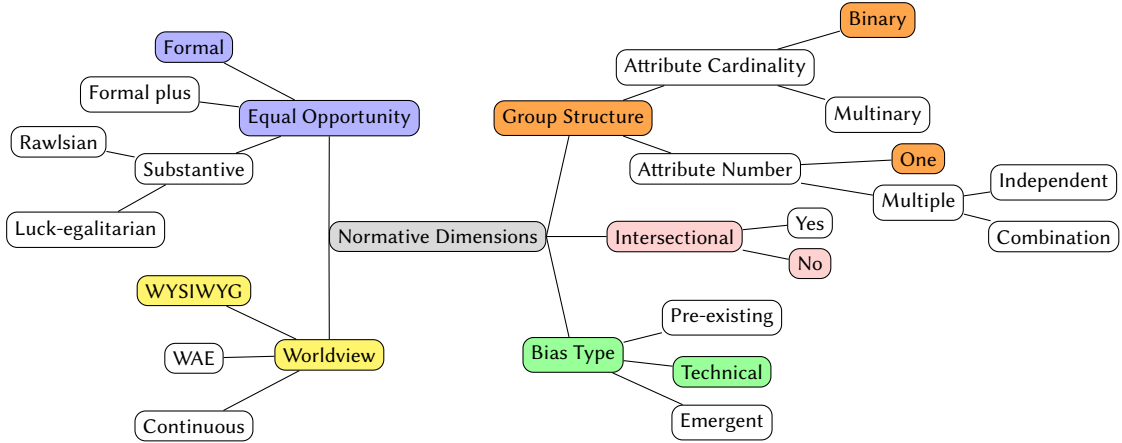
\begin{figure}[h]
    \centering
    \begin{tikzpicture}[scale=0.9,align=center]
    \tikzset{
        root/.style = {rectangle, rounded corners, draw=black, fill=gray!30, font=\small\sffamily},
        child/.style = {rectangle, rounded corners,draw=black, font=\small\sffamily},
        orangec/.style = {rectangle, rounded corners, draw=black, fill=orange!70, font=\small\sffamily},
        greenc/.style = {rectangle, rounded corners, draw=black, fill=green!40, font=\small\sffamily},
        bluec/.style = {rectangle, rounded corners, draw=black, fill=blue!30, font=\small\sffamily},
        yellowc/.style = {rectangle, rounded corners, draw=black, fill=yellow!70, font=\small\sffamily},
        pinkc/.style = {rectangle, rounded corners, draw=black, fill=pink!70, font=\small\sffamily},
    }
    \node (ND) at (-0.2,0) [root] {Normative Dimensions};    
    \node (GS) at (1.8,1.75) [orangec] {Group Structure};    
    \node (BT) at (1.8,-1.75) [greenc] {Bias Type};    
    \node (EO) at (-1.8,1.75) [bluec] {Equal Opportunity};    
    \node (WV) at (-1.8,-1.75) [yellowc] {Worldview};  
    \node (IT) at (3.3,0.0) [pinkc] {Intersectional};
    
    \node (AC) at (3.9,2.5) [child] {Attribute Cardinality};
    \node (ACB) at (6.3,3.2) [orangec] {Binary};
    \node (ACM) at (6.3,1.8) [child] {Multinary};
    
    \node (AN) at (3.9,1.0) [child] {Attribute Number};
    \node (ANO) at (6.9,1.15) [orangec] {One};
    \node (ANM) at (6.9,0.5) [child] {Multiple};
    
    \node (ANMI) at (8.9,0.8) [child] {Independent};
    \node (ANMC) at (8.9,0.0) [child] {Combination};
    
    \node (ITY) at (5.3,0.2) [child] {Yes};
    \node (ITN) at (5.3,-0.4) [pinkc] {No};
    
    \node (BTP) at (4.0,-1.2) [child] {Pre-existing};
    \node (BTT) at (4.0,-2.0) [greenc] {Technical};
    \node (BTE) at (4.0,-2.75) [child] {Emergent};
    
    \node (EOF) at (-3.8,2.8) [bluec] {Formal};
    \node (EOFP) at (-4.6,2.1) [child] {Formal plus};
    \node (EOS) at (-3.8,1.0) [child] {Substantive};
    
    \node (EOSR) at (-5.8,1.4) [child] {Rawlsian};
    \node (EOSL) at (-5.4,0.1) [child] {Luck-egalitarian};

    \node (WVI) at (-3.8,-0.8) [yellowc] {WYSIWYG};
    \node (WVE) at (-3.8,-1.8) [child] {WAE};
    \node (WVC) at (-3.8,-2.8) [child] {Continuous};
    
    \draw (ND.east) -- (GS);
    \draw (ND.east) -- (BT);
    \draw (ND.east) -- (IT);
    \draw (ND.west) -- (EO);
    \draw (ND.west) -- (WV);
    
    \draw (GS) -- (AC);
    \draw (GS) -- (AN);
    
    \draw (AC) -- (ACB);
    \draw (AC) -- (ACM);
    
    \draw (AN) -- (ANO);
    \draw (AN) -- (ANM);
    
    \draw (IT) -- (ITY);
    \draw (IT) -- (ITN);
    
    \draw (ANM) -- (ANMI);
    \draw (ANM) -- (ANMC);
    
    \draw (BT) -- (BTP);
    \draw (BT) -- (BTT);
    \draw (BT) -- (BTE);
    
    \draw (EO) -- (EOF);
    \draw (EO) -- (EOFP);
    \draw (EO) -- (EOS);
    
    \draw (EOS) -- (EOSR);
    \draw (EOS) -- (EOSL);
    
    \draw (WV) -- (WVI);
    \draw (WV) -- (WVE);
    \draw (WV) -- (WVC);
\end{tikzpicture}
    \caption{\revv{Summary of the normative values encoded by Fair-PG-Rank (\citet{singh2019policy}). This translates into normative choices that are implicitly taken when applying the method.}}
    \label{fig:in-proc:fairpgrank}
\end{figure}

\spara{Insights. } As already mentioned, the method is concerned explicitly with the inherent technical bias of a ranking that arises from \revv{showing ranked results} in a one-dimensional list.
It assumes a WYSIWYG world in which a document's merit in $OS$ truly reflects its merit in $CS$, and no effort is made to reduce any errors that might have been introduced by the mapping function $g$. This method allocates outcomes (visibility) on the basis of merit (utility), and thereby codifies formal \eop. 

This method is explicitly concerned with equity of exposure. This means that documents being ranked will only receive as much exposure as they ``deserve'' based on their relevance.
Returning to our college admission example (Figure~\ref{fig:ir_example}, page~\pageref{fig:ir_example}), we see that in the training data (white background lines), all women have worse scores than men, as reflected in the relative ranking of the two groups.
A standard LTR algorithm is therefore likely to treat $A_1$ as an important decision criterion and assign a high weight to it. 
Because Fair-PG-Rank optimizes for equity of exposure, this approach ensures that documents receive equal exposure among those that  ``deserve'' it based on their relevance.
As such, Fair-PG-Rank ensures that a model is not discriminating based on a sensitive attribute, while assuming that relevance is computed correctly and does embed any patterns of discrimination.
Note, that, similarly to \deltr, this method works for binary groups, and that it cannot handle multiple protected attributes.
Figure~\ref{fig:in-proc:fairpgrank} and Table~\ref{tbl:method-summary_ir} summarize our analysis.

\spara{Experiments.}
Experiments are done on a synthetic dataset, \revv{German credit~\cite{GermanCreditData} with the group fairness definition in Eq.~\ref{eq:fpgr:ind} as part of the learning objective, and on the Yahoo! LTR dataset~\cite{YahooData}, with the individual fairness definition in Eq.~\ref{eq:fpgr:grp} as part of the learning objective. }
The synthetic dataset has two features for each document, \revv{one of which} is corrupted for the minority group $G_1$.
The results show that, with increasing values of $\lambda$ (see Eq.~\ref{eq:fpgr:opt}), the weight of the corrupted feature is decreasing. 
With the real world datasets, the authors show that the method works in a real world setting, but they do not discuss its implications. 
\revv{The authors compare their method to \deltr~\cite{zehlike2018reducing}, but, because both their measure of disparate exposure and the type bias on which they are focusing are vastly different, it is not clear whether meaningful conclusions can be drawn from such a comparison.}
Further research is needed to understand \revv{whether and how} methods with opposing worldviews, conflicting understandings of equality of opportunity, and different \revv{bias types can be compared}.

\subsubsection{Pairwise Fairness for Rankings, \citet{Beutel:2019:FRR:3292500.3330745}}
\label{subsubsec:beutel}
\spara{Fairness Definition.}
This method is the first one to introduce a pairwise fairness metric for ranking predictions. 
The authors setup their method as a component of a cascading recommender system, however, it only considers the final ranking of items. 
We therefore present the method here, rather than in Section~\ref{sec:fair_recsys}, where we talk about fairness in recommendation systems.

The framework is formalized as follows: Each query consists of user features $\mathbf{U}_i$ for user $i$ and context features  $\mathbf{C}$.
Each ranking candidate $a$ is described by a feature vector $\featOfCand{a}$.
Then, a ranker $\hat{f}$ is trained to predict user engagement, which relates to clicks $\hat{y}$ \emph{and} interaction after a click $\hat{z}$ (such as purchase, ratings, etc.),
which are then mapped to a scalar value to rank items: $\hat{f}(\featureSet) = \predScore$.

The focus of the fairness definition is on the risk for groups of ranked candidates to be under-recommended, \revv{under binary group membership (\ie  $A_a \in {0,1}$ for candidate $a$)}.
For this, the authors first define pairwise accuracy, which describes the probability that a clicked candidate is ranked above another relevant unclicked candidate, for the same query:
\[P\left(\hat f (\featOfCand{a}) > \hat f(\featOfCand{b}) \; | \; \score{a} > \score{b}\right)\]
Pairwise fairness asks if the pairwise accuracy is the same across the two groups: 
\[P\left(\hat f (\featOfCand{a}) > \hat f(\featOfCand{b}) \; | \; \score{a} > \score{b}, \sensAttr_a=0\right) = P\left(\hat f (\featOfCand{a}) > \hat f(\featOfCand{b}) \; | \; \score{a} > \score{b}, \sensAttr_a=1\right)\]

To account for user engagement $z$, the definition is extended to compare only those candidates with each other that receive the same amount of engagement $\Tilde{z}$: 
\begin{align*}
    &P\left(\hat f (\featOfCand{a}) > \hat f(\featOfCand{b}) \; | \; \score{a} > \score{b}, \sensAttr_a=0, z_a=\Tilde{z}\right) = \\
    &P\left(\hat f (\featOfCand{a}) > \hat f(\featOfCand{b}) \; | \; \score{a} > \score{b}, \sensAttr_a=1, z_a=\Tilde{z}\right) \forall \Tilde{z}
\end{align*}
The authors further extend the definition to also consider group exposure in rankings, because two rankings could have the same pairwise accuracy across groups, while systematically putting candidates of one group to lower ranks of the list.
To account for this, they split the definition into \emph{intra-group} pairwise fairness:
\begin{align*}
    &P\left(\hat f (\featOfCand{a}) > \hat f(\featOfCand{b}) \; | \; \score{a} > \score{b}, \sensAttr_a=\sensAttr_b=0, z_a=\Tilde{z}\right) = \\
    &P\left(\hat f (\featOfCand{a}) > \hat f(\featOfCand{b}) \; | \; \score{a} > \score{b}, \sensAttr_a=\sensAttr_b=1, z_a=\Tilde{z}\right) \forall \Tilde{z}
\end{align*}
and \emph{inter-group} pairwise fairness:
\begin{align*}\label{eq:pair:inter}
    &P\left(\hat f (\featOfCand{a}) > \hat f(\featOfCand{b}) \; | \; \score{a} > \score{b}, \sensAttr_a=0, \sensAttr_b=1, z_a=\Tilde{z}\right) = \\
    &P\left(\hat f (\featOfCand{a}) > \hat f(\featOfCand{b}) \; | \; \score{a} > \score{b}, \sensAttr_a=1, \sensAttr_b=0, z_a=\Tilde{z}\right) \forall \Tilde{z}
\end{align*}
Intra-group fairness indicates whether, across candidates from the same group, those that are more likely to be clicked are ranked above those that are less likely to be clicked. On the other hand, inter-group fairness describes whether mistakes of the ranker are at the cost of one particular group.

\spara{Insights.} The framework as is is not clearly classifiable into WAE or WYSIWYG, because the authors talk about click probability and user engagement without further specifying what these two are composed off. 
\revv{Particularly they do not specify whether a candidate's merit is part of the click probability.}
Click through rate (CTR) is usually defined to contain some measure of relevance~\cite{richardson2007predicting}.  If that is the case here, then  $\score{}$ would contain a component that encodes merit, which is measured in $OS$, and hence the framework would correspond to WYSIWYG.
The authors briefly mention that they assume the final ranking model $\hat f$ \revv{to only operate on relevant documents, which supports the assumption that the underlying worldview of this framework is WYSIWYG.}
We face the same uncertainty when thinking about which \eop framework this work corresponds to. 
\revv{Without knowledge of the actual underlying estimation of click probability and user engagement, and without a statement on an individual's effort, it is not clear with which \eop framework the fairness definition is consistent.}
\revv{However, because the authors state that they adopted a definition of ``equal opportunity'', and because their inter-group fairness objective balances error rates across groups, we map this method to formal-plus \eop.} 
An identification of the addressed bias is also not possible without the CTR definition. 

\begin{figure}[h]
    \centering
    \begin{tikzpicture}[scale=0.9,align=center]
    \tikzset{
        root/.style = {rectangle, rounded corners, draw=black, fill=gray!30, font=\small\sffamily},
        child/.style = {rectangle, rounded corners,draw=black, font=\small\sffamily},
        orangec/.style = {rectangle, rounded corners, draw=black, fill=orange!70, font=\small\sffamily},
        greenc/.style = {rectangle, rounded corners, draw=black, fill=green!40, font=\small\sffamily},
        bluec/.style = {rectangle, rounded corners, draw=black, fill=blue!30, font=\small\sffamily},
        yellowc/.style = {rectangle, rounded corners, draw=black, fill=yellow!70, font=\small\sffamily},
        pinkc/.style = {rectangle, rounded corners, draw=black, fill=pink!70, font=\small\sffamily},
    }
    \node (ND) at (-0.2,0) [root] {Normative Dimensions};    
    \node (GS) at (1.8,1.75) [orangec] {Group Structure};    
    \node (BT) at (1.8,-1.75) [greenc] {Bias Type};    
    \node (EO) at (-1.8,1.75) [bluec] {Equal Opportunity};    
    \node (WV) at (-1.8,-1.75) [yellowc] {Worldview};  
    \node (IT) at (3.3,0.0) [pinkc] {Intersectional};
    
    \node (AC) at (3.9,2.5) [child] {Attribute Cardinality};
    \node (ACB) at (6.3,3.2) [orangec] {Binary};
    \node (ACM) at (6.3,1.8) [child] {Multinary};
    
    \node (AN) at (3.9,1.0) [child] {Attribute Number};
    \node (ANO) at (6.9,1.15) [orangec] {One};
    \node (ANM) at (6.9,0.5) [child] {Multiple};
    
    \node (ANMI) at (8.9,0.8) [child] {Independent};
    \node (ANMC) at (8.9,0.0) [child] {Combination};
    
    \node (ITY) at (5.3,0.2) [child] {Yes};
    \node (ITN) at (5.3,-0.4) [pinkc] {No};
    
    \node (BTP) at (4.0,-1.2) [child] {Pre-existing};
    \node (BTT) at (4.0,-2.0) [child] {Technical};
    \node (BTE) at (4.0,-2.75) [child] {Emergent};
    
    \node (EOF) at (-3.8,2.8) [child] {Formal};
    \node (EOFP) at (-4.6,2.1) [bluec] {Formal plus};
    \node (EOS) at (-3.8,1.0) [child] {Substantive};
    
    \node (EOSR) at (-5.8,1.4) [child] {Rawlsian};
    \node (EOSL) at (-5.4,0.1) [child] {Luck-egalitarian};

    \node (WVI) at (-3.8,-0.8) [child] {WYSIWYG};
    \node (WVE) at (-3.8,-1.8) [child] {WAE};
    \node (WVC) at (-3.8,-2.8) [child] {Continuous};
    
    \draw (ND.east) -- (GS);
    \draw (ND.east) -- (BT);
    \draw (ND.east) -- (IT);
    \draw (ND.west) -- (EO);
    \draw (ND.west) -- (WV);
    
    \draw (GS) -- (AC);
    \draw (GS) -- (AN);
    
    \draw (AC) -- (ACB);
    \draw (AC) -- (ACM);
    
    \draw (AN) -- (ANO);
    \draw (AN) -- (ANM);
    
    \draw (IT) -- (ITY);
    \draw (IT) -- (ITN);
    
    \draw (ANM) -- (ANMI);
    \draw (ANM) -- (ANMC);
    
    \draw (BT) -- (BTP);
    \draw (BT) -- (BTT);
    \draw (BT) -- (BTE);
    
    \draw (EO) -- (EOF);
    \draw (EO) -- (EOFP);
    \draw (EO) -- (EOS);
    
    \draw (EOS) -- (EOSR);
    \draw (EOS) -- (EOSL);
    
    \draw (WV) -- (WVI);
    \draw (WV) -- (WVE);
    \draw (WV) -- (WVC);
\end{tikzpicture}
    \caption{\rev{Summary of the normative values encoded by \revv{Pairwise fariness for rankings (\citet{Beutel:2019:FRR:3292500.3330745})}.}} 
    \label{fig:in-proc:beutel}
\end{figure}

\spara{Experiments.}
The experiments study the performance of the ranker with respect to the \revv{protected subgroup} of candidates in the synthetic dataset, comparing the performance of this subgroup to the rest of the data, denoted by ``not subgroup.'' \revv{(We will refer to these ``not subgroup'' candidates as ``privileged'' for ease of exposition.)}
The \revv{protected group} represents approximately 0.2\% of all items.
The authors compare two versions of their approach: a model without any pairwise fairness constraint and one with an inter-group fairness constraint.
\revv{User engagement is grouped into four levels.}  Performance measures (pairwise accuracy) are aggregated across user engagement levels and averaged.
Then, \revv{the ratio of the accuracy for the protected vs. privileged groups is computed, with a ratio of 1 corresponding to perfect fairness, a value above 1 corresponding to higher accuracy for the protected group, and a value below 1 corresponding to higher accuracy for the privileged group.}

\revv{The overall pairwise fairness evaluation shows that the system under-ranks protected group candidates when the level of engagement is low, but, interestingly, it slightly over-ranks protected group candidates when the level of engagement is high.}
Intra-group pairwise fairness evaluation shows that, across all levels of engagement, the model has more difficulty selecting the clicked candidate when comparing protected group candidates than when comparing privileged group candidates. 
This is partly because the protected group is small and less diverse.
Inter-group fairness evaluation shows that, across all levels of engagement, protected group candidates are significantly under-ranked relative to the privileged candidates.
\revv{Further, the results show that the pairwise accuracy for the protected group in inter-group pairs is notably higher than in intra-group pairs, suggesting that protected group candidates, even when of interest to the user, are ranked below the privileged group candidates.}
\revv{When optimizing for inter-group fairness, these disparities are mitigated and protected group candidates  receive more exposure (measure as the probability that candidate $a$ is ranked above candidate $b$).}

\subsection{Post-Processing Methods: Re-Ordering Ranked Items}

Post-processing algorithms assume that a ranking model has already been trained.
A predicted ranking is handed to the algorithm, which re-orders items to improve fairness. 
\revv{Most algorithms operate on a notion of group membership, where certain groups are denoted as protected, while one group is denoted as non-protected (or privileged).}
\revv{As such, post-processing methods often model fairness constraints similarly to  score-based fair ranking methods.}
\revv{We will reuse Figure~\ref{fig:ad_example} from Section~\ref{sec:intro:example} as our running example in this section, but will omit the non-sensitive features $X_1, X_2$ and $X_3$ because they are unimportant here.}

\newcommand{\postProcDatatab}{
    \small
	\begin{tabular}{|c||c|c||c||c|}
		\hline
		\rowcolor[HTML]{C0C0C0} 
		candidate & $A_1$ (sex) & $A_2$ (race) & $\predScore{}$ & $\mathbf v$  \\ \hline
		\rowcolor[HTML]{FFFFFF}
		\val{b}  & \val{male}      & \val{White}    & 9  & 3.32 \\ \hline
		\val{c}  & \val{male}      & \val{Asian}    &  8 & 2.10\\ \hline
		\val{d}  & \val{male}      & \val{White}   & 7 & 1.66\\ \hline
		\val{e}  & \val{male}      & \val{White}     & 6 & 1.43\\ \hline
		\val{f}  & \val{female}       & \val{Asian}  & 5 & 1.28\\ \hline
		\val{k}  & \val{female}       & \val{Black}  & 4  & 1.18 \\ \hline
		\val{l}  & \val{female}      & \val{Black}     & 3 & 1.11  \\ \hline
		\val{o}  & \val{female}       & \val{White}  & 2 & 1.04\\ \hline
	\end{tabular}
}
\newcommand{\postProcRankrawtab}{
    \small
	\begin{tabular}{|c|}
		\hline
		\rowcolor[HTML]{C0C0C0} 
		$\hat \btau_1$ \\ \hline
		\rowcolor[HTML]{CBCEFB} 
		b                    \\ \hline
		\rowcolor[HTML]{CBCEFB} 
		c                    \\ \hline
		\rowcolor[HTML]{CBCEFB} 
		d                    \\ \hline
		\rowcolor[HTML]{CBCEFB} 
		e                    \\ \hline \hline
		\rowcolor[HTML]{FFCE93} 
		f                    \\ \hline
		\rowcolor[HTML]{FFCE93} 
		k                    \\ \hline
		\rowcolor[HTML]{FFCE93} 
		l                    \\ \hline
		\rowcolor[HTML]{FFCE93} 
		o                    \\ \hline
	\end{tabular}
}
\newcommand{\postProcRankpreftab}{
    \small
	\begin{tabular}{|c|}
		\hline
		\rowcolor[HTML]{C0C0C0} 
		$\hat \btau_2$ \\ \hline
		\rowcolor[HTML]{CBCEFB} 
		\val{b}   \\ \hline
		\rowcolor[HTML]{FFCE93} 
		\val{f}   \\ \hline 
		\rowcolor[HTML]{CBCEFB} 
		\val{c}   \\ \hline
		\rowcolor[HTML]{FFCE93} 
		\val{k}   \\ \hline
		\hline
		\rowcolor[HTML]{CBCEFB} 
		\val{d}   \\ \hline
		\rowcolor[HTML]{FFCE93} 
		\val{l}   \\ \hline
		\rowcolor[HTML]{CBCEFB} 
		\val{e}   \\ \hline
		\rowcolor[HTML]{FFCE93} 
		\val{o}   \\ \hline
	\end{tabular}
}
\newcommand{\postProcRankproptab}{
    \small
	\begin{tabular}{|c|}
		\hline
		\rowcolor[HTML]{C0C0C0} 
		$\hat \btau_3$ \\ \hline
		\rowcolor[HTML]{CBCEFB} 
		\val{b}   \\ \hline
		\rowcolor[HTML]{CBCEFB} 
		\val{c}   \\ \hline
		\rowcolor[HTML]{CBCEFB} 
		\val{d}   \\ \hline
		\rowcolor[HTML]{FFCE93} 
		\val{l}   \\ \hline
		\hline
		\rowcolor[HTML]{CBCEFB} 
		\val{e}   \\ \hline
		\rowcolor[HTML]{CBCEFB} 
		\val{f}   \\ \hline 
		\rowcolor[HTML]{CBCEFB} 
		\val{k}   \\ \hline
		\rowcolor[HTML]{FFCE93} 
		\val{o}   \\ \hline
	\end{tabular}
}
\newcommand{\postProcRankSinghTab}{
    \small
	\begin{tabular}{|c|}
		\hline
		\rowcolor[HTML]{C0C0C0} 
		$\hat \btau_4$ \\ \hline
		\rowcolor[HTML]{CBCEFB} 
		\val{b}   \\ \hline
		\rowcolor[HTML]{FFCE93} 
		\val{f}   \\ \hline 
		\rowcolor[HTML]{FFCE93} 
		\val{k}   \\ \hline
		\rowcolor[HTML]{FFCE93} 
		\val{l}   \\ \hline
		\hline
		\rowcolor[HTML]{FFCE93} 
		\val{o}   \\ \hline
		\rowcolor[HTML]{CBCEFB} 
		\val{d}   \\ \hline
		\rowcolor[HTML]{CBCEFB} 
		\val{c}   \\ \hline
		\rowcolor[HTML]{CBCEFB} 
		\val{e}   \\ \hline
	\end{tabular}
}
\newcommand{\oldtablefive}{
    \small
	\begin{tabular}{r|cccccccccccc}
		\diaghead{som ext}%
		{p}{k}
		& 1 & 2 & 3 & 4 & 5 & 6 & 7 & 8 & 9 & 10 & 11 & 12 \\ \midrule
		0.1      & 0 & 0 & 0 & 0 & 0 & 0 & 0 & 0 & 0 & 0  &  0 &  0 \\
		0.3      & 0 & 0 & 0 & 0 & 0 & 0 & 1 & 1 & 1 & 1  &  1 &  2 \\
		0.4      & 0 & 0 & 0 & 0 & 1 & 1 & 1 & 1 & 2 & 2  &  2 &  3 \\
		0.5      & 0 & 0 & 0 & 1 & 1 & 1 & 2 & 2 & 3 & 3  &  3 &  4 \\
		0.6      & 0 & 0 & 1 & 1 & 2 & 2 & 3 & 3 & 4 & 4  &  5 &  5 \\
		0.7      & 0 & 1 & 1 & 2 & 2 & 3 & 3 & 4 & 5 & 5  &  6 &  6 \\
		\bottomrule
	\end{tabular}
}
\setlength{\tabcolsep}{0.25em}
\begin{table*}
    \begin{tabular}{cc}
        \begin{minipage}{0.6\textwidth}
            \centering
           \begin{minipage}{0.55\textwidth}
                \postProcDatatab
           \end{minipage}
           \begin{minipage}{0.05\textwidth}
                \postProcRankrawtab
           \end{minipage}
           \hspace{1.1em}
           \begin{minipage}{0.05\textwidth}
                \postProcRankpreftab
           \end{minipage}
           \hspace{1.1em}
           \begin{minipage}{0.05\textwidth}
                \postProcRankproptab
           \end{minipage}
           \hspace{1.1em}
           \begin{minipage}{0.05\textwidth}
                \postProcRankSinghTab
           \end{minipage}
        \end{minipage}
        &
        \begin{minipage}{0.4\textwidth}
            \centering
            \oldtablefive
            \vspace{3em}
        \end{minipage}
     \\
    \begin{minipage}[t]{0.65\textwidth}
        \captionof{figure}{A dataset $\candidateSet{}$ of college applicants. Score $\predScore{}$ is predicted by a learning-to-rank model. $\mathbf v$ is the position bias of positions 1~--~8. Ranking $\hat \btau_1$ is predicted based on $\predScore{}$, and the top-$4$ candidates will be admitted. \revv{Note that no female applicants are admitted in this scenario.} A fair ranking $\hat \btau_2$ is produced according to the ranked group fairness condition of the \algofair algorithm \revv{(\citet{zehlike2017fa})}, with $p=0.7$ and $\alpha = 0.1$.
        Ranking $\hat \btau_3$ is produced by the algorithm CFA$\theta$ with $\theta=1$ if the dataset was changed to candidate $f$ and $k$ being male. Note that CFA$\theta$ cannot achieve a ranking where the only two female candidates $l$ and $o$ are both admitted. Ranking $\hat \btau_4$ with equal exposure across groups is produced by the method of~\citet{singh2018fairness}.
        }
        \label{fig:postProc_example}
    \end{minipage}
     & 
    \begin{minipage}[t]{0.3\textwidth}
        \vspace{1.2em}
        \caption{Example values of the minimum number of protected items that must appear in the top-$k$ positions, to pass the binomial ranked group fairness test with $\alpha=0.1$. Table reporduced from~\citet{zehlike2017fa}.}
	    \label{tbl:ranked_group_fairness_table}
    \end{minipage}
    \end{tabular}
\end{table*}

\noindent General advantages of post-processing methods are: 
\begin{itemize}
    \item Many of them provide a guaranteed share of visibility for the protected group in the ranking.
    \item The effect the algorithms have on the ranked output is easy to visualize and understand, because the original ranking before the application of the fairness method can be compared to its result, in terms of how the items are re-ordered and, in some cases, in terms of the loss in a \revv{utility} metric such as NDCG. 
\end{itemize}
General disadvantages are:
\begin{itemize}
    \item The use of post-processing inherently suggests that fairness comes at the expense of accuracy, because the scores of a previously trained ranking model are taken as ``the true anchor point.'' Depending on the properties of pre-existing bias in the training data, however, ranking models may incorporate biases that decrease accuracy as shown by~\citet{zehlike2018reducing}, which renders any measurement of accuracy loss obsolete.
    \item Assuming biases are small, algorithms with a fixed fairness constraint \cite{zehlike2017fa,zehlike2017matching,singh2018fairness} may cause a substantial loss in performance w.r.t. the original ranking. \revv{This may be the case if the score distribution of the protected group is very different than that of the privileged group, leading to systematically lower scores for protected group members.} 
\end{itemize}

\subsubsection{\algofair,~\citet{zehlike2017fa, zehlike2022fair}}
\label{subsubsec:FAIR}
\spara{Fairness Definition.}
\revv{This work builds on~\citet{yang2017measuring}, discussed in Section~\ref{sec:fair_db:prop}, and adopts a fairness definition} based on the assumption that rankings are fair when the decisions on candidate placement are drawn from a Bernoulli distribution (coin tosses) \revv{that is not impacted by the candidate's sensitive attributes.} 
\revv{\algofair~\cite{zehlike2017fa} ensures} that the number of protected candidates does not fall far below a required minimum percentage $p$ at any point in the ranking, by formulating this fairness as a statistical significance test of whether a ranking was likely to have been produced by a Bernoulli process.
In~\citet{zehlike2022fair}, the authors extend the mathematical framework of~\cite{zehlike2017fa} to a multinomial distribution (roll of a dice), and provide a vector of minimum proportions $p_{\{\group{}\}}$, containing one $p$ for each group $\group{}$.
This way the extended algorithm can handle more than two groups at the same time, while the original operates in a binary group setting.
Both methods are concerned with disparate impact, as they do not take any notion of merit into account for their re-ranking strategy.
\revv{A ranking prefix of length $k$ is considered to fairly represent the protected group if the proportion of protected group members $\alpha$ 
does not fall 
below the minimum target proportion $p_{\{\group{}\}}$: $F\left(\btau_{\{\group{}\}}\; k, p_{\{\group{}\}}\right)$, 
with $F$ corresponding to the} multinomial cumulative distribution function.
If this condition holds for each prefix $k=1, \ldots, n$, then the entire ranking is considered to be fair. 

\begin{figure}[h]
    \centering
    \begin{tikzpicture}[scale=0.9,align=center]
    \tikzset{
        root/.style = {rectangle, rounded corners, draw=black, fill=gray!30, font=\small\sffamily},
        child/.style = {rectangle, rounded corners,draw=black, font=\small\sffamily},
        orangec/.style = {rectangle, rounded corners, draw=black, fill=orange!70, font=\small\sffamily},
        greenc/.style = {rectangle, rounded corners, draw=black, fill=green!40, font=\small\sffamily},
        bluec/.style = {rectangle, rounded corners, draw=black, fill=blue!30, font=\small\sffamily},
        yellowc/.style = {rectangle, rounded corners, draw=black, fill=yellow!70, font=\small\sffamily},
        pinkc/.style = {rectangle, rounded corners, draw=black, fill=pink!70, font=\small\sffamily},
    }
    \node (ND) at (-0.2,0) [root] {Normative Dimensions};    
    \node (GS) at (1.8,1.75) [orangec] {Group Structure};    
    \node (BT) at (1.8,-1.75) [greenc] {Bias Type};    
    \node (EO) at (-1.8,1.75) [bluec] {Equal Opportunity};    
    \node (WV) at (-1.8,-1.75) [yellowc] {Worldview};  
    \node (IT) at (3.3,0.0) [pinkc] {Intersectional};
    
    \node (AC) at (3.9,2.5) [child] {Attribute Cardinality};
    \node (ACB) at (6.3,3.2) [child] {Binary};
    \node (ACM) at (6.3,1.8) [orangec] {Multinary};
    
    \node (AN) at (3.9,1.0) [child] {Attribute Number};
    \node (ANO) at (6.9,1.15) [child] {One};
    \node (ANM) at (6.9,0.5) [child] {Multiple};
    
    \node (ANMI) at (8.9,0.8) [child] {Independent};
    \node (ANMC) at (8.9,0.0) [orangec] {Combination};
    
    \node (ITY) at (5.3,0.2) [child] {Yes};
    \node (ITN) at (5.3,-0.4) [pinkc] {No};
    
    \node (BTP) at (4.0,-1.2) [greenc] {Pre-existing};
    \node (BTT) at (4.0,-2.0) [child] {Technical};
    \node (BTE) at (4.0,-2.75) [child] {Emergent};
    
    \node (EOF) at (-3.8,2.8) [child] [bluec] {Formal};
    \node (EOFP) at (-4.6,2.1) [child] {Formal plus};
    \node (EOS) at (-3.8,1.0) [child] {Substantive};
    
    \node (EOSR) at (-5.8,1.4) [child] {Rawlsian};
    \node (EOSL) at (-5.4,0.1) [bluec] {Luck-egalitarian};

    \node (WVI) at (-3.8,-0.8) [child] {WYSIWYG};
    \node (WVE) at (-3.8,-1.8) [child] {WAE};
    \node (WVC) at (-3.8,-2.8) [yellowc] {Continuous};
    
    \draw (ND.east) -- (GS);
    \draw (ND.east) -- (BT);
    \draw (ND.east) -- (IT);
    \draw (ND.west) -- (EO);
    \draw (ND.west) -- (WV);
    
    \draw (GS) -- (AC);
    \draw (GS) -- (AN);
    
    \draw (AC) -- (ACB);
    \draw (AC) -- (ACM);
    
    \draw (AN) -- (ANO);
    \draw (AN) -- (ANM);
    
    \draw (IT) -- (ITY);
    \draw (IT) -- (ITN);
    
    \draw (ANM) -- (ANMI);
    \draw (ANM) -- (ANMC);
    
    \draw (BT) -- (BTP);
    \draw (BT) -- (BTT);
    \draw (BT) -- (BTE);
    
    \draw (EO) -- (EOF);
    \draw (EO) -- (EOFP);
    \draw (EO) -- (EOS);
    
    \draw (EOS) -- (EOSR);
    \draw (EOS) -- (EOSL);
    
    \draw (WV) -- (WVI);
    \draw (WV) -- (WVE);
    \draw (WV) -- (WVC);
\end{tikzpicture}
    \caption{\rev{Summary of the normative values encoded by \algofair \revv{(\citet{zehlike2017fa, zehlike2022fair})}. 
    }}
    \label{fig:post-proc:FAIRDimensions}
\end{figure}
\spara{Insights.} The existence of minimum proportions for a protected group suggests a WAE worldview, \revv{and a focus on pre-existing bias.} 
However, the WYSIWYG worldview can also be adopted for a particular group, if needed: the respective $p$ can be set to a low value, meaning that only very few candidates from that group need to be in the ranking.
This permits a gradual transition from WYSIWYG to WAE. 
However, this is a mere side effect of the method, and, \revv{in contrast to~\citet{zehlike2017matching}, this is not to be seen as a desideratum of \algofair.}
Critically, if $p$ is chosen too high, the framework will rank a lot more protected candidates in the highest positions, than non-protected ones. 
\revv{
This would be justified only 
if one assumes \emph{either} that the protected group is actually a lot more relevant (their scores are very high in the construct space $CS$), but the measurements in observable space are extremely biased;  \emph{or} that the scoring model produces inverted predictions for the protected group in decision space $DS$. 
While such cases do exist, it is questionable whether one should attempt to correct such a flawed ranking, rather than disregarding it altogether.}

\algofair assumes that differences between groups in the relevance distribution are an artifact of different circumstances for each group.  \revv{Under the assumption that the goal of the fairness-enhancing intervention is to equalize access to opportunity over a lifetime, this method is consistent with substantive \eop.  Furthermore, because the proposed method conditions on morally irrelevant characteristics, we classify it as luck-egalitarian \eop. Figure~\ref{fig:post-proc:FAIRDimensions} and Table~\ref{tbl:method-summary_ir} summarize the mapping of the methods in this section to normative frameworks.}
%
%
%

\spara{Algorithm.}
For performance improvements, the binomial algorithm~\cite{zehlike2017fa} pre-computes a table that contains the minimum number of protected candidates from each group at each position in the ranking given a minimum proportion $p$ (see Table~\ref{tbl:ranked_group_fairness_table}). 
%
This is done by computing $F^{-1}\left(\alpha; k, p\right)$, the percent point function of the binomial CDF.
Then, the two groups are ranked separately by decreasing scores, and are then merged into a single ranking according to the table. 
Whenever the ranking by score violates the minimum number requirement, a protected item is put at the respective position.
For multiple protected groups the algorithm receives a minimum proportion for each group, and the data structure to hold the minimum numbers of candidates of each position becomes a tree instead of a table.
Thus multiple ways are possible to order candidates \revv{to produce a ranking that is fair} according to a multinomial distribution process.

%
As an example, consider Figure~\ref{fig:postProc_example} where a model has predicted relevance scores $\predScore{}$, based on which ranking $\hat \btau_1$ in Figure~\ref{fig:postProc_example} is produced. 
This ranking is fair according to the binomial \algofair algorithm, if the input is $p<0.5, \alpha=0.1$.
If $p \geq 0.5$ the candidates have to be reordered. 
Ranking $\hat \btau_2$ in Figure~\ref{fig:postProc_example} is produced by \algofair with $p=0.7, \alpha=0.1$.

\spara{Experiments.} 
The experimental evaluation in~\cite{zehlike2017fa} is done on the three real datasets, COMPAS~\cite{COMPASData}, German credit~\cite{GermanCreditData}, and SAT~\cite{SATData}, and shows the effects of the methods on performance in terms of NDCG and maximum rank loss. 
In~\cite{zehlike2022fair}, experiments are performed on COMPAS, German Credit, and LSAC~\cite{wightman1998lsac}.
All scenarios under consideration relate to questions of distributive fairness, where certain benefits (scholarship, visibility, pardon) are to be distributed fairly across two groups.

\subsubsection{Fairness-Aware Ranking at LinkedIn~\cite{geyik2019fairness}}
\label{subsubsec:LinkedIn}
\spara{Fairness Definition and Algorithm.} 
In this work, fairness is defined through the minimum and the maximum number of candidates at each position in the ranking. 
A proportion $p_{\group{}}$ for each group $\group{}$  is given as input, 
and a ranking $\btau$ is declared as fair for all groups if the following conditions hold:
\begin{align*}
\forall k \leq |\btau|, \; \forall \group{} &: \sum_{i = 0}^k \btau(i) \in \group{} \leq \lceil p_{\group{} \rceil \cdot } \\
\forall k \leq |\btau|, \; \forall \group{} &: \sum_{i = 0}^k \btau(i) \in \group{} \geq \lfloor p_{\group{} \rfloor \cdot k}
\end{align*}
To create a fair ranking, the algorithm first checks whether any group has not yet met its minimum number requirement (first inequality), and \revv{, if so, adds protected candidate to the ranking.} %
If more than one group \revv{does not yet meet the requirement,} the candidate with the highest utility score $\predScore$ among all eligible ones is chosen.
If all minimum requirements are met, the algorithm takes the highest-scoring candidate among all that have not yet exceeded their maximum count (second inequality).
As soon as more than three groups are present, this can easily lead to an infeasible state, where the requirements can no longer be met, no matter which candidate is chosen.
The authors therefore propose three additional ranking algorithms, each of them improving the likelihood to create a fair ranking by changing the strategy to choose the next candidate.

\begin{figure}[b!]
    \centering
    \begin{tikzpicture}[scale=0.9,align=center]
    \tikzset{
        root/.style = {rectangle, rounded corners, draw=black, fill=gray!30, font=\small\sffamily},
        child/.style = {rectangle, rounded corners,draw=black, font=\small\sffamily},
        orangec/.style = {rectangle, rounded corners, draw=black, fill=orange!70, font=\small\sffamily},
        greenc/.style = {rectangle, rounded corners, draw=black, fill=green!40, font=\small\sffamily},
        bluec/.style = {rectangle, rounded corners, draw=black, fill=blue!30, font=\small\sffamily},
        yellowc/.style = {rectangle, rounded corners, draw=black, fill=yellow!70, font=\small\sffamily},
        pinkc/.style = {rectangle, rounded corners, draw=black, fill=pink!70, font=\small\sffamily},
    }
    \node (ND) at (-0.2,0) [root] {Normative Dimensions};    
    \node (GS) at (1.8,1.75) [orangec] {Group Structure};    
    \node (BT) at (1.8,-1.75) [greenc] {Bias Type};    
    \node (EO) at (-1.8,1.75) [bluec] {Equal Opportunity};    
    \node (WV) at (-1.8,-1.75) [yellowc] {Worldview};  
    \node (IT) at (3.3,0.0) [pinkc] {Intersectional};
    
    \node (AC) at (3.9,2.5) [child] {Attribute Cardinality};
    \node (ACB) at (6.3,3.2) [child] {Binary};
    \node (ACM) at (6.3,1.8) [orangec] {Multinary};
    
    \node (AN) at (3.9,1.0) [child] {Attribute Number};
    \node (ANO) at (6.9,1.15) [child] {One};
    \node (ANM) at (6.9,0.5) [child] {Multiple};
    
    \node (ANMI) at (8.9,0.8) [child] {Independent};
    \node (ANMC) at (8.9,0.0) [orangec] {Combination};
    
    \node (ITY) at (5.3,0.2) [child] {Yes};
    \node (ITN) at (5.3,-0.4) [pinkc] {No};
    
    \node (BTP) at (4.0,-1.2) [greenc] {Pre-existing};
    \node (BTT) at (4.0,-2.0) [child] {Technical};
    \node (BTE) at (4.0,-2.75) [child] {Emergent};
    
    \node (EOF) at (-3.8,2.8) [child] {Formal};
    \node (EOFP) at (-4.6,2.1) [child] {Formal plus};
    \node (EOS) at (-3.8,1.0) [child] {Substantive};
    
    \node (EOSR) at (-5.8,1.4) [child] {Rawlsian};
    \node (EOSL) at (-5.4,0.1) [bluec] {Luck-egalitarian (1 sensitive attribute)};

    \node (WVI) at (-3.8,-0.8) [child] {WYSIWYG};
    \node (WVE) at (-3.8,-1.8) [child] {WAE};
    \node (WVC) at (-3.8,-2.8) [yellowc] {Continuous};
    
    \draw (ND.east) -- (GS);
    \draw (ND.east) -- (BT);
    \draw (ND.east) -- (IT);
    \draw (ND.west) -- (EO);
    \draw (ND.west) -- (WV);
    
    \draw (GS) -- (AC);
    \draw (GS) -- (AN);
    
    \draw (AC) -- (ACB);
    \draw (AC) -- (ACM);
    
    \draw (AN) -- (ANO);
    \draw (AN) -- (ANM);
    
    \draw (IT) -- (ITY);
    \draw (IT) -- (ITN);
    
    \draw (ANM) -- (ANMI);
    \draw (ANM) -- (ANMC);
    
    \draw (BT) -- (BTP);
    \draw (BT) -- (BTT);
    \draw (BT) -- (BTE);
    
    \draw (EO) -- (EOF);
    \draw (EO) -- (EOFP);
    \draw (EO) -- (EOS);
    
    \draw (EOS) -- (EOSR);
    \draw (EOS) -- (EOSL);
    
    \draw (WV) -- (WVI);
    \draw (WV) -- (WVE);
    \draw (WV) -- (WVC);
\end{tikzpicture}
    \caption{\rev{Summary of the normative values encoded by the fair LinkedIn method (\revv{\citet{geyik2019fairness}}). This translates into normative choices that are implicitly taken when applying the method. \revv{Note, that in contrast to \algofair, this method is not compliant with substantive \eop for more than 1 sensitive attribute, even though the methods appear to be very similar.} This shows how a small technical detail can lead to profound differences in underlying values.}}
    \label{fig:post-proc:geyik}
\end{figure}
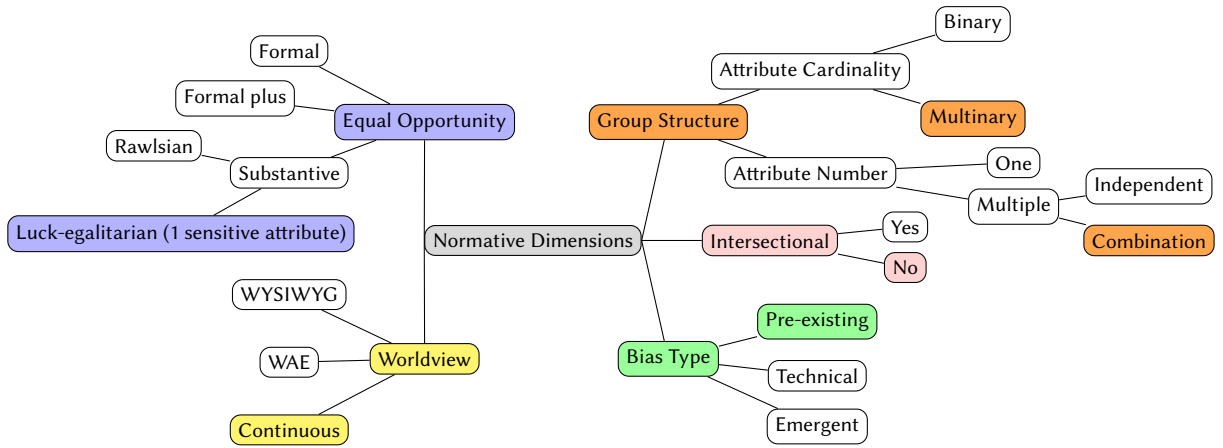

\spara{Insights.} The algorithm can handle multinary protected attributes, and is concerned with both pre-existing and technical bias.  As in \algofair, the proposed methods operate on minimum proportions for protected groups, and the fairness constraints are not incorporating a notion of merit.
The values of $p_{\group{}}$ can also be chosen freely and, therefore, \revv{at first glance}, a spectrum of worldviews and \eop frameworks can be adopted.
%
%
However, there is an important detail in the algorithm that makes it inconsistent with WAE and with substantive \eop: From all possible candidates to be selected, the algorithm always chooses the highest-scoring one. This is essentially the same design choice as was made by~\citet{celis2018ranking} in their Constrained ranking maximization paper (discussed in Section ~\ref{sec:fair_db:celis2018}) when the number of sensitive attributes exceeds 1.

%
This choice leads to an implicit integration of merit into the fairness objective, and it assumes that scores are comparable across groups. For this reason, we classify this method as being compatible with substantive \eop, and, specifically, with the luck-egalitarian \eop doctrine, for \emph{one} sensitive group, but deem it inconsistent with any \eop framework for multiple sensitive groups.  
It also ignores the fact that groups facing intersectional discrimination commonly show larger group skews and biases when their true merit from $\os$ is translated into scores in $\ds$.  
Figure~\ref{fig:post-proc:geyik} and Table~\ref{tbl:method-summary_ir} summarize our analysis.


\spara{Experiments.} Experiments are performed on synthetic datasets with different numbers of sensitive attributes.
The evaluation measures report fairness in terms of normalized discounted KL divergence~\cite{yang2017measuring}, and performance in terms of NDCG~\cite{jarvelin2002cumulated}.
Additionally, A/B-tests are performed after implementing the method as part of the LinkedIn Recruiter product.
The minimum proportions are set to match the distribution of all genders among the relevant candidates. 
The results show significant increases in terms of 
representativeness of gender among the top-$k$, but no decreases in performance measures, which is why the algorithm was implemented permanently in the product.

\subsubsection{Continuous Fairness with Optimal Transport, ~\citet{zehlike2017matching}}
\spara{Fairness Definition and Algorithm.}
This work defines a mathematical framework, CFA$\theta$, to continuously interpolate between the WYSIWYG and WAE worldviews.
The authors argue that, legally speaking, WYSIWYG is consistent with individual fairness, while WAE is consistent with current anti-discrimination law that defines group fairness in terms of \emph{statistical parity of outcomes}.
The answer to the question of what a fair distribution of outcome is, depends on the estimated extend of indirect discrimination and pre-existing bias in the scoring model. 
Interestingly, this means that any fairness definitions departing from group fairness as statistical parity, and individual fairness as meritocratic scores, do not yet have any legal meaning.
Furthermore, the authors state that the current rulings on anti-discrimination cases only involve actual unfair decisions, and not ``softer'' disadvantages such as reduced visibility in a ranking. 

As we discussed earlier, generally speaking, the WYSIWYG worldview corresponds to the meritocratic ideal and, hence, is consistent with formal \eop, while WAE corresponds to substantive \eop.  
This is because an individual's measurable effort (here, their raw score) is seen to be drawn from different distributions $\mu_k$ per group in $OS$, while in $CS$, there exists only one $\nu$, meaning that all groups have essentially the same distribution of true effort. 

The framework can handle multiple \revv{sensitive attributes}, each defining a \revv{partitioning} over all individuals: $\featureSet = \bigcup_{k\in\{0,1\}^N} \groupFunc^{-1}(k)$, where $\groupFunc: \featureSet \rightarrow \{0,1\}^N$ is a mapping that returns 1 if an individual carries a certain trait from the set of $N$ features. 
The $k$-th group is therefore $\group{k} := \groupFunc^{-1} (k)$.
The authors explicitly include all features in the group definition instead of only taking certain attributes that are legally protected into account. 
This broad definition has the advantage that it can handle non-sensitive features that serve as proxies for the sensitive features. 

The framework assumes that a potentially biased scoring function $S$ is given, and maps from the space of individual traits $\featureSet$ to an $n$-dimensional vector $S: \featureSet \rightarrow \mathds{R}^n$, and that each group's score forms a probability distribution $\mu_k$.
The combined score distribution is called $\mu$ and corresponds to a metric from $OS$, which, as usual, is prone to pre-existing bias and other systematic errors with different group skews.

The authors then define a score distribution $\nu_k = \mu_k \circ T^{-1}_k$ to be the fair score representation of group $k$ obtained by an optimal transport map $T_k: \mathds{R}^n \rightarrow \mathds{R}^n$. 
Depending on the worldview, $\nu$ is defined differently: In WYSIWYG $\mu = \nu_k$ for all $k$ groups, while in WAE any differences between the $\mu_k$ are solely the product of bias, and, there, there exists a single $\nu$ that is the same for all groups.
In the former case, the optimal transport matrix is the identity matrix.
In the latter, the WAE fair representation distribution $\nu$ that satisfies statistical parity is defined as the barycenter in Wasserstein space of the $\mu_k$, and a $T_k$ has to be found for each group to transform $\mu_k$ into $\nu$ while minimizing violations against decision maker utility and individual fairness.
The framework further defines a displacement interpolation with a fairness parameter $\theta \in [0, 1]$, which allows to transform $\mu_k$ into any distribution $\mu_k^\theta = \mu_k \circ (T^\theta)^{-1}$ between the WYSIWYG (or individual fairness) policy $\mu$ and the WAE (or group fairness) policy $\nu$.
This means that a high $\theta$ corresponds to more emphasis on group fairness, \revv{while a low $\theta$ corresponds to more individual fairness, with $\mu^0 = \mu$ and $\mu^1 = \nu$.}

A notable advantage of this approach is that it does not rely on the existence of a distance metric between individuals, in contrast to many other methods~\cite{dwork2012fairness, lahoti2019ifair}. 
This is important because it is not clear how such a distance metric, \revv{if it were actually} available in $OS$, would be less prone to biases and errors \revv{than any other commonly used} optimization metric. 

\begin{figure}[t]
    \centering
    \begin{tikzpicture}[scale=0.9,align=center]
    \tikzset{
        root/.style = {rectangle, rounded corners, draw=black, fill=gray!30, font=\small\sffamily},
        child/.style = {rectangle, rounded corners,draw=black, font=\small\sffamily},
        orangec/.style = {rectangle, rounded corners, draw=black, fill=orange!70, font=\small\sffamily},
        greenc/.style = {rectangle, rounded corners, draw=black, fill=green!40, font=\small\sffamily},
        bluec/.style = {rectangle, rounded corners, draw=black, fill=blue!30, font=\small\sffamily},
        yellowc/.style = {rectangle, rounded corners, draw=black, fill=yellow!70, font=\small\sffamily},
        pinkc/.style = {rectangle, rounded corners, draw=black, fill=pink!70, font=\small\sffamily},
    }
    \node (ND) at (-0.2,0) [root] {Normative Dimensions};    
    \node (GS) at (1.8,1.75) [orangec] {Group Structure};    
    \node (BT) at (1.8,-1.75) [greenc] {Bias Type};    
    \node (EO) at (-1.8,1.75) [bluec] {Equal Opportunity};    
    \node (WV) at (-1.8,-1.75) [yellowc] {Worldview};  
    \node (IT) at (3.3,0.0) [pinkc] {Intersectional};
    
    \node (AC) at (3.9,2.5) [child] {Attribute Cardinality};
    \node (ACB) at (6.3,3.2) [child] {Binary};
    \node (ACM) at (6.3,1.8) [orangec] {Multinary};
    
    \node (AN) at (3.9,1.0) [child] {Attribute Number};
    \node (ANO) at (6.9,1.15) [child] {One};
    \node (ANM) at (6.9,0.5) [child] {Multiple};
    
    \node (ANMI) at (8.9,0.8) [child] {Independent};
    \node (ANMC) at (8.9,0.0) [orangec] {Combination};
    
    \node (ITY) at (5.3,0.2) [child] {Yes};
    \node (ITN) at (5.3,-0.4) [pinkc] {No};
    
    \node (BTP) at (4.0,-1.2) [greenc] {Pre-existing};
    \node (BTT) at (4.0,-2.0) [child] {Technical};
    \node (BTE) at (4.0,-2.75) [child] {Emergent};
    
    \node (EOF) at (-3.8,2.8) [bluec] {Formal};
    \node (EOFP) at (-4.6,2.1) [child] {Formal plus};
    \node (EOS) at (-3.8,1.0) [bluec] {Substantive};
    
    \node (EOSR) at (-5.8,1.4) [child] {Rawlsian};
    \node (EOSL) at (-5.4,0.1) [child] {Luck-egalitarian};

    \node (WVI) at (-3.8,-0.8) [child] {WYSIWYG};
    \node (WVE) at (-3.8,-1.8) [child] {WAE};
    \node (WVC) at (-3.8,-2.8) [yellowc] {Continuous};
    
    \draw (ND.east) -- (GS);
    \draw (ND.east) -- (BT);
    \draw (ND.east) -- (IT);
    \draw (ND.west) -- (EO);
    \draw (ND.west) -- (WV);
    
    \draw (GS) -- (AC);
    \draw (GS) -- (AN);
    
    \draw (AC) -- (ACB);
    \draw (AC) -- (ACM);
    
    \draw (AN) -- (ANO);
    \draw (AN) -- (ANM);
    
    \draw (IT) -- (ITY);
    \draw (IT) -- (ITN);
    
    \draw (ANM) -- (ANMI);
    \draw (ANM) -- (ANMC);
    
    \draw (BT) -- (BTP);
    \draw (BT) -- (BTT);
    \draw (BT) -- (BTE);
    
    \draw (EO) -- (EOF);
    \draw (EO) -- (EOFP);
    \draw (EO) -- (EOS);
    
    \draw (EOS) -- (EOSR);
    \draw (EOS) -- (EOSL);
    
    \draw (WV) -- (WVI);
    \draw (WV) -- (WVE);
    \draw (WV) -- (WVC);
\end{tikzpicture}
    \caption{\rev{Summary of the normative values encoded by CFA$\theta$ \revv{(\citet{zehlike2017matching})}. 
    The paper is a rare example of explicit statements on the incorporated values.}}
    \label{fig:post-proc:cfa}
\end{figure}
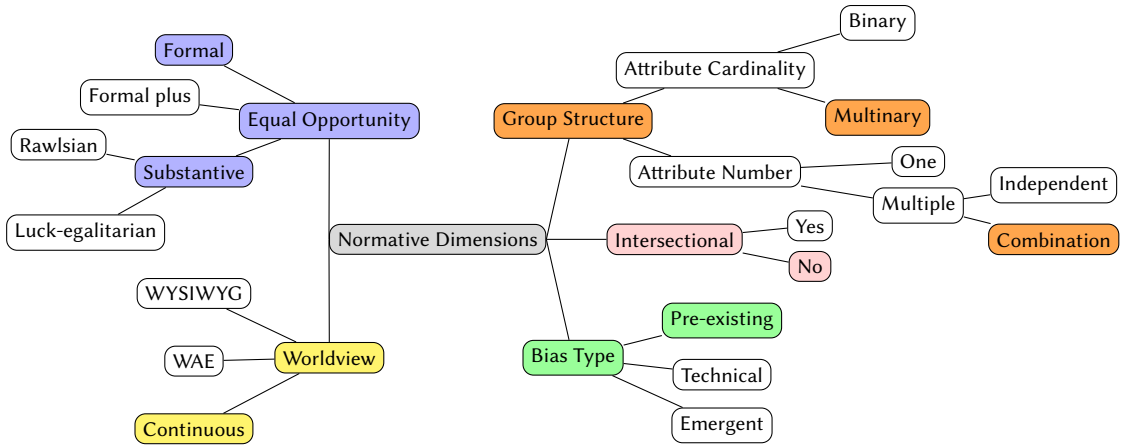

\spara{Insights.}
As CFA$\theta$ moves the group distributions of predicted scores closer to each other, this means that when setting $\theta=1$ the algorithm achieves statistical parity for each group throughout the ranking, \emph{but not more so} as compared to \algofair, which can push candidates from the protected group even higher.
Such a setting would result in the same ranking $\hat \btau_2$ as shown in Figure~\ref{fig:postProc_example}, given that there are 50\% of \revv{male and female candidates} in the dataset.
However if there where only 25\% \revv{female candidates} in the dataset (\eg if candidates $f$ and $k$ were male), then the algorithm would produce the ranking $\hat \btau_3$ as shown in Figure~\ref{fig:postProc_example} when given $\theta=1$ as input.
Note that with such a dataset the method cannot return a ranking in which the now only two female candidates are ranked among those admitted (\ie in the top-$4$).
Figure~\ref{fig:post-proc:cfa} and Table~\ref{tbl:method-summary_ir} summarize our analysis.

\spara{Experiments.}
The experimental evaluation is done on a synthetic dataset with 100,000 data points, a score feature and two sensitive features. 
Group membership for this experiment is defined by all combinations of the values of the sensitive features.
Rankings are produced based on the score column and the performance of a fair ranking with different $\theta$ is measured in terms of NDCG.
The fairness of a ranking is measured as the share of each group at each $n$ positions, with $n$ ranging from 10 to 1000. 

A second experiment is conducted on the LSAC dataset~\cite{LSACData}.  
The experiments confirm the general properties of post-processing methods: it is clearly visible how groups are distributed more evenly across all positions with increasing values of $\theta$. 
However, depending on the differences between the $\mu_k$, a processed ranking based on the fair representation $\nu$ can show significant declines in performance measures w.r.t. the raw score ranking.

\subsubsection{Fairness of Exposure,~\citet{singh2018fairness}}
\label{subsubsec:FairnessOfExposure}
\spara{Fairness Definition. }
The fairness objective of this work is set as a linear combination $\mathbf{a}^T\mathbf{P}^{\btau}_{a,i}\mathbf{v}=h$ with $\mathbf{a}$ being a vector to encode group membership, $\mathbf{P}^{\btau}_{a,i}$ as the probability that $\hat f$ places candidate $a$ at rank $i$ in $\btau$, and $\mathbf{v}$ reflecting the importance of a position in a ranking. 
This equation is solved under three different group fairness constraints based on a definition of exposure that a candidate $a$ receives under $\mathbf P^{\btau}$:
\begin{equation*}
    \operatorname{Exposure}(\featOfCand{a}|\mathbf{P}^{\btau}) = \sum^{k}_{i=1} \mathbf{P}^{\btau}_{a,i}\posBias{i}
\end{equation*}
with $\posBias{i}$ being the position bias of position $i$ in ranking $\btau$.
The average exposure of a group $\group{}$ is defined as follows:
\begin{equation*}
    \operatorname{Exposure}(\group{}|\mathbf{P}^{\btau}) = \frac{1}{|\group{}|}\sum_{\featOfCand{a} \in \group{}}\operatorname{Exposure}(\featOfCand{a}|\mathbf{P}^{\btau})
\end{equation*}
The goal is to distribute exposure fairly between groups $\group{0}$ and $\group{1}$ using the following three definitions:
\begin{enumerate}
    \item \textbf{Demographic Parity} states that the average exposure of groups shall be equal $\operatorname{Exposure}(\group{0}|\mathbf{P}^{\btau}) = \operatorname{Exposure}(\group{1}|\mathbf{P}^{\btau})$. 
    \item \textbf{Disparate Treatment} requires equity of exposure across groups (\ie the average exposure in relation to their average utility should be equal across groups):
    \begin{equation}
        \frac{\operatorname{Exposure}(\group{0}|\mathbf{P}^{\btau})}{\utilityThreePara{}{\btau}{\group{0}}} = \frac{\operatorname{Exposure}(\group{1}|\mathbf{P}^{\btau})}{\utilityThreePara{}{\btau}{\group{1}}}
    \end{equation}
    \item \textbf{Disparate Impact} \revv{is measured in terms of disparate click through rates (CTR)~\cite{richardson2007predicting} across group.  The goal is to equalize CTR across groups, given the groups average utility:}
    \begin{equation*}
        \operatorname{CTR}(\group{}|\mathbf{P}^{\btau}) = \frac{1}{|\group{}|}\sum_{a \in \group{}} \sum^{k}_{i=1}\mathbf{P}_{a,i}^{\btau} \score{a} \posBias{i}
    \end{equation*}
    with $\score{a}$ being the relevance of candidate $a$.
    \begin{equation}
        \frac{\operatorname{CTR}(\group{0}|\mathbf{P}^{\btau})}{\utilityThreePara{}{\btau}{\group{0}}} = \frac{\operatorname{CTR}(\group{1}|\mathbf{P}^{\btau})}{\utilityThreePara{}{\btau}{\group{1}}}
    \end{equation}
\end{enumerate}

\begin{figure}[b!]
    \centering
    \begin{tikzpicture}[scale=0.9,align=center]
    \tikzset{
        root/.style = {rectangle, rounded corners, draw=black, fill=gray!30, font=\small\sffamily},
        child/.style = {rectangle, rounded corners,draw=black, font=\small\sffamily},
        orangec/.style = {rectangle, rounded corners, draw=black, fill=orange!70, font=\small\sffamily},
        greenc/.style = {rectangle, rounded corners, draw=black, fill=green!40, font=\small\sffamily},
        bluec/.style = {rectangle, rounded corners, draw=black, fill=blue!30, font=\small\sffamily},
        yellowc/.style = {rectangle, rounded corners, draw=black, fill=yellow!70, font=\small\sffamily},
        pinkc/.style = {rectangle, rounded corners, draw=black, fill=pink!70, font=\small\sffamily},
    }
    \node (ND) at (-0.2,0) [root] {Normative Dimensions};    
    \node (GS) at (1.8,1.75) [orangec] {Group Structure};    
    \node (BT) at (1.8,-1.75) [greenc] {Bias Type};    
    \node (EO) at (-1.8,1.75) [bluec] {Equal Opportunity};    
    \node (WV) at (-1.8,-1.75) [yellowc] {Worldview};  
    \node (IT) at (3.3,0.0) [pinkc] {Intersectional};
    
    \node (AC) at (3.9,2.5) [child] {Attribute Cardinality};
    \node (ACB) at (6.3,3.2) [orangec] {Binary};
    \node (ACM) at (6.3,1.8) [child] {Multinary};
    
    \node (AN) at (3.9,1.0) [child] {Attribute Number};
    \node (ANO) at (6.9,1.15) [orangec] {One};
    \node (ANM) at (6.9,0.5) [child] {Multiple};
    
    \node (ANMI) at (8.9,0.8) [child] {Independent};
    \node (ANMC) at (8.9,0.0) [child] {Combination};
    
    \node (ITY) at (5.3,0.2) [child] {Yes};
    \node (ITN) at (5.3,-0.4) [pinkc] {No};
    
    \node (BTP) at (4.0,-1.2) [greenc] {Pre-existing};
    \node (BTT) at (4.0,-2.0) [child] {Technical};
    \node (BTE) at (4.0,-2.75) [child] {Emergent};
    
    \node (EOF) at (-3.8,2.8) [child] {Formal};
    \node (EOFP) at (-4.6,2.1) [child] {Formal plus};
    \node (EOS) at (-3.8,1.0) [child] {Substantive};
    
    \node (EOSR) at (-5.8,1.4) [child] {Rawlsian};
    \node (EOSL) at (-5.4,0.1) [bluec] {Luck-egalitarian};

    \node (WVI) at (-3.8,-0.8) [child] {WYSIWYG};
    \node (WVE) at (-3.8,-1.8) [yellowc] {WAE};
    \node (WVC) at (-3.8,-2.8) [child] {Continuous};
    
    \draw (ND.east) -- (GS);
    \draw (ND.east) -- (BT);
    \draw (ND.east) -- (IT);
    \draw (ND.west) -- (EO);
    \draw (ND.west) -- (WV);
    
    \draw (GS) -- (AC);
    \draw (GS) -- (AN);
    
    \draw (AC) -- (ACB);
    \draw (AC) -- (ACM);
    
    \draw (AN) -- (ANO);
    \draw (AN) -- (ANM);
    
    \draw (IT) -- (ITY);
    \draw (IT) -- (ITN);
    
    \draw (ANM) -- (ANMI);
    \draw (ANM) -- (ANMC);
    
    \draw (BT) -- (BTP);
    \draw (BT) -- (BTT);
    \draw (BT) -- (BTE);
    
    \draw (EO) -- (EOF);
    \draw (EO) -- (EOFP);
    \draw (EO) -- (EOS);
    
    \draw (EOS) -- (EOSR);
    \draw (EOS) -- (EOSL);
    
    \draw (WV) -- (WVI);
    \draw (WV) -- (WVE);
    \draw (WV) -- (WVC);
\end{tikzpicture}
    \caption{\rev{Summary of the normative values encoded by the definition of demographic parity of Singh and Joachims~\cite{singh2018fairness}.}}
    \label{fig:post-proc:singhOne}
\end{figure}
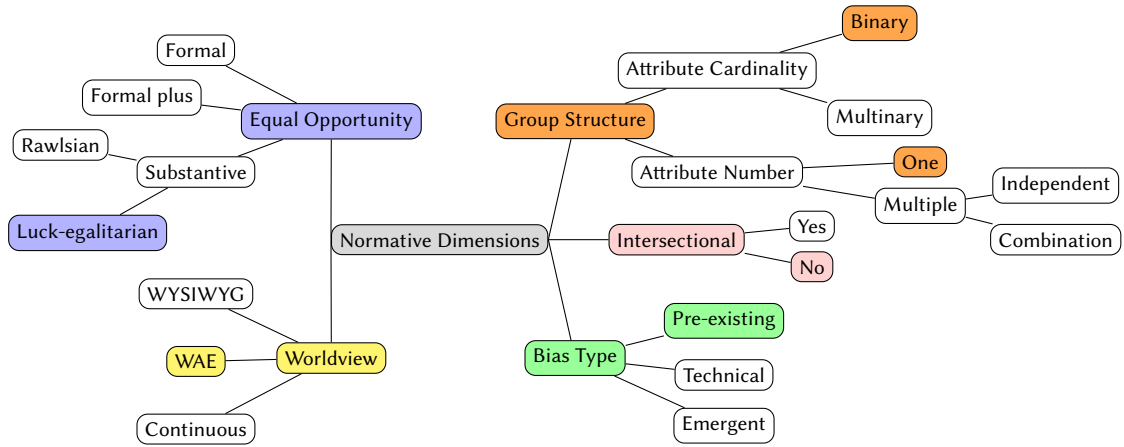
\spara{Insights.} The definition of demographic parity addresses the problem of pre-existing bias and corresponds to the WAE framework, as it tries to balance visibility across groups independently of their performance in $OS$ (note, however, that a group's exposure depends on $\mathbf{P}^{\btau}$, and may thus indirectly depend on a utility measure, if $\mathbf{P}^{\btau}$ is calculated based on document utility). 
The definition assumes that $CS \nsim OS$ and therefore that an individuals true effort is different from the measured effort, \revv{accounted for by the demographic parity definition.}
\revv{Under the assumption that the goal of the competition is to equalize opportunity over a lifetime, these methods are consistent with substantive \eop.  Further, because of conditioning on group membership, we map these methods to luck-egalitarian \eop.}
%
%
Figure~\ref{fig:post-proc:singhOne} and Table~\ref{tbl:method-summary_ir} summarize our analysis.

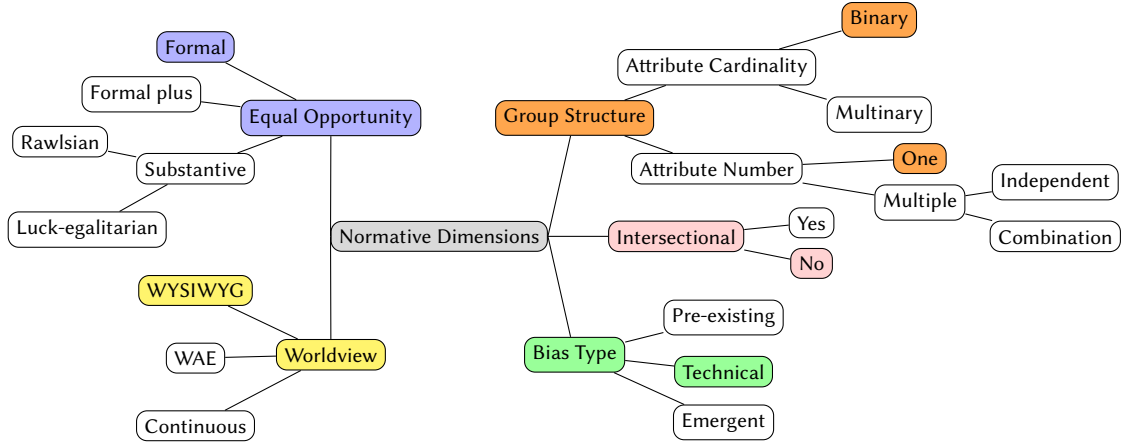
\begin{figure}[t!]
    \centering
    \begin{tikzpicture}[scale=0.9,align=center]
    \tikzset{
        root/.style = {rectangle, rounded corners, draw=black, fill=gray!30, font=\small\sffamily},
        child/.style = {rectangle, rounded corners,draw=black, font=\small\sffamily},
        orangec/.style = {rectangle, rounded corners, draw=black, fill=orange!70, font=\small\sffamily},
        greenc/.style = {rectangle, rounded corners, draw=black, fill=green!40, font=\small\sffamily},
        bluec/.style = {rectangle, rounded corners, draw=black, fill=blue!30, font=\small\sffamily},
        yellowc/.style = {rectangle, rounded corners, draw=black, fill=yellow!70, font=\small\sffamily},
        pinkc/.style = {rectangle, rounded corners, draw=black, fill=pink!70, font=\small\sffamily},
    }
    \node (ND) at (-0.2,0) [root] {Normative Dimensions};    
    \node (GS) at (1.8,1.75) [orangec] {Group Structure};    
    \node (BT) at (1.8,-1.75) [greenc] {Bias Type};    
    \node (EO) at (-1.8,1.75) [bluec] {Equal Opportunity};    
    \node (WV) at (-1.8,-1.75) [yellowc] {Worldview};  
    \node (IT) at (3.3,0.0) [pinkc] {Intersectional};
    
    \node (AC) at (3.9,2.5) [child] {Attribute Cardinality};
    \node (ACB) at (6.3,3.2) [orangec] {Binary};
    \node (ACM) at (6.3,1.8) [child] {Multinary};
    
    \node (AN) at (3.9,1.0) [child] {Attribute Number};
    \node (ANO) at (6.9,1.15) [orangec] {One};
    \node (ANM) at (6.9,0.5) [child] {Multiple};
    
    \node (ANMI) at (8.9,0.8) [child] {Independent};
    \node (ANMC) at (8.9,0.0) [child] {Combination};
    
    \node (ITY) at (5.3,0.2) [child] {Yes};
    \node (ITN) at (5.3,-0.4) [pinkc] {No};
    
    \node (BTP) at (4.0,-1.2) [child] {Pre-existing};
    \node (BTT) at (4.0,-2.0) [greenc] {Technical};
    \node (BTE) at (4.0,-2.75) [child] {Emergent};
    
    \node (EOF) at (-3.8,2.8) [bluec] {Formal};
    \node (EOFP) at (-4.6,2.1) [child] {Formal plus};
    \node (EOS) at (-3.8,1.0) [child] {Substantive};
    
    \node (EOSR) at (-5.8,1.4) [child] {Rawlsian};
    \node (EOSL) at (-5.4,0.1) [child] {Luck-egalitarian};

    \node (WVI) at (-3.8,-0.8) [yellowc] {WYSIWYG};
    \node (WVE) at (-3.8,-1.8) [child] {WAE};
    \node (WVC) at (-3.8,-2.8) [child] {Continuous};
    
    \draw (ND.east) -- (GS);
    \draw (ND.east) -- (BT);
    \draw (ND.east) -- (IT);
    \draw (ND.west) -- (EO);
    \draw (ND.west) -- (WV);
    
    \draw (GS) -- (AC);
    \draw (GS) -- (AN);
    
    \draw (AC) -- (ACB);
    \draw (AC) -- (ACM);
    
    \draw (AN) -- (ANO);
    \draw (AN) -- (ANM);
    
    \draw (IT) -- (ITY);
    \draw (IT) -- (ITN);
    
    \draw (ANM) -- (ANMI);
    \draw (ANM) -- (ANMC);
    
    \draw (BT) -- (BTP);
    \draw (BT) -- (BTT);
    \draw (BT) -- (BTE);
    
    \draw (EO) -- (EOF);
    \draw (EO) -- (EOFP);
    \draw (EO) -- (EOS);
    
    \draw (EOS) -- (EOSR);
    \draw (EOS) -- (EOSL);
    
    \draw (WV) -- (WVI);
    \draw (WV) -- (WVE);
    \draw (WV) -- (WVC);
\end{tikzpicture}
    \caption{\revv{Summary of the normative values encoded by the definitions of ``disparate treatment'' and  ``disparate impact'' of~\citet{singh2018fairness}.}}
    \label{fig:post-proc:singhTwo}
\end{figure}

The definition of disparate treatment explicitly addresses the technical bias of a ranking, also known as position bias, by ensuring that all documents of the same utility receive equal visibility. 
This is consistent with formal \eop framework.  
%
%
The method corresponds to the WYSIWYG worldview because document utility is measured through features from observable space without taking into account that a biased observation process may exist, hence $CS \sim OS$. 
Note that this does not necessarily correspond to individual fairness, because utility and exposure are averaged across individuals of a group, and can lead to a downgrading of high scoring individuals in otherwise badly performing groups. 
\revv{Figure~\ref{fig:post-proc:singhTwo} and Table~\ref{tbl:method-summary_ir} summarize our analysis.}
%

The definition of disparate impact is misleading does not comply with the \emph{legal} definition of disparate impact, which is described solely in terms of the deviation from statistical parity.
\revv{Referring to this definition as ``disparate impact'' is misleading, because the click through rate contains a notion of document relevance. } 
Statistical parity, in contrast, does not consider any relevance measure whatsoever, precisely because it assumes that these very measurements are subject to pre-existing biases and a biased mapping from $CS$ to $OS$. 
As the given definition mostly corresponds to formal \eop and a WYSIWYG worldview, it would be more appropriate to label it as a different version of disparate treatment that is concerned with click through rate instead of exposure. 
Since the two definitions correspond to the same value frameworks, we have summarized both of them in Figure~\ref{fig:post-proc:singhTwo}, as well as in Table~\ref{tbl:method-summary_ir}.

\spara{Algorithm.}
The algorithmic framework is implemented as an ILP that maximizes ranking utility given one of the above constraints translated into a scalar $h$. \revv{Note that this is in contrast to~\citet{biega2018equity}, to be discussed in Section~\ref{subsubsec:equityOfAttention}, who instead constraint quality and optimize for disparate treatment}:
\begin{argmaxi}|l|
    {\mathbf{P}}{\score{}^T\mathbf{Pv}}{}{}
    \addConstraint{\mathds{1}^T\mathbf{P}}{=\mathds{1}^T}{}
    \addConstraint{\mathbf{P}\mathds{1}}{=\mathds{1}}{}
    \addConstraint{0 \leq \mathbf{P}_{a,i}}{\leq 1}{}
    \addConstraint{\mathbf{a}^T\mathbf{Pv}}{=h}{}
\end{argmaxi}

Depending on the respective definition of fairness, the outcome rankings will look quite differently.
Let us assume that the model is absolutely sure about where to place each candidate, \revv{such that $\mathbf{P}$ becomes the unit matrix and exposure of a group is calculated as the sum of each group member's position bias in the ranking.}
This gives us a group exposure of 2.13 for the male group and 1.15 for the female group for the ranking $\hat \btau_1$ in Figure~\ref{fig:postProc_example}.
\revv{Let us consider the demographic parity objective, that requires to equalize exposure for both groups.}
As the position bias $\mathbf v$ is a constant value, the algorithm will modify the scores $\predScore{}$ until the parity objective is met.
A possible solution is ranking $\hat \btau_4$, shown in Figure~\ref{fig:postProc_example}, with group exposure of 1.66 for the male group and 1.62 for the female group.
%

%
The other two objectives work in an similar manner, \revv{except that they take utility of a ranking into account.}

\spara{Experiments.}
The experiments are framed within three different scenarios of unfairness: biased allocation of opportunity (in job candidate rankings), misrepresentation of real-world distributions (biased Google image search for CEO), and fairness as freedom of speech (equality of voice within news media channels like YouTube or Twitter). 
For these, the authors create a synthetic dataset with 100,000 entries and a binary protected attribute. 
Furthermore they \revv{use the real YOW news} recommendation dataset~\cite{YowData}.

\subsubsection{Equity of Attention, \citet{biega2018equity}}
\label{subsubsec:equityOfAttention}
\spara{Fairness Definition.}
%
Each position in a ranked list carries an inherent position bias that increases with the position number, meaning that even if all items had the same relevance, those at the top of the ranking would receive a lot more attention compared to items at lower ranks.
\citet{biega2018equity} frame this discrepancy as a problem of distributive individual fairness, \revv{aiming to achieve} equity of attention on the level of individuals, \revv{and postulating that} the attention an item gets from users should be proportional to their relevance to the query.
Assuming that relevance decreases linearly and attention decreases geometrically, \revv{there is necessarily a discrepancy between the attention loss of a candidates and their decrease in relevance.}

This work proposes a post-processing algorithm that optimizes equity of user attention with a constrained overall relevance loss.  \revv{(Note that this is in contrast to~\citet{singh2018fairness}, discussed in Section~\ref{subsubsec:FairnessOfExposure}, that constrains fairness and optimizes for relevance.)}
Because a single ranking cannot be fair according to their definition, the authors propose an approach in which unfairness is mitigated \emph{over time}. \revv{For a pair of well-suited candidates $a$ and $b$, attention enjoyed by them should be equalized over $m$ rankings $\btau_{1 \ldots m}$: }
%
    \begin{equation}
        \frac{\sum^{m}_{i=1} \attention{\btau_i}{a}}{\sum^{m}_{i=1} \utilityThreePara{}{\btau_i}{a}} = \frac{\sum^{m}_{i=1} \attention{\btau_i}{b}}{\sum^{m}_{i=1} \utilityThreePara{}{\btau_i}{b}}
    \end{equation}
Hence, unfairness is measured as the accumulated difference between the attention enjoyed by \revv{the candidates} and their relevance:
    \begin{equation}
        \operatorname{unfairness}(\btau_1, \ldots, \btau_m) = \sum^{n}_{a=1} \left| \sum^{m}_{i=1} \attention{\btau_i}{a} - \sum^{m}_{i=1}  \utilityThreePara{}{\btau_i}{a} \right|
    \end{equation}
    
\begin{figure}[h]
    \centering
    \begin{tikzpicture}[scale=0.9,align=center]
    \tikzset{
        root/.style = {rectangle, rounded corners, draw=black, fill=gray!30, font=\small\sffamily},
        child/.style = {rectangle, rounded corners,draw=black, font=\small\sffamily},
        orangec/.style = {rectangle, rounded corners, draw=black, fill=orange!70, font=\small\sffamily},
        greenc/.style = {rectangle, rounded corners, draw=black, fill=green!40, font=\small\sffamily},
        bluec/.style = {rectangle, rounded corners, draw=black, fill=blue!30, font=\small\sffamily},
        yellowc/.style = {rectangle, rounded corners, draw=black, fill=yellow!70, font=\small\sffamily},
        pinkc/.style = {rectangle, rounded corners, draw=black, fill=pink!70, font=\small\sffamily},
    }
    \node (ND) at (-0.2,0) [root] {Normative Dimensions};    
    \node (GS) at (1.8,1.75) [orangec] {Group Structure};    
    \node (BT) at (1.8,-1.75) [greenc] {Bias Type};    
    \node (EO) at (-1.8,1.75) [bluec] {Equal Opportunity};    
    \node (WV) at (-1.8,-1.75) [yellowc] {Worldview};  
    \node (IT) at (3.3,0.0) [pinkc] {Intersectional};
    
    \node (AC) at (3.9,2.5) [child] {Attribute Cardinality};
    \node (ACB) at (6.3,3.2) [child] {Binary};
    \node (ACM) at (6.3,1.8) [orangec] {Multinary};
    
    \node (AN) at (3.9,1.0) [child] {Attribute Number};
    \node (ANO) at (6.9,1.15) [child] {One};
    \node (ANM) at (6.9,0.5) [child] {Multiple};
    
    \node (ANMI) at (8.9,0.8) [orangec] {Independent};
    \node (ANMC) at (8.9,0.0) [child] {Combination};
    
    \node (ITY) at (5.3,0.2) [child] {Yes};
    \node (ITN) at (5.3,-0.4) [pinkc] {No};
    
    \node (BTP) at (4.0,-1.2) [child] {Pre-existing};
    \node (BTT) at (4.0,-2.0) [greenc] {Technical};
    \node (BTE) at (4.0,-2.75) [greenc] {Emergent};
    
    \node (EOF) at (-3.8,2.8) [bluec] {Formal};
    \node (EOFP) at (-4.6,2.1) [child] {Formal plus};
    \node (EOS) at (-3.8,1.0) [child] {Substantive};
    
    \node (EOSR) at (-5.8,1.4) [child] {Rawlsian};
    \node (EOSL) at (-5.4,0.1) [child] {Luck-egalitarian};

    \node (WVI) at (-3.8,-0.8) [yellowc] {WYSIWYG};
    \node (WVE) at (-3.8,-1.8) [child] {WAE};
    \node (WVC) at (-3.8,-2.8) [child] {Continuous};
    
    \draw (ND.east) -- (GS);
    \draw (ND.east) -- (BT);
    \draw (ND.east) -- (IT);
    \draw (ND.west) -- (EO);
    \draw (ND.west) -- (WV);
    
    \draw (GS) -- (AC);
    \draw (GS) -- (AN);
    
    \draw (AC) -- (ACB);
    \draw (AC) -- (ACM);
    
    \draw (AN) -- (ANO);
    \draw (AN) -- (ANM);
    
    \draw (IT) -- (ITY);
    \draw (IT) -- (ITN);
    
    \draw (ANM) -- (ANMI);
    \draw (ANM) -- (ANMC);
    
    \draw (BT) -- (BTP);
    \draw (BT) -- (BTT);
    \draw (BT) -- (BTE);
    
    \draw (EO) -- (EOF);
    \draw (EO) -- (EOFP);
    \draw (EO) -- (EOS);
    
    \draw (EOS) -- (EOSR);
    \draw (EOS) -- (EOSL);
    
    \draw (WV) -- (WVI);
    \draw (WV) -- (WVE);
    \draw (WV) -- (WVC);
\end{tikzpicture}
    \caption{\rev{Summary of the normative values encoded by~\citet{biega2018equity}. This is the only method \revv{in this survey} that explicitly addresses emergent bias.}}
    \label{fig:post-proc:biega}
\end{figure}
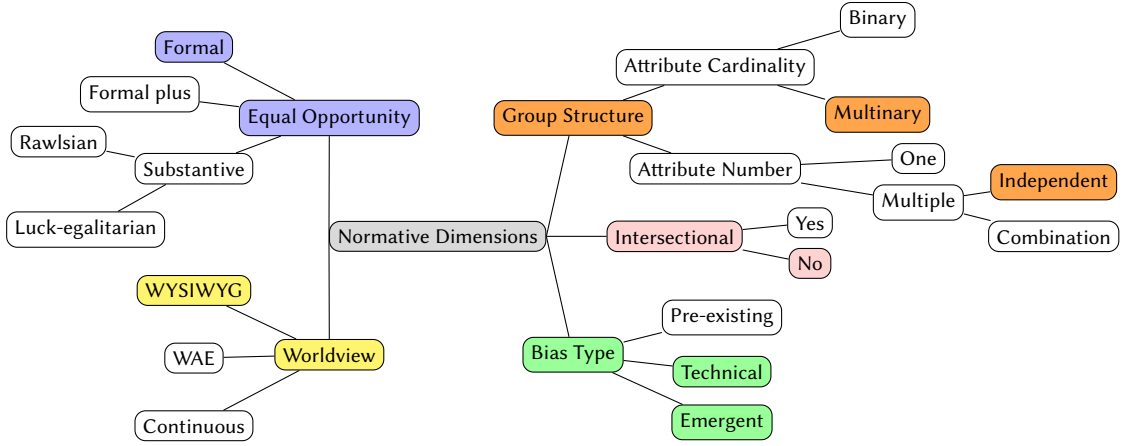

\spara{Insights.} The authors relate their work to the notion of individual fairness by~\citet{dwork2012fairness} and treat group fairness, \revv{which they call ``equality of attention,''} as a special case, \revv{when} utility distributions are uniform across all rankings: $ \utilityThreePara{}{\btau}{a} =  \utilityThreePara{}{\btau}{b}, \forall a, b$.
Note that this understanding of group fairness can not account for biases and errors in the data that are manifesting differently across groups, meaning that the error for the protected group might be high, while the data for the \revv{privileged group may be error-free.}  This method is consistent with formal \eop.
This work's definition of group fairness corresponds therefore explicitly with a WYSIWYG worldview and contrasts with most of the other definitions in the literature.
\revv{In addition to an explicit focus on technical bias, this method can also address} emergent bias, which may result from online learning algorithms that learn through user feedback and click data.
Figure~\ref{fig:post-proc:biega} and Table~\ref{tbl:method-summary_ir} summarize our analysis.

\spara{Algorithm.}
The fairness definition is implemented as a constrained optimization problem (which is then translated into an Integer Linear Program) in which unfairness is minimized subject to constraints on the maximum NDCG-loss in an online manner.
This means that the algorithms allows unfairness minimization over time, and candidates can enter and leave the system at any point.
The algorithm reorders a given ranking such that unfairness is minimized given the cumulative attention and relevance distribution seen so far:
\begin{mini}|l|
    {}{\sum_a \left| \sum^{l-1}_{i=1} \attention{\btau_i}{a} + \attention{\btau_l}{a} - \left(\sum^{l-1}_{i=1}  \utilityThreePara{}{\btau_i}{a} +  \utilityThreePara{}{\btau_l}{a}\right) \right| }{}{}
    \addConstraint{\operatorname{NDCG@k}(\btau_l, \btau_{l^*}) \geq \theta}{}{}
\end{mini}
where $\btau_l$ is the current,  $\btau_{l^*}$ is the reordered ranking, and $\theta$ is the quality constraint, \revv{with higher values corresponding to more fairness}.

As this method measures the attention each item received over time given their relevance, we have to consider several rankings to illustrate its effect. 
As attention is defined through position bias too, just as in~\cite{singh2018fairness}, let us come back to Figure~\ref{fig:postProc_example}.
It is likely that the first few rankings will look like the ranking $\hat \btau_1$ in Figure~\ref{fig:postProc_example}, \revv{with candidate $f$ ranked right below candidate $e$}.
However, at some point, the attention received by candidate $f$ will be much lower than attention received by its neighbor $e$, because the relevance of $f$ decreases linearly, while its attention decreases geometrically.
If this happens, then candidates $e$ and $f$ will be swapped in the next ranking produced by the model.

\spara{Experiments.} The experiments are run on two real datasets, AirBnB~\cite{AirBnBData} and StackExchange~\cite{StackData}, and a synthetic dataset, each with two different models for attention gains by position: (1) geometric attention decrease by position, and (2) the first position gets all attention.
\begin{itemize}
    \item \textbf{Synthetic Data:} Experiments are set up with three different relevance score distributions (uniform, linear, exponential) and the aforementioned two attention models. 
    In all cases, the algorithm shows periodic behavior under lab conditions, meaning that every $x$ rounds, it brings unfairness to 0.  Furthermore, fairness is not a steady state but will grow with each new ranking for each individual item.
    \item \textbf{AirBnB:} For these experiments rankings were created from AirBnB apartment listings in Hongkong (4529 items), Boston (3944 items) and Geneva (1728 items) under two scenarios --- (1) always using  the same query, (2) using different queries.
    The results verify that the difference between the distributions of attention and relevance is substantial in real world datasets, and whether unfairness can be effectively mitigated is highly dependant on the dataset at hand.
    In all cases, unfairness did not increase over time only when no utility constraint was given (\it $\theta = 0$.
    \item \textbf{StackExchange:} The experiments show that individual subjects appear in relatively few result rankings, leading to an extended fairness amortization time.
\end{itemize}

\subsection{Discussion of the Normative Framework Mapping}
In this section, we have seen how technical decisions lead to different value frameworks that are supported by a method.
In some cases these differences are quite obvious: those method that explicitly incorporate a notion of utility \emph{into their fairness objective}, namely \citet{lahoti2019ifair, singh2019policy, biega2018equity}, and the disparate treatment and disparate impact definition of~\citet{singh2018fairness}, generally lean towards the WYSIWYG worldview, with $\cs \approx \os$, and are consistent with formal \eop.
In contrast, methods that explicitly exclude a utility measure from the fairness definition (\citet{zehlike2018reducing, zehlike2017fa, zehlike2017matching, geyik2019fairness}, and the demographic parity definition of~\citet{singh2018fairness}), generally lean towards the WAE worldview and substantive \eop.
Additionally, some methods explicitly allow continuous interpolation between these two worldviews, either by introducing a sliding parameter, or by allowing a range of values for the fairness constraints (\citet{zehlike2017fa, zehlike2018reducing, zehlike2017matching, geyik2019fairness}).

However, the devil is in the details, and even though some methods appear to make \revv{very similar technical assumptions}, a minor difference in design choices can lead to tremendous differences in their underlying value frameworks.
Let us take \algofair (Sec.~\ref{subsubsec:FAIR}) and LinkedIn (Sec.~\ref{subsubsec:LinkedIn}) as examples.
The technical choices of these two algorithms appear to be extremely similar: both receive a vector of minimum proportions for the protected groups, and reorder a ranking such that  candidates from all groups are shown throughout the ranking according to these minimum proportions.
Neither method incorporates utility into the fairness constraints, and both seem to continuously adopt worldviews between WYSIWYG and WAE. 
However, \citet{geyik2019fairness} explicitly decided to sort all protected candidates based on their utility and always pick the highest-scoring candidate whenever ties in the proportion would allow different choices.
In the same situation, \citet{zehlike2017fa} explicitly pick the next candidate based on a probability distribution, which again is not based on any utility measure.
\citet{geyik2019fairness} incorporate a utility notion for protected candidates through a backdoor, which is why we classified it as leaning towards WYSIWYG and formal \eop, even though it is technically very similar to \algofair. 
This design choice also has tremendous consequences for individuals who experience intersectional discrimination.
%
%

Being able to map mathematical formulas to normative dimensions, and thus to identify normative value frameworks is crucial for several reasons: 
First, it facilitates the decision on which method to choose under different application scenarios. Obviously one wants to use methods whose value frameworks match the ones of the application at hand.
However, we have seen that some methods are not easy to analyse for their values because important details are missing.
This was particularly true for the method of~\citet{Beutel:2019:FRR:3292500.3330745}, where an explicit formula on how to calculate the click through rate is needed to come up with a mapping.
Second, our comprehensive analysis identified research gaps in terms of value frameworks. 
For example, we could only identify one method, ~\citet{biega2017privacy}, which explicitly addresses the problem of emergent bias.
At the same time, the method is consistent with WYSIWYG and with formal \eop.
Thus, to the best of our knowledge, there is currently no method that explicitly handles emergent bias and is consistent with supports WAE. 
%
%
%
Finally, in contrast to the work presented in Section~\ref{sec:fair_db}, \revv{no methods  discussed in this part of the survey explicitly handle intersectional discrimination.  To the best of our knowledge, no such methods have been developed to-date in the learning-to-rank literature.}

\section{Fairness in Other Domains: Recommendation and Matching}
\label{sec:fair_recsys}
Today many web applications with search functionality also implement a recommendation system that provides personalized search results to the user with items dedicated precisely to their interest. 
The main objective of recommender systems is to facilitate transactions between multiple stakeholders in which personalization plays a key role and therefore fairness issues for more than one group of participants have to be considered.
RecSys tasks can be grouped into three different types: finding good items that meet a user's interest, optimizing the utility of users, and predicting ratings of items for a user~\cite{kamishima2018recommendation}.
They often consist of multiple models, must balance multiple goals and are difficult to evaluate due to extreme and skewed sparsity~\cite{Beutel:2019:FRR:3292500.3330745}.
Examples for such systems are user recommendations in web shops or on streaming platforms.
\begin{figure}[t]
    \centering
    \includegraphics[width=0.8\textwidth]{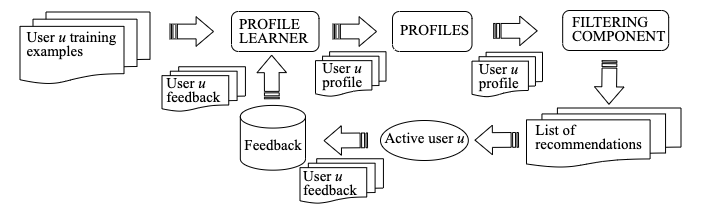}
    \caption{Principle components of a recommender system. A profile learner determines user profiles from training examples (upper right) which are stored in a profile database. When the user issues a query, the filtering component processes the general search result, and returns a list of user specific recommendations. The interaction of the user are collected as feedback and stored in a feedback database. This data is used in turn to update the model of the profile learner. Image taken from~\citet{de2015semantics}.}
    \label{fig:recsys:functional}
\end{figure}
Figure~\ref{fig:recsys:functional} shows the functional principle of a recommender system. 
A profile learner determines user profiles from training examples (upper right) which are stored in a profile database. 
When the user issues a query, the filtering component processes the general search result (and additional items to be displayed), and returns a list of user specific recommendations. 
The interaction of the user with the result are collected as feedback and stored in a feedback database. 
This data is used in turn to update the model of the profile learner.

\citet{burke2017multisided} identifies three different stakeholders in recommender systems, namely the consumers, producers and the platform itself.
They further identify three different types of recommender systems, distinguished by the respective stakeholders that are considered for fairness.
When considering fairness for the recommended items, they speak of producer-fairness, while fairness for the users of the system is considered under the term consumer-fairness.
In a system that satisfies consumer fairness the disparate impact of recommendation on protected classes of consumers has to be taken into account, while the fairness of outcomes is not considered for producers.
Producer fairness regards the producer side of the system, but not the consumer side, e.g. to ensure market diversity and avoid monopoly domination.
At the same time the producers are a passive system component that do not seek out recommendation opportunities, but instead wait for users to request recommendations from the platform.
The third fairness constraint simultaneously takes producers and consumers into account and has to be applied when protected groups exist among both stakeholders.

In our paper's language consumers would be called users $\userSet{}$ und producers would be called candidates $\candidateSet{}$.

\rev{Recommender systems often use rankings to present the most suitable candidates to their users, and the question of the fairness for all stakeholders has been raised from different perspectives.  We therefore give a brief overview on the topic of fairness for recommender systems, and refer the reader to several recent tutorials and surveys of fairness in recommendation and matchmaking systems for additional details: \citet{ekstrand2019fairness} work out which algorithmic fairness concepts from classification and scoring do and do not translate to information access scenarios. \citet{li2021tutorial} summarize foundations and algorithms for fairness in recommendation systems.  \citet{chen2020bias} outline seven types of biases in recommendation that typically stems from user behavioral data, and provide a taxonomy of existing work on de-biasing. \citet{gao2020counteracting} bridge socio-technical terminologies and metrics to important recommendation concepts such as relevance, novelty, diversity, bias, and fairness.}

While maintaining the same level of detail as in other parts of this survey, we will discuss three papers that highlight different directions within fairness in recommender systems.  In Section~\ref{sec:recsys:indep}, we discuss~\citet{kamishima2018recommendation} to showcase a theoretical contribution.  In Section~\ref{sec:recsys:familiar}, we discuss \citet{mehrotra2018towards} to demonstrate a countermeasure against the familiarity bias feedback loop in the music streaming platform Spotify.  Finally, in Section~\ref{sec:recsys:twosided} we discuss \citet{suhr2019two}, as an example of work on two-sided fairness for repeated matches.

Before diving in, we also briefly discuss other notable work on fairness in recommendation and matching that do not strongly connect to ranking.
An early work on user fairness in recommendations by~\citet{leonhardt2018user} shows that methods that only focus on diversification on results can lead to discrimination among users.
\citet{rastegarpanah2019fighting} present a bias mitigation approach by augmenting the input to an unfair recommendation model with additional ``antidote data,'' as an analogy to work studying data poisoning.
\citet{wang2021practical} study the compositionality of fairness definitions for recommender systems and provide a set of conditions under which fairness of individual models does indeed compose.
\citet{sonboli2019localized} examine fairness in situations where membership in protected groups is not given a priori, but must be derived from the data itself. Additionally, the authors propose a localized understanding of fairness in cases where global system properties are insufficient to identify protected groups.
\citet{deldjoo2021flexible} propose a probabilistic framework based on generalized cross entropy to measure fairness for a given recommendation model, instead of understanding fairness as the divergence of a system from some form equality.
\citet{liu2019personalized} propose a fairness method to produce recommendations in the microlending domain that are fair for the borrowers and attentive to individual lender preferences. 
\citet{ge2021towards} account for the dynamic nature of recommender systems and consider long-term fairness by proposing a fairness-constrained reinforcement learning algorithm, which models the recommendation problem as a constrained Markov decision process.
\citet{sonboli2021fairness} describe the results of an exploratory interview study that investigates user perspectives on fairness-aware recommender systems and on techniques for enhancing their transparency.
\citet{zhu2018fairness} introduce a tensor-based fair recommendation system that preserves the benefits of using matrix factorization for recommendation, while increasing fairness by excluding sensitive features and their latent influence on non-sensitive ones.
\citet{li2021user} study unfair recommendation performance disparities for active vs. inactive users, since those who actively interact with the platform produce larger amounts of data for collaborative filtering approaches.

In this section we will use the terms users and consumers, as well as candidates and producers interchangeably to demonstrate that these concepts are analogous to each other, and to illustrate the similarities between the previously described methods, and those that follow in this section.

All  methods presented in this section can be considered in-processing approaches.

\subsection{Recommendation Independence}
\label{sec:recsys:indep}
\spara{Fairness Definition.} 
The work of~\citet{kamishima2018recommendation} introduces a concept of fairness as \emph{recommendation independence}: an unconditional statistical independence between a recommendation outcome and a specified piece of information. In other words, predictions of ratings should not be based on some previously specified feature.
This is formalized as a regularization-based approach that can deal with the encoding of sensitive information in the first and second moments of the distributions, which means that independence shall be given in terms of the mean \emph{and} the standard deviation of the distribution of predicted user ratings.
A binary sensitive feature will be specified by user or manager.

\revv{Let us assume that the random variable $\featOfUser{}$ represents users, $\featureSet$ represents items, $\sensFeatSet$ represents sensitive features, and $\score{}$  represents ground truth ratings.}
The $i$-th instance of the training dataset $\recsysTrainSet{train}$ is a 4-tuple $(\featOfUser{i}, \featOfCand{i}, \sensFeatSet_i, \score{i})$. 
The rating predictions $\predScore{}$ are calculated using a modified loss function in which an independence term shall be maximized, meaning that the higher $\operatorname{ind}(\predScore{}, \sensFeatSet)$, the less statistically dependent are $\predScore{}$ and $\sensFeatSet$:
\[\sum_{(\featOfUser{i}, \featOfCand{i}, \sensFeatSet_i, \score{i}) \in \recsysTrainSet{train}} \operatorname{loss}(\score{}, \predScore(\featOfUser{i}, \featOfCand{i}, \sensFeatSet_i)) - \eta \operatorname{ind}(\predScore{}, \sensFeatSet) + \operatorname{reg}(\Theta)\]
The method can therefore be seen as an in-processing approach.

\citet{kamishima2018recommendation} give three ways in which independence can be measured, all aiming to produce a rating model in which the distributions of predicted ratings are indistinguishable \revv{for different values of the sensitive feature}:
\begin{enumerate}
    \item \textbf{Mean Matching:} The means of two normal distributions shall match
    \[-\left(\frac{\mathbb{S}^{(0)}}{N^{(0)}} - \frac{\mathbb{S}^{(1)}}{N^{(1)}}\right)\]
    where $\mathbb{S}^{(\group{})}$ is the sum of predicted ratings per group, and $N^{(\group{})}$ is the number of training items per group.
    \item \textbf{Distribution Matching:} The similarity between two distributions $Pr[\predScore{}|\sensAttr=0]$ and $Pr[\predScore{}|\sensAttr=1]$ is measured in terms of the negative Bhattacharyya-distance:
    \[\frac{1}{2}\ln{\left(\frac{2\sqrt{\mathbb{V}^{(0)}\mathbb{V}^{(1)}}}{\mathbb{V}^{(0)} + \mathbb{V}^{(1)}}\right)}
    -
    \frac{\left(\frac{\mathbb{S}^{(0)}}{N^{(0)}} - \frac{\mathbb{S}^{(1)}}{N^{(1)}}\right)^2}
    {4(\mathbb{V}^{(0)} + \mathbb{V}^{(1)})}\]
    where $\mathbb{V}^{(\group{})}$ is the variance of the training items per group.
    \item \textbf{Mutual information:} \revv{The degree of statistical independence is quantified by} a differential entropy function for normal distributions:
    \[-I(\predScore{};\sensFeatSet) = -(H(\predScore{}) - H(\predScore{}|\sensFeatSet))\]
    where $H(\predScore{}) = \frac{1}{2}\ln 2 \pi e \mathbb{V}$ and $H(\predScore{}|\sensFeatSet) = \frac{1}{2}\ln 2 \pi e \mathbb{V}^{(s)}$.
\end{enumerate}

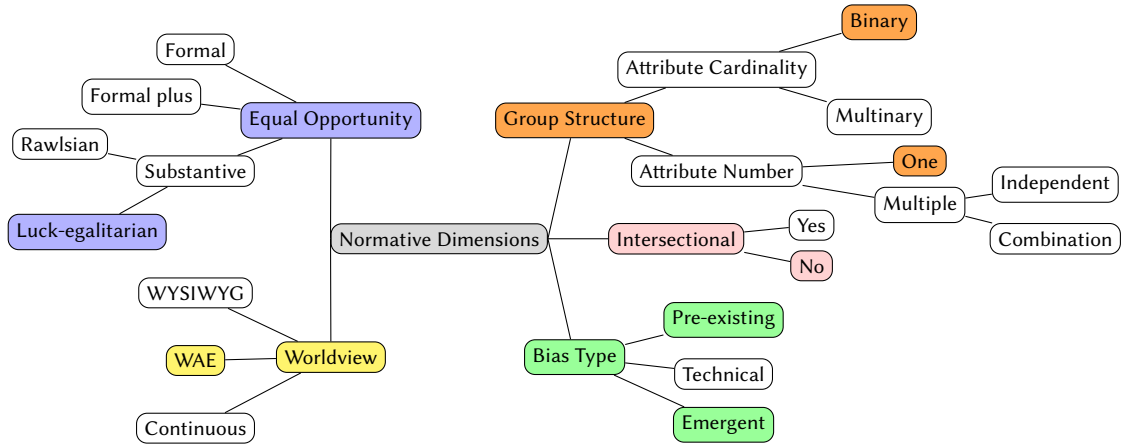
\begin{figure}[h]
    \centering
    \begin{tikzpicture}[scale=0.9,align=center]
    \tikzset{
        root/.style = {rectangle, rounded corners, draw=black, fill=gray!30, font=\small\sffamily},
        child/.style = {rectangle, rounded corners,draw=black, font=\small\sffamily},
        orangec/.style = {rectangle, rounded corners, draw=black, fill=orange!70, font=\small\sffamily},
        greenc/.style = {rectangle, rounded corners, draw=black, fill=green!40, font=\small\sffamily},
        bluec/.style = {rectangle, rounded corners, draw=black, fill=blue!30, font=\small\sffamily},
        yellowc/.style = {rectangle, rounded corners, draw=black, fill=yellow!70, font=\small\sffamily},
        pinkc/.style = {rectangle, rounded corners, draw=black, fill=pink!70, font=\small\sffamily},
    }
    \node (ND) at (-0.2,0) [root] {Normative Dimensions};    
    \node (GS) at (1.8,1.75) [orangec] {Group Structure};    
    \node (BT) at (1.8,-1.75) [greenc] {Bias Type};    
    \node (EO) at (-1.8,1.75) [bluec] {Equal Opportunity};    
    \node (WV) at (-1.8,-1.75) [yellowc] {Worldview};  
    \node (IT) at (3.3,0.0) [pinkc] {Intersectional};
    
    \node (AC) at (3.9,2.5) [child] {Attribute Cardinality};
    \node (ACB) at (6.3,3.2) [orangec] {Binary};
    \node (ACM) at (6.3,1.8) [child] {Multinary};
    
    \node (AN) at (3.9,1.0) [child] {Attribute Number};
    \node (ANO) at (6.9,1.15) [orangec] {One};
    \node (ANM) at (6.9,0.5) [child] {Multiple};
    
    \node (ANMI) at (8.9,0.8) [child] {Independent};
    \node (ANMC) at (8.9,0.0) [child] {Combination};
    
    \node (ITY) at (5.3,0.2) [child] {Yes};
    \node (ITN) at (5.3,-0.4) [pinkc] {No};
    
    \node (BTP) at (4.0,-1.2) [greenc] {Pre-existing};
    \node (BTT) at (4.0,-2.0) [child] {Technical};
    \node (BTE) at (4.0,-2.75) [greenc] {Emergent};
    
    \node (EOF) at (-3.8,2.8) [child] {Formal};
    \node (EOFP) at (-4.6,2.1) [child] {Formal plus};
    \node (EOS) at (-3.8,1.0) [child] {Substantive};
    
    \node (EOSR) at (-5.8,1.4) [child] {Rawlsian};
    \node (EOSL) at (-5.4,0.1) [bluec] {Luck-egalitarian};

    \node (WVI) at (-3.8,-0.8) [child] {WYSIWYG};
    \node (WVE) at (-3.8,-1.8) [yellowc] {WAE};
    \node (WVC) at (-3.8,-2.8) [child] {Continuous};
    
    \draw (ND.east) -- (GS);
    \draw (ND.east) -- (BT);
    \draw (ND.east) -- (IT);
    \draw (ND.west) -- (EO);
    \draw (ND.west) -- (WV);
    
    \draw (GS) -- (AC);
    \draw (GS) -- (AN);
    
    \draw (AC) -- (ACB);
    \draw (AC) -- (ACM);
    
    \draw (AN) -- (ANO);
    \draw (AN) -- (ANM);
    
    \draw (IT) -- (ITY);
    \draw (IT) -- (ITN);
    
    \draw (ANM) -- (ANMI);
    \draw (ANM) -- (ANMC);
    
    \draw (BT) -- (BTP);
    \draw (BT) -- (BTT);
    \draw (BT) -- (BTE);
    
    \draw (EO) -- (EOF);
    \draw (EO) -- (EOFP);
    \draw (EO) -- (EOS);
    
    \draw (EOS) -- (EOSR);
    \draw (EOS) -- (EOSL);
    
    \draw (WV) -- (WVI);
    \draw (WV) -- (WVE);
    \draw (WV) -- (WVC);
\end{tikzpicture}
    \caption{\rev{Summary of the normative values encoded in the fairness definitions by~\citet{kamishima2018recommendation}.}}
    \label{fig:recsys:kamishima}
\end{figure}

\spara{Insights.} \revv{Although this is not explicitly stated,} the goal of equalizing rating distributions is consistent with the WAE worldview.  \revv{Further, since the fairness definitions do not contain a measure of utility or merit, and under the assumption that the goal is to equalize access to opportunity over a lifetime, the approach is consistent with substantive \eop.  Because of explicit conditioning on group membership, we map this approach to luck-egalitarian \eop.}
%
%
%
%
The definition implicitly addresses the problem of pre-existing biases. 
Depending on the choice of the protected feature (\eg popularity), it may also be capable to address emergent bias.
Figure~\ref{fig:recsys:kamishima} summarizes this analysis.

\spara{Algorithm.} 
The algorithm is implemented using probabilistic matrix factorization to predict ratings:
    \[\predScore(\featOfUser{}, \featOfCand{}, \sensFeatSet) = \mu^{(\group{})} + b^{(\group{})}_{\featOfUser{}} + c^{(\group{})}_{\featOfCand{}} + \mathbf{p}^{(\group{})\top}_{\featOfUser{}}\mathbf{q}^{(\group{})}_{\featOfCand{}}\]
where $\mu, b_{\featOfUser{}}$ and $c_{\featOfCand{}}$ are global, per-user and per-item bias parameters respectively and $\mathbf{p}_{\featOfUser{}}$ and $\mathbf{q}_{\featOfCand{}}$ are $K$-dimensional parameter vectors, which represent the cross effects between users and items.
Said parameters are learned by minimizing the loss function, where the loss is expressed as a $L_2$-norm:
    \[\sum_{(\featOfUser{i}, \featOfCand{i}, \sensFeatSet_i, \score{i}) \in \recsysTrainSet{train}} (\score{i} - \predScore(\featOfUser{i}, \featOfCand{i}, \sensFeatSet_i))^2 - \eta \operatorname{ind}(\score{}, \sensFeatSet) + \operatorname{reg}(\Theta)\]
All independence measures are differentiable and can hence be optimized efficiently using conjugate gradient methods.

\spara{Experiments. }
The experiments are done on three datasets with different sensitive features:
\begin{itemize}
    \item The ML1M \textbf{Movielens dataset}~\cite{harper2015movielens} contains 1M movie items. Two sensitive features where chosen for two separate settings: \revv{(1)} whether the movie's release year was before 1990, and \revv{(2)} the user's gender. 
    The first setting investigates fairness on the side of the producer, whereas the second relates to a fairness concern w.r.t. the users of the system.
    \item The \textbf{Flixster dataset}~\cite{jamali2010matrix} is also a dataset on movie recommendations and contains almost 10M entries.
    The popularity of an item was chosen as the protected feature: movies that received the most user ratings (top 1\%) belong to one group, while all movies that received \revv{fewer} ratings belong to the other group.
    \item The \textbf{Sushi dataset}~\cite{kamishima2003nantonac} contains around 50,000 data points of users and their preferences when ordering Sushi from 25 different restaurants.
    Three different choices of sensitive features where investigated: \revv{(1) whether the user was a teen, (2) gender: male or female, and (3) whether the type of sushi was seafood.}
\end{itemize}

The performance evaluation uses mean absolute error, while the independence (\ie fairness) evaluation uses the Kolmogorov-Smirnoff test.
The latter evaluates the area between two empirical cumulative distributions and shall be close to zero for high fairness.
The experiments compare all three independence measures to each other\revv{, and show that independence can be achieved in all cases, with a small cost in accuracy.}


\subsection{Familiarity Bias and Superstar Economics}
\label{sec:recsys:familiar}
\spara{Fairness Definition.} \revv{The work of~\citet{mehrotra2018towards}} addresses the problem of disparities in exposure of items to users due to the combination of two factors: the pre-existing familiarity of the user with certain items (someone who likes action movies will likely know Tom Cruise) and the current recommendation strategies of two-sided markets (``superstar economics'').
\revv{The fact that recommendation systems optimize for relevance can lead} to a lock-in of popular and relevant suppliers, and thus cause \revv{many} suppliers at the tail end of the exposure distribution to struggle to attract consumers.
This may\revv{, in turn, lead to a dissatisfaction of such suppliers} with the marketplace.

\revv{\citet{mehrotra2018towards} aim} to understand the interplay \revv{between} relevance, fairness and satisfaction, and investigate consumer relevance and supplier fairness on a music streaming platform.
\revv{The paper} introduces a notion of multinomial group fairness, which requires that the content shown to users be spread well across the wide long-tailed popularity spectrum, rather than focusing on a small subset of popular artists.
From the popularity distribution of all artists, $K$ bins of equal size are created and artists are grouped into these bins depending on their popularity:
    \[\Psi(s) = \sum_{i=1}^{K} \sqrt{|t_j|_{\forall t_j \in P_i \cap T(s)}}\]
where $P_i$ is the set of artists that belong to popularity bin $i$, $s$ is the recommended set, and $T(s)$ is the collection of artists in the set $s$ with $t_j$ being the $j$-th track in $s$.
This definition rewards sets that are \emph{diverse} in terms of \revv{the represented popularity bins} and, as per the current definition, \emph{fair} to different popularity bins of suppliers. 
There is more benefit to selecting an artist from a popularity bin that is not yet represented. 
As soon as at least one artist is selected from a bin, other artists from the same bin start having diminishing gain owing to the square root function.

The paper then presents three different policies with $\phi(.)$ being a relevance and $\psi(.)$ a fairness measure:
\begin{enumerate}
    \item A weighted combination of relevance and fairness:
    \[s^*_u = \argmax_{s \in S_u} (1 - \beta) \phi (u, s) + \beta \psi(s)\]
    where $S_u$ is the collection of all sets pertinent to the user $u$.
    \item A probabilistic combination of relevance and fairness, where the weighting factor $\beta \in [0, 1]$ decides on whether to recommend a set based on relevance or fairness:
    \begin{equation*}
        s^*_u =
        \begin{cases}
          \argmax_{s \in S_u} \psi(s), & \text{if}\ p < \beta \\
          \argmax_{s \in S_u} \phi(u, s), & \text{otherwise}
        \end{cases}
    \end{equation*}
    where $p \in [0, 1]$ is a randomly generated number.
    \item A guaranteed relevance term to ensure that the minimum relevance is $\beta$: 
    \[s^*_u = \argmax_{s \in S_u} \psi(s) \;\; s.t.\;\; \phi(u,s) \geq \beta\]
\end{enumerate}
They further investigate different affinities of users to fairness; for example, that some users only want to listen to one particular artist, while other users are more flexible.
This affinity is measured as the difference between the satisfactions of a user when recommended relevant content vs.  fair content.
The paper states that a fairness recommendation policy should be adaptive to this affinity $\xi_u$, and therefore redefine the second fairness policy as:
    \begin{equation*}
        s^*_u =
        \begin{cases}
          \argmax_{s \in S_u} \psi(s), & \text{if}\ \xi_u \geq 0 \\
          \argmax_{s \in S_u} \phi(u, s), & \text{otherwise}
        \end{cases}
    \end{equation*}
Another version redefines the first policy to use the z-scored affinity, denoted as $\hat{\xi}_u$: 
    \[s^*_u = \argmax_{s \in S_u} (1 - \hat{\xi}_u) \phi (u, s) + \hat{\xi}_u \psi(s)\]
Because all definitions are modifying the objective function of the learning algorithm, the method can be classified as in-processing approach.

\spara{Insights. } The method is addressing a combination of pre-existing and technical bias, where the popularity bias can be seen as pre-existing, and its reinforcement by the recommender system as technical bias.
Furthermore, because of the temporal nature of recommender systems, proposed methods may also address emergent bias w.r.t. popularity shifts in the future.

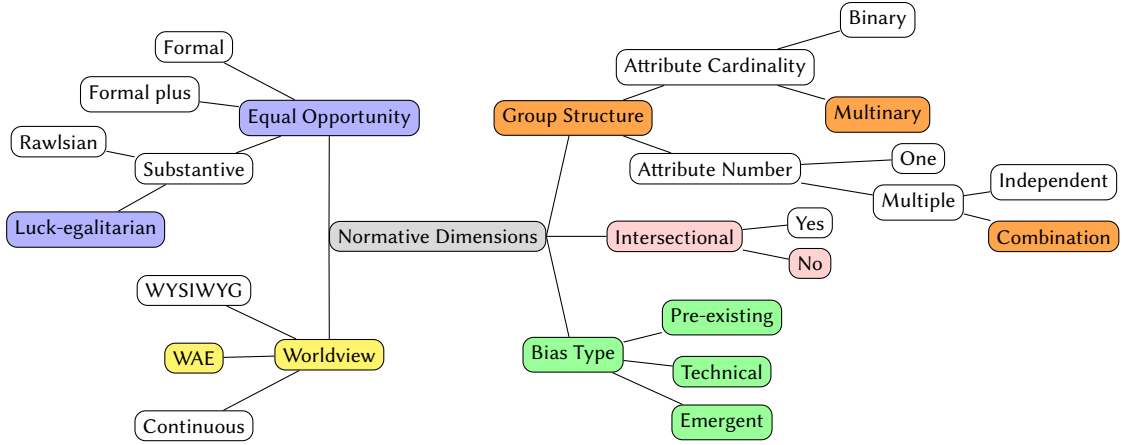
\begin{figure}[h]
    \centering
    \begin{tikzpicture}[scale=0.9,align=center]
    \tikzset{
        root/.style = {rectangle, rounded corners, draw=black, fill=gray!30, font=\small\sffamily},
        child/.style = {rectangle, rounded corners,draw=black, font=\small\sffamily},
        orangec/.style = {rectangle, rounded corners, draw=black, fill=orange!70, font=\small\sffamily},
        greenc/.style = {rectangle, rounded corners, draw=black, fill=green!40, font=\small\sffamily},
        bluec/.style = {rectangle, rounded corners, draw=black, fill=blue!30, font=\small\sffamily},
        yellowc/.style = {rectangle, rounded corners, draw=black, fill=yellow!70, font=\small\sffamily},
        pinkc/.style = {rectangle, rounded corners, draw=black, fill=pink!70, font=\small\sffamily},
    }
    \node (ND) at (-0.2,0) [root] {Normative Dimensions};    
    \node (GS) at (1.8,1.75) [orangec] {Group Structure};    
    \node (BT) at (1.8,-1.75) [greenc] {Bias Type};    
    \node (EO) at (-1.8,1.75) [bluec] {Equal Opportunity};    
    \node (WV) at (-1.8,-1.75) [yellowc] {Worldview};  
    \node (IT) at (3.3,0.0) [pinkc] {Intersectional};
    
    \node (AC) at (3.9,2.5) [child] {Attribute Cardinality};
    \node (ACB) at (6.3,3.2) [child] {Binary};
    \node (ACM) at (6.3,1.8) [orangec] {Multinary};
    
    \node (AN) at (3.9,1.0) [child] {Attribute Number};
    \node (ANO) at (6.9,1.15) [child] {One};
    \node (ANM) at (6.9,0.5) [child] {Multiple};
    
    \node (ANMI) at (8.9,0.8) [child] {Independent};
    \node (ANMC) at (8.9,0.0) [orangec] {Combination};
    
    \node (ITY) at (5.3,0.2) [child] {Yes};
    \node (ITN) at (5.3,-0.4) [pinkc] {No};
    
    \node (BTP) at (4.0,-1.2) [greenc] {Pre-existing};
    \node (BTT) at (4.0,-2.0) [greenc] {Technical};
    \node (BTE) at (4.0,-2.75) [greenc] {Emergent};
    
    \node (EOF) at (-3.8,2.8) [child] {Formal};
    \node (EOFP) at (-4.6,2.1) [child] {Formal plus};
    \node (EOS) at (-3.8,1.0) [child] {Substantive};
    
    \node (EOSR) at (-5.8,1.4) [child] {Rawlsian};
    \node (EOSL) at (-5.4,0.1) [bluec] {Luck-egalitarian};

    \node (WVI) at (-3.8,-0.8) [child] {WYSIWYG};
    \node (WVE) at (-3.8,-1.8) [yellowc] {WAE};
    \node (WVC) at (-3.8,-2.8) [child] {Continuous};
    
    \draw (ND.east) -- (GS);
    \draw (ND.east) -- (BT);
    \draw (ND.east) -- (IT);
    \draw (ND.west) -- (EO);
    \draw (ND.west) -- (WV);
    
    \draw (GS) -- (AC);
    \draw (GS) -- (AN);
    
    \draw (AC) -- (ACB);
    \draw (AC) -- (ACM);
    
    \draw (AN) -- (ANO);
    \draw (AN) -- (ANM);
    
    \draw (IT) -- (ITY);
    \draw (IT) -- (ITN);
    
    \draw (ANM) -- (ANMI);
    \draw (ANM) -- (ANMC);
    
    \draw (BT) -- (BTP);
    \draw (BT) -- (BTT);
    \draw (BT) -- (BTE);
    
    \draw (EO) -- (EOF);
    \draw (EO) -- (EOFP);
    \draw (EO) -- (EOS);
    
    \draw (EOS) -- (EOSR);
    \draw (EOS) -- (EOSL);
    
    \draw (WV) -- (WVI);
    \draw (WV) -- (WVE);
    \draw (WV) -- (WVC);
\end{tikzpicture}
    \caption{\rev{Summary of the normative values encoded by~\citet{mehrotra2018towards}.}}
    \label{fig:recsys:mehrota}
\end{figure}

The work does not explicitly express the authors' beliefs about the mapping $g$ from construct space $CS$ to observable space $OS$.
That being said, the fairness definition aims to equalize the exposure of different artist independent of their popularity and can therefore be associated with the WAE worldview.
Further, because of a natural connection to emergent bias, the assumption of equalizing opportunity over the lifetime is also very natural.  For these reasons, the methods can be associated to substantive \eop, and specifically with luck-egalitarian \eop, which would acknowledge different distributions of popularity in construct space. %
However, it is not clear how an artist's popularity relates to their true underlying effort at all, as it is mainly driven by the \textit{users} of the system, rather than by the artists themselves. 
%
%
%
%
Figure~\ref{fig:recsys:mehrota} summarizes our analysis.

\spara{Algorithm.} The algorithm is implemented as a combinatorial contextual bandit problem with the following consecutive interactions between the customers and the recommendation system:
\begin{enumerate}
    \item The system observes context $\featureSet$ drawn from a distribution of contexts $\mathbf P(\featureSet)$.
    \item Based on $\featureSet$, the system chooses the sets $s$ to recommend to the user.
    \item A reward function $\score{} \in [0,1]$ is drawn from the distribution $\mathbf P(\score{}|\featureSet,s)$ that expresses user satisfaction.
    \item The system maximizes $\score{}$ under different fairness policies.
\end{enumerate}
As estimating user satisfaction is not easy, because it relies on user feedback which is hard to estimate offline, the recommender system is modeled as a stochastic policy $\pi$ that specifies a conditional distribution $\pi(a|\featureSet)$.
The value $V(\pi)$ of a policy is the expected reward (\ie the user's satisfaction) if an action $a$ is chosen under that policy.
This value is to be estimated for a new policy $\pi^*$ given logged training data, using the inverse propensity score (IPS) estimator, which is provably unbiased:
\[V_{\operatorname{offline}} = \sum_{\forall (\featureSet, a, \score{a}, p_a)} \frac{\score{a} \mathds{1}(\pi^*(\featureSet) = a)}{p_a}\]
with $p_a$ being the propensity scores for an action $a$ that was randomly chosen from the space of all possible actions under context $\featureSet$,
and $\mathds{1}(\pi^*(x) = a)$ being the set indicator function that evaluates to 1 if the action selected by the target policy matches the logged training data.
With this, a metric of user satisfaction can be computed as:
\[\score{\pi^*(x)} = \mathds{E}_a \left[ \frac{\score{a} \mathds{1}(\pi^*(\featureSet) = a)}{p_a} \right]\]

\spara{Experiments.} 
The experimental part contains a trade-off analysis between user relevance and fairness of sets, which shows that only very few highly relevant sets also achieve high scores on fairness.
This means that a recommendation system that solely optimizes for user relevance will not automatically lead to fair and diverse sets for the suppliers, and, in turn, that two-sided marketplaces have to optimize for both users and suppliers to satisfy both.

\revv{Evaluation is conducted} on a proprietary dataset from Spotify with 400K users, 49K artists, and 5K sets (\ie playlists).
When maximizing relevance $\phi(u,s)$ only, the results show the highest user satisfaction, whereas when optimizing for fairness only, user satisfaction drops by 35\%.
When optimizing for both, user satisfaction increases with an increase of the weight that is given to the relevance objective.
However satisfaction does not drop significantly up to a fairness weight of 20\%, such that policies could easily increase fairness up to that point without trading user satisfaction significantly. 
The guaranteed relevance policy yields the best satisfaction results in absolute numbers, and shows a linear trend of user satisfaction with increasing weights for relevance.
However, for the same levels of $\beta$, this policy achieves the least average fairness scores compared to the other two policies. 
This means that the usage of the interpolating fairness policy and the probabilistic fairness policy is preferable in situations where less fairness shall be traded for higher relevance values.
When evaluating the adaptive policies, the results highlight that personalizing the recommendation policy and adapting based on user level affinity is better than globally balancing relevance and fairness.
Interestingly, adaptive policies lead to a relatively high fairness mean compared to global policies, while at the same time increasing the overall user satisfaction.
This is another example where the general assumption that a trade-off between fairness and relevance is a necessary evil, is \revv{shown not to hold}.


\subsection{Fairness in Two-sided Markets}
\label{sec:recsys:twosided}

\spara{Fairness Definition.} \revv{The work of~\citet{suhr2019two}} gives a case study of a two-sided matching platform, namely, a ride-hailing platform such as Uber.
\revv{This paper} discusses fairness in a platform performing repeated matches of \revv{providers (drivers) and consumers (riders)} over time.
%
%
Fairness is seen in terms of fair distribution of \revv{driver's income, given their} active time in the system. 
It is explicitly considered over time, because a single matching does not have a significant long-term effect on the life of the people that are matched, and because income equity can be amortized over a longer period, such as a week or a month.
Note that the paper speaks explicitly about \revv{riders} and drivers, but in a broader sense those can be seen as users $\userSet{}$ of the system and candidates $\candidateSet{}$ to be matched to them.
%

Customer utility $U_{\userSet{}}$ is described as a customer $\userSet{b}$'s waiting time, which is approximated using the negative distance $d$ of a driver $\candidateSet{a}$ to them:
\[U_{\userSet{}}(b, a) = -d(\userSet{b}, \candidateSet{a})\]
Driver utility is described as the income a driver receives from transporting a customer, which is approximated using the distance from the customer's pick-up location to their destination, reduced by the distance the driver has to travel in order to arrive at the respective pick-up location:
    \[U_{\candidateSet{}}(a, b) = d(\userSet{b}, \operatorname{dest}(\userSet{b}^t)) - d(\userSet{b}, \candidateSet{a})\]

The paper then defines two fairness concepts. 
First, the authors introduce \textbf{parity fairness}: over time, the sum of received utility shall be (almost) equal for all drivers in $\candidateSet{}$:
\[\sum_a \sum_b | U^T_{\candidateSet{a}} - U^T_{\candidateSet{b}}| < \epsilon \]
with $U^T_{\candidateSet{a}}$ being the total utility that driver $a$ received until matching round $T$.
Second, the authors define \textbf{proportional fairness}: over time, the sum of received utility normalized by their active driving time shall be equal for all drivers 
\[\sum_a \sum_b \left| \frac{U^T_{\candidateSet{a}}}{\Lambda^T_{\candidateSet{a}}}  - \frac{U^T_{\candidateSet{b}}}{\Lambda^T_{\candidateSet{b}}}\right| < \epsilon\]
where $\Lambda^T_D(j)$ is the total amount of time a driver has been active on the platform until $T$.

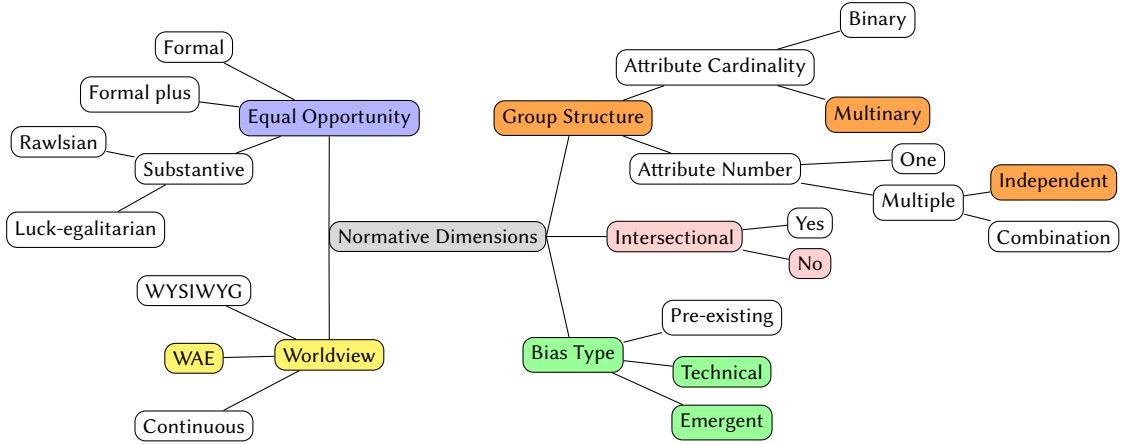
\begin{figure}[h]
    \centering
    \begin{tikzpicture}[scale=0.9,align=center]
    \tikzset{
        root/.style = {rectangle, rounded corners, draw=black, fill=gray!30, font=\small\sffamily},
        child/.style = {rectangle, rounded corners,draw=black, font=\small\sffamily},
        orangec/.style = {rectangle, rounded corners, draw=black, fill=orange!70, font=\small\sffamily},
        greenc/.style = {rectangle, rounded corners, draw=black, fill=green!40, font=\small\sffamily},
        bluec/.style = {rectangle, rounded corners, draw=black, fill=blue!30, font=\small\sffamily},
        yellowc/.style = {rectangle, rounded corners, draw=black, fill=yellow!70, font=\small\sffamily},
        pinkc/.style = {rectangle, rounded corners, draw=black, fill=pink!70, font=\small\sffamily},
    }
    \node (ND) at (-0.2,0) [root] {Normative Dimensions};    
    \node (GS) at (1.8,1.75) [orangec] {Group Structure};    
    \node (BT) at (1.8,-1.75) [greenc] {Bias Type};    
    \node (EO) at (-1.8,1.75) [bluec] {Equal Opportunity};    
    \node (WV) at (-1.8,-1.75) [yellowc] {Worldview};  
    \node (IT) at (3.3,0.0) [pinkc] {Intersectional};
    
    \node (AC) at (3.9,2.5) [child] {Attribute Cardinality};
    \node (ACB) at (6.3,3.2) [child] {Binary};
    \node (ACM) at (6.3,1.8) [orangec] {Multinary};
    
    \node (AN) at (3.9,1.0) [child] {Attribute Number};
    \node (ANO) at (6.9,1.15) [child] {One};
    \node (ANM) at (6.9,0.5) [child] {Multiple};
    
    \node (ANMI) at (8.9,0.8) [orangec] {Independent};
    \node (ANMC) at (8.9,0.0) [child] {Combination};
    
    \node (ITY) at (5.3,0.2) [child] {Yes};
    \node (ITN) at (5.3,-0.4) [pinkc] {No};
    
    \node (BTP) at (4.0,-1.2) [child] {Pre-existing};
    \node (BTT) at (4.0,-2.0) [greenc] {Technical};
    \node (BTE) at (4.0,-2.75) [greenc] {Emergent};
    
    \node (EOF) at (-3.8,2.8) [child] {Formal};
    \node (EOFP) at (-4.6,2.1) [child] {Formal plus};
    \node (EOS) at (-3.8,1.0) [child] {Substantive};
    
    \node (EOSR) at (-5.8,1.4) [child] {Rawlsian};
    \node (EOSL) at (-5.4,0.1) [child] {Luck-egalitarian};

    \node (WVI) at (-3.8,-0.8) [child] {WYSIWYG};
    \node (WVE) at (-3.8,-1.8) [yellowc] {WAE};
    \node (WVC) at (-3.8,-2.8) [child] {Continuous};
    
    \draw (ND.east) -- (GS);
    \draw (ND.east) -- (BT);
    \draw (ND.east) -- (IT);
    \draw (ND.west) -- (EO);
    \draw (ND.west) -- (WV);
    
    \draw (GS) -- (AC);
    \draw (GS) -- (AN);
    
    \draw (AC) -- (ACB);
    \draw (AC) -- (ACM);
    
    \draw (AN) -- (ANO);
    \draw (AN) -- (ANM);
    
    \draw (IT) -- (ITY);
    \draw (IT) -- (ITN);
    
    \draw (ANM) -- (ANMI);
    \draw (ANM) -- (ANMC);
    
    \draw (BT) -- (BTP);
    \draw (BT) -- (BTT);
    \draw (BT) -- (BTE);
    
    \draw (EO) -- (EOF);
    \draw (EO) -- (EOFP);
    \draw (EO) -- (EOS);
    
    \draw (EOS) -- (EOSR);
    \draw (EOS) -- (EOSL);
    
    \draw (WV) -- (WVI);
    \draw (WV) -- (WVE);
    \draw (WV) -- (WVC);
\end{tikzpicture}
    \caption{\rev{Summary of the normative values encoded by~\citet{suhr2019two}.}}
    \label{fig:recsys:suehr}
\end{figure}

\spara{Insights. } The work does not talk about any protected groups and instead tries to equalize the (hourly) wage of all drivers in the system. 
Also, the drivers' effort is not considered other than in the sense of their active time, and, as such, it does not make sense to associate this method with a particular equality of opportunity framework. 
The assumption behind this work is that driving skills are essentially the same and therefore all drivers should be paid equally. It is an extreme case of the WAE worldview in which not only do all groups have the same qualification distribution, but the absolute qualification values are the same for each individual. 
The biases addressed are of a technical and emergent nature.
Emergent bias would result in the sense that the drivers' locations are shaped by the platform, and driver may be concentrated in certain hot spots, while some locations remain less frequented, \revv{potentially disadvantaging riders who which to be picked up near those locations.} 
Figure~\ref{fig:recsys:suehr} summarizes our analysis.

\spara{Algorithm.} The method defines a two-sided optimization objective to minimize the difference of the utilities of drivers (resp. customers) as compared to the maximum utility gained by any driver (resp. customer) up until the previous matching round:
    \begin{equation*}
    \begin{aligned}
    \sum_{a} \sum_{b} & \text{   }
     \lambda \cdot \left| \max_{a'}U^{T-1}_{\candidateSet{a'}} - \left( U^{T-1}_{\candidateSet{a}} + M^T_{a,b} \cdot U_{\candidateSet{}}^T(a,b) \right) \right| \\
     &  + (1 - \lambda)\cdot \left| \max_{b'}U^{T-1}_{\userSet{b'}} - \left( U^{T-1}_{\userSet{b}} + M^T_{a,b} \cdot U^{T}_{\userSet{}}(b,a) \right) \right|
    \end{aligned}
    \label{eq:objective2}
    \end{equation*}
where $M^T_{a,b}$ is 1 if driver $\candidateSet{a}$ is matched to customer $\userSet{b}$ in round $T$ and 0 otherwise, and $\lambda$ is a hyper-parameter to continuously interpolate between producer and consumer fairness.
This objective is translated into an integer linear program.

\spara{Experiments.} 
%
\revv{Experiments} are performed on a dataset from a ride hailing platform in an Asian city consisting of 1462 registered drivers, \revv{and measure income inequality using the generalized entropy index (GEI).}. 
As passengers are not registered, every request is handled using a unique job id, with 231,268 jobs in total.
The analysis of the dataset shows that supply exceeds demand by an order of magnitude, and driver income is, hence, a scarce resource.
Then different matching strategies are compared to each other w.r.t. their effects on driver and customer utility:
\begin{itemize}
    \item \textbf{Nearest driver first} is a simple objective for low customer waiting times. 
    It yields the highest \emph{average} driver utility, but has the largest discrepancies in terms of driver income (\ie a high GEI). 
    \item \textbf{Worst-off driver first} yields almost equal driver income, but lowest customer utility.
    It also shows undesired effects, such as lowering hourly wages for drivers that have long active periods, because whenever a new driver joins the system, they have the lowest possible income so far, namely, zero.
    \item \textbf{Two-sided optimization} achieves better results \revv{both} in terms of income equality for drivers \revv{and} in terms of customer waiting time, because drivers happen to be placed in better positions for \revv{subsequent} requests. 
    \revv{However, it is not clear whether this result is an artifact of the dataset, or if it points to a general property of the strategy.}
    \revv{In any case, this finding confirms that, just as is the case for rankings~\cite{zehlike2018reducing} and for two-sided platforms~\cite{mehrotra2018towards}, fairness does not always come at the cost of utility.  In fact, optimizing for fairness can sometimes increase system utility.}
\end{itemize}

\section{Frameworks and Benchmarks}
\label{sec:eval}
In this section, we survey several frameworks and benchmarks that focus on fairness in ranking, and are relevant to both score-based ranking and to supervised learning methods.

\subsection{Fair Search}
FairSearch~\cite{zehlike2020fairsearch} is an open source API to provide fair search results, which is designed as stand-alone libraries (supported in Python and Java) and plugins of Elasticsearch (supported in Java). 
Users can run FairSearch together with their own datasets once these have been formatted as required. FairSearch is designed for learning-to-rank and information retrieval tasks, where the input is queries and documents with relevance scores to each query. For an unseen query, FairSearch outputs a ranked list of relevant documents for it.

FairSearch implements two algorithms to guarantee fair ranking as search results: an in-processing technique named as DELTR (see details in Section~\ref{subsubsec:DELTR}) and a post-processing one called FA*IR (details in Section~\ref{subsubsec:FAIR}) For the support of DELTR, FairSearch provides an off-line wrapper to train a fair ranking model using DELTR that is later uploaded and stored in Elasticsearch as plugins. For FA*IR, FairSearch applies FA*IR algorithm to rerank the search results provided by Elasticsearch and presents it as final search results to users. The implementation of FairSearch can be found in \url{https://github.com/fair-search}. 

\subsection{TREC Fair Ranking Track}
The Text REtrieval Conference (TREC)\footnote{\url{https://trec.nist.gov}}, an annual conference operated by the U.S. National Institute of Standards and Technology (NIST) to support information retrieval research, has a track for fair ranking that evaluates systems according to how well they fairly rank documents, starting from 2019. Each year's Fair Ranking track defines 1-2 information retrieval tasks that ask participants to build systems to address. The tasks are provided with a corpus of documents as training data and a set of queries for evaluation. The participants submit their system's responses to these evaluation queries, and the task organizers score each submission based on some pre-defined measurements that are also provided to the participants with the tasks. Thus, the Fair Ranking track could be seen as a benchmark that allows the comparison of different methods to guarantee fairness in information retrieval tasks.

The Fair Ranking track 2019~\cite{trec-fair-ranking-2019} focuses on re-ranking academic abstracts given a query. The objective is to fairly represent relevant authors from several, undisclosed group definitions. Note that these groups can be defined in a variety of ways, and the definition of groups is not provided by the task organizers and will be a part of evaluation to test the robustness of the results for various group definitions. The Fair Ranking track 2020 also focuses on scholarly search and fairly ranking academic abstracts and papers from authors belonging to different groups. The 2021 focused on fairly prioritising Wikimedia\footnote{\url{https://www.wikimedia.org}} articles for editing to provide a fair exposure to articles from different groups.


%

\subsection{Ranking Facts}
Ranking Facts~\cite{yang2018nutritional} is a Web-based application that generates a “nutritional label” for rankings. The nutritional label is made up of a collection of visual widgets that implement research results on fairness, stability, and transparency for rankings, and that communicate details of the ranking methodology, or of the output, to the end user. Ranking Facts is designed for score-based rankings, where rankers are specified by users as input. 

Ranking Facts supports tabular datasets as inputs. Users are asked to upload a dataset as in CSV format and specify a score-based ranking function. For fairness and diversity concerns, users are also required to specify an attribute in the uploaded data as the sensitive one to define group membership. Then, Ranking Facts generates a nutritional label of the generated ranking. The fairness widget in a produced label implements three definitions of group fairness: whether a fair ranking shows proportional representation, encodes pairwise comparison, or satisfies the definition by FA*IR (see details in Section~\ref{subsubsec:FAIR}). Ranking Facts can be accessed at \url{https://dataresponsibly.github.io/tools/}. The code is available at \url{https://github.com/DataResponsibly/RankingFacts}.

\section{Discussion and Recommendations}
\label{sec:discuss}
In the previous three sections, we discussed the principal functioning of fair ranking methods, and made explicit the technical choices they make and the value frameworks they encode. In this section we discuss our insights and draw a set of recommendations for the evaluation of fair ranking methods. 

Our recommendations are aimed at data science researchers and practitioners.  With these recommendations we aim to establish best practices for the development, evaluation, and deployment of fair ranking algorithms, and to avoid potentially harmful uninformed transfer of methods from one application domain to another. 

\paragraph{Recommendation 1: Make Values and Context of Use Explicit}
Different application scenarios require different value frameworks.
The classification frameworks we presented in this paper are meaningful if the application scenario is concerned with aspects of \emph{distributive justice}.
However, even if a situation requires distributive justice to be taken into account, the goods or benefits to be distributed play a key role in determining which framework should be applied.

For example, college admissions (educational opportunity) may require a different interpretation of fairness than hiring or user rating prediction in online shops (economic opportunity).
To avoid algorithmic solutionism, users of fair rankers must first clearly articulate their own moral goals for a ranking task, and choose a fairness-enhancing method that is consistent with their goals.  To facilitate this choice, the fairness concepts behind a fair rankers must be made explicit by their creators. 

Values are rarely made explicit in fair ranking papers.  
A reader looking to adopt a method to their application context will often use the experiments section of a paper to decide whether the method will ``work'' for them.  
We often see experimental sections in which all available datasets, corresponding to vastly different contexts of use---from recidivism risk prediction, to credit scoring, to college admissions, to matching platforms like AirBnB---are used to show performance of a method, but without an explanation as to why the dataset was selected, other than that it was available and items in it have scores on which to rank.  
For example, the COMPAS dataset~\cite{angwin2016machine} is often used in experiments for papers that propose methods for distributive justice~\cite{zehlike2017fa, yang2017measuring}, yet the dataset was collected for a legal decision making task and so this use is out of scope. 
We caution against this practice and recommend that, when designing their experiments, the authors should carefully substantiate the appropriateness of using a proposed method on a particular dataset in the context of a specific ranking task.  
This substantiation should be made by mapping the method and the task to a value framework.

\paragraph{Recommendation 2: Surface Normative Consequences of Technical Choices}
Algorithmic rankers are complex technical objects, and many implicit and explicit choices go into their design.  
In this paper we discussed an important technical dimension of ranker design, namely, the  representation of group structure: how many sensitive attributes a ranker handles, and whether these are binary or multinary. 
This technical choice in turn impacts what type of discrimination a fair ranker can help address (\eg on one or on several sensitive attributes), and  whether it can address intersectional concerns and, if so, what specific concerns are in scope (\eg representation constraints on intersectional groups, differences in score distributions, or imbalanced loss in utility).  
We deliberately discussed intersectional discrimination under the heading of mitigation objectives, rather than presenting it as a purely technical choice, and we recommend that designers of fair rankers explicitly discuss both their technical choices, and what consequences they have for applicability.

Another important technical dimension is where in the processing pipeline bias mitigation is applied (recall Figure~\ref{fig:ir-flowchart} on page~\pageref{fig:ir-flowchart}).  
Pre-processing methods have the advantage of early intervention on pre-existing bias.  The advantage of in-processing methods is that they do not allow a biased model to be trained. 
The advantage of post-processing methods is that they provide a guaranteed share of visibility for protected groups. 
However, post-processing methods are be subject to legal debates in the US because of due process concerns that may make it illegal to intervene at the decision stage (\eg Ricci v. DeStefano~\cite{ricci}).
In the EU, post-processing methods can be used if other methods fail to comply with EU anti-discrimination law. 
We recommend that the designers of fair rankers substantiate the appropriateness of their technical choice based on the context of use, on for which their method was designed, as well as on the region of use to avoid legal pitfalls. 

\paragraph{Recommendation 3: Draw Meaningful Comparisons}
Additional research is needed to understand how methods with opposing worldviews, conflicting understandings of equality of opportunity, and different addressed biases can be compared in a meaningful way.  
For example, in their experiments the authors of \fairpg~\cite{singh2019policy} compare its results to those of \deltr~\cite{zehlike2018reducing}, but as their measure of disparate exposure is vastly different and the bias they are focusing on is not the same, it is not clear what conclusion to draw from such comparisons.  Making the values and the context of use explicit will go a long way towards helping design meaningful experimental comparisons between methods, rather than mechanically comparing apples to oranges.

\section{Conclusions}
\label{sec:conc}
In this survey we gave an extensive overview of the state-of-the-art literature of fair ranking in the domains of score-based and supervised learning-based ranking. We introduced important dimensions along which to classify fair ranking methods, mapping their assumptions and design choices to the normative values they incorporate. We outlined the technical details of all methods, presented \revv{commonalities and differences}, and categorized each technical choice by its implicit values within the normative dimensions. We discussed implications of normative choices and gave recommendations for researchers on how to make such choices in their work explicit.

Most fair ranking methods are concerned with the concept of \emph{distributive justice}, as they aim to fairly distribute the visibility in a ranking among the candidates. Our focus on distributive justice allowed for the mapping between the worldviews and the equality of opportunity concepts in the framework we proposed. However, this mapping is only meaningful in a distributive context and most likely cannot be transferred to a different setting. In the future we hope to see work that relates to other concepts of justice, such as \emph{procedural justice}, which is concerned with the fairness and transparency of a decision making \emph{process}, and is therefore particularly important in legal decision making. It will be interesting to study whether fairness-enhancing methods designed for concerns of distributive justice can be transferred to the context of procedural justice in a meaningful way.

Another interesting direction to classify fair ranking methods within distributive justice contexts is to understand the properties of ranking scores with respect to different indexes of advantage.
Commonly, the score models a candidate's \textit{potential utility} to the user of the ranking, which stems from welfarism/utilitarianism and, as such, incorporates an idea of satisfaction and preference~\cite{sen1980equality}. In contrast, Rawls judges the goodness of a distribution in terms of so-called \textit{primary goods}. It is important to explore the implications of these different conceptions of advantage, and, crucially,  understand whether they can be combined in a common fairness objective.
\section{Acknowledgements}
\label{sec:ack}

We are grateful to Falaah Arif Khan for her input on equality of opportunity (\eop) frameworks, and on the mapping of specific methods to \eop doctrines. Additionally we want to thank the authors of various methods we presented here for their discussions on the mathematical and fairness details. This research was supported in part by NSF Awards No. 1934464, 1916505, and 1922658. 

\balance

{
\bibliographystyle{ACM-Reference-Format}
\bibliography{main.bib}


\begin{thebibliography}{122}


\ifx \showCODEN    \undefined \def \showCODEN     #1{\unskip}     \fi
\ifx \showDOI      \undefined \def \showDOI       #1{#1}\fi
\ifx \showISBNx    \undefined \def \showISBNx     #1{\unskip}     \fi
\ifx \showISBNxiii \undefined \def \showISBNxiii  #1{\unskip}     \fi
\ifx \showISSN     \undefined \def \showISSN      #1{\unskip}     \fi
\ifx \showLCCN     \undefined \def \showLCCN      #1{\unskip}     \fi
\ifx \shownote     \undefined \def \shownote      #1{#1}          \fi
\ifx \showarticletitle \undefined \def \showarticletitle #1{#1}   \fi
\ifx \showURL      \undefined \def \showURL       {\relax}        \fi
\providecommand\bibfield[2]{#2}
\providecommand\bibinfo[2]{#2}
\providecommand\natexlab[1]{#1}
\providecommand\showeprint[2][]{arXiv:#2}

\bibitem[\protect\citeauthoryear{AirBnB}{AirBnB}{[n.\,d.]}]%
        {AirBnBData}
\bibfield{author}{\bibinfo{person}{AirBnB}.}
  \bibinfo{year}{[n.\,d.]}\natexlab{}.
\newblock \bibinfo{title}{AirBnB}.
\newblock
\newblock
\urldef\tempurl%
\url{https://insideairbnb.com}
\showURL{%
\tempurl}


\bibitem[\protect\citeauthoryear{Angwin, Larson, Mattu, and Kirchner}{Angwin
  et~al\mbox{.}}{2016}]%
        {angwin2016machine}
\bibfield{author}{\bibinfo{person}{Julia Angwin}, \bibinfo{person}{Jeff
  Larson}, \bibinfo{person}{Surya Mattu}, {and} \bibinfo{person}{Lauren
  Kirchner}.} \bibinfo{year}{2016}\natexlab{}.
\newblock \showarticletitle{Machine bias. ProPublica}.
\newblock \bibinfo{journal}{\emph{See https://www. propublica.
  org/article/machine-bias-risk-assessments-in-criminal-sentencing}}
  (\bibinfo{year}{2016}).
\newblock


\bibitem[\protect\citeauthoryear{Arneson}{Arneson}{2018}]%
        {Arneson2018FourCO}
\bibfield{author}{\bibinfo{person}{Richard~J. Arneson}.}
  \bibinfo{year}{2018}\natexlab{}.
\newblock \showarticletitle{Four Conceptions of Equal Opportunity}.
\newblock \bibinfo{journal}{\emph{Wiley-Blackwell: Economic Journal}}
  (\bibinfo{year}{2018}).
\newblock


\bibitem[\protect\citeauthoryear{Asudeh, Jagadish, Stoyanovich, and Das}{Asudeh
  et~al\mbox{.}}{2019}]%
        {asudeh2019designing}
\bibfield{author}{\bibinfo{person}{Abolfazl Asudeh}, \bibinfo{person}{HV
  Jagadish}, \bibinfo{person}{Julia Stoyanovich}, {and} \bibinfo{person}{Gautam
  Das}.} \bibinfo{year}{2019}\natexlab{}.
\newblock \showarticletitle{Designing fair ranking schemes}. In
  \bibinfo{booktitle}{\emph{Proceedings of the 2019 International Conference on
  Management of Data}}. ACM, \bibinfo{pages}{1259--1276}.
\newblock


\bibitem[\protect\citeauthoryear{Asudeh and Jagadish}{Asudeh and
  Jagadish}{2020}]%
        {vldb20tutorial}
\bibfield{author}{\bibinfo{person}{Abolfazl Asudeh} {and}
  \bibinfo{person}{H.~V. Jagadish}.} \bibinfo{year}{2020}\natexlab{}.
\newblock \showarticletitle{Fairly Evaluating and Scoring Items in a Data Set}.
\newblock \bibinfo{journal}{\emph{Proc. {VLDB} Endow.}} \bibinfo{volume}{13},
  \bibinfo{number}{12} (\bibinfo{year}{2020}), \bibinfo{pages}{3445--3448}.
\newblock
\urldef\tempurl%
\url{https://doi.org/10.14778/3415478.3415566}
\showDOI{\tempurl}


\bibitem[\protect\citeauthoryear{Babaioff, Immorlica, Kempe, and
  Kleinberg}{Babaioff et~al\mbox{.}}{2008}]%
        {DBLP:journals/sigecom/BabaioffIKK08}
\bibfield{author}{\bibinfo{person}{Moshe Babaioff}, \bibinfo{person}{Nicole
  Immorlica}, \bibinfo{person}{David Kempe}, {and} \bibinfo{person}{Robert
  Kleinberg}.} \bibinfo{year}{2008}\natexlab{}.
\newblock \showarticletitle{Online auctions and generalized secretary
  problems}.
\newblock \bibinfo{journal}{\emph{SIGecom Exchanges}} \bibinfo{volume}{7},
  \bibinfo{number}{2} (\bibinfo{year}{2008}).
\newblock
\urldef\tempurl%
\url{https://doi.org/10.1145/1399589.1399596}
\showDOI{\tempurl}


\bibitem[\protect\citeauthoryear{Baeza{-}Yates}{Baeza{-}Yates}{2018}]%
        {DBLP:journals/cacm/Baeza-Yates18}
\bibfield{author}{\bibinfo{person}{Ricardo Baeza{-}Yates}.}
  \bibinfo{year}{2018}\natexlab{}.
\newblock \showarticletitle{Bias on the web}.
\newblock \bibinfo{journal}{\emph{Commun. {ACM}}} \bibinfo{volume}{61},
  \bibinfo{number}{6} (\bibinfo{year}{2018}), \bibinfo{pages}{54--61}.
\newblock
\urldef\tempurl%
\url{https://doi.org/10.1145/3209581}
\showDOI{\tempurl}


\bibitem[\protect\citeauthoryear{Barocas, Hardt, and Narayanan}{Barocas
  et~al\mbox{.}}{2017}]%
        {fairMLbook}
\bibfield{author}{\bibinfo{person}{Solon Barocas}, \bibinfo{person}{Moritz
  Hardt}, {and} \bibinfo{person}{Arvind Narayanan}.}
  \bibinfo{year}{2017}\natexlab{}.
\newblock \showarticletitle{Fairness in machine learning}.
\newblock \bibinfo{journal}{\emph{Nips tutorial}}  \bibinfo{volume}{1}
  (\bibinfo{year}{2017}), \bibinfo{pages}{2}.
\newblock


\bibitem[\protect\citeauthoryear{Baswana, Chakrabarti, Kanoria, Patange, and
  Chandran}{Baswana et~al\mbox{.}}{2019}]%
        {DBLP:journals/corr/abs-1904-06698}
\bibfield{author}{\bibinfo{person}{Surender Baswana}, \bibinfo{person}{P.~P.
  Chakrabarti}, \bibinfo{person}{Yashodhan Kanoria}, \bibinfo{person}{Utkarsh
  Patange}, {and} \bibinfo{person}{Sharat Chandran}.}
  \bibinfo{year}{2019}\natexlab{}.
\newblock \showarticletitle{Joint Seat Allocation 2018: An algorithmic
  perspective}.
\newblock \bibinfo{journal}{\emph{CoRR}}  \bibinfo{volume}{abs/1904.06698}
  (\bibinfo{year}{2019}).
\newblock
\showeprint[arxiv]{1904.06698}
\urldef\tempurl%
\url{http://arxiv.org/abs/1904.06698}
\showURL{%
\tempurl}


\bibitem[\protect\citeauthoryear{Bellamy, Dey, Hind, Hoffman, Houde, Kannan,
  Lohia, Martino, Mehta, Mojsilovic, Nagar, Ramamurthy, Richards, Saha,
  Sattigeri, Singh, Varshney, and Zhang}{Bellamy et~al\mbox{.}}{2018}]%
        {bellamy2018ai}
\bibfield{author}{\bibinfo{person}{Rachel K.~E. Bellamy},
  \bibinfo{person}{Kuntal Dey}, \bibinfo{person}{Michael Hind},
  \bibinfo{person}{Samuel~C. Hoffman}, \bibinfo{person}{Stephanie Houde},
  \bibinfo{person}{Kalapriya Kannan}, \bibinfo{person}{Pranay Lohia},
  \bibinfo{person}{Jacquelyn Martino}, \bibinfo{person}{Sameep Mehta},
  \bibinfo{person}{Aleksandra Mojsilovic}, \bibinfo{person}{Seema Nagar},
  \bibinfo{person}{Karthikeyan~Natesan Ramamurthy}, \bibinfo{person}{John~T.
  Richards}, \bibinfo{person}{Diptikalyan Saha}, \bibinfo{person}{Prasanna
  Sattigeri}, \bibinfo{person}{Moninder Singh}, \bibinfo{person}{Kush~R.
  Varshney}, {and} \bibinfo{person}{Yunfeng Zhang}.}
  \bibinfo{year}{2018}\natexlab{}.
\newblock \showarticletitle{{AI} Fairness 360: An Extensible Toolkit for
  Detecting, Understanding, and Mitigating Unwanted Algorithmic Bias}.
\newblock \bibinfo{journal}{\emph{CoRR}}  \bibinfo{volume}{abs/1810.01943}
  (\bibinfo{year}{2018}).
\newblock
\showeprint[arxiv]{1810.01943}
\urldef\tempurl%
\url{http://arxiv.org/abs/1810.01943}
\showURL{%
\tempurl}


\bibitem[\protect\citeauthoryear{Beutel, Chen, Doshi, Qian, Wei, Wu, Heldt,
  Zhao, Hong, Chi, and Goodrow}{Beutel et~al\mbox{.}}{2019}]%
        {Beutel:2019:FRR:3292500.3330745}
\bibfield{author}{\bibinfo{person}{Alex Beutel}, \bibinfo{person}{Jilin Chen},
  \bibinfo{person}{Tulsee Doshi}, \bibinfo{person}{Hai Qian},
  \bibinfo{person}{Li Wei}, \bibinfo{person}{Yi Wu}, \bibinfo{person}{Lukasz
  Heldt}, \bibinfo{person}{Zhe Zhao}, \bibinfo{person}{Lichan Hong},
  \bibinfo{person}{Ed~H. Chi}, {and} \bibinfo{person}{Cristos Goodrow}.}
  \bibinfo{year}{2019}\natexlab{}.
\newblock \showarticletitle{Fairness in Recommendation Ranking Through Pairwise
  Comparisons}. In \bibinfo{booktitle}{\emph{Proceedings of the 25th ACM SIGKDD
  International Conference on Knowledge Discovery \& Data Mining}} (Anchorage,
  AK, USA) \emph{(\bibinfo{series}{KDD '19})}. \bibinfo{publisher}{ACM},
  \bibinfo{address}{New York, NY, USA}, \bibinfo{pages}{2212--2220}.
\newblock
\showISBNx{978-1-4503-6201-6}
\urldef\tempurl%
\url{https://doi.org/10.1145/3292500.3330745}
\showDOI{\tempurl}


\bibitem[\protect\citeauthoryear{Biega, Gummadi, and Weikum}{Biega
  et~al\mbox{.}}{2018}]%
        {biega2018equity}
\bibfield{author}{\bibinfo{person}{Asia~J Biega}, \bibinfo{person}{Krishna~P
  Gummadi}, {and} \bibinfo{person}{Gerhard Weikum}.}
  \bibinfo{year}{2018}\natexlab{}.
\newblock \showarticletitle{Equity of attention: Amortizing individual fairness
  in rankings}. In \bibinfo{booktitle}{\emph{The 41st International ACM SIGIR
  Conference on Research \& Development in Information Retrieval}}. ACM,
  \bibinfo{pages}{405--414}.
\newblock


\bibitem[\protect\citeauthoryear{Biega, Saha~Roy, and Weikum}{Biega
  et~al\mbox{.}}{2017}]%
        {biega2017privacy}
\bibfield{author}{\bibinfo{person}{Asia~J Biega}, \bibinfo{person}{Rishiraj
  Saha~Roy}, {and} \bibinfo{person}{Gerhard Weikum}.}
  \bibinfo{year}{2017}\natexlab{}.
\newblock \showarticletitle{Privacy through solidarity: A
  user-utility-preserving framework to counter profiling}. In
  \bibinfo{booktitle}{\emph{Proceedings of the 40th International ACM SIGIR
  Conference on Research and Development in Information Retrieval}}.
  \bibinfo{pages}{675--684}.
\newblock


\bibitem[\protect\citeauthoryear{Bradley and Terry}{Bradley and Terry}{1952}]%
        {bradley1952rank}
\bibfield{author}{\bibinfo{person}{Ralph~Allan Bradley} {and}
  \bibinfo{person}{Milton~E Terry}.} \bibinfo{year}{1952}\natexlab{}.
\newblock \showarticletitle{Rank analysis of incomplete block designs: I. The
  method of paired comparisons}.
\newblock \bibinfo{journal}{\emph{Biometrika}} \bibinfo{volume}{39},
  \bibinfo{number}{3/4} (\bibinfo{year}{1952}), \bibinfo{pages}{324--345}.
\newblock


\bibitem[\protect\citeauthoryear{Burges}{Burges}{2010}]%
        {burges2010ranknet}
\bibfield{author}{\bibinfo{person}{Christopher~JC Burges}.}
  \bibinfo{year}{2010}\natexlab{}.
\newblock \showarticletitle{From ranknet to lambdarank to lambdamart: An
  overview}.
\newblock \bibinfo{journal}{\emph{Learning}} \bibinfo{volume}{11},
  \bibinfo{number}{23-581} (\bibinfo{year}{2010}), \bibinfo{pages}{81}.
\newblock


\bibitem[\protect\citeauthoryear{Burke}{Burke}{2017}]%
        {burke2017multisided}
\bibfield{author}{\bibinfo{person}{Robin Burke}.}
  \bibinfo{year}{2017}\natexlab{}.
\newblock \showarticletitle{Multisided fairness for recommendation}.
\newblock \bibinfo{journal}{\emph{arXiv preprint arXiv:1707.00093}}
  (\bibinfo{year}{2017}).
\newblock


\bibitem[\protect\citeauthoryear{Cao, Qin, Liu, Tsai, and Li}{Cao
  et~al\mbox{.}}{2007}]%
        {cao2007learning}
\bibfield{author}{\bibinfo{person}{Zhe Cao}, \bibinfo{person}{Tao Qin},
  \bibinfo{person}{Tie-Yan Liu}, \bibinfo{person}{Ming-Feng Tsai}, {and}
  \bibinfo{person}{Hang Li}.} \bibinfo{year}{2007}\natexlab{}.
\newblock \showarticletitle{Learning to rank: from pairwise approach to
  listwise approach}. In \bibinfo{booktitle}{\emph{Proceedings of the 24th
  international conference on Machine learning}}. ACM,
  \bibinfo{pages}{129--136}.
\newblock


\bibitem[\protect\citeauthoryear{Castillo}{Castillo}{2019}]%
        {castillo2019fairness}
\bibfield{author}{\bibinfo{person}{Carlos Castillo}.}
  \bibinfo{year}{2019}\natexlab{}.
\newblock \showarticletitle{Fairness and transparency in ranking}. In
  \bibinfo{booktitle}{\emph{ACM SIGIR Forum}}, Vol.~\bibinfo{volume}{52}. ACM
  New York, NY, USA, \bibinfo{pages}{64--71}.
\newblock


\bibitem[\protect\citeauthoryear{Celis, Mehrotra, and Vishnoi}{Celis
  et~al\mbox{.}}{2020}]%
        {celis2020interventions}
\bibfield{author}{\bibinfo{person}{L~Elisa Celis}, \bibinfo{person}{Anay
  Mehrotra}, {and} \bibinfo{person}{Nisheeth~K Vishnoi}.}
  \bibinfo{year}{2020}\natexlab{}.
\newblock \showarticletitle{Interventions for ranking in the presence of
  implicit bias}. In \bibinfo{booktitle}{\emph{Proceedings of the 2020
  Conference on Fairness, Accountability, and Transparency}}.
  \bibinfo{pages}{369--380}.
\newblock


\bibitem[\protect\citeauthoryear{Celis, Straszak, and Vishnoi}{Celis
  et~al\mbox{.}}{2018}]%
        {celis2018ranking}
\bibfield{author}{\bibinfo{person}{L~Elisa Celis}, \bibinfo{person}{Damian
  Straszak}, {and} \bibinfo{person}{Nisheeth~K Vishnoi}.}
  \bibinfo{year}{2018}\natexlab{}.
\newblock \showarticletitle{Ranking with Fairness Constraints}. In
  \bibinfo{booktitle}{\emph{45th International Colloquium on Automata,
  Languages, and Programming (ICALP 2018)}}. Schloss Dagstuhl-Leibniz-Zentrum
  fuer Informatik.
\newblock


\bibitem[\protect\citeauthoryear{Chen, Dong, Wang, Feng, Wang, and He}{Chen
  et~al\mbox{.}}{2020}]%
        {chen2020bias}
\bibfield{author}{\bibinfo{person}{Jiawei Chen}, \bibinfo{person}{Hande Dong},
  \bibinfo{person}{Xiang Wang}, \bibinfo{person}{Fuli Feng},
  \bibinfo{person}{Meng Wang}, {and} \bibinfo{person}{Xiangnan He}.}
  \bibinfo{year}{2020}\natexlab{}.
\newblock \showarticletitle{Bias and debias in recommender system: A survey and
  future directions}.
\newblock \bibinfo{journal}{\emph{arXiv preprint arXiv:2010.03240}}
  (\bibinfo{year}{2020}).
\newblock


\bibitem[\protect\citeauthoryear{Chouldechova and Roth}{Chouldechova and
  Roth}{2020}]%
        {DBLP:journals/cacm/ChouldechovaR20}
\bibfield{author}{\bibinfo{person}{Alexandra Chouldechova} {and}
  \bibinfo{person}{Aaron Roth}.} \bibinfo{year}{2020}\natexlab{}.
\newblock \showarticletitle{A snapshot of the frontiers of fairness in machine
  learning}.
\newblock \bibinfo{journal}{\emph{Commun. {ACM}}} \bibinfo{volume}{63},
  \bibinfo{number}{5} (\bibinfo{year}{2020}), \bibinfo{pages}{82--89}.
\newblock
\urldef\tempurl%
\url{https://doi.org/10.1145/3376898}
\showDOI{\tempurl}


\bibitem[\protect\citeauthoryear{Cohen, Cohen, and Banthin}{Cohen
  et~al\mbox{.}}{2009}]%
        {cohen2009medical}
\bibfield{author}{\bibinfo{person}{Joel~W Cohen}, \bibinfo{person}{Steven~B
  Cohen}, {and} \bibinfo{person}{Jessica~S Banthin}.}
  \bibinfo{year}{2009}\natexlab{}.
\newblock \showarticletitle{The medical expenditure panel survey: a national
  information resource to support healthcare cost research and inform policy
  and practice}.
\newblock \bibinfo{journal}{\emph{Medical care}} (\bibinfo{year}{2009}),
  \bibinfo{pages}{S44--S50}.
\newblock


\bibitem[\protect\citeauthoryear{Collective Learning~Group}{Collective
  Learning~Group}{[n.\,d.]}]%
        {PantheonData}
\bibfield{author}{\bibinfo{person}{MIT Collective Learning~Group}.}
  \bibinfo{year}{[n.\,d.]}\natexlab{}.
\newblock \bibinfo{title}{Pantheon}.
\newblock
\newblock
\urldef\tempurl%
\url{https://github.com/DataResponsibly/Datasets}
\showURL{%
\tempurl}


\bibitem[\protect\citeauthoryear{Collins}{Collins}{2007}]%
        {rooney}
\bibfield{author}{\bibinfo{person}{Brian Collins}.}
  \bibinfo{year}{2007}\natexlab{}.
\newblock \showarticletitle{Tackling Unconscious Bias in Hiring Practices: The
  Plight of the Rooney Rule}.
\newblock \bibinfo{journal}{\emph{NYU Law Review}} \bibinfo{volume}{82},
  \bibinfo{number}{3} (\bibinfo{year}{2007}).
\newblock


\bibitem[\protect\citeauthoryear{Collins}{Collins}{2002}]%
        {collins2002black}
\bibfield{author}{\bibinfo{person}{Patricia~Hill Collins}.}
  \bibinfo{year}{2002}\natexlab{}.
\newblock \bibinfo{booktitle}{\emph{Black feminist thought: Knowledge,
  consciousness, and the politics of empowerment}}.
\newblock \bibinfo{publisher}{routledge}.
\newblock


\bibitem[\protect\citeauthoryear{Coston, Ramamurthy, Wei, Varshney, Speakman,
  Mustahsan, and Chakraborty}{Coston et~al\mbox{.}}{2019}]%
        {coston2019fair}
\bibfield{author}{\bibinfo{person}{Amanda Coston},
  \bibinfo{person}{Karthikeyan~Natesan Ramamurthy}, \bibinfo{person}{Dennis
  Wei}, \bibinfo{person}{Kush~R Varshney}, \bibinfo{person}{Skyler Speakman},
  \bibinfo{person}{Zairah Mustahsan}, {and} \bibinfo{person}{Supriyo
  Chakraborty}.} \bibinfo{year}{2019}\natexlab{}.
\newblock \showarticletitle{Fair Transfer Learning with Missing Protected
  Attributes}. In \bibinfo{booktitle}{\emph{Proceedings of the 2019 {AAAI/ACM}
  Conference on AI, Ethics, and Society, {AIES}}}. \bibinfo{publisher}{{ACM}}.
\newblock


\bibitem[\protect\citeauthoryear{Crenshaw}{Crenshaw}{1990}]%
        {crenshaw1990mapping}
\bibfield{author}{\bibinfo{person}{Kimberle Crenshaw}.}
  \bibinfo{year}{1990}\natexlab{}.
\newblock \showarticletitle{Mapping the margins: Intersectionality, identity
  politics, and violence against women of color}.
\newblock \bibinfo{journal}{\emph{Stan. L. Rev.}}  \bibinfo{volume}{43}
  (\bibinfo{year}{1990}), \bibinfo{pages}{1241}.
\newblock


\bibitem[\protect\citeauthoryear{{CS Rankings}}{{CS Rankings}}{[n.\,d.]}]%
        {CSData}
\bibfield{author}{\bibinfo{person}{{CS Rankings}}.}
  \bibinfo{year}{[n.\,d.]}\natexlab{}.
\newblock \bibinfo{title}{CSRankings: Computer Science Rankings}.
\newblock
\newblock
\urldef\tempurl%
\url{https://csrankings.org}
\showURL{%
\tempurl}


\bibitem[\protect\citeauthoryear{De~Gemmis, Lops, Musto, Narducci, and
  Semeraro}{De~Gemmis et~al\mbox{.}}{2015}]%
        {de2015semantics}
\bibfield{author}{\bibinfo{person}{Marco De~Gemmis}, \bibinfo{person}{Pasquale
  Lops}, \bibinfo{person}{Cataldo Musto}, \bibinfo{person}{Fedelucio Narducci},
  {and} \bibinfo{person}{Giovanni Semeraro}.} \bibinfo{year}{2015}\natexlab{}.
\newblock \showarticletitle{Semantics-aware content-based recommender systems}.
\newblock In \bibinfo{booktitle}{\emph{Recommender systems handbook}}.
  \bibinfo{publisher}{Springer}, \bibinfo{pages}{119--159}.
\newblock


\bibitem[\protect\citeauthoryear{Deldjoo, Anelli, Zamani, Bellogin, and
  Di~Noia}{Deldjoo et~al\mbox{.}}{2021}]%
        {deldjoo2021flexible}
\bibfield{author}{\bibinfo{person}{Yashar Deldjoo},
  \bibinfo{person}{Vito~Walter Anelli}, \bibinfo{person}{Hamed Zamani},
  \bibinfo{person}{Alejandro Bellogin}, {and} \bibinfo{person}{Tommaso
  Di~Noia}.} \bibinfo{year}{2021}\natexlab{}.
\newblock \showarticletitle{A flexible framework for evaluating user and item
  fairness in recommender systems}.
\newblock \bibinfo{journal}{\emph{User Modeling and User-Adapted Interaction}}
  (\bibinfo{year}{2021}), \bibinfo{pages}{1--55}.
\newblock


\bibitem[\protect\citeauthoryear{D'Ignazio and Klein}{D'Ignazio and
  Klein}{2020}]%
        {d2020data}
\bibfield{author}{\bibinfo{person}{Catherine D'Ignazio} {and}
  \bibinfo{person}{Lauren~F Klein}.} \bibinfo{year}{2020}\natexlab{}.
\newblock \bibinfo{booktitle}{\emph{Data feminism}}.
\newblock \bibinfo{publisher}{MIT Press}.
\newblock


\bibitem[\protect\citeauthoryear{Drosou, Jagadish, Pitoura, and
  Stoyanovich}{Drosou et~al\mbox{.}}{2017}]%
        {drosou2017diversity}
\bibfield{author}{\bibinfo{person}{Marina Drosou}, \bibinfo{person}{HV
  Jagadish}, \bibinfo{person}{Evaggelia Pitoura}, {and} \bibinfo{person}{Julia
  Stoyanovich}.} \bibinfo{year}{2017}\natexlab{}.
\newblock \showarticletitle{Diversity in {B}ig {D}ata: A review}.
\newblock \bibinfo{journal}{\emph{Big Data}} \bibinfo{volume}{5},
  \bibinfo{number}{2} (\bibinfo{year}{2017}), \bibinfo{pages}{73--84}.
\newblock


\bibitem[\protect\citeauthoryear{Dwork, Hardt, Pitassi, Reingold, and
  Zemel}{Dwork et~al\mbox{.}}{2012}]%
        {dwork2012fairness}
\bibfield{author}{\bibinfo{person}{Cynthia Dwork}, \bibinfo{person}{Moritz
  Hardt}, \bibinfo{person}{Toniann Pitassi}, \bibinfo{person}{Omer Reingold},
  {and} \bibinfo{person}{Richard Zemel}.} \bibinfo{year}{2012}\natexlab{}.
\newblock \showarticletitle{Fairness through awareness}. In
  \bibinfo{booktitle}{\emph{Proceedings of the 3rd innovations in theoretical
  computer science conference}}. ACM, \bibinfo{pages}{214--226}.
\newblock


\bibitem[\protect\citeauthoryear{Dworkin}{Dworkin}{1981}]%
        {dworkin_1981}
\bibfield{author}{\bibinfo{person}{Ronald Dworkin}.}
  \bibinfo{year}{1981}\natexlab{}.
\newblock \showarticletitle{What is Equality? Part 1: Equality of Welfare}.
\newblock \bibinfo{journal}{\emph{Philosophy and Public Affairs}}
  \bibinfo{volume}{10}, \bibinfo{number}{3} (\bibinfo{year}{1981}),
  \bibinfo{pages}{185--246}.
\newblock
\showISSN{00483915, 10884963}
\urldef\tempurl%
\url{http://www.jstor.org/stable/2264894}
\showURL{%
\tempurl}


\bibitem[\protect\citeauthoryear{Dynkin}{Dynkin}{1963}]%
        {Dynkin:1963}
\bibfield{author}{\bibinfo{person}{E.B. Dynkin}.}
  \bibinfo{year}{1963}\natexlab{}.
\newblock \showarticletitle{The optimum choice of the instant for stopping a
  Markov process}.
\newblock \bibinfo{journal}{\emph{Sov. Math. Dokl.}}  \bibinfo{volume}{4}
  (\bibinfo{year}{1963}).
\newblock


\bibitem[\protect\citeauthoryear{Ekstrand, Burke, and Diaz}{Ekstrand
  et~al\mbox{.}}{2019a}]%
        {DBLP:conf/recsys/EkstrandBD19}
\bibfield{author}{\bibinfo{person}{Michael~D. Ekstrand}, \bibinfo{person}{Robin
  Burke}, {and} \bibinfo{person}{Fernando Diaz}.}
  \bibinfo{year}{2019}\natexlab{a}.
\newblock \showarticletitle{Fairness and discrimination in recommendation and
  retrieval}. In \bibinfo{booktitle}{\emph{Proceedings of the 13th {ACM}
  Conference on Recommender Systems, RecSys 2019, Copenhagen, Denmark,
  September 16-20, 2019}}, \bibfield{editor}{\bibinfo{person}{Toine Bogers},
  \bibinfo{person}{Alan Said}, \bibinfo{person}{Peter Brusilovsky}, {and}
  \bibinfo{person}{Domonkos Tikk}} (Eds.). \bibinfo{publisher}{{ACM}},
  \bibinfo{pages}{576--577}.
\newblock
\urldef\tempurl%
\url{https://doi.org/10.1145/3298689.3346964}
\showDOI{\tempurl}


\bibitem[\protect\citeauthoryear{Ekstrand, Burke, and Diaz}{Ekstrand
  et~al\mbox{.}}{2019b}]%
        {ekstrand2019fairness}
\bibfield{author}{\bibinfo{person}{Michael~D. Ekstrand}, \bibinfo{person}{Robin
  Burke}, {and} \bibinfo{person}{Fernando Diaz}.}
  \bibinfo{year}{2019}\natexlab{b}.
\newblock \showarticletitle{Fairness and Discrimination in Retrieval and
  Recommendation}. In \bibinfo{booktitle}{\emph{Proceedings of the 42nd
  International ACM SIGIR Conference on Research and Development in Information
  Retrieval}} (Paris, France) \emph{(\bibinfo{series}{SIGIR’19})}.
  \bibinfo{publisher}{Association for Computing Machinery},
  \bibinfo{address}{New York, NY, USA}, \bibinfo{pages}{1403–1404}.
\newblock
\showISBNx{9781450361729}
\urldef\tempurl%
\url{https://doi.org/10.1145/3331184.3331380}
\showDOI{\tempurl}


\bibitem[\protect\citeauthoryear{Ferguson}{Ferguson}{1989}]%
        {ferguson1989}
\bibfield{author}{\bibinfo{person}{Thomas~S. Ferguson}.}
  \bibinfo{year}{1989}\natexlab{}.
\newblock \showarticletitle{Who Solved the Secretary Problem?}
\newblock \bibinfo{journal}{\emph{Statist. Sci.}} \bibinfo{volume}{4},
  \bibinfo{number}{3} (\bibinfo{date}{08} \bibinfo{year}{1989}),
  \bibinfo{pages}{282--289}.
\newblock
\urldef\tempurl%
\url{https://doi.org/10.1214/ss/1177012493}
\showDOI{\tempurl}


\bibitem[\protect\citeauthoryear{Fishkin}{Fishkin}{2014}]%
        {Fishkin2014Bottlenecks}
\bibfield{author}{\bibinfo{person}{Joseph Fishkin}.}
  \bibinfo{year}{2014}\natexlab{}.
\newblock \bibinfo{booktitle}{\emph{Bottlenecks: A New Theory of Equal
  Opportunity}}.
\newblock \bibinfo{publisher}{Oup Usa}.
\newblock


\bibitem[\protect\citeauthoryear{Forbes}{Forbes}{[n.\,d.]}]%
        {ForbesRichesData}
\bibfield{author}{\bibinfo{person}{Forbes}.}
  \bibinfo{year}{[n.\,d.]}\natexlab{}.
\newblock \bibinfo{title}{Forbes Richest Americans}.
\newblock
\newblock
\urldef\tempurl%
\url{https://github.com/DataResponsibly/Datasets}
\showURL{%
\tempurl}


\bibitem[\protect\citeauthoryear{Friedler, Scheidegger, and
  Venkatasubramanian}{Friedler et~al\mbox{.}}{2016}]%
        {friedler2016possibility}
\bibfield{author}{\bibinfo{person}{Sorelle~A Friedler}, \bibinfo{person}{Carlos
  Scheidegger}, {and} \bibinfo{person}{Suresh Venkatasubramanian}.}
  \bibinfo{year}{2016}\natexlab{}.
\newblock \showarticletitle{On the (im) possibility of fairness}.
\newblock \bibinfo{journal}{\emph{arXiv preprint arXiv:1609.07236}}
  (\bibinfo{year}{2016}).
\newblock


\bibitem[\protect\citeauthoryear{Friedman and Nissenbaum}{Friedman and
  Nissenbaum}{1996}]%
        {DBLP:journals/tois/FriedmanN96}
\bibfield{author}{\bibinfo{person}{Batya Friedman} {and} \bibinfo{person}{Helen
  Nissenbaum}.} \bibinfo{year}{1996}\natexlab{}.
\newblock \showarticletitle{Bias in Computer Systems}.
\newblock \bibinfo{journal}{\emph{{ACM} Trans. Inf. Syst.}}
  \bibinfo{volume}{14}, \bibinfo{number}{3} (\bibinfo{year}{1996}),
  \bibinfo{pages}{330--347}.
\newblock
\urldef\tempurl%
\url{https://doi.org/10.1145/230538.230561}
\showDOI{\tempurl}


\bibitem[\protect\citeauthoryear{Gao and Shah}{Gao and Shah}{2020}]%
        {gao2020counteracting}
\bibfield{author}{\bibinfo{person}{Ruoyuan Gao} {and} \bibinfo{person}{Chirag
  Shah}.} \bibinfo{year}{2020}\natexlab{}.
\newblock \showarticletitle{Counteracting Bias and Increasing Fairness in
  Search and Recommender Systems}. In \bibinfo{booktitle}{\emph{Fourteenth ACM
  Conference on Recommender Systems}}. \bibinfo{pages}{745--747}.
\newblock


\bibitem[\protect\citeauthoryear{Ge, Liu, Gao, Xian, Li, Zhao, Pei, Sun, Ge,
  Ou, et~al\mbox{.}}{Ge et~al\mbox{.}}{2021}]%
        {ge2021towards}
\bibfield{author}{\bibinfo{person}{Yingqiang Ge}, \bibinfo{person}{Shuchang
  Liu}, \bibinfo{person}{Ruoyuan Gao}, \bibinfo{person}{Yikun Xian},
  \bibinfo{person}{Yunqi Li}, \bibinfo{person}{Xiangyu Zhao},
  \bibinfo{person}{Changhua Pei}, \bibinfo{person}{Fei Sun},
  \bibinfo{person}{Junfeng Ge}, \bibinfo{person}{Wenwu Ou}, {et~al\mbox{.}}}
  \bibinfo{year}{2021}\natexlab{}.
\newblock \showarticletitle{Towards Long-term Fairness in Recommendation}. In
  \bibinfo{booktitle}{\emph{Proceedings of the 14th ACM International
  Conference on Web Search and Data Mining}}. \bibinfo{pages}{445--453}.
\newblock


\bibitem[\protect\citeauthoryear{Geyik, Ambler, and Kenthapadi}{Geyik
  et~al\mbox{.}}{2019}]%
        {geyik2019fairness}
\bibfield{author}{\bibinfo{person}{Sahin~Cem Geyik}, \bibinfo{person}{Stuart
  Ambler}, {and} \bibinfo{person}{Krishnaram Kenthapadi}.}
  \bibinfo{year}{2019}\natexlab{}.
\newblock \showarticletitle{Fairness-aware ranking in search \& recommendation
  systems with application to linkedin talent search}. In
  \bibinfo{booktitle}{\emph{Proceedings of the 25th acm sigkdd international
  conference on knowledge discovery \& data mining}}.
  \bibinfo{pages}{2221--2231}.
\newblock


\bibitem[\protect\citeauthoryear{Grari, Ruf, Lamprier, and Detyniecki}{Grari
  et~al\mbox{.}}{2019}]%
        {grari2019fairnessaware}
\bibfield{author}{\bibinfo{person}{Vincent Grari}, \bibinfo{person}{Boris Ruf},
  \bibinfo{person}{Sylvain Lamprier}, {and} \bibinfo{person}{Marcin
  Detyniecki}.} \bibinfo{year}{2019}\natexlab{}.
\newblock \bibinfo{title}{Fairness-Aware Neural R\'eyni Minimization for
  Continuous Features}.
\newblock
\newblock
\showeprint[arxiv]{1911.04929}~[cs.LG]


\bibitem[\protect\citeauthoryear{Hajian, Bonchi, and Castillo}{Hajian
  et~al\mbox{.}}{2016}]%
        {hajian2016algorithmic}
\bibfield{author}{\bibinfo{person}{Sara Hajian}, \bibinfo{person}{Francesco
  Bonchi}, {and} \bibinfo{person}{Carlos Castillo}.}
  \bibinfo{year}{2016}\natexlab{}.
\newblock \showarticletitle{Algorithmic bias: From discrimination discovery to
  fairness-aware data mining}. In \bibinfo{booktitle}{\emph{Proceedings of the
  22nd ACM SIGKDD international conference on knowledge discovery and data
  mining}}. ACM, \bibinfo{pages}{2125--2126}.
\newblock


\bibitem[\protect\citeauthoryear{Hardt, Price, and Srebro}{Hardt
  et~al\mbox{.}}{2016}]%
        {hardt2016equality}
\bibfield{author}{\bibinfo{person}{Moritz Hardt}, \bibinfo{person}{Eric Price},
  {and} \bibinfo{person}{Nati Srebro}.} \bibinfo{year}{2016}\natexlab{}.
\newblock \showarticletitle{Equality of opportunity in supervised learning}. In
  \bibinfo{booktitle}{\emph{Advances in neural information processing
  systems}}. \bibinfo{pages}{3315--3323}.
\newblock


\bibitem[\protect\citeauthoryear{Harper and Konstan}{Harper and
  Konstan}{2015}]%
        {harper2015movielens}
\bibfield{author}{\bibinfo{person}{F~Maxwell Harper} {and}
  \bibinfo{person}{Joseph~A Konstan}.} \bibinfo{year}{2015}\natexlab{}.
\newblock \showarticletitle{The movielens datasets: History and context}.
\newblock \bibinfo{journal}{\emph{Acm transactions on interactive intelligent
  systems (tiis)}} \bibinfo{volume}{5}, \bibinfo{number}{4}
  (\bibinfo{year}{2015}), \bibinfo{pages}{1--19}.
\newblock


\bibitem[\protect\citeauthoryear{Heidari, Loi, Gummadi, and Krause}{Heidari
  et~al\mbox{.}}{2019}]%
        {heidari2019moral}
\bibfield{author}{\bibinfo{person}{Hoda Heidari}, \bibinfo{person}{Michele
  Loi}, \bibinfo{person}{Krishna~P Gummadi}, {and} \bibinfo{person}{Andreas
  Krause}.} \bibinfo{year}{2019}\natexlab{}.
\newblock \showarticletitle{A moral framework for understanding fair ml through
  economic models of equality of opportunity}. In
  \bibinfo{booktitle}{\emph{Proceedings of the Conference on Fairness,
  Accountability, and Transparency}}. ACM, \bibinfo{pages}{181--190}.
\newblock


\bibitem[\protect\citeauthoryear{Jamali and Ester}{Jamali and Ester}{2010}]%
        {jamali2010matrix}
\bibfield{author}{\bibinfo{person}{Mohsen Jamali} {and} \bibinfo{person}{Martin
  Ester}.} \bibinfo{year}{2010}\natexlab{}.
\newblock \showarticletitle{A matrix factorization technique with trust
  propagation for recommendation in social networks}. In
  \bibinfo{booktitle}{\emph{Proceedings of the fourth ACM conference on
  Recommender systems}}. \bibinfo{pages}{135--142}.
\newblock


\bibitem[\protect\citeauthoryear{J{\"a}rvelin and
  Kek{\"a}l{\"a}inen}{J{\"a}rvelin and Kek{\"a}l{\"a}inen}{2002}]%
        {jarvelin2002cumulated}
\bibfield{author}{\bibinfo{person}{Kalervo J{\"a}rvelin} {and}
  \bibinfo{person}{Jaana Kek{\"a}l{\"a}inen}.} \bibinfo{year}{2002}\natexlab{}.
\newblock \showarticletitle{Cumulated gain-based evaluation of IR techniques}.
\newblock \bibinfo{journal}{\emph{ACM Transactions on Information Systems
  (TOIS)}} \bibinfo{volume}{20}, \bibinfo{number}{4} (\bibinfo{year}{2002}),
  \bibinfo{pages}{422--446}.
\newblock


\bibitem[\protect\citeauthoryear{Joachims, Granka, Pan, Hembrooke, and
  Gay}{Joachims et~al\mbox{.}}{2017}]%
        {joachims2017accurately}
\bibfield{author}{\bibinfo{person}{Thorsten Joachims}, \bibinfo{person}{Laura
  Granka}, \bibinfo{person}{Bing Pan}, \bibinfo{person}{Helene Hembrooke},
  {and} \bibinfo{person}{Geri Gay}.} \bibinfo{year}{2017}\natexlab{}.
\newblock \showarticletitle{Accurately interpreting clickthrough data as
  implicit feedback}. In \bibinfo{booktitle}{\emph{ACM SIGIR Forum}},
  Vol.~\bibinfo{volume}{51}. Acm New York, NY, USA, \bibinfo{pages}{4--11}.
\newblock


\bibitem[\protect\citeauthoryear{Kamishima}{Kamishima}{2003}]%
        {kamishima2003nantonac}
\bibfield{author}{\bibinfo{person}{Toshihiro Kamishima}.}
  \bibinfo{year}{2003}\natexlab{}.
\newblock \showarticletitle{Nantonac collaborative filtering: recommendation
  based on order responses}. In \bibinfo{booktitle}{\emph{Proceedings of the
  ninth ACM SIGKDD international conference on Knowledge discovery and data
  mining}}. \bibinfo{pages}{583--588}.
\newblock


\bibitem[\protect\citeauthoryear{Kamishima, Akaho, Asoh, and Sakuma}{Kamishima
  et~al\mbox{.}}{2018}]%
        {kamishima2018recommendation}
\bibfield{author}{\bibinfo{person}{Toshihiro Kamishima},
  \bibinfo{person}{Shotaro Akaho}, \bibinfo{person}{Hideki Asoh}, {and}
  \bibinfo{person}{Jun Sakuma}.} \bibinfo{year}{2018}\natexlab{}.
\newblock \showarticletitle{Recommendation independence}. In
  \bibinfo{booktitle}{\emph{Conference on Fairness, Accountability and
  Transparency}}. \bibinfo{pages}{187--201}.
\newblock


\bibitem[\protect\citeauthoryear{Khan, Manis, and Stoyanovich}{Khan
  et~al\mbox{.}}{2021}]%
        {fairfriends}
\bibfield{author}{\bibinfo{person}{Falaah~Arif Khan}, \bibinfo{person}{Eleni
  Manis}, {and} \bibinfo{person}{Julia Stoyanovich}.}
  \bibinfo{year}{2021}\natexlab{}.
\newblock \showarticletitle{Translation Tutorial: Fairness and Friends}. In
  \bibinfo{booktitle}{\emph{Proceedings of the ACM Conference on Fairness,
  Accountability, and Transparency}}.
\newblock


\bibitem[\protect\citeauthoryear{Kleinberg and Raghavan}{Kleinberg and
  Raghavan}{2018}]%
        {kleinberg_et_al:LIPIcs:2018:8323}
\bibfield{author}{\bibinfo{person}{Jon Kleinberg} {and} \bibinfo{person}{Manish
  Raghavan}.} \bibinfo{year}{2018}\natexlab{}.
\newblock \showarticletitle{{Selection Problems in the Presence of Implicit
  Bias}}. In \bibinfo{booktitle}{\emph{9th Innovations in Theoretical Computer
  Science Conference (ITCS 2018)}} \emph{(\bibinfo{series}{Leibniz
  International Proceedings in Informatics (LIPIcs)},
  Vol.~\bibinfo{volume}{94})}, \bibfield{editor}{\bibinfo{person}{Anna~R.
  Karlin}} (Ed.). \bibinfo{publisher}{Schloss Dagstuhl--Leibniz-Zentrum fuer
  Informatik}, \bibinfo{address}{Dagstuhl, Germany},
  \bibinfo{pages}{33:1--33:17}.
\newblock
\showISBNx{978-3-95977-060-6}
\showISSN{1868-8969}
\urldef\tempurl%
\url{https://doi.org/10.4230/LIPIcs.ITCS.2018.33}
\showDOI{\tempurl}


\bibitem[\protect\citeauthoryear{Kumar and Vassilvitskii}{Kumar and
  Vassilvitskii}{2010}]%
        {DBLP:conf/www/KumarV10}
\bibfield{author}{\bibinfo{person}{Ravi Kumar} {and} \bibinfo{person}{Sergei
  Vassilvitskii}.} \bibinfo{year}{2010}\natexlab{}.
\newblock \showarticletitle{Generalized distances between rankings}. In
  \bibinfo{booktitle}{\emph{Proceedings of the 19th International Conference on
  World Wide Web, {WWW} 2010, Raleigh, North Carolina, USA, April 26-30,
  2010}}, \bibfield{editor}{\bibinfo{person}{Michael Rappa},
  \bibinfo{person}{Paul Jones}, \bibinfo{person}{Juliana Freire}, {and}
  \bibinfo{person}{Soumen Chakrabarti}} (Eds.). \bibinfo{publisher}{{ACM}},
  \bibinfo{pages}{571--580}.
\newblock
\urldef\tempurl%
\url{https://doi.org/10.1145/1772690.1772749}
\showDOI{\tempurl}


\bibitem[\protect\citeauthoryear{Lahoti, Gummadi, and Weikum}{Lahoti
  et~al\mbox{.}}{2019a}]%
        {lahoti2019ifair}
\bibfield{author}{\bibinfo{person}{Preethi Lahoti}, \bibinfo{person}{Krishna~P
  Gummadi}, {and} \bibinfo{person}{Gerhard Weikum}.}
  \bibinfo{year}{2019}\natexlab{a}.
\newblock \showarticletitle{ifair: Learning individually fair data
  representations for algorithmic decision making}. In
  \bibinfo{booktitle}{\emph{2019 IEEE 35th International Conference on Data
  Engineering (ICDE)}}. IEEE, \bibinfo{pages}{1334--1345}.
\newblock


\bibitem[\protect\citeauthoryear{Lahoti, Gummadi, and Weikum}{Lahoti
  et~al\mbox{.}}{2019b}]%
        {lahoti2019operationalizing}
\bibfield{author}{\bibinfo{person}{Preethi Lahoti}, \bibinfo{person}{Krishna~P
  Gummadi}, {and} \bibinfo{person}{Gerhard Weikum}.}
  \bibinfo{year}{2019}\natexlab{b}.
\newblock \showarticletitle{Operationalizing individual fairness with pairwise
  fair representations}.
\newblock \bibinfo{journal}{\emph{arXiv preprint arXiv:1907.01439}}
  (\bibinfo{year}{2019}).
\newblock


\bibitem[\protect\citeauthoryear{Leonhardt, Anand, and Khosla}{Leonhardt
  et~al\mbox{.}}{2018}]%
        {leonhardt2018user}
\bibfield{author}{\bibinfo{person}{Jurek Leonhardt}, \bibinfo{person}{Avishek
  Anand}, {and} \bibinfo{person}{Megha Khosla}.}
  \bibinfo{year}{2018}\natexlab{}.
\newblock \showarticletitle{User fairness in recommender systems}. In
  \bibinfo{booktitle}{\emph{Companion Proceedings of the The Web Conference
  2018}}. \bibinfo{pages}{101--102}.
\newblock


\bibitem[\protect\citeauthoryear{Li}{Li}{2014}]%
        {DBLP:series/synthesis/2014Li}
\bibfield{author}{\bibinfo{person}{Hang Li}.} \bibinfo{year}{2014}\natexlab{}.
\newblock \bibinfo{booktitle}{\emph{Learning to Rank for Information Retrieval
  and Natural Language Processing, Second Edition}}.
\newblock \bibinfo{publisher}{Morgan {\&} Claypool Publishers}.
\newblock
\urldef\tempurl%
\url{https://doi.org/10.2200/S00607ED2V01Y201410HLT026}
\showDOI{\tempurl}


\bibitem[\protect\citeauthoryear{Li, Chen, Fu, Ge, and Zhang}{Li
  et~al\mbox{.}}{2021a}]%
        {li2021user}
\bibfield{author}{\bibinfo{person}{Yunqi Li}, \bibinfo{person}{Hanxiong Chen},
  \bibinfo{person}{Zuohui Fu}, \bibinfo{person}{Yingqiang Ge}, {and}
  \bibinfo{person}{Yongfeng Zhang}.} \bibinfo{year}{2021}\natexlab{a}.
\newblock \showarticletitle{User-oriented Fairness in Recommendation}. In
  \bibinfo{booktitle}{\emph{Proceedings of the Web Conference 2021}}.
  \bibinfo{pages}{624--632}.
\newblock


\bibitem[\protect\citeauthoryear{Li, Ge, and Zhang}{Li et~al\mbox{.}}{2021b}]%
        {li2021tutorial}
\bibfield{author}{\bibinfo{person}{Yunqi Li}, \bibinfo{person}{Yingqiang Ge},
  {and} \bibinfo{person}{Yongfeng Zhang}.} \bibinfo{year}{2021}\natexlab{b}.
\newblock \showarticletitle{Tutorial on Fairness of Machine Learning in
  Recommender Systems}. SIGIR.
\newblock


\bibitem[\protect\citeauthoryear{Lichman}{Lichman}{2013}]%
        {lichman_2013_uci}
\bibfield{author}{\bibinfo{person}{M. Lichman}.}
  \bibinfo{year}{2013}\natexlab{}.
\newblock \bibinfo{title}{{UCI} Machine Learning Repository}.
\newblock
\newblock


\bibitem[\protect\citeauthoryear{Lindley}{Lindley}{1961}]%
        {Lindley:1961}
\bibfield{author}{\bibinfo{person}{D.~V. Lindley}.}
  \bibinfo{year}{1961}\natexlab{}.
\newblock \showarticletitle{Dynamic Programming and Decision Theory}.
\newblock \bibinfo{journal}{\emph{Journal of the Royal Statistical Society}}
  \bibinfo{volume}{10}, \bibinfo{number}{1} (\bibinfo{date}{March}
  \bibinfo{year}{1961}), \bibinfo{pages}{39--51}.
\newblock


\bibitem[\protect\citeauthoryear{Liu, Guo, Sonboli, Burke, and Zhang}{Liu
  et~al\mbox{.}}{2019}]%
        {liu2019personalized}
\bibfield{author}{\bibinfo{person}{Weiwen Liu}, \bibinfo{person}{Jun Guo},
  \bibinfo{person}{Nasim Sonboli}, \bibinfo{person}{Robin Burke}, {and}
  \bibinfo{person}{Shengyu Zhang}.} \bibinfo{year}{2019}\natexlab{}.
\newblock \showarticletitle{Personalized fairness-aware re-ranking for
  microlending}. In \bibinfo{booktitle}{\emph{Proceedings of the 13th ACM
  Conference on Recommender Systems}}. \bibinfo{pages}{467--471}.
\newblock


\bibitem[\protect\citeauthoryear{Makkonen}{Makkonen}{2002}]%
        {makkonen2002multiple}
\bibfield{author}{\bibinfo{person}{Timo Makkonen}.}
  \bibinfo{year}{2002}\natexlab{}.
\newblock \showarticletitle{Multiple, compound and intersectional
  discrimination: Bringing the experiences of the most marginalized to the
  fore}.
\newblock \bibinfo{journal}{\emph{Institute for Human Rights, {\AA}bo Akademi
  University}} (\bibinfo{year}{2002}).
\newblock


\bibitem[\protect\citeauthoryear{Manning, Raghavan, and Sch{\"u}tze}{Manning
  et~al\mbox{.}}{2008}]%
        {manning2008evaluation}
\bibfield{author}{\bibinfo{person}{Christopher~D Manning},
  \bibinfo{person}{Prabhakar Raghavan}, {and} \bibinfo{person}{Hinrich
  Sch{\"u}tze}.} \bibinfo{year}{2008}\natexlab{}.
\newblock \showarticletitle{Evaluation in information retrieval}.
\newblock \bibinfo{journal}{\emph{Introduction to information retrieval}}
  \bibinfo{volume}{1} (\bibinfo{year}{2008}), \bibinfo{pages}{188--210}.
\newblock


\bibitem[\protect\citeauthoryear{Mary, Calauz{\`e}nes, and Karoui}{Mary
  et~al\mbox{.}}{2019}]%
        {pmlr-v97-mary19a}
\bibfield{author}{\bibinfo{person}{Jeremie Mary}, \bibinfo{person}{Cl{\'e}ment
  Calauz{\`e}nes}, {and} \bibinfo{person}{Noureddine~El Karoui}.}
  \bibinfo{year}{2019}\natexlab{}.
\newblock \showarticletitle{Fairness-Aware Learning for Continuous Attributes
  and Treatments}. In \bibinfo{booktitle}{\emph{Proceedings of the 36th
  International Conference on Machine Learning}}
  \emph{(\bibinfo{series}{Proceedings of Machine Learning Research},
  Vol.~\bibinfo{volume}{97})}, \bibfield{editor}{\bibinfo{person}{Kamalika
  Chaudhuri} {and} \bibinfo{person}{Ruslan Salakhutdinov}} (Eds.).
  \bibinfo{publisher}{PMLR}, \bibinfo{pages}{4382--4391}.
\newblock
\urldef\tempurl%
\url{https://proceedings.mlr.press/v97/mary19a.html}
\showURL{%
\tempurl}


\bibitem[\protect\citeauthoryear{Mehrabi, Morstatter, Saxena, Lerman, and
  Galstyan}{Mehrabi et~al\mbox{.}}{2021}]%
        {DBLP:journals/csur/MehrabiMSLG21}
\bibfield{author}{\bibinfo{person}{Ninareh Mehrabi}, \bibinfo{person}{Fred
  Morstatter}, \bibinfo{person}{Nripsuta Saxena}, \bibinfo{person}{Kristina
  Lerman}, {and} \bibinfo{person}{Aram Galstyan}.}
  \bibinfo{year}{2021}\natexlab{}.
\newblock \showarticletitle{A Survey on Bias and Fairness in Machine Learning}.
\newblock \bibinfo{journal}{\emph{{ACM} Comput. Surv.}} \bibinfo{volume}{54},
  \bibinfo{number}{6} (\bibinfo{year}{2021}), \bibinfo{pages}{115:1--115:35}.
\newblock
\urldef\tempurl%
\url{https://doi.org/10.1145/3457607}
\showDOI{\tempurl}


\bibitem[\protect\citeauthoryear{Mehrotra, McInerney, Bouchard, Lalmas, and
  Diaz}{Mehrotra et~al\mbox{.}}{2018}]%
        {mehrotra2018towards}
\bibfield{author}{\bibinfo{person}{Rishabh Mehrotra}, \bibinfo{person}{James
  McInerney}, \bibinfo{person}{Hugues Bouchard}, \bibinfo{person}{Mounia
  Lalmas}, {and} \bibinfo{person}{Fernando Diaz}.}
  \bibinfo{year}{2018}\natexlab{}.
\newblock \showarticletitle{Towards a fair marketplace: Counterfactual
  evaluation of the trade-off between relevance, fairness \& satisfaction in
  recommendation systems}. In \bibinfo{booktitle}{\emph{Proceedings of the 27th
  acm international conference on information and knowledge management}}.
  \bibinfo{pages}{2243--2251}.
\newblock


\bibitem[\protect\citeauthoryear{NASA}{NASA}{[n.\,d.]}]%
        {NASAData}
\bibfield{author}{\bibinfo{person}{NASA}.} \bibinfo{year}{[n.\,d.]}\natexlab{}.
\newblock \bibinfo{title}{Astronauts}.
\newblock
\newblock
\urldef\tempurl%
\url{https://github.com/DataResponsibly/Datasets}
\showURL{%
\tempurl}


\bibitem[\protect\citeauthoryear{Noble}{Noble}{2018}]%
        {noble2018algorithms}
\bibfield{author}{\bibinfo{person}{Safiya~Umoja Noble}.}
  \bibinfo{year}{2018}\natexlab{}.
\newblock \bibinfo{booktitle}{\emph{Algorithms of oppression: How search
  engines reinforce racism}}.
\newblock \bibinfo{publisher}{nyu Press}.
\newblock


\bibitem[\protect\citeauthoryear{of~Health and Services}{of~Health and
  Services}{[n.\,d.]}]%
        {MEPSData}
\bibfield{author}{\bibinfo{person}{Department of Health} {and}
  \bibinfo{person}{Human Services}.} \bibinfo{year}{[n.\,d.]}\natexlab{}.
\newblock \bibinfo{title}{{Medical Expenditure Panel Survey}}.
\newblock
\newblock
\urldef\tempurl%
\url{https://meps.ahrq.gov/mepsweb/}
\showURL{%
\tempurl}


\bibitem[\protect\citeauthoryear{of~Transportation~Statistics}{of~Transportation~Statistics}{[n.\,d.]}]%
        {DOTData}
\bibfield{author}{\bibinfo{person}{Bureau of Transportation~Statistics}.}
  \bibinfo{year}{[n.\,d.]}\natexlab{}.
\newblock \bibinfo{title}{{National Summary of U.S. Flights}}.
\newblock
\newblock
\urldef\tempurl%
\url{https://www.transtats.bts.gov}
\showURL{%
\tempurl}


\bibitem[\protect\citeauthoryear{Pan, Hembrooke, Joachims, Lorigo, Gay, and
  Granka}{Pan et~al\mbox{.}}{2007}]%
        {pan2007google}
\bibfield{author}{\bibinfo{person}{Bing Pan}, \bibinfo{person}{Helene
  Hembrooke}, \bibinfo{person}{Thorsten Joachims}, \bibinfo{person}{Lori
  Lorigo}, \bibinfo{person}{Geri Gay}, {and} \bibinfo{person}{Laura Granka}.}
  \bibinfo{year}{2007}\natexlab{}.
\newblock \showarticletitle{In Google we trust: Users’ decisions on rank,
  position, and relevance}.
\newblock \bibinfo{journal}{\emph{Journal of computer-mediated communication}}
  \bibinfo{volume}{12}, \bibinfo{number}{3} (\bibinfo{year}{2007}),
  \bibinfo{pages}{801--823}.
\newblock


\bibitem[\protect\citeauthoryear{Pessach and Shmueli}{Pessach and
  Shmueli}{2020}]%
        {DBLP:journals/corr/abs-2001-09784}
\bibfield{author}{\bibinfo{person}{Dana Pessach} {and} \bibinfo{person}{Erez
  Shmueli}.} \bibinfo{year}{2020}\natexlab{}.
\newblock \showarticletitle{Algorithmic Fairness}.
\newblock \bibinfo{journal}{\emph{CoRR}}  \bibinfo{volume}{abs/2001.09784}
  (\bibinfo{year}{2020}).
\newblock
\showeprint[arXiv]{2001.09784}
\urldef\tempurl%
\url{https://arxiv.org/abs/2001.09784}
\showURL{%
\tempurl}


\bibitem[\protect\citeauthoryear{ProPublica}{ProPublica}{[n.\,d.]}]%
        {COMPASData}
\bibfield{author}{\bibinfo{person}{ProPublica}.}
  \bibinfo{year}{[n.\,d.]}\natexlab{}.
\newblock \bibinfo{title}{Correctional Offender Management Profiling for
  Alternative Sanctions}.
\newblock
\newblock
\urldef\tempurl%
\url{https://github.com/propublica/compas-analysis}
\showURL{%
\tempurl}


\bibitem[\protect\citeauthoryear{Rastegarpanah, Gummadi, and
  Crovella}{Rastegarpanah et~al\mbox{.}}{2019}]%
        {rastegarpanah2019fighting}
\bibfield{author}{\bibinfo{person}{Bashir Rastegarpanah},
  \bibinfo{person}{Krishna~P Gummadi}, {and} \bibinfo{person}{Mark Crovella}.}
  \bibinfo{year}{2019}\natexlab{}.
\newblock \showarticletitle{Fighting fire with fire: Using antidote data to
  improve polarization and fairness of recommender systems}. In
  \bibinfo{booktitle}{\emph{Proceedings of the Twelfth ACM International
  Conference on Web Search and Data Mining}}. \bibinfo{pages}{231--239}.
\newblock


\bibitem[\protect\citeauthoryear{Rawls}{Rawls}{1971}]%
        {Rawls}
\bibfield{author}{\bibinfo{person}{John Rawls}.}
  \bibinfo{year}{1971}\natexlab{}.
\newblock \bibinfo{booktitle}{\emph{A theory of justice}}.
\newblock \bibinfo{publisher}{Harvard University Press}.
\newblock


\bibitem[\protect\citeauthoryear{Reeves and Halikias}{Reeves and
  Halikias}{2017}]%
        {brookings_race}
\bibfield{author}{\bibinfo{person}{Richard~V. Reeves} {and}
  \bibinfo{person}{Dimitrios Halikias}.} \bibinfo{year}{2017}\natexlab{}.
\newblock \showarticletitle{Race gaps in SAT scores highlight inequality and
  hinder upward mobility}.
\newblock  (\bibinfo{year}{2017}).
\newblock
\urldef\tempurl%
\url{https://www.brookings.edu/research/race-gaps-in-sat-scores-highlight-inequality-and-hinder-upward-mobility}
\showURL{%
\tempurl}


\bibitem[\protect\citeauthoryear{Repository}{Repository}{[n.\,d.]}]%
        {GermanCreditData}
\bibfield{author}{\bibinfo{person}{UCI Machine~Learning Repository}.}
  \bibinfo{year}{[n.\,d.]}\natexlab{}.
\newblock \bibinfo{title}{{German Credit}}.
\newblock
\newblock
\urldef\tempurl%
\url{https://archive.ics.uci.edu/ml/machine-learning-databases/statlog/german/}
\showURL{%
\tempurl}


\bibitem[\protect\citeauthoryear{Richardson, Dominowska, and Ragno}{Richardson
  et~al\mbox{.}}{2007}]%
        {richardson2007predicting}
\bibfield{author}{\bibinfo{person}{Matthew Richardson}, \bibinfo{person}{Ewa
  Dominowska}, {and} \bibinfo{person}{Robert Ragno}.}
  \bibinfo{year}{2007}\natexlab{}.
\newblock \showarticletitle{Predicting clicks: estimating the click-through
  rate for new ads}. In \bibinfo{booktitle}{\emph{Proceedings of the 16th
  international conference on World Wide Web}}. \bibinfo{pages}{521--530}.
\newblock


\bibitem[\protect\citeauthoryear{Roemer}{Roemer}{2002}]%
        {Roemer2002}
\bibfield{author}{\bibinfo{person}{John~E. Roemer}.}
  \bibinfo{year}{2002}\natexlab{}.
\newblock \showarticletitle{Equality of opportunity: a progress report}.
\newblock \bibinfo{journal}{\emph{Social Choice and Welfare}}
  \bibinfo{volume}{19}, \bibinfo{number}{2} (\bibinfo{year}{2002}),
  \bibinfo{pages}{405--471}.
\newblock


\bibitem[\protect\citeauthoryear{SAT}{SAT}{[n.\,d.]}]%
        {SATData}
\bibfield{author}{\bibinfo{person}{SAT}.} \bibinfo{year}{[n.\,d.]}\natexlab{}.
\newblock \bibinfo{title}{SAT}.
\newblock
\newblock
\urldef\tempurl%
\url{https://www.qsleap.com/sat/resources/sat-2014-percentiles}
\showURL{%
\tempurl}


\bibitem[\protect\citeauthoryear{Scholar}{Scholar}{[n.\,d.]}]%
        {SSORCData}
\bibfield{author}{\bibinfo{person}{Semantic Scholar}.}
  \bibinfo{year}{[n.\,d.]}\natexlab{}.
\newblock \bibinfo{title}{Semantic Scholar Open Research Corpus}.
\newblock
\newblock
\urldef\tempurl%
\url{https://api.semanticscholar.org/corpus/}
\showURL{%
\tempurl}


\bibitem[\protect\citeauthoryear{Sen}{Sen}{1980}]%
        {sen1980equality}
\bibfield{author}{\bibinfo{person}{Amartya Sen}.}
  \bibinfo{year}{1980}\natexlab{}.
\newblock \showarticletitle{Equality of what?}
\newblock \bibinfo{journal}{\emph{The Tanner lecture on human values}}
  \bibinfo{volume}{1} (\bibinfo{year}{1980}), \bibinfo{pages}{197--220}.
\newblock


\bibitem[\protect\citeauthoryear{Series}{Series}{[n.\,d.]}]%
        {LSACData}
\bibfield{author}{\bibinfo{person}{Law School Admission Council Research~Report
  Series}.} \bibinfo{year}{[n.\,d.]}\natexlab{}.
\newblock \bibinfo{title}{LSAC national longitudinal bar passage study}.
\newblock
\newblock
\urldef\tempurl%
\url{https://github.com/MilkaLichtblau/DELTR-Experiments/tree/master/data/LawStudents}
\showURL{%
\tempurl}


\bibitem[\protect\citeauthoryear{Shields}{Shields}{2008}]%
        {shields2008gender}
\bibfield{author}{\bibinfo{person}{Stephanie~A Shields}.}
  \bibinfo{year}{2008}\natexlab{}.
\newblock \showarticletitle{Gender: An intersectionality perspective}.
\newblock \bibinfo{journal}{\emph{Sex roles}} \bibinfo{volume}{59},
  \bibinfo{number}{5-6} (\bibinfo{year}{2008}), \bibinfo{pages}{301--311}.
\newblock


\bibitem[\protect\citeauthoryear{Singh and Joachims}{Singh and
  Joachims}{2018}]%
        {singh2018fairness}
\bibfield{author}{\bibinfo{person}{Ashudeep Singh} {and}
  \bibinfo{person}{Thorsten Joachims}.} \bibinfo{year}{2018}\natexlab{}.
\newblock \showarticletitle{Fairness of exposure in rankings}. In
  \bibinfo{booktitle}{\emph{Proceedings of the 24th ACM SIGKDD International
  Conference on Knowledge Discovery \& Data Mining}}. ACM,
  \bibinfo{pages}{2219--2228}.
\newblock


\bibitem[\protect\citeauthoryear{Singh and Joachims}{Singh and
  Joachims}{2019}]%
        {singh2019policy}
\bibfield{author}{\bibinfo{person}{Ashudeep Singh} {and}
  \bibinfo{person}{Thorsten Joachims}.} \bibinfo{year}{2019}\natexlab{}.
\newblock \showarticletitle{Policy Learning for Fairness in Ranking}.
\newblock \bibinfo{journal}{\emph{arXiv preprint arXiv:1902.04056}}
  (\bibinfo{year}{2019}).
\newblock


\bibitem[\protect\citeauthoryear{Sonboli and Burke}{Sonboli and Burke}{2019}]%
        {sonboli2019localized}
\bibfield{author}{\bibinfo{person}{Nasim Sonboli} {and} \bibinfo{person}{Robin
  Burke}.} \bibinfo{year}{2019}\natexlab{}.
\newblock \showarticletitle{Localized fairness in recommender systems}. In
  \bibinfo{booktitle}{\emph{Adjunct Publication of the 27th Conference on User
  Modeling, Adaptation and Personalization}}. \bibinfo{pages}{295--300}.
\newblock


\bibitem[\protect\citeauthoryear{Sonboli, Smith, Cabral~Berenfus, Burke, and
  Fiesler}{Sonboli et~al\mbox{.}}{2021}]%
        {sonboli2021fairness}
\bibfield{author}{\bibinfo{person}{Nasim Sonboli}, \bibinfo{person}{Jessie~J
  Smith}, \bibinfo{person}{Florencia Cabral~Berenfus}, \bibinfo{person}{Robin
  Burke}, {and} \bibinfo{person}{Casey Fiesler}.}
  \bibinfo{year}{2021}\natexlab{}.
\newblock \showarticletitle{Fairness and transparency in recommendation: The
  users’ perspective}. In \bibinfo{booktitle}{\emph{Proceedings of the 29th
  ACM Conference on User Modeling, Adaptation and Personalization}}.
  \bibinfo{pages}{274--279}.
\newblock


\bibitem[\protect\citeauthoryear{Spearman}{Spearman}{1961}]%
        {spearman1961proof}
\bibfield{author}{\bibinfo{person}{Charles Spearman}.}
  \bibinfo{year}{1961}\natexlab{}.
\newblock \showarticletitle{The proof and measurement of association between
  two things.}
\newblock  (\bibinfo{year}{1961}).
\newblock


\bibitem[\protect\citeauthoryear{StackExchange}{StackExchange}{[n.\,d.]}]%
        {StackData}
\bibfield{author}{\bibinfo{person}{StackExchange}.}
  \bibinfo{year}{[n.\,d.]}\natexlab{}.
\newblock \bibinfo{title}{StackExchange}.
\newblock
\newblock
\urldef\tempurl%
\url{https://stackexchange.com/}
\showURL{%
\tempurl}


\bibitem[\protect\citeauthoryear{Stoyanovich, Yang, and Jagadish}{Stoyanovich
  et~al\mbox{.}}{2018}]%
        {DBLP:conf/edbt/StoyanovichYJ18}
\bibfield{author}{\bibinfo{person}{Julia Stoyanovich}, \bibinfo{person}{Ke
  Yang}, {and} \bibinfo{person}{H.~V. Jagadish}.}
  \bibinfo{year}{2018}\natexlab{}.
\newblock \showarticletitle{Online Set Selection with Fairness and Diversity
  Constraints}. In \bibinfo{booktitle}{\emph{Proceedings of the 21th
  International Conference on Extending Database Technology, {EDBT} 2018,
  Vienna, Austria, March 26-29, 2018}}. \bibinfo{pages}{241--252}.
\newblock
\urldef\tempurl%
\url{https://doi.org/10.5441/002/edbt.2018.22}
\showDOI{\tempurl}


\bibitem[\protect\citeauthoryear{students}{students}{[n.\,d.]}]%
        {EngineeringData}
\bibfield{author}{\bibinfo{person}{Engineering students}.}
  \bibinfo{year}{[n.\,d.]}\natexlab{}.
\newblock \bibinfo{title}{Engineering students}.
\newblock
\newblock
\urldef\tempurl%
\url{https://github.com/MilkaLichtblau/DELTR-Experiments/tree/master/data/EngineeringStudents}
\showURL{%
\tempurl}


\bibitem[\protect\citeauthoryear{S{\"u}hr, Biega, Zehlike, Gummadi, and
  Chakraborty}{S{\"u}hr et~al\mbox{.}}{2019}]%
        {suhr2019two}
\bibfield{author}{\bibinfo{person}{Tom S{\"u}hr}, \bibinfo{person}{Asia~J
  Biega}, \bibinfo{person}{Meike Zehlike}, \bibinfo{person}{Krishna~P Gummadi},
  {and} \bibinfo{person}{Abhijnan Chakraborty}.}
  \bibinfo{year}{2019}\natexlab{}.
\newblock \showarticletitle{Two-sided fairness for repeated matchings in
  two-sided markets: A case study of a ride-hailing platform}. In
  \bibinfo{booktitle}{\emph{Proceedings of the 25th ACM SIGKDD International
  Conference on Knowledge Discovery \& Data Mining}}. ACM,
  \bibinfo{pages}{3082--3092}.
\newblock


\bibitem[\protect\citeauthoryear{{Supreme Court of the United States}}{{Supreme
  Court of the United States}}{2009}]%
        {ricci}
\bibfield{author}{\bibinfo{person}{{Supreme Court of the United States}}.}
  \bibinfo{year}{2009}\natexlab{}.
\newblock \bibinfo{title}{{Ricci v. DeStefano (Nos. 07-1428 and 08-328), 530 F.
  3d 87, reversed and remanded}}.
\newblock
  \bibinfo{howpublished}{\url{https://www.law.cornell.edu/supct/html/07-1428.ZO.html}}.
\newblock


\bibitem[\protect\citeauthoryear{Technology}{Technology}{[n.\,d.]}]%
        {IITJEEData}
\bibfield{author}{\bibinfo{person}{Indian Institute~Of Technology}.}
  \bibinfo{year}{[n.\,d.]}\natexlab{}.
\newblock \bibinfo{title}{{IIT-JEE}}.
\newblock
\newblock
\urldef\tempurl%
\url{https://indiankanoon.org/doc/1955304/}
\showURL{%
\tempurl}


\bibitem[\protect\citeauthoryear{{The College Board}}{{The College
  Board}}{2014}]%
        {sat_2014}
\bibfield{author}{\bibinfo{person}{{The College Board}}.}
  \bibinfo{year}{2014}\natexlab{}.
\newblock \bibinfo{title}{{SAT} Percentile Ranks}.
\newblock
\newblock


\bibitem[\protect\citeauthoryear{TREC}{TREC}{[n.\,d.]}]%
        {W3CData}
\bibfield{author}{\bibinfo{person}{TREC}.} \bibinfo{year}{[n.\,d.]}\natexlab{}.
\newblock \bibinfo{title}{W3C Experts}.
\newblock
\newblock
\urldef\tempurl%
\url{https://github.com/MilkaLichtblau/DELTR-Experiments/tree/master/data/TREC}
\showURL{%
\tempurl}


\bibitem[\protect\citeauthoryear{Wang, Thain, Sinha, Prost, Chi, Chen, and
  Beutel}{Wang et~al\mbox{.}}{2021}]%
        {wang2021practical}
\bibfield{author}{\bibinfo{person}{Xuezhi Wang}, \bibinfo{person}{Nithum
  Thain}, \bibinfo{person}{Anu Sinha}, \bibinfo{person}{Flavien Prost},
  \bibinfo{person}{Ed~H Chi}, \bibinfo{person}{Jilin Chen}, {and}
  \bibinfo{person}{Alex Beutel}.} \bibinfo{year}{2021}\natexlab{}.
\newblock \showarticletitle{Practical Compositional Fairness: Understanding
  Fairness in Multi-Component Recommender Systems}. In
  \bibinfo{booktitle}{\emph{Proceedings of the 14th ACM International
  Conference on Web Search and Data Mining}}. \bibinfo{pages}{436--444}.
\newblock


\bibitem[\protect\citeauthoryear{Wightman and Ramsey}{Wightman and
  Ramsey}{1998}]%
        {wightman1998lsac}
\bibfield{author}{\bibinfo{person}{Linda~F Wightman} {and}
  \bibinfo{person}{Henry Ramsey}.} \bibinfo{year}{1998}\natexlab{}.
\newblock \bibinfo{booktitle}{\emph{LSAC national longitudinal bar passage
  study}}.
\newblock \bibinfo{publisher}{Law School Admission Council}.
\newblock


\bibitem[\protect\citeauthoryear{Wu, Zhang, and Wu}{Wu et~al\mbox{.}}{2018}]%
        {DBLP:conf/kdd/WuZW18}
\bibfield{author}{\bibinfo{person}{Yongkai Wu}, \bibinfo{person}{Lu Zhang},
  {and} \bibinfo{person}{Xintao Wu}.} \bibinfo{year}{2018}\natexlab{}.
\newblock \showarticletitle{On Discrimination Discovery and Removal in Ranked
  Data using Causal Graph}. In \bibinfo{booktitle}{\emph{Proceedings of the
  24th {ACM} {SIGKDD} International Conference on Knowledge Discovery {\&} Data
  Mining, {KDD} 2018, London, UK, August 19-23, 2018}}.
  \bibinfo{pages}{2536--2544}.
\newblock
\urldef\tempurl%
\url{https://doi.org/10.1145/3219819.3220087}
\showDOI{\tempurl}


\bibitem[\protect\citeauthoryear{Xing}{Xing}{[n.\,d.]}]%
        {XINGData}
\bibfield{author}{\bibinfo{person}{Xing}.} \bibinfo{year}{[n.\,d.]}\natexlab{}.
\newblock \bibinfo{title}{XING}.
\newblock
\newblock
\urldef\tempurl%
\url{https://github.com/MilkaLichtblau/xing_dataset}
\showURL{%
\tempurl}


\bibitem[\protect\citeauthoryear{Yahoo}{Yahoo}{[n.\,d.]}]%
        {YahooData}
\bibfield{author}{\bibinfo{person}{Yahoo}.}
  \bibinfo{year}{[n.\,d.]}\natexlab{}.
\newblock \bibinfo{title}{The Yahoo Webscope Program}.
\newblock
\newblock
\urldef\tempurl%
\url{https://webscope.sandbox.yahoo.com/}
\showURL{%
\tempurl}


\bibitem[\protect\citeauthoryear{Yang, Gkatzelis, and Stoyanovich}{Yang
  et~al\mbox{.}}{2019}]%
        {DBLP:conf/ijcai/YangGS19}
\bibfield{author}{\bibinfo{person}{Ke Yang}, \bibinfo{person}{Vasilis
  Gkatzelis}, {and} \bibinfo{person}{Julia Stoyanovich}.}
  \bibinfo{year}{2019}\natexlab{}.
\newblock \showarticletitle{Balanced Ranking with Diversity Constraints}. In
  \bibinfo{booktitle}{\emph{Proceedings of the Twenty-Eighth International
  Joint Conference on Artificial Intelligence, {IJCAI} 2019, Macao, China,
  August 10-16, 2019}}. \bibinfo{pages}{6035--6042}.
\newblock
\urldef\tempurl%
\url{https://doi.org/10.24963/ijcai.2019/836}
\showDOI{\tempurl}


\bibitem[\protect\citeauthoryear{Yang, Loftus, and Stoyanovich}{Yang
  et~al\mbox{.}}{2021}]%
        {yang2020causal}
\bibfield{author}{\bibinfo{person}{Ke Yang}, \bibinfo{person}{Joshua~R.
  Loftus}, {and} \bibinfo{person}{Julia Stoyanovich}.}
  \bibinfo{year}{2021}\natexlab{}.
\newblock \showarticletitle{Causal intersectionality and fair ranking}. In
  \bibinfo{booktitle}{\emph{Symposium on Foundations of Responsible Computing
  {(FORC)}}}.
\newblock
\urldef\tempurl%
\url{https://doi.org/10.4230/LIPIcs.FORC.2021.7}
\showDOI{\tempurl}


\bibitem[\protect\citeauthoryear{Yang and Stoyanovich}{Yang and
  Stoyanovich}{2017}]%
        {yang2017measuring}
\bibfield{author}{\bibinfo{person}{Ke Yang} {and} \bibinfo{person}{Julia
  Stoyanovich}.} \bibinfo{year}{2017}\natexlab{}.
\newblock \showarticletitle{Measuring fairness in ranked outputs}. In
  \bibinfo{booktitle}{\emph{Proceedings of the 29th International Conference on
  Scientific and Statistical Database Management}}. ACM, \bibinfo{pages}{22}.
\newblock


\bibitem[\protect\citeauthoryear{Yang, Stoyanovich, Asudeh, Howe, Jagadish, and
  Miklau}{Yang et~al\mbox{.}}{2018}]%
        {yang2018nutritional}
\bibfield{author}{\bibinfo{person}{Ke Yang}, \bibinfo{person}{Julia
  Stoyanovich}, \bibinfo{person}{Abolfazl Asudeh}, \bibinfo{person}{Bill Howe},
  \bibinfo{person}{HV Jagadish}, {and} \bibinfo{person}{Gerome Miklau}.}
  \bibinfo{year}{2018}\natexlab{}.
\newblock \showarticletitle{A nutritional label for rankings}. In
  \bibinfo{booktitle}{\emph{Proceedings of the 2018 International Conference on
  Management of Data}}. ACM, \bibinfo{pages}{1773--1776}.
\newblock


\bibitem[\protect\citeauthoryear{Zehlike, Bonchi, Castillo, Hajian, Megahed,
  and Baeza-Yates}{Zehlike et~al\mbox{.}}{2017a}]%
        {zehlike2017fa}
\bibfield{author}{\bibinfo{person}{Meike Zehlike}, \bibinfo{person}{Francesco
  Bonchi}, \bibinfo{person}{Carlos Castillo}, \bibinfo{person}{Sara Hajian},
  \bibinfo{person}{Mohamed Megahed}, {and} \bibinfo{person}{Ricardo
  Baeza-Yates}.} \bibinfo{year}{2017}\natexlab{a}.
\newblock \showarticletitle{Fa* ir: A fair top-k ranking algorithm}. In
  \bibinfo{booktitle}{\emph{Proceedings of the 2017 ACM on Conference on
  Information and Knowledge Management}}. ACM, \bibinfo{pages}{1569--1578}.
\newblock


\bibitem[\protect\citeauthoryear{Zehlike and Castillo}{Zehlike and
  Castillo}{2018}]%
        {zehlike2018reducing}
\bibfield{author}{\bibinfo{person}{Meike Zehlike} {and} \bibinfo{person}{Carlos
  Castillo}.} \bibinfo{year}{2018}\natexlab{}.
\newblock \showarticletitle{Reducing disparate exposure in ranking: A learning
  to rank approach}.
\newblock \bibinfo{journal}{\emph{arXiv preprint arXiv:1805.08716}}
  (\bibinfo{year}{2018}).
\newblock


\bibitem[\protect\citeauthoryear{Zehlike, Hacker, and Wiedemann}{Zehlike
  et~al\mbox{.}}{2017b}]%
        {zehlike2017matching}
\bibfield{author}{\bibinfo{person}{Meike Zehlike}, \bibinfo{person}{Philipp
  Hacker}, {and} \bibinfo{person}{Emil Wiedemann}.}
  \bibinfo{year}{2017}\natexlab{b}.
\newblock \showarticletitle{Matching code and law: achieving algorithmic
  fairness with optimal transport}.
\newblock \bibinfo{journal}{\emph{Data Mining and Knowledge Discovery}}
  (\bibinfo{year}{2017}), \bibinfo{pages}{1--38}.
\newblock


\bibitem[\protect\citeauthoryear{Zehlike, S{\"u}hr, Baeza-Yates, Bonchi,
  Castillo, and Hajian}{Zehlike et~al\mbox{.}}{2022}]%
        {zehlike2022fair}
\bibfield{author}{\bibinfo{person}{Meike Zehlike}, \bibinfo{person}{Tom
  S{\"u}hr}, \bibinfo{person}{Ricardo Baeza-Yates}, \bibinfo{person}{Francesco
  Bonchi}, \bibinfo{person}{Carlos Castillo}, {and} \bibinfo{person}{Sara
  Hajian}.} \bibinfo{year}{2022}\natexlab{}.
\newblock \showarticletitle{Fair Top-k Ranking with multiple protected groups}.
\newblock \bibinfo{journal}{\emph{Information Processing \& Management}}
  \bibinfo{volume}{59}, \bibinfo{number}{1} (\bibinfo{year}{2022}),
  \bibinfo{pages}{102707}.
\newblock


\bibitem[\protect\citeauthoryear{Zehlike, S{\"u}hr, Castillo, and
  Kitanovski}{Zehlike et~al\mbox{.}}{2020}]%
        {zehlike2020fairsearch}
\bibfield{author}{\bibinfo{person}{Meike Zehlike}, \bibinfo{person}{Tom
  S{\"u}hr}, \bibinfo{person}{Carlos Castillo}, {and} \bibinfo{person}{Ivan
  Kitanovski}.} \bibinfo{year}{2020}\natexlab{}.
\newblock \showarticletitle{FairSearch: A Tool For Fairness in Ranked Search
  Results}. In \bibinfo{booktitle}{\emph{Companion Proceedings of the Web
  Conference 2020}}. \bibinfo{pages}{172--175}.
\newblock


\bibitem[\protect\citeauthoryear{Zemel, Wu, Swersky, Pitassi, and Dwork}{Zemel
  et~al\mbox{.}}{2013}]%
        {DBLP:conf/icml/ZemelWSPD13}
\bibfield{author}{\bibinfo{person}{Richard~S. Zemel}, \bibinfo{person}{Yu Wu},
  \bibinfo{person}{Kevin Swersky}, \bibinfo{person}{Toniann Pitassi}, {and}
  \bibinfo{person}{Cynthia Dwork}.} \bibinfo{year}{2013}\natexlab{}.
\newblock \showarticletitle{Learning Fair Representations}. In
  \bibinfo{booktitle}{\emph{International Conference on Machine Learning}}.
\newblock
\urldef\tempurl%
\url{http://jmlr.org/proceedings/papers/v28/zemel13.html}
\showURL{%
\tempurl}


\bibitem[\protect\citeauthoryear{Zhang}{Zhang}{[n.\,d.]}]%
        {YowData}
\bibfield{author}{\bibinfo{person}{Yi Zhang}.}
  \bibinfo{year}{[n.\,d.]}\natexlab{}.
\newblock \bibinfo{title}{Yow News Recommendation}.
\newblock
\newblock
\urldef\tempurl%
\url{https://www.younow.com}
\showURL{%
\tempurl}


\bibitem[\protect\citeauthoryear{Zhang}{Zhang}{2005}]%
        {zhang2005bayesian}
\bibfield{author}{\bibinfo{person}{Yi Zhang}.} \bibinfo{year}{2005}\natexlab{}.
\newblock \emph{\bibinfo{title}{Bayesian Graphical Model for Adaptive
  Information Filtering}}.
\newblock \bibinfo{thesistype}{Ph.\,D. Dissertation}. \bibinfo{school}{Carnegie
  Mellon University}.
\newblock


\bibitem[\protect\citeauthoryear{Zhu, Hu, and Caverlee}{Zhu
  et~al\mbox{.}}{2018}]%
        {zhu2018fairness}
\bibfield{author}{\bibinfo{person}{Ziwei Zhu}, \bibinfo{person}{Xia Hu}, {and}
  \bibinfo{person}{James Caverlee}.} \bibinfo{year}{2018}\natexlab{}.
\newblock \showarticletitle{Fairness-aware tensor-based recommendation}. In
  \bibinfo{booktitle}{\emph{Proceedings of the 27th ACM International
  Conference on Information and Knowledge Management}}.
  \bibinfo{pages}{1153--1162}.
\newblock


\end{thebibliography}
}
\appendix

\end{document}